\documentclass[Master, ngerman, USenglish, UKenglish, openany]{scrbook}
\usepackage[T1]{fontenc}
\usepackage{cmap}
\usepackage[texlive=2020, shownotes, cleveref, feynmp, txfonts]{ubonn-thesis}
\usepackage{ubonn-biblatex}

\usepackage{thesis_defs}
\graphicspath{%
  {figs/}%
  {figs/cover/}%
}

\ifthenelse{\equal{\ThesisVersion}{Submit}}{%
  \usepackage{background}
  \ifthenelse{\texlive < 2013}{%
    \SetBgContents{DRAFT}
    \SetBgColor{blue!30}
  }{%
    \backgroundsetup{contents=DRAFT, color=blue!30}
  }
}

\begin{document}
%\input{ubonn-thesis-7.0/cover/Master_Cover}
% Cover page of thesis - this is only needed for the printed final
% version to be submitted to the department library
% Do not use this page for thesis submission to the Prüfungsamt or Promotionsbüro!
\ifthenelse{\equal{\ThesisVersion}{PILibrary}}{%
  \typeout{Document \jobname, Info: PI library version of thesis}
  \input{cover/\ThesisType_Cover}
}{}

% Start counting pages from the title page
\frontmatter
% Dedication has to come before \maketitle
% \dedication{Dedicated to no one}

% Select the correct title page(s)
\ifthenelse{\equal{\ThesisType}{Unknown}}{%
  \typeout{Document \jobname, Error: Unknown thesis type - no title page printed}
}{%
  % Bachelor thesis only has one title page
  \ifthenelse{\equal{\ThesisType}{Bachelor}}{%
    \typeout{Document \jobname, Info: Bachelor thesis}
    \input{cover/\ThesisType_Title}
  }{%
    \ifthenelse{\equal{\ThesisVersion}{Final} \OR \equal{\ThesisVersion}{PILibrary}}{%
      % Final and PI library versions
      \typeout{Document \jobname, Info: Final version of a \ThesisType  thesis}
      \input{cover/\ThesisType_Final_Title}
    }{% Submission and draft versions
      \input{cover/\ThesisType_Submit_Title}
      \typeout{Document \jobname, Info: Draft/submission version of a \ThesisType  thesis}
    }
  }
}

\pagestyle{scrplain}

%------------------------------------------------------------------------------
% You can add your acknowledgements here - don't forget to also add
% them to \includeonly above
%==============================================================================
\chapter*{Abstract}
\label{sec:abstract}
%==============================================================================

The main purpose of this work is to study the performance of the APV25 and VMM3a readout chips with regards to using them in the front-end readout system of a Triple-GEM detector. For the APV25, the SNR\footnote{Signal-to-Noise Ratio} was estimated with a cosmics measurement around 84.29$\pm$0.12. For the VMM3a, the noise of the chip was characterised in terms of the ENC\footnote{Equivalent Noise Charge} in the full parameter space and with increasing input capacitance. For the highest VMM gain and peaktime settings, the calculated ENC was about 300 electrons $+$ 8 electrons/pF. Finally, a comparison was made between the two chips, based on their noise, while also the setup for future cosmics measurements with a VMM readout was prepared in order to also derive the VMM SNR ratio. \\

%\textbf{Keywords:}  APV25, VMM3a, APV, VMM, ASIC, GEM detectors.

%------------------------------------------------------------------------------
\chapter*{Acknowledgements}
\label{sec:ack}
%------------------------------------------------------------------------------

I would like to thank:

\begin{itemize}
    \item Michael Lupberger for his invaluable help and resources provided during the implementation of this thesis project;
    \item my colleagues from the AG Ketzer research group and specifically: Michael Hösgen for his input and help with the GEM-APV measurement setup; Markus Ball for his constant guidance; Henri Pekeler for his help in setting up a remote connection with the oscilloscope; Karl Jonathan Flöthner, Christian Honisch, Dimitri Schaab, Philip Hauer and Jan Paschek;
    \item Prof. Bernhard Ketzer for openly accepting me in his group and allowing me to work on an interesting project, as well as for his constructive criticism and guidance.
    \item Patrick Schwäbig and Finn Jaekel from the Physics Institute (PI) for their help with the VMM.
    \item Francesco Cozzi, because his previous work with the APV at the HISKP provided a helpful reference for my own work; 
    \item Steven D. Sharples, because his open source code for the VXI-11 Ethernet Protocol was immensely helpful in setting up the oscilloscope part of the VMM automation framework; 
    \item my international friends from Bonn, who played a huge part in me enjoying my time during the entirety of this Master's Program;
    \item my dad, my mom and my sister for their support and encouragement, as well as their understanding and patience, despite the difficulties presented from my extended leave abroad.
\end{itemize}

%%% Local Variables: 
%%% mode: latex
%%% TeX-master: "../mythesis"
%%% End: 

\tableofcontents

\mainmatter
\pagestyle{scrheadings}

% Turn off DRAFT for the following pages
\ifthenelse{\equal{\ThesisVersion}{Submit}}{%
  \ifthenelse{\texlive < 2013}{%
    \SetBgContents{}
  }{%
    \backgroundsetup{contents={ }}
  }
}{}

%------------------------------------------------------------------------------
% Add your chapters here - don't forget to also add them to \includeonly above
% !TEX root = mythesis.tex

%==============================================================================
\chapter{Introduction}
\label{sec:intro}
%==============================================================================

The advancement and characterisation of detector readouts is essential in order to study the physics of the microcosm. The high-energy physics requirements set by experiments such as the ATLAS\footnote{A Toroidal LHC ApparatuS \cite{aad2008atlas}} and CMS\footnote{Compact Muon Solenoid \cite{collaboration2008cms}} at CERN have guided the development of detector technologies over the years. 

In particular, the design of gaseous detectors has been upgraded through time in order to achieve major goals, such as: 1) minimize the dimensions of the readout systems so as to perform better particle tracking, 2) incorporate schemes which will keep discharge and electrostatic effects to the minimum so as to protect the electronics from damage, 3) provide sufficient amplification of a particle's initial signal, and 4) eliminate dead readout times between physics events in the detector medium. The latter is necessary in order to make detectors able to withstand high-rate particle environments. 

Micro-pattern gaseous detectors (MPGDs) are designed to include all of the above mentioned features. One of them is the Gas Electron Multiplier (GEM), designed by F. Sauli \cite{sauli1997gem} and upgraded for the purposes of the COMPASS\footnote{Common Muon and Proton Apparatus for Structure and Spectroscopy \cite{abbon2007compass}} experiment at CERN to incorporate a triple-stage gain amplification. Triple-GEM detectors have so far made use of the APV25 readout ASIC. Initially designed for the silicon strip detectors in the CMS tracker \cite{jones1999apv25}, some of the key features of this readout chip is an analog pipeline readout, radiation tolerance and low-power consumption. 

In view of the GEM detector upgrade of the COMPASS++/AMBER \cite{compassamber} experiment, plans to change the main electronic readout component of the Triple-GEM detectors were set in motion. One of the requirements of the experiment is a self-triggered readout. For this reason, the VMM3a ASIC seemed like a viable option for the upgrade of the Triple-GEM readout system. The VMM chip was designed for the ATLAS New Small Wheel upgrade \cite{polychronakos2018vmm3a} and implemented in the Scalable Readout System \cite{lupberger2018implementation}. Some of its main features are a self-triggering system, adaptive gain and neighbour-clustering logic.   

The main objective of this thesis is to compare the APV25 and VMM3a readout chips regarding their performance on a standard-size Triple-GEM detector readout system. In the case of the APV25, this is approached by acquiring its signal-to-noise ratio from measurements with cosmic rays. For the VMM3a, a direct signal-to-noise measurement was not possible so far, as individual care needed to be taken regarding the chip's noise levels. In order to accomplish this, the ENC\footnote{Equivalent Noise Charge} was measured for several VMM chips with all parameter settings and with increasing input capacitance. 

In \cref{sec:theory} of this work, there is a brief overview of the fundamental processes and elements that we will encounter in this study; in \cref{sec:apv25}, the noise measurements of the APV25 chip are listed and explained in detail, while \cref{sec:vmmstg1,,sec:script,,sec:vmmstg2} provide the full spectrum of noise measurements regarding the VMM3a chip. Finally, in \cref{sec:vmmsn}, a brief comparison between the two chips is provided.  

%%% Local Variables: 
%%% mode: latex
%%% TeX-master: "mythesis"
%%% End: 

%==============================================================================
\chapter{Fundamentals}
\label{sec:theory}
%==============================================================================
\noindent
Interactions of particles with matter are the working principles around which particle detectors are designed. This chapter aims to give a brief insight into these interactions and how the various gaseous detectors take advantage of them, as well as introduce the triple GEM detector and its readout electronics which are used in this study. Finally, electronic noise sources in detectors are a necessary knowledge background in order to understand the main material of this thesis. The first two sections of this chapter are largely based on studies of Ref. \cite{groom2000passage},\cite{leotechniques}, \cite{grupen2005astroparticle}, \cite{virdee1999experimental} and \cite{stapnes2006instrumentation}.
%------------------------------------------------------------------------------
\section{Particle Detection}%
\label{sec:theory:particles}
%------------------------------------------------------------------------------

\begin{figure}[htbp]
  \centering
  \includegraphics[width=9cm]{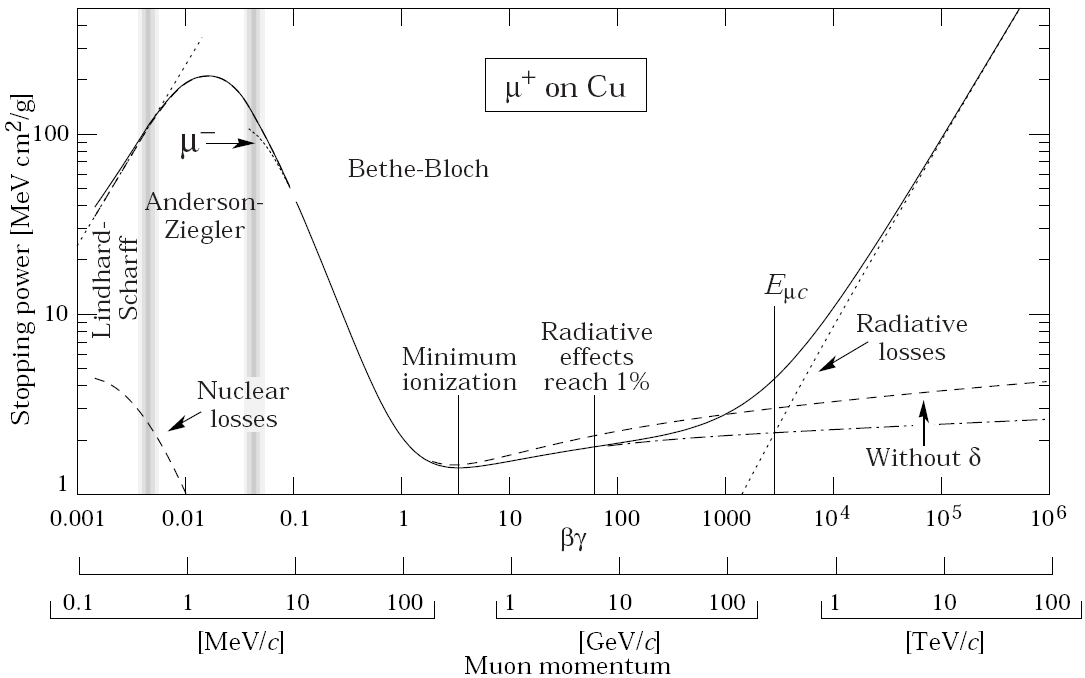} 
  \caption{Energy loss of charged particles in matter as described by the Bethe-Bloch formula \cite{groom2000passage}}%
  \label{fig:bethebloch}
\end{figure}

Depending on their mass, energy and charge, particles have a variety of interactions with matter through which they deposit energy in the medium. The energy loss of charged particles is represented by the Bethe-Bloch formula, shown graphically in \cref{fig:bethebloch}. The formula describes the mean energy loss or stopping power of a heavy charged particle in a medium according to its velocity or momentum. The main source of energy deposition is through ionization and excitation of the atoms of the matter (creation of electron-ion pairs). It presents a minimum at $\beta\gamma=3.5$, which rises slightly for higher particle momenta, unless radiative losses take effect. 
The latter mainly affect electrons, positrons and high-energy muons. The Bethe-Bloch formula may also be integrated to derive the range $R$ of the incident particle in the medium. Electromagnetic interactions not represented in the Bethe-Bloch formula are namely the Cherenkov radiation and transition radiation and the interactions of photons. Information for the first two can obtained from \cite{leotechniques} and exceed the relevance of this work.

\begin{figure}
    \centering
    \includegraphics[width=6.7cm]{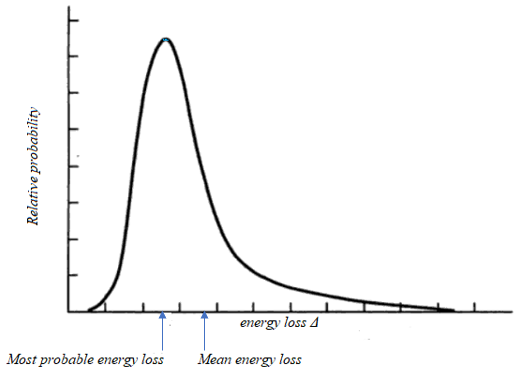}
    \caption{The Landau Distribution \cite{landauglen}.}
    \label{fig:landau}
\end{figure}

While the Bethe-Bloch shows the mean energy loss, the fluctuations of the energy loss of particles in a thin layer of material is described by a Landau distribution (\cref{fig:landau}). The fluctuations are due to a small number of collisions leading to high energy transfers. In thicker materials, the distribution becomes Gaussian due to the increase in the number of such interactions.

The energy loss of photons is dominated by three main processes. Photons can be entirely absorbed from the medium they traverse and produce either an electron (photoelectric effect) or an electron-positron pair (pair production). Additionally, they can be unelastically scattered from free electrons (Compton scattering). 

%The photoelectric effect is most notably taken advantage of in scintillator-photomultiplier (PMT) systems. Incident radiation in the scintillator releases scintillation light or photons which hit the photocathode of the PMT. Electrons released from the photocathode are then 

In order to understand how these interactions can be utilized in gaseous detector systems, which is the main focus in this thesis, it is essential to first explain the kinematic properties of charges in an electrified gas, i.e. a gas, on which an electric field is applied. Namely, the phenomena of drift and amplification.

%\begin{figure}[htbp]
 % \centering
  %\includegraphics[width=8cm]{figs/theory/ionizationchamber.png}
  %\caption{A simple gaseous ionization chamber (no gas multiplication) %\cite{masciocchi2017gaseous}.}%
  %\label{fig:ionizationchamber}
%\end{figure}

The electrons and ions liberated during ionization of the electrified gas medium are accelerated in the electric field, however, their motion is hindered by collisions with the atomic nuclei of the matter. The normal speeding movement of the charges is thus attained to a drift with an average drift velocity of: 
\begin{equation}
    \label{eq:vdrift}
    u_D = \frac{e}{m}E\tau 
  \end{equation}
where $E$ is the electric field intensity, $\tau$ is the mean free path between collisions, $m$ is the electron mass and $e$ its charge. Due to their much smaller mass, electrons can attain speeds $10^2$ times higher than ions. 

\begin{figure}[htbp]
  \centering
  \includegraphics[width=6cm]{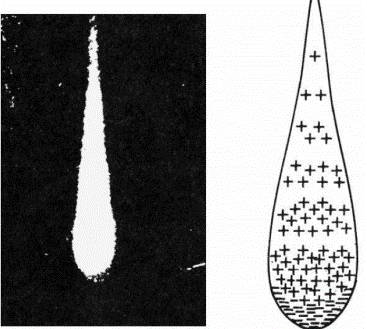}
  \caption{The drop-like shape of an avalanche formation. On the left, a cloud chamber picture. On the right, a schematic view. Electrons drift faster than ions and thus they group together at the head of the drop, while the ions slowly drift backwards \cite{garutti2012gaseous}.}
  \label{fig:avalanche}
\end{figure}

With a high enough electric field, the drifting electrons gain enough energy to cause secondary, tertiary and further ionizations, which gradually results in the formation of a so-called avalanche. Due to their faster movement, the electrons concentrate at the forefront, while the positive ions slowly trail backwards, producing the avalanche's known shape of a drop, shown in \cref{fig:avalanche}.

The charge amplification is characterized by the ratio of produced electrons $n$ to the number of primary electrons $n_0$, or the gas gain $G$. In order to calculate this, one needs to take into account the mean free path $\lambda$ between collisions. So after a distance $dx$ in the medium, the new number of electrons will be:
\begin{equation}
    \label{eq:dndx}
    dn =  \frac{n}{\lambda} dx
  \end{equation}
In uniform electric fields, the mean free path can be expressed as the inverse Townsend coefficient $a^{-1}$. Hence, integrating \cref{eq:dndx} yields: 
\begin{equation}
    \label{eq:nx}
    n =  n_0e^{\alpha x} 
  \end{equation}
Therefore, the gas gain $G$ can be finally expressed as:
\begin{equation}
    \label{eq:gasgain}
    G = \frac{n}{n_0}=e^{\alpha x} 
  \end{equation}
However, for non-uniform fields, $\alpha$ is a function of the distance $x$:

\begin{equation}
    \label{eq:townsend}
    \alpha = \int_{x_1}^x \alpha(x)dx 
  \end{equation}

The gas gain shouldn't exceed the Raether limit of $10^8$ as this would lead to a continuous discharge and subsequently, the breakdown of the supplied voltage. In reality, due to the statistical nature of the multiplication process, this breakdown can occur already at gain $10^6$ \cite{venditti2018avalanche}. Spontaneous increases in the gas gain can be caused by a variety of factors. One such is the emission of UV photons from the decaying of excited atoms that weren't ionized during their collision. These photons can induce new avalanches and lead to a continuous discharge. To avoid this in gaseous detector systems, usually a quench gas is added to the primary gas. This is usually a polyatomic gas with many vibrational and rotational states, suitable for absorbing the energy from the excited atoms, allowing a stable gain in the detector medium. 

%------------------------------------------------------------------------------
\section{Gaseous Detectors}%
\label{sec:theory:gaseous}
%------------------------------------------------------------------------------

In gaseous detectors, the ionizing effect of incoming particles is utilized by applying an electric field on a gas medium, and collecting the liberated charges on a readout anode. The electric field is created by setting varying voltages on two encompassing electrodes, the cathode and the anode. At low voltages, this produces at most a weak electric signal. Increasing the operating voltage, however, provides amplification of the primary charge inside the gas, causing the detector to enter its proportional mode. This mode is generally preferred for gaseous detectors as the output signal is directly proportional to the initial energy deposited and hence, by measuring the energy loss, it is possible in some cases to export information about the incident particle, such as its species. 

\begin{figure}[htbp]
    \centering
    \includegraphics[width=7cm]{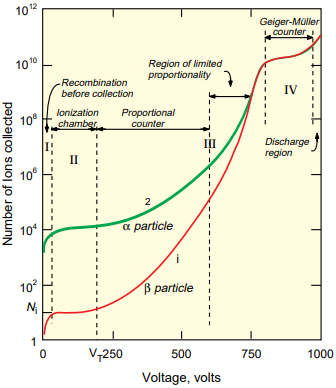}
    \caption{Operational modes of a gaseous detector \cite{virdee1999experimental}.}
    \label{fig:gasmodes}
\end{figure}

A more complete overview of the operating modes of a gaseous detector is provided in \cref{fig:gasmodes}. The main ones are: 1) the recombination mode, where the operating voltage is too low and thus charge pairs 'recombine' into atoms before their collection, 2) the ionization mode, where all the primary charge is collected and no amplification stage is provided, 3) the proportional mode which is split to high and limited proportionality regions, and 4) the Geiger-Müller mode with the highest possible operating voltage before continuous discharge takes effect, where the gain has reached saturation and the signal is independent of the initial deposited energy.

\begin{figure}[htbp]
  \centering
  \begin{tabular}{cc}
  \includegraphics[width=3.35cm]{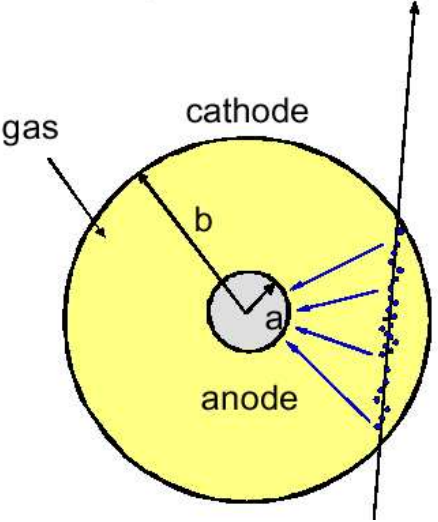} &
  \includegraphics[width=9.2cm]{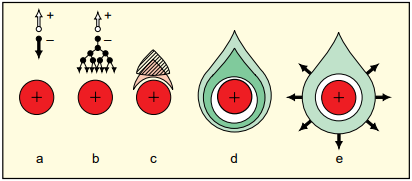} \\
  (a) & (b)
  \end{tabular}
  \caption{(a) Cross section of a cylindrical proportional counter \cite{stapnes2006instrumentation}, (b) Development of an avalanche in a proportional wire chamber.}
  \label{fig:pcounter}
\end{figure}

The original and most simple design of a gaseous proportional detector is the proportional counter, a cross section of which is shown in \cref{fig:pcounter}a. Its main characteristics are a cathode container, upon which negative voltage is applied, and an anode wire with positive voltage along its axis. Due to its cylindrical geometry, the electric field intensity is naturally higher near the anode wire ($E\propto 1/r$), hence gas multiplication occurs. \cref{fig:pcounter}b shows a schematic of the time development of an avalanche in a proportional wire chamber. The multiplication process lasts about $\sim$ \SI{1}{ns} \cite{virdee1999experimental}. Normally, the electrons are collected rapidly while the ions slowly drift toward the cathode. The actual signal on the anode is induced by the movement of charges. Hence, due to the longer distance that the ions have to traverse, it is the ion drift that determines the size and time of the signal. 

However, the proportional counter lacks the possibility of particle position tracking, which is a desired characteristic for gaseous detectors.

\subsection{Multiwire Proportional Chambers}

\begin{figure}[htbp]
  \centering  
  \begin{tabular}{cc}
  \includegraphics[width=7cm]{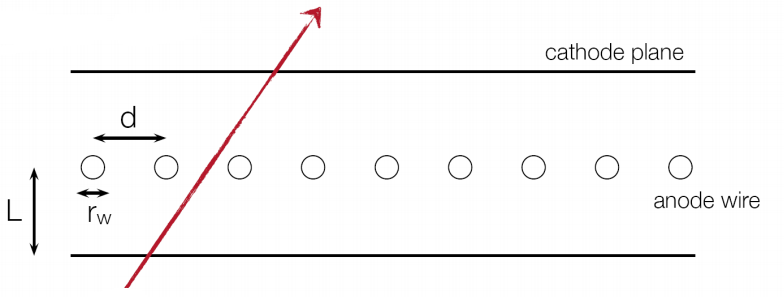} &
  \includegraphics[width=6cm]{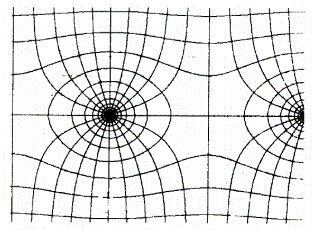} \\
  (a) & (b)
  \end{tabular}
  \caption{(a) Schematic view of an MWPC. (b) Equipotential and field lines of an MWPC \cite{garutti2012gaseous}.}
  \label{fig:wirechamber}
\end{figure}

Multiple adjacent anode wires in a planar chamber can act as individual proportional counters. If they are each connected to a distinct amplifier, they can also offer position tracking. This was the idea for the Multiwire Proportional Chambers (MWPC), invented by Charpak in 1968. A schematic view is provided in \cref{fig:wirechamber}a. With the cathodes at a negative voltage and the wires grounded, an almost homogeneous electric field is created in the drift regions with a radial field near the wires, as shown in \cref{fig:wirechamber}b. In specific, the applied field is of several kV, the wire pitch $d$ is 1-\SI{4}{mm} with a wire diameter $r_w$ of 20-\SI{25}{\micro m}, while the drift distance $L$ on each side of the wires is 3-\SI{6}{mm}. The spatial resolution is limited by the minimum possible distance between wires being $\sim$ \SI{1}{mm}, since smaller wire distances were technologically difficult to achieve and they would also leave the setup vulnerable to electrostatic effects. Hence, the readout resolution is limited to $\sigma_x = d/\sqrt{12} \approx$ \SI{300}{\micro m}. 

\begin{figure}[htbp]
  \centering
  \includegraphics[width=7cm]{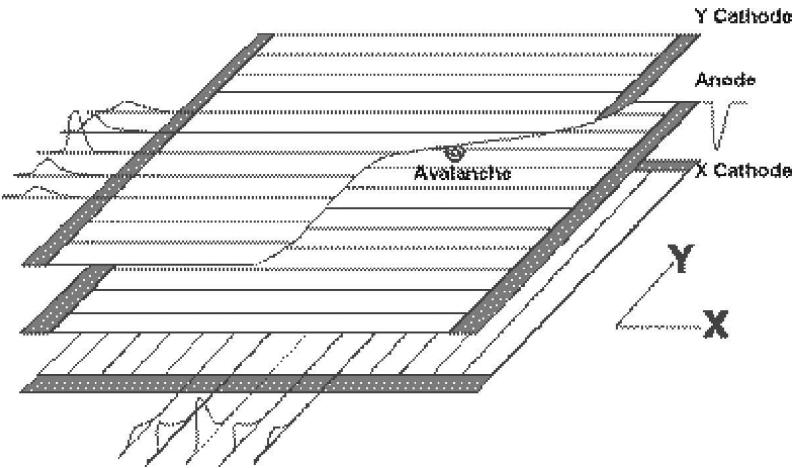}
  \caption{A 2D MWPC with segmented cathode planes. Signals induced on the two cathode planes are used to reconstruct the X-Y coordinates of the incident particle \cite{stapnes2006instrumentation}.}
  \label{fig:mwpcx3}
\end{figure}

The spatial resolution can be improved by having a segmented cathode plane readout, as shown in \cref{fig:mwpcx3}. This allows determination of the X and Y coordinates of the incident particle and provides information about the exact hit position. Additionally, having multiple planes of anode wires placed in varying angles could provide full reconstruction of the particle track.

One of the major disadvantages of the MWPCs was the fact that if a single wire is slightly displaced, it distorts the field shape. Therefore, high mechanical precision is required for their construction. Additionally, the long distance between the anode and cathode planes extends the ion drift time, which limits the rate capability of the detector.

Other ideas for particle tracking were implemented in the drift chambers--similar to proportional chambers, only with the added feature of measurement of the drift time, which can be extrapolated to the origin of the electrons in the chamber. Alternatively, the straw-detector unit is composed of proportional chambers of small diameter stacked together, hence a triggered signal from several straw tubes can help reconstruct the particle track in space. Time projection chambers (TPCs), on the other hand, combined planar MWPCs and drift time measurement in order to provide three-dimensional particle tracking. 

\subsection{Micro-Pattern Gaseous Detectors}

Modern applications for detectors prioritized high-rate capability and improved spatial resolution, which are limited in the previous detectors due to the long ion drift time to the cathode in the former case and due to the limitations in the construction technology in the latter case. This brought forward a new generation of detectors aided by the photo-lithographic technology: the micro-pattern gaseous detectors (MPGDs). 

\begin{figure}[htbp]
  \centering
  \begin{tabular}{cc}
  \includegraphics[width=9cm]{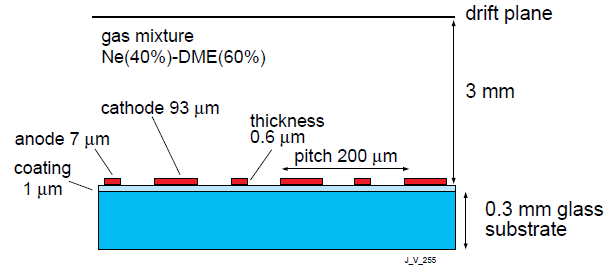} &
  \includegraphics[width=4cm]{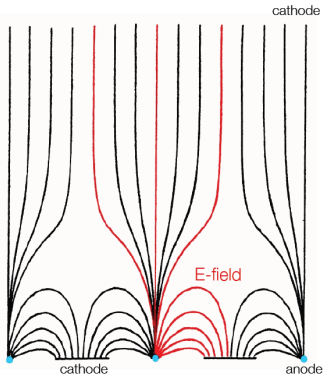} \\
  (a) & (b)
  \end{tabular}
  \caption{(a) Schematic view of an MSGC \cite{virdee1999experimental}. (b) Field structure of the MSGC \cite{garutti2012gaseous}.}
  \label{fig:msgc}
\end{figure}

The first design in this category was the micro-strip gaseous chambers (MSGCs) introduced by A. Oed \cite{oed1988position}. As shown in \cref{fig:msgc}a, this type of detector consists of alternating 0.6-\SI{}{\micro\meter}-thick anode and cathode strips imprinted on an 0.3-mm-thick glass substrate for electrostatic insulation, while the electric field is provided from a drift electrode placed at \SI{3}{mm} distance from the substrate. The small gap between the anode and cathode allows a fast evacuation of ions, which makes higher rates achievable for the detector. The typical pitch between the anodes is \SI{200}{\micro m} with an anode width of $\sim$ \SI{7}{\micro m} and cathode width of $\sim$ \SI{93}{\micro m}. The intrinsic spatial resolution is $\sim$ \SI{30}{\micro m}, while the multi-track resolution is $\sim$ \SI{250}{\micro m}. The field structure can be seen in \cref{fig:msgc}b. Similarly to the MWPCs, a high field is created directly above the anodes.

A major disadvantage of this structure, however, was the fragility of the electrodes, which were pliable to discharges, due to the high electric fields developed in their proximity. This accelerated the aging of the detectors. A solution to this was the use of an intermediate structure, which is the main feature of the latest Micromegas and GEM detectors. 

\subsection{The Gas Electron Multiplier}

\begin{figure}[htbp]
  \centering
  \begin{tabular}{cc}
  \includegraphics[width=6.5cm]{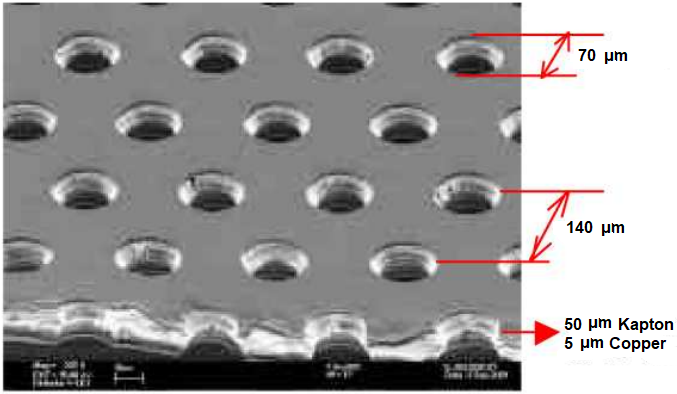} &
  \includegraphics[width=6.5cm]{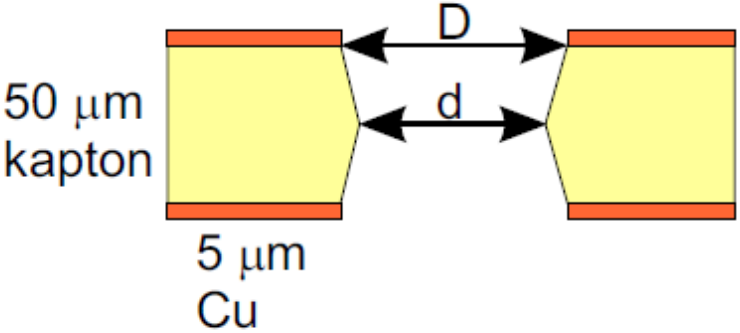} \\
  (a) & (b)
  \end{tabular}
  \caption{(a) Electron microscope view of a GEM foil, modified from \cite{stapnes2006instrumentation}. (b) Cut view of a GEM hole \cite{hoesgen2017tracking}.}
  \label{fig:gemfoil}
\end{figure}

The \textbf{G}as \textbf{E}lectron \textbf{M}ultiplier (GEM), invented by F. Sauli in 1997 \cite{sauli1997gem}, multiplies the charge in a stage prior to it reaching the anode. This is done in a thin metal-coated polymer foil, etched with an aggregation of holes. In standard designs, the foil is \SI{50}{\micro m} thick kapton, while the metal coating is \SI{5}{\micro m} thick copper. The holes are of double-conical shape with an outer diameter $D$ of \SI{70}{\micro m} and an inner diameter $d$ of \SI{65}{\micro m}, set at a \SI{140}{\micro m} pitch. Applying a potential difference on the two sides of the holes creates a high enough field ($\sim$ tens of kV/cm) to allow charge multiplication. The back-drifting ions are then easily collected by the foil. \cref{fig:gemfoil}a shows a microscopic view of the foil, while the double-conical shape of the gem hole can be better seen in \cref{fig:gemfoil}b.

\begin{figure}[htbp]
  \centering
  \begin{tabular}{cc}
  \includegraphics[width=6cm]{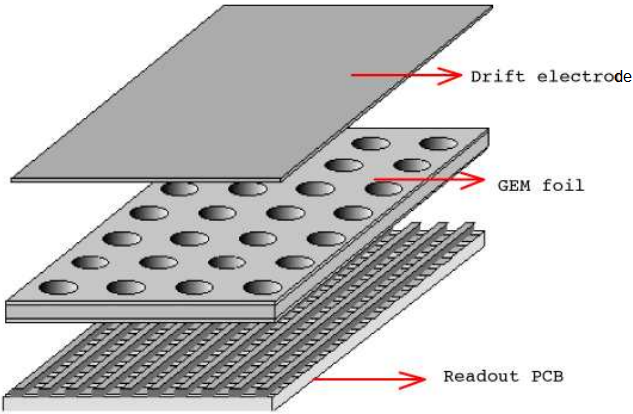} &
  \includegraphics[width=4cm]{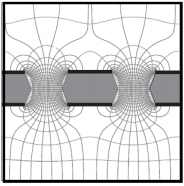} \\
  (a) & (b)
  \end{tabular}
  \caption{(a) Schematic view of a GEM detector. The GEM foil is where avalanches occur. It separates the detector in a drift region (above) and an induction region (below) \cite{stapnes2006instrumentation}. (b) Field structure of a GEM. The asymmetric field allows high electron extraction and fast ion collection \cite{garutti2012gaseous}.}
  \label{fig:gem}
\end{figure}

The GEM foil is set between a cathode and anode electrode, dividing the detector in a drift region and an induction region, the latter leading to the readout electronics. \cref{fig:gem}a shows a schematic view of the GEM detector. In \cref{fig:gem}b, one can see the field structure in the GEM holes, the drift region (above the GEM foil) and the induction region (below the GEM foil). Separating the amplification region from the readout electronics limits the probability of discharges propagating to the electronic circuits and hence, greatly improves the performance of the GEM detector compared to the MSGCs. A typical spatial resolution is \SI{45}{\micro m} with a time resolution around \SI{12}{ns}. In many cases, multiple GEM foils can be placed in cascade in order to increase the gain sufficiently to detect minimum ionizing particles.

%------------------------------------------------------------------------------
\subsection{Triple GEM Detectors}%
\label{sec:theory:triple-gem}
%------------------------------------------------------------------------------

It has been found that using a triple GEM structure, as opposed to using a double GEM, reduces the probability of a discharge at the presence of heavily ionizing particles by one magnitude of order \cite{ketzer2002triple}. %This probability is further reduced when setting the top GEM foil at higher gain and decreasing it in the successive GEM foils.

\begin{figure}[htbp]
  \centering
  \includegraphics[width=8cm]{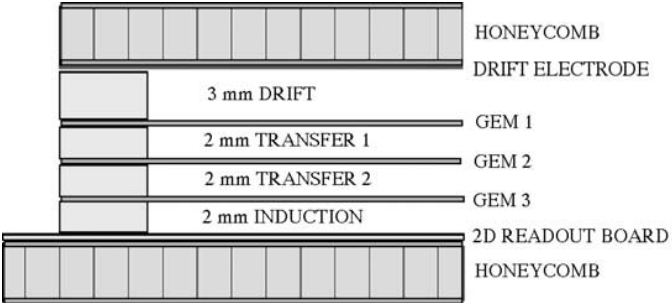}
  \caption{Schematic view of a triple GEM structure \cite{altunbas2002construction}.}
  \label{fig:triplegem}
\end{figure}

The architecture of a triple GEM as used in this study is shown in \cref{fig:triplegem}. The design closely follows that of the GEM detectors used at the COMPASS experiment at CERN \cite{abbon2007compass}. The three GEM foils are set between copper electrodes, forming a \SI{3}{mm} drift volume between the drift electrode and the first GEM foil, two gain-transfer areas of \SI{2}{mm} in between the GEM foils, and finally a \SI{2}{mm} induction area between the last GEM foil and the readout electronics. The whole structure is sandwiched between two honeycomb plates which provide mechanical stiffness and gas tightness. 

Discharge probability is further reduced in the tripe GEM detector by the use of an asymmetric gain across the three GEM foils. In order to distribute voltages throughout the GEM foils, a resistor chain is used. This simplifies the operation of the detector, since only a single high voltage needs to be applied. Similarly to the COMPASS GEM design, the resistor chain includes a \SI{1}{M}$\Omega$ resistor for each gap (drift, transfer and induction regions), and resistors that vary between \SI{0.5}{M}$\Omega$-\SI{0.7}{M}$\Omega$ for the GEM foils \cite{hoesgen2017tracking}. 

Regarding the gas, an Ar:CO$_2$ (70:30) mixture is used. For this gas mixture, according to previous studies of the detector gain \cite{cozzi2019asic,,hoesgen2017tracking}, the detector must be powered with a $\sim$ \SI{4}{kV} high voltage.

%In order to reduce the energy released in a discharge, each GEM foil is segmented into 12 sectors of $\sim$\SI{5}{nF} capacitance each, distinctly connected to the voltage supply. The distance between sectors is \SI{200}{\micro m}, which causes only a slight deviation in the uniformity of the detector performance. The center is additionally segmented in the shape of a \SI{50}{mm}-diameter disc. This region is powered separately and is used to kill the primary non-interacting beam during normal operation of the detector, when such a detector is used at the COMPASS~\cite{abbon2007compass} experiment at CERN. This partitioning only concerns one side of the foil, while the other remains in constant voltage to prevent discharge propagation. \cref{fig:gemfoil}b shows the segmented side of the GEM foil described above.

\begin{figure}[htbp]
  \centering
  \includegraphics[width=6cm]{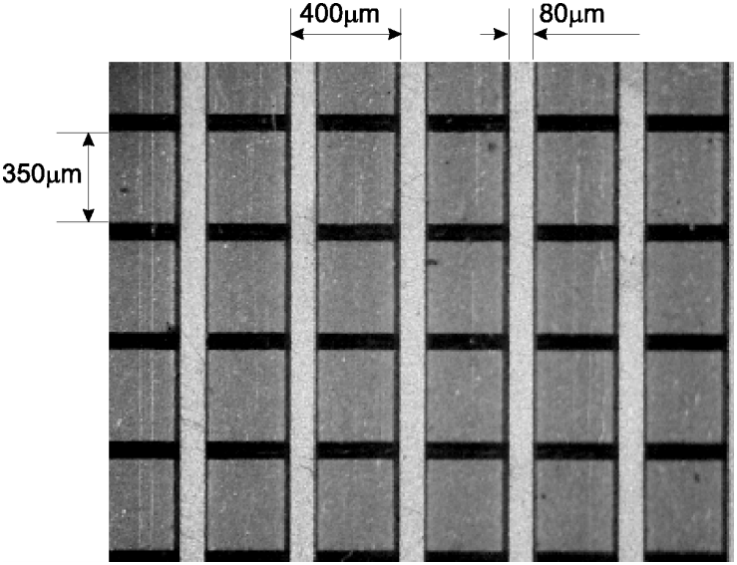}
  \caption{Microscope view of the readout PCB. The strips are \SI{80}{\micro m} and \SI{350}{\micro m} wide, pitched at \SI{400}{\micro m} and separated between the top and bottom layer with \SI{50}{\micro m} thick kapton ridges \cite{altunbas2002construction}.}
  \label{fig:gemreadout}
\end{figure}

The readout printed circuit board (PCB) is organized in two orthogonal layers of copper strips separated by \SI{50}{\micro m} thick kapton ridges, as shown in \cref{fig:gemreadout}. The widths of the strips on each layer have been optimized for an almost equal charge sharing. The top layer consists of \SI{80}{\micro m} wide strips and is conventionally used as the 'X' coordinate, while the bottom layer with \SI{350}{\micro m} wide strips serves as the 'Y' coordinate. Both planes have a total of 512 strips (256 strips per plane) set at a \SI{400}{\micro m} pitch. The total active area is $10\times10$ cm$^2$.

The charge collected from the strips of a detector readout is initially too small to resolve the position of the interaction. Therefore, additional amplification is required. For this reason, the front-end readout of a detector includes application-specific integrated circuits (ASIC), or readout chips, which implement such pre-amplifiers in a small area for each individual detector channel. This allows a channel-independent readout and leads to higher rate capabilities. 

%------------------------------------------------------------------------------
\section{Readout Electronics}%
\label{sec:theory:readout}
%------------------------------------------------------------------------------

\begin{figure}[htbp]
  \centering
  \includegraphics[width=9.5cm]{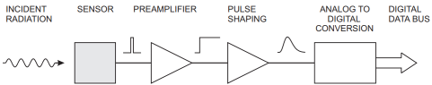}
  \caption{General schematic of a readout ASIC \cite{spieler2003analog}.}
  \label{fig:readoutschem}
\end{figure}

A general schematic of a detector readout chip is shown in \cref{fig:readoutschem}. The main parts are the aforementioned pre-amplifier, a shaping filter that determines the pulse shape and time and as such determines the overall bandwidth and electronic noise contribution of the system. The shaped data go through read and write sequences, before they are sent out to be digitized (analog to digital conversion). After this stage, they may be compared to an external trigger signal, which determines the importance of the acquired data in terms of being a real physics event or not. 

\subsection{The APV25 Readout Chip}
\label{sec:theory:apv25}

\begin{figure}[htbp]
  \centering
  \includegraphics[width=12cm]{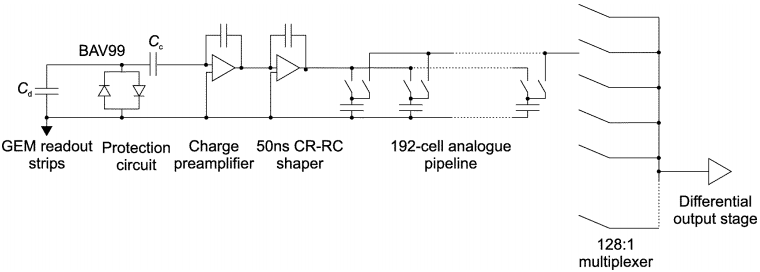}
  \caption{Schematics of the channel read-out electronics, with the diode protection and APV25 analogue pipeline \cite{altunbas2002construction}.}%
  \label{fig:apv25schematic}
\end{figure}

The APV25 readout chip consists of 128 channels, where input charges are pre-amplified at gain \SI{4.6}{mV}/fC\footnote{The APV gain is stated as \SI{18.7}{mV}/mip ($\sim$ 25000 electrons) in this source: \cite{jones1999apv25}. The 25000 electrons is a charge of \SI{4}{fC}, so the gain is 18.7/\SI{4}{mV}/fC $=$ \SI{4.6}{mV}/fC.} and shaped by a CR-RC filter of \SI{50}{ns} peaking time \cite{french2001design}. High-speed grounded signal diodes (BAV99) and a decoupling \SI{220}{pF} capacitor serve as a protection circuit to prevent leakage currents and discharges in the detector from entering the preamplifier. Shaped data are then sampled and stored into a 192-columns long pipeline composed of switched capacitor elements. A schematic of the APV25 circuit is shown in \cref{fig:apv25schematic}.

The writer pointer samples the shaper output for a few samples after an external trigger. The sampled pipeline data is then read out to a Firt-In/First-Out (FIFO) buffer, which can store up to 32 samples. There are three sampling modes upon a trigger signal: 1) the peak mode, 2) the deconvolution mode, and 3) the multi-mode. In the multi-mode operation, which is the one used in this thesis, three consecutive samples are stored for each trigger with the third sample placed on the peak of the received signal, and the other two samples on the rising edge at a \SI{25}{ns} and \SI{50}{ns} distance from the peak. This allows for precise timing information which prevents saturation at high-rate beam environments. A key feature in this process is the trigger latency, which determines which buffer ID elements will be reserved for the trigger signal and hence the delay between a trigger signal and a physics event. The latency is set in values of \SI{25}{ns}. Due to the pipeline depth, the latency can be programmed up to 160 clock cycles, which amounts to about \SI{4}{\micro s} \footnote{The other 32 locations are used as buffers for up to 10 events.}.

\begin{figure}[htbp]
    \centering
    \begin{tabular}{cc}
       \includegraphics[width=8cm]{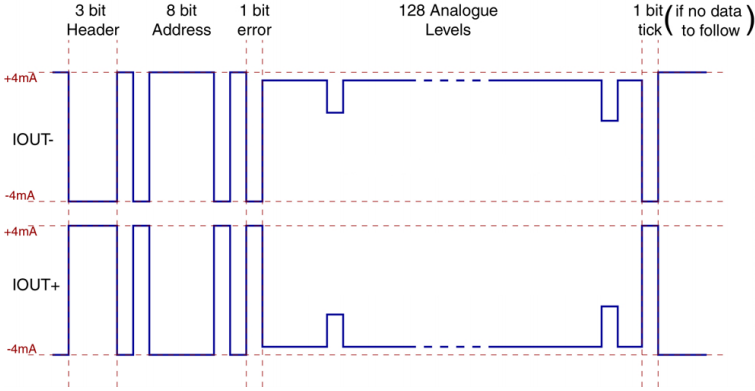}  &  \includegraphics[width=6cm]{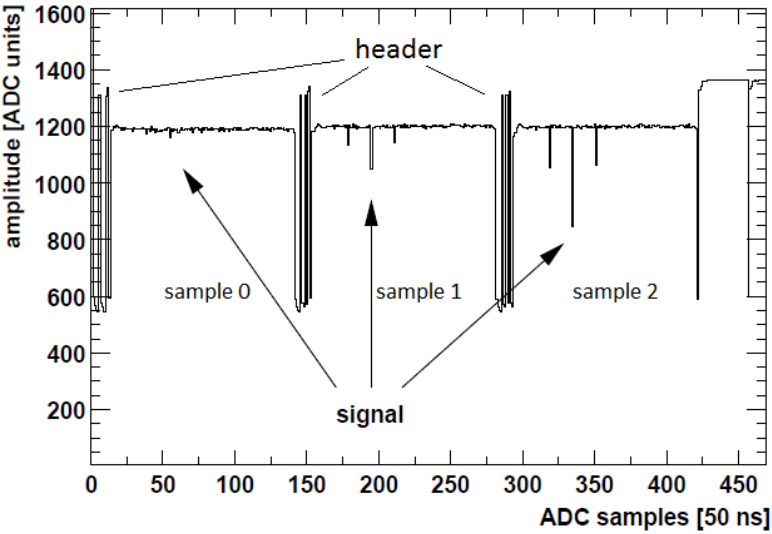}\\
        (a) & (b)
    \end{tabular}
    \caption{(a) Serial format of the data output of the APV. A signal affecting three channels can be seen \cite{simon2001commissioning}. (b) An analogue signal is sent out in two inverted data lines and consists of a header, address, error flag, the 128 channels of the sample and a synchronization tick (if all samples are sent out) \cite{jones1999apv25}.}
    \label{fig:apvdataformat}
\end{figure}

The sampled data are then read out separately in an analogue multiplexer which produces an analogue signal containing all 128 channels for each sample and converts it to a differential output in order to send it to the analog-to-digital converter (ADC). The conversion factor between the shaped and analogue differential signal can be adjusted through an I2C pipeline and its value is critical to the baseline that can be derived later from the APV chip. As shown in \cref{fig:apvdataformat}a, the data is sent in serial format, i.e. the samples are set in increasing order in the data output, sample 2 in the figure being the sample on the signal peak. More specifically, the data output format of the multiplexer consists of two signal lines, inverted to each other, as shown in \cref{fig:apvdataformat}b. This way, when subtracted, these signal lines can be used to reconstruct the real signal and also cancel out some of the pick-up electric noise. Each analogue signal/sample is preceded by a 3-bit header, an 8-bit address and an 1-bit error flag. A 1-bit synchronization tick also follows the data output if all three samples have been sent out and there is no more data to record.

\begin{figure}[htbp]
  \centering
  \includegraphics[width=8cm]{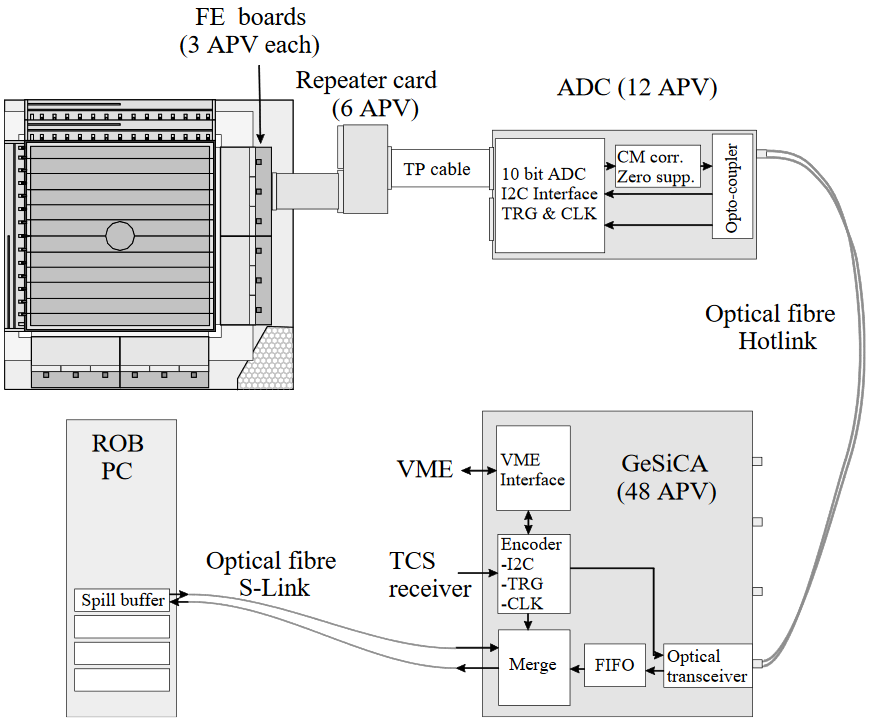}
  \caption{Schematic representation of the readout electronics \cite{altunbas2002construction}}%
  \label{fig:readoutschematic}
\end{figure}

As shown in \cref{fig:readoutschematic}, the multiplexed data from each APV chip are transmitted through a \SI{2}{m} twisted-pair flat cable to a 10-bit ADC, which digitizes the signal at a rate of \SI{20}{MHz}. The ADC has two modes of operation: 1) the Latch-All mode, during which pedestals are recorded (baseline mean and RMS), and 2) the Sparse mode for physics events. Regarding the noise that can be recorded during the first mode, this can be categorized into an intrinsic noise, which is independent for each channel, but also a so-called common-mode-shift noise, which is caused by external noise sources like electronic pick-up noise and grounding issues; this noise may affect groups of adjacent channels or even the whole chip. During the Sparse mode, the ADC removes the recorded pedestal values and makes use of a field programmable gate array (FPGA) in order to: 1) correct for the common mode shift, 2) perform zero-suppression if required, that is, send only those channels with an amplitude higher than a noise threshold. For channel $i$, this threshold amplitude $\alpha_i$ is determined from the baseline $b_i$, the baseline RMS $s_i$ and a suppression factor $k_s$ which is an ADC setting (usually set to 3) \cite{hoesgen2017tracking}. So only the signals with amplitude higher than: 
\begin{equation}
    \label{eq:apvthres}
    \alpha_i = b_i + k_s \cdot s_i
\end{equation}
can be recorded.

ADC data are sent through a bi-directional optical fiber to a GEM and Silicon Control and Acquisition module (GeSiCA) using the Hotlink protocol. The GeSiCA module hosts a VME interface, a buffer and merger of the ADC data and an encoder to process trigger and clock signals from the Trigger Control System (TCS) as well as slow control sequences for the ADC and APV25 chips which are sent back to the front-end electronics. Finally, the buffered and merged data are transmitted through optical fibers using the S-Link protocol to the data acquisition (DAQ) system.

\subsection{The VMM Readout Chip}
\label{sec:theory:vmm}

Initially designed for the Micromegas detectors of the ATLAS New Small Wheel (NSW) upgrade, the VMM chip was implemented in the Scalable Readout System (SRS) to replace the APV25 chip \cite{lupberger2018implementation}. Though it only carries half of the channels of the APV, the VMM ASIC presents features that promote its usage in various applications with Micro-Pattern Gaseous Detectors (MPGDs). Mainly, a peak and a time detector along with a sub-hysteresis discriminator which provide internal triggering. This reduces data pile-up and allows for use in high-rate environments with a readout frequency of up to 4 MHz/channel \cite{polychronakos2018vmm3a}.

\begin{figure}[htbp]
  \centering
  \includegraphics[width=12cm]{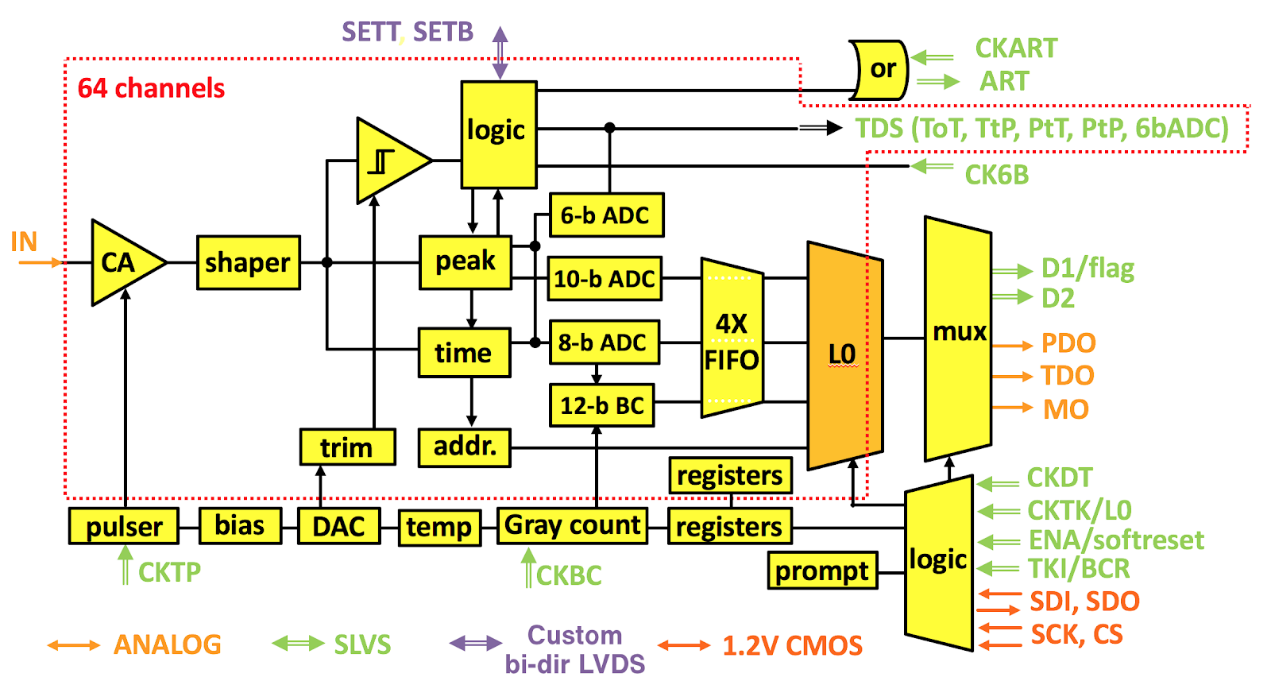}
  \caption{Architecture of the VMM3a \cite{polychronakos2018vmm3a}.}%
  \label{fig:vmmschematic}
\end{figure}

A schematic of the design of a VMM channel is shown in Fig. \ref{fig:vmmschematic}. Each channel processes an incoming signal through a low-noise charge pre-amplifier (CA), a shaper, the aforementioned peak and time detector and discriminator, and two of three available ADCs. The CA is designed with adaptive feedback, a test capacitor of either 0.3 pF or 3 pF capacitance and two different polarities, while the gain can be chosen from eight different values (0.5 - 16 mV/fC).

The shaper has four available peaking times (\SI{25}{ns} - \SI{200}{ns}) that change the time of charge integration over the signal. It is designed with a high analog dynamic range, which allows for high resolution at input capacitances of less than 200 pF. An internal pulser circuit connected to the injection capacitor of the CA makes it possible to also send test pulses through the signal-processing components. A 10-bit Digital to Analog Converter (DAC) allows global threshold adjustment that can be trimmed by a 5-bit DAC per channel. The discriminator then filters the signal through the channel’s threshold. Neighbour logic provides the reading of channels next to the hit one regardless of their signal threshold. Channels of adjacent chips are also included via a bidirectional signal.

%The VMM3a has two readout modes, a two phase analog mode and a continuous simultaneous read-write mode. Timing diagrams of both these modes are shown in \cref{fig:readoutmodes}. %In the two-phase analog mode, the VMM switches between acquisition and readout mode. The data acquisition is triggered by an ena signal and an event is processed into amplitude and time voltages in the peak (PDO) and time (TDO) detector respectively and gets stored in their analog memories. The ena signal is then set to low and the ASIC switches to readout phase. The PDO and TDO outputs are digitised, while the channel address is serialized. Acquisition is then re-enabled.  

At the crossing of the peak of a detector signal or test pulse, the peak detector stores the peak amplitude first into analog memory. This triggers the linear voltage ramp of the time-to-amplitude converter (TAC), which can also be set to start at threshold crossing and stops at the time of the next bunch crossing (BC) clock. The BC clock cycle lasts 25 ns. The TAC value is saved into analog memory and the ramp duration can be chosen from four values (60 ns - 650 ns). The BC clock count (BCID) indicates the coarse time of the VMM, while the time detector output (TDO) amplitude constitutes the fine timing, which has to be converted from an 8-bit ADC value to nanoseconds. Depending on the readout mode, the analog signal amplitude is processed through a 6-bit or 8-bit ADC. 

\begin{figure}[htbp]
  \centering
  \begin{tabular}{cc}
  \includegraphics[width=7cm]{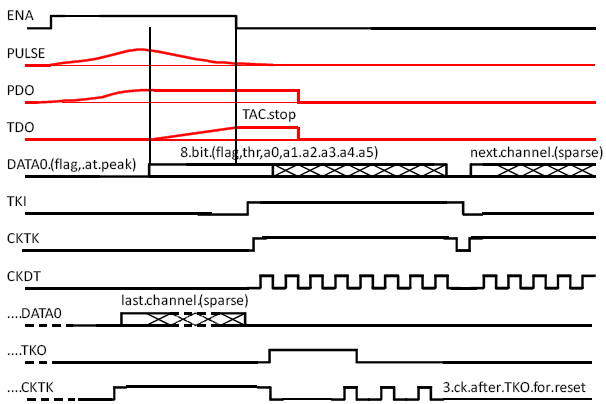} &
  \includegraphics[width=7cm]{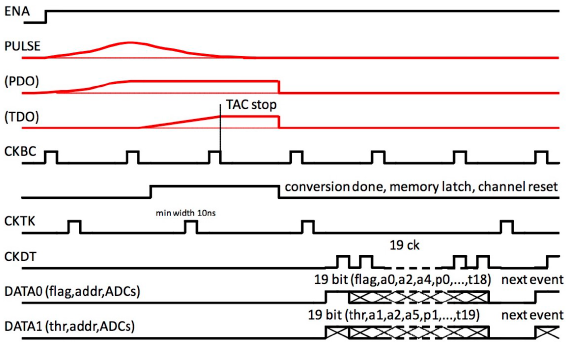}\\
  (a) & (b)
  \end{tabular}
  \caption{Readout modes of the VMM ASIC. (a) Two-Phase Analog Mode. (b) Continuous Mode \cite{polychronakos2018vmm3a}.}%
  \label{fig:readoutmodes}
\end{figure}

The readout mode can be a two phase analog mode, where the VMM switches between acquisition and readout mode, or a continuous mode, where data are read and written simultaneously, eliminating dead times. Timing diagrams of both these modes are shown in \cref{fig:readoutmodes}. 

In the continuous mode, which is used in this study, the peak amplitude is digitised in a 10-bit ADC, while the time amplitude in an 8-bit ADC \cite{iakovidis2018vmm3}. A 6-bit ADC value can also be obtained as a fast and direct digital output signal. For each event, a total of 38 bits are saved into a 4-hit-deep FIFO buffer. Under these conditions, a hit rate of up to 4MHz/channel can be achieved. The event is read out to external devices using a token (CKTK) scheme and multiplexed into two data outputs.

Additional VMM features also include a temperature sensor and an analog monitoring output (MO) where the shaper output can be monitored. Multiple signals can be routed to the MO via configuration of the VMM: the global and individual channel threshold, the analog channel baseline, the temperature, the pulser step and test pulses. 

\begin{figure}[htbp]
  \centering
  \begin{tabular}{cc}
  \includegraphics[width=6cm]{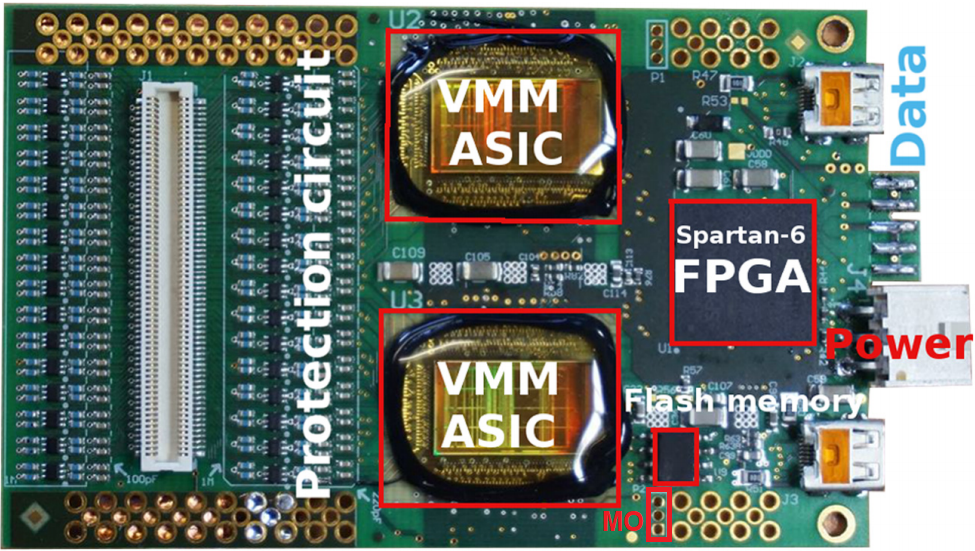} &
  \includegraphics[width=5cm]{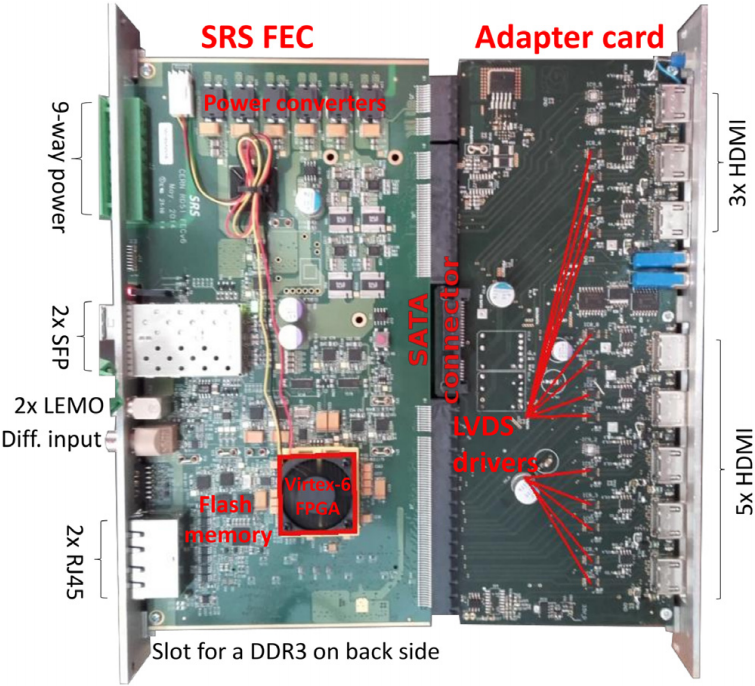}\\
  (a) & (b)
  \end{tabular}
  \caption{Implementation of the VMM in the SRS system. (a) VMM3 hybrid, modified from \cite{lupberger2018implementation}. (b) From left to right, the SRS FEC and Adapter card, also from \cite{lupberger2018implementation}.}
  \label{fig:vmmsrs}
\end{figure}

As mentioned before, the VMM chip was also implemented in the SRS system \cite{lupberger2018implementation} and this is the way it was used in the experimental setup. Developed by the RD51 collaboration in 2009 \cite{martoiu2013development}, the SRS constitutes a universal method of implementing detector front-end readouts. More specifically, with the use of a common back-end and an application-specific front-end adapter board, it is possible to use the system for a variety of detector chips. 

Most of the general-purpose components of the readout sit on the main part of the back-end, the Front-End Concentrator (FEC). This includes power converters, flash memory, programmable logic and high-speed devices. Specifically, the FEC is based on a Virtex-6 FPGA, integrating two small form-factor pluggable transceivers (SFP) that provide Gigabit Ethernet connection to a computer for data transfer, a DDR3 memory chip, clock and trigger capabilities (RJ45, LEMO 50-$\Omega$), a differential input and a 9-way power connector. Regarding scalability, multiple FECs can be placed side by side in a Eurocrate.

Developments toward including the VMM in the SRS included the design of a hybrid, a new FPGA firmware on the SRS FEC and an adapter card, all shown in \cref{fig:vmmsrs}b.

The VMM was wire-bonded on a hybrid, which similarly to the APV25 chip is directly connected to the detector readout. In order to match the 128 channel number of the APV, two VMMs were placed on a single hybrid. The signals of both VMMs are collected by an FPGA on the hybrid, which controls the general operation of the VMMs. As shown in \cref{fig:vmmsrs}a, the hybrid features include a flash memory to configure the FPGA, a mini HDMI with a powering option, detector and external power connectors. The latter connect to Low-Dropout Regulators (LDOs) which require two supply voltages, one for the FPGA operation (\SI{2.5}{V}) and one for the VMMs (\SI{1.2}{V}). Inter-ASIC signals on the hybrid help apply the neighbour logic across the entirety of the detector. In addition, the MO pad of the ASIC's periphery was wire-bonded to the hybrid PCB, providing an external test point. This way the MO test point can be connected to eg. an oscilloscope, to provide measurement of the signals that can be configured to the shaper output.

\begin{figure}[htbp]
  \centering
  \includegraphics[width=11cm]{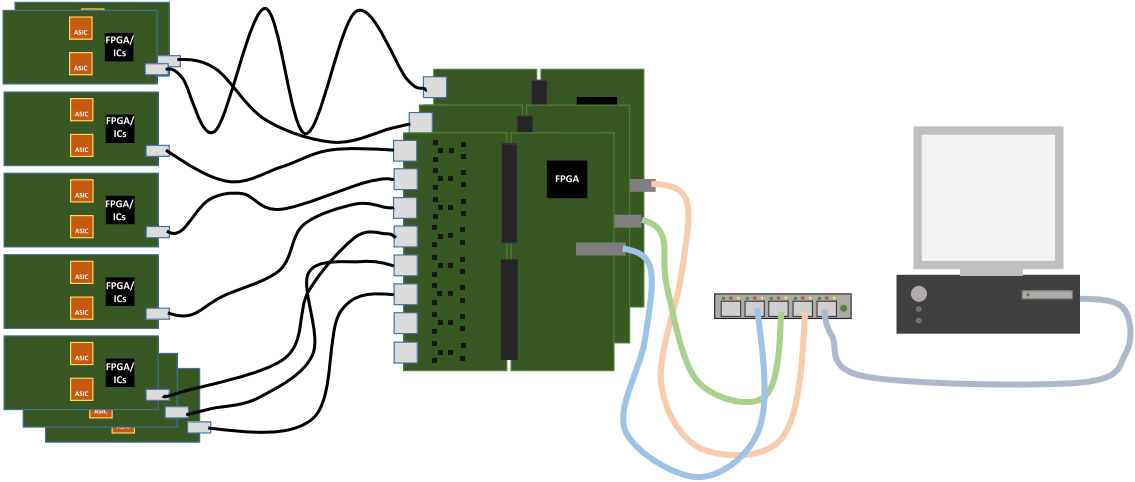}
  \caption{Schematic of the SRS VMM readout \cite{lupberger2018implementation}.}
  \label{fig:vmmconnection}
\end{figure}

The adapter card hosts 8 HDMI sockets, where the hybrids can be connected. An option to power the hybrids via the HDMI connection is provided by a Serial AT Attachment (SATA) connector between the FEC and the adapter card. The necessary voltages can then be configured in the power converters of the FEC. The SRS FEC can be connected via Ethernet connection to a data acquisition (DAQ) computer and it provides both control and configuration of the adapter card and hybrid as well as data reading from the hybrid. Additionally, a decoder is included to handle the 8b10b decoded data sent between the hybrid and the FEC. A schematic of a connection with HDMI-powered hybrids is shown in \cref{fig:vmmconnection}.

\begin{figure}[htbp]
  \centering
  \includegraphics[width=11cm]{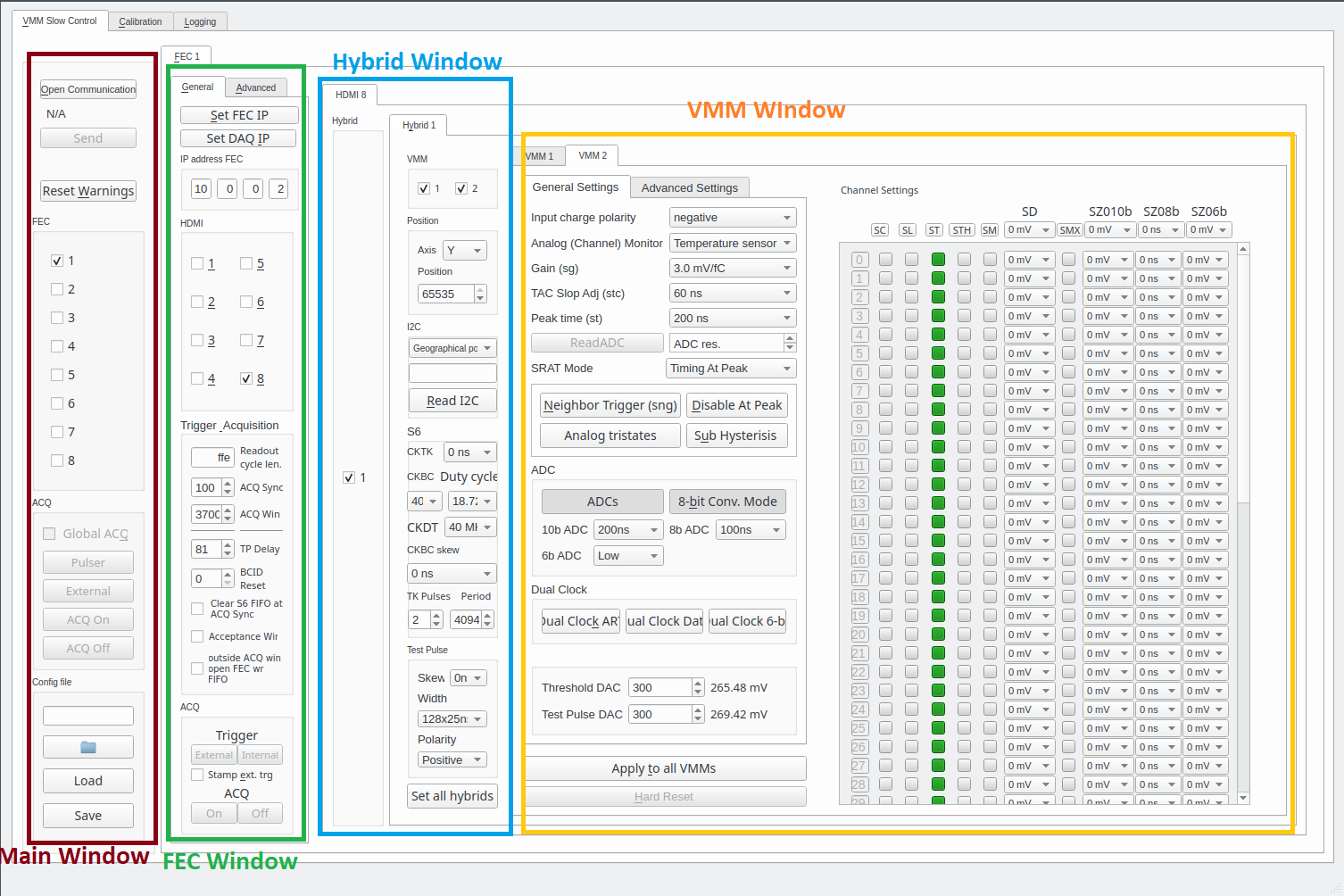}
  \caption{Main window of the VMM Slow Control Interface.}
  \label{fig:vscimain}
\end{figure}

The software used to control the SRS-VMM readout is the VMM Slow Control Interface (VSCI), based on \texttt{C++/Qt} \cite{guth2017software}. It was adapted from the corresponding software used in the ATLAS NSW upgrade. It has a main graphical window which categorizes the hardware and provides options to directly handle the FEC and the hybrid with the VMMs. As can be seen in \cref{fig:vscimain}, one can choose multiple FECs, HDMIS, Hybrids and VMMs to activate simultaneously. The VMM controls are also all reflected; the gain and peaktime can be adjusted, as well as the global and single-channel threshold, the pulser step and the polarity. A direct reading of the monitoring output is provided and it can be chosen to display the temperature, the analog baseline of a channel, the pulser step, the threshold DAC and a test pulse. One can choose to mask channels in order to study an individual one, use the internal pulser circuit to send a test pulse and adjust the internal test capacitance. Data acquisition can be manually switched off and on.

\begin{figure}[htbp]
  \centering
  \includegraphics[width=11cm]{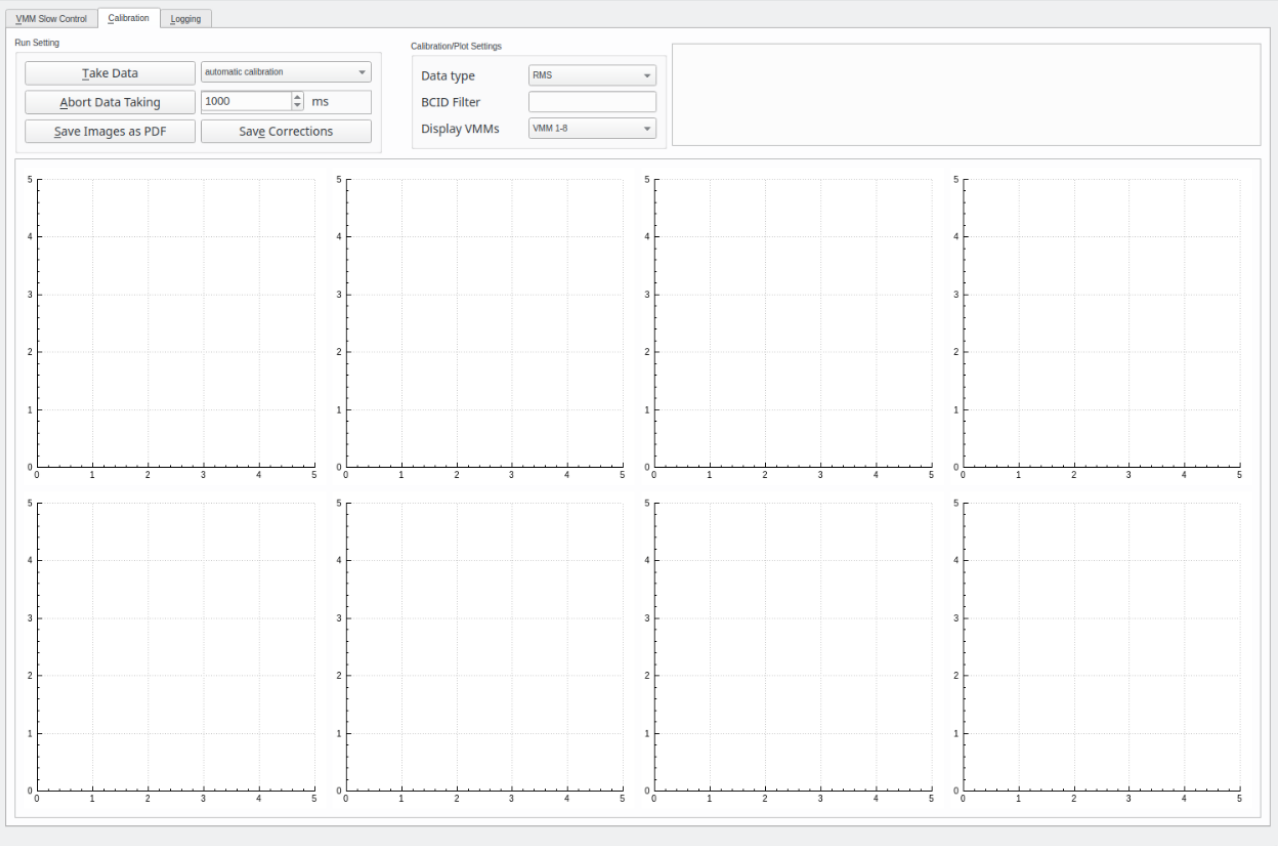}
  \caption{Calibration window of the VMM Slow Control Interface.}
  \label{fig:vscical}
\end{figure}

Aside from the main window, the slow control provides extensive calibration modules which can be edited and adjusted. \cref{fig:vscical} shows the calibration window. Eight graph canvases are provided for direct output of the calibration processes. For instance, one such calibration can refer to the baseline of all channels of a VMM chip, so the baseline of the VMM would be displayed on one of the graphs (Baseline (ADC Chn) vs. channel). The third visible window at the top of the control panel is the Logging window, where tracking and error-displaying of the commands used is provided.

%------------------------------------------------------------------------------
\subsection{Noise Sources in Electronic Circuits}%
\label{sec:theory:noise}
%------------------------------------------------------------------------------

Since the main focus of this work is the measurement of the noise in a readout ASIC, it is essential to know some of the major noise contributions. This noise is intrinsic to the devices that make up the circuit and it is due to the random motion of charge carriers\footnote{The information given in this section is largely based on studies of \cite{spieler2002pulse,,spieler2003analog,,flicker,,fish2017electronic}}. There are two basic noise mechanisms: 

\begin{itemize}
    \item thermal noise, which is induced by the random motion of conduction electrons, or velocity fluctuations
    \item shot noise, which is due to the independent passage of electrons over a potential barrier, eg. a semiconductor diode. Because of this, it allows leakage currents through circuit elements.
    \item excess or "$1/f$" noise
\end{itemize}

Both thermal and shot noise are "white noise" sources, i.e. they have a spectral density, or power per unit bandwidth, which is constant. Thermal noise depends on the bandwidth of a system, hence since this is usually determined by the shaping time of a shaper filter, small shaping times decrease this type of noise. Shot noise is the passage of current through
circuit elements.

On the other hand, the $1/f$ is excess noise (other than thermal or shot noise) that is generated when a current passes through a resistor or semiconductor and it occurs in all electronic devices. It has a variety of causes which are not entirely understood, but for example, in semiconductors, which is the diodes used in the pre-amplifiers of electronic circuits, it is due to fluctuations in the number of charge carriers that arrive from a signal source (eg. detector). These fluctuations in turn may arise from the random release of electrons that were trapped in the gas medium of a detector due to impurities in the gas. Its power density decreases with increasing frequency from the signal and hence it follows a $1/f$ characteristic. Because it is most visible in low frequencies it is also often referred to as a low-frequency noise. 

Adding a capacitance to the input of the ASIC, such as one from the strips of a detector or from other external circuits, adds extra noise depending on the capacitance value. This dependence should be linear 

The overall noise (thermal, shot, excess) has a Gaussian distribution, since it is caused by a large number of random events and it appears as a baseline noise which characterizes electronics. Superimposing a constant amplitude signal on the baseline noise will result in a Gaussian amplitude distribution, whose width is equal to the noise level.

%==============================================================================
\chapter{APV25 Noise Performance}
\label{sec:apv25}
%==============================================================================

The APV25 noise performance tests include pedestal measurements, setting up an external cosmics trigger, determining the trigger latency value and finally, the signal-to-noise ratio evaluation. 

%------------------------------------------------------------------------------
\section{Pedestal measurements}%
\label{sec:apv25:pedestals}
%------------------------------------------------------------------------------

\begin{figure}[htbp]
  \centering
  \begin{tabular}{cc}
  \includegraphics[width=6.8cm]{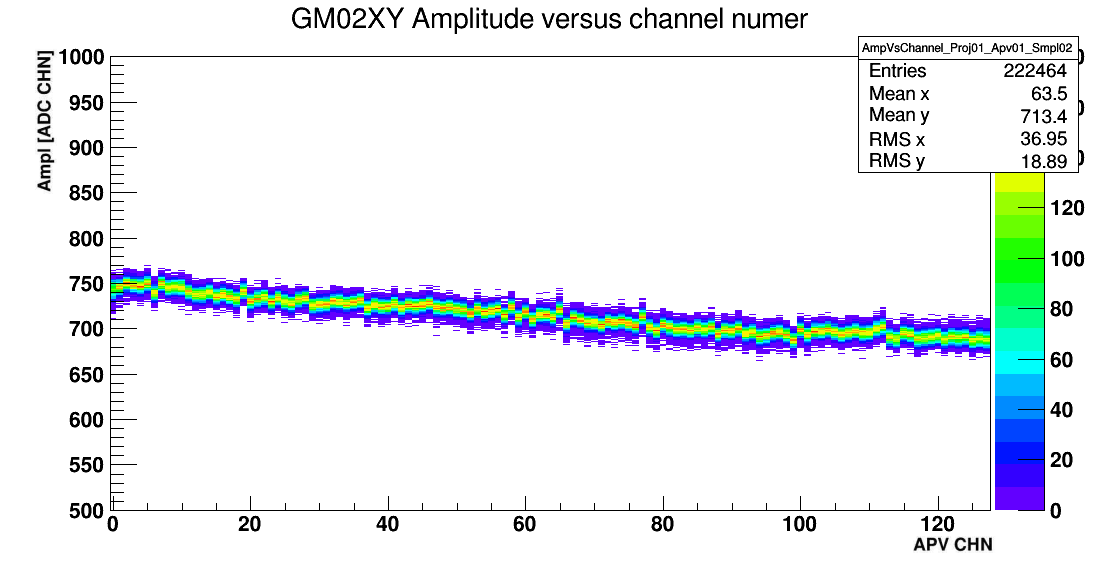} &
  \includegraphics[width=6.8cm]{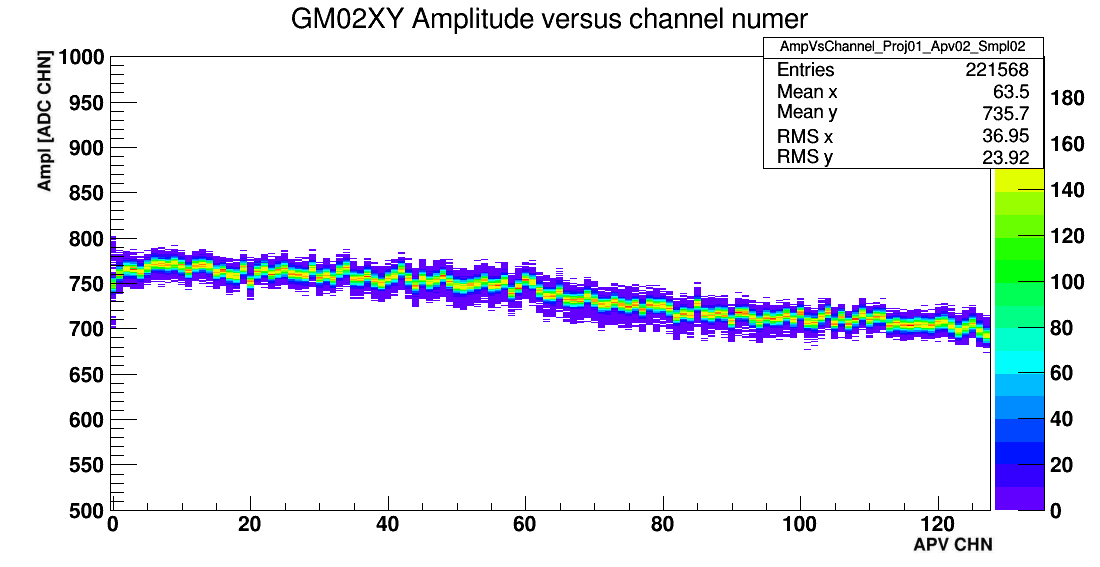} \\
  \includegraphics[width=6.8cm]{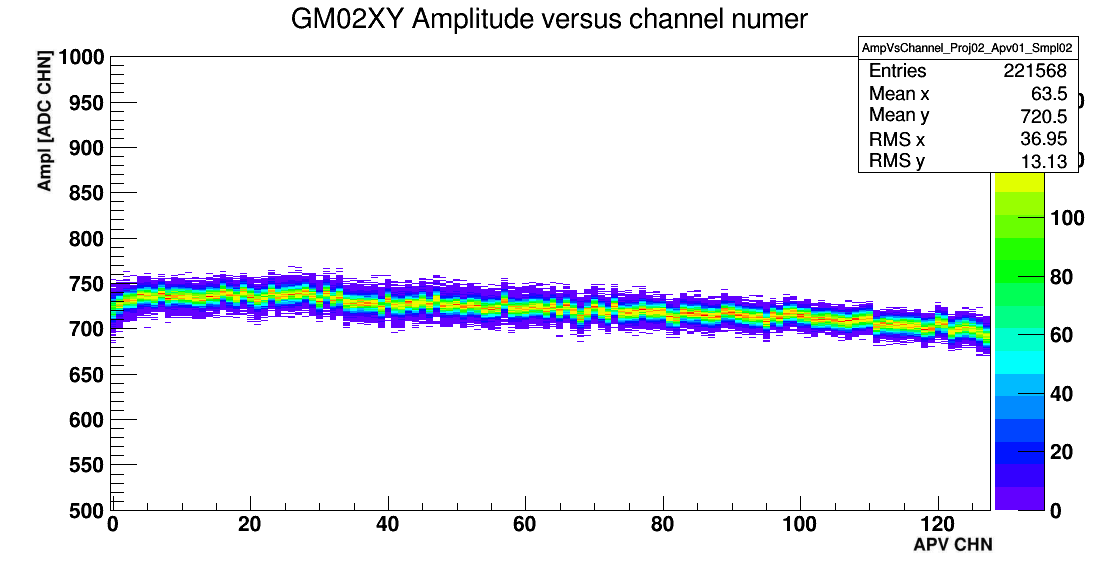} &
  \includegraphics[width=6.8cm]{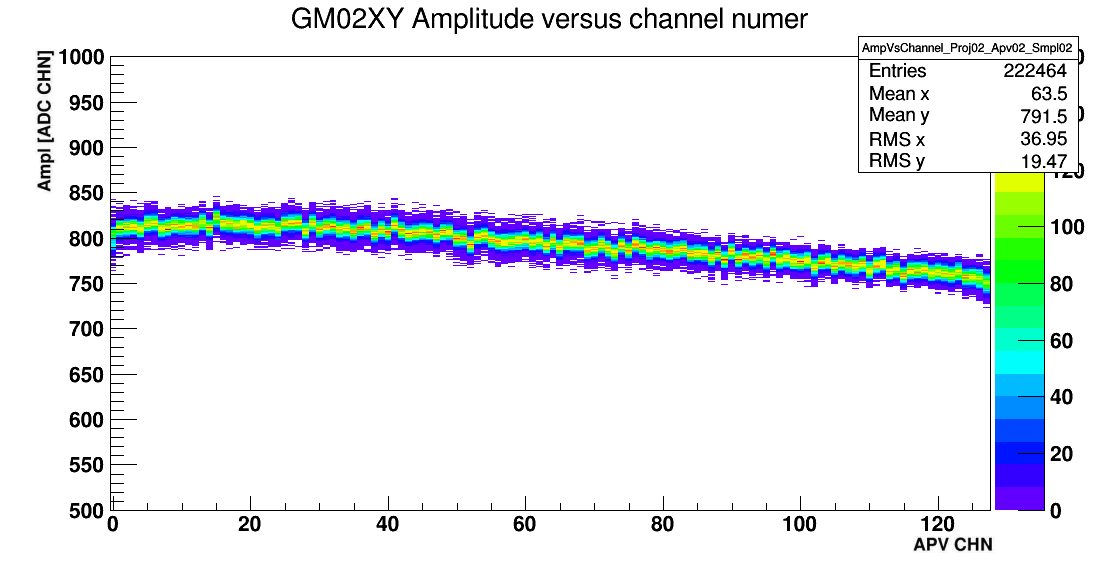} 
  \end{tabular}
  \caption{Pedestals measurement of both chips at the X strips (top) and the Y strips (bottom).}
  \label{fig:apvbase}
\end{figure}

\begin{figure}[htbp]
  \centering
  \begin{tabular}{cc}
  \includegraphics[width=7cm]{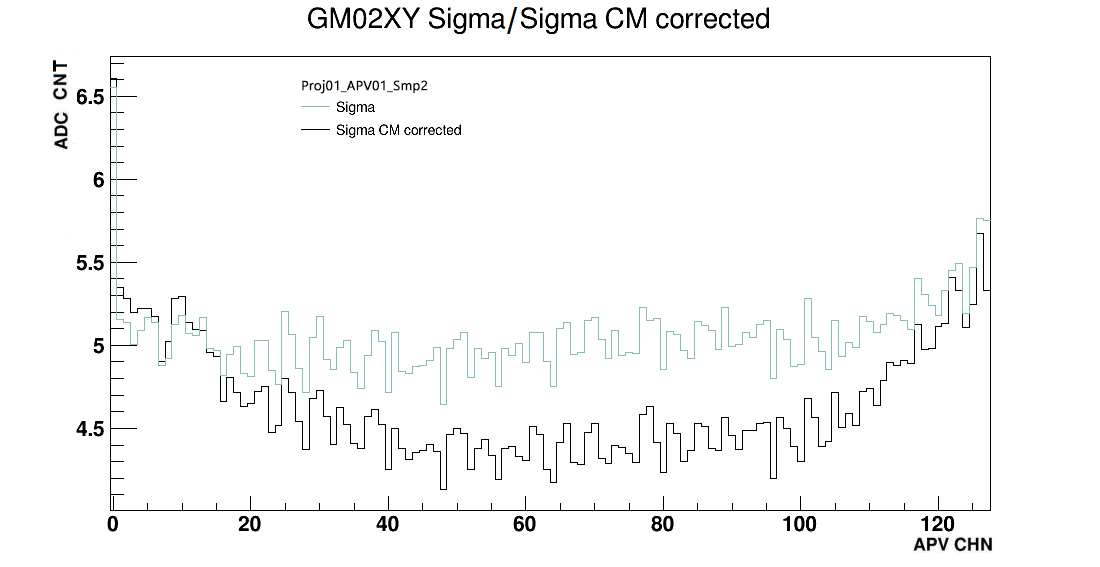} &
  \includegraphics[width=7cm]{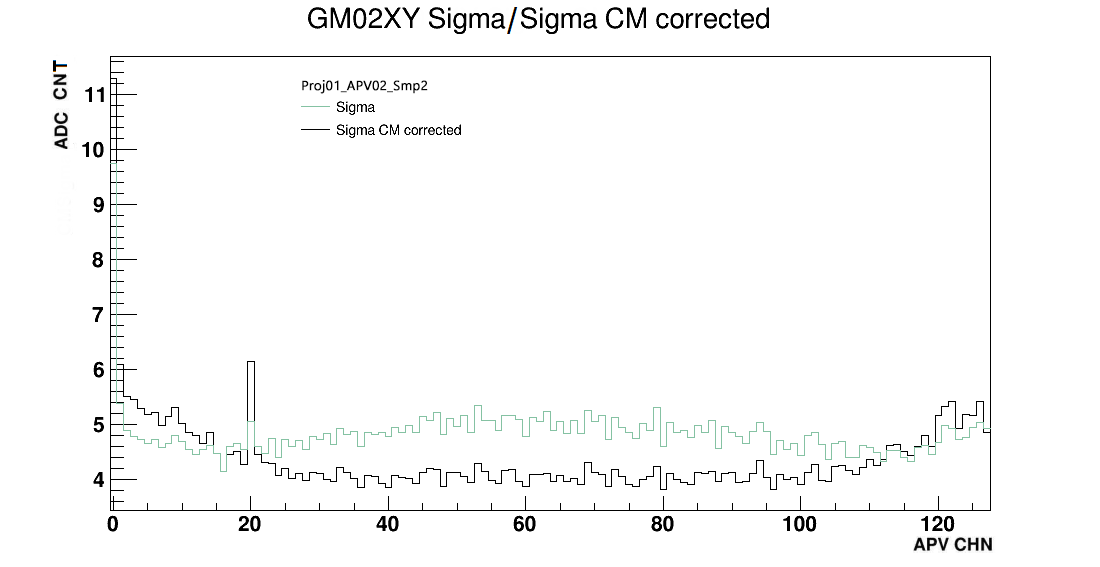} \\
  \includegraphics[width=7cm]{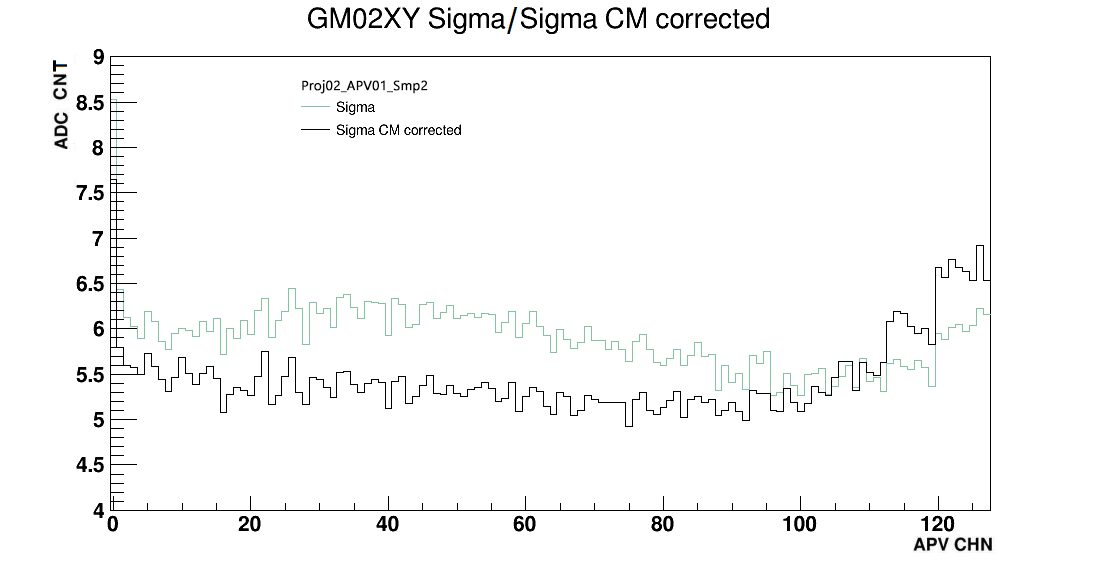} &
  \includegraphics[width=7cm]{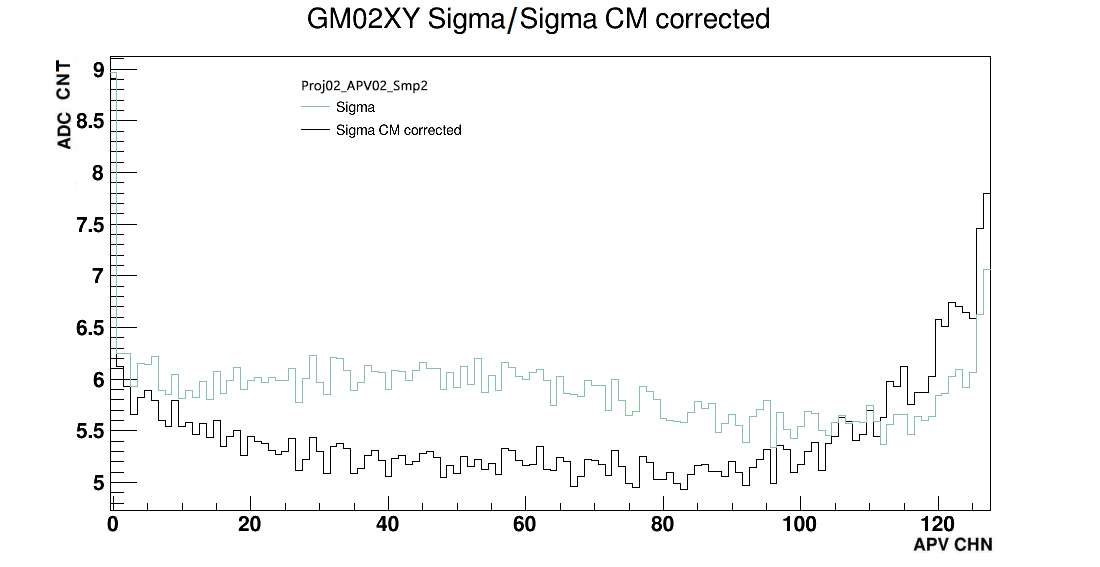} 
  \end{tabular}
  \caption{Sigma (green) and CM Sigma (black) of the two chips at the X strips (top) and the ones at the Y strips (bottom).}
  \label{fig:apvsigmacm}
\end{figure}

In order to measure the pedestals, or the baseline, of the APV25 chip, the ADC is set to operate in the Latch-All mode, while a simple periodic trigger is used. As explained in \cref{sec:theory:apv25}, the baseline RMS or sigma can be split down to an intrinsic noise, which is different for each channel, and a common-mode-shift noise, which affects several channels at once. The ADC can correct for this common noise, providing the common-mode-corrected sigma (CM sigma). First measurements of the baseline, sigma and CM sigma are shown in \cref{fig:apvbase} and \cref{fig:apvsigmacm}. The sigma and CM sigma are both displayed in \cref{fig:apvsigmacm} in green and black respectively, in order to better realize the common-mode-shift correction applied by the ADC.

The mean of the baseline and its RMS are used to perform zero suppression during a cosmics measurement, so by looking at the plots of \cref{fig:apvbase} and according to \cref{eq:apvthres}, at the maximum signals of amplitude less than $\sim$ \SI{850}{ADC CHN} are filtered out from the final data (this number is roughly calculated from the baseline mean and RMS on the second plot at the bottom of the figure, so roughly 790$+3\cdot20$.)

Regarding the sigma plots, it can be observed the common-mode correction omits the outer channels and in some occasions the CM-corrected sigma is higher than the sigma. This could be due to a variety of reasons such as possible edge effects of the chip or of the readout coupling to the detector edges.  

%------------------------------------------------------------------------------
\section{Setting up a Cosmics Trigger}%
\label{sec:apv25:cosmicstrigger}
%------------------------------------------------------------------------------

\begin{figure}[htbp]
  \centering
  \includegraphics[width=10cm]{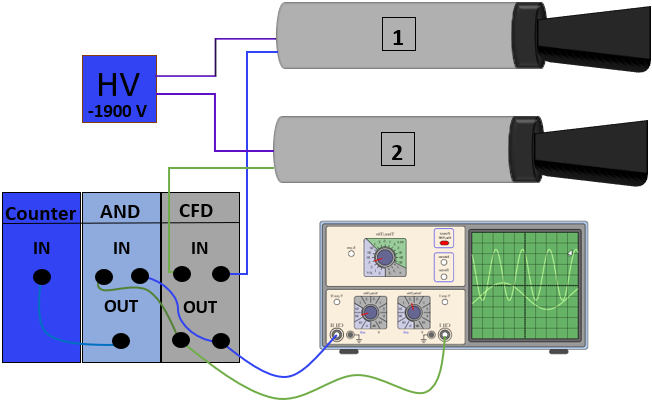}
  \caption{Schematic of the setup used to establish an external cosmics trigger for the APV25.}
  \label{fig:sci-signal-setup}
\end{figure}

Due to an absence of an internal trigger on the APV25, it is necessary to set up one externally. This was done by using the setup shown in \cref{fig:sci-signal-setup}. The main element providing the trigger is the scintillators. Scintillators are made of luminescent materials that once traversed by ionizing radiation, they emit scintillation light. This light needs to be amplified in order to be detected and so scintillators are connected to photomultipliers (PMTs)--devices which consist of a photocathode, where the light hits and releases photo-electrons via the photoelectric effect; these electrons are then multiplied on a cascade of dynodes set in an electric field and finally they are collected on a readout anode. For the rest of the work, the PMT-scintillator detector is simply referred to as a scintillator. For simplicity, the top scintillator will be referred to as Scintillator 1, while the bottom will be referred to as Scintillator 2.

\begin{figure}[htbp]
  \centering
  \begin{tabular}{cc}
  \includegraphics[width=6cm]{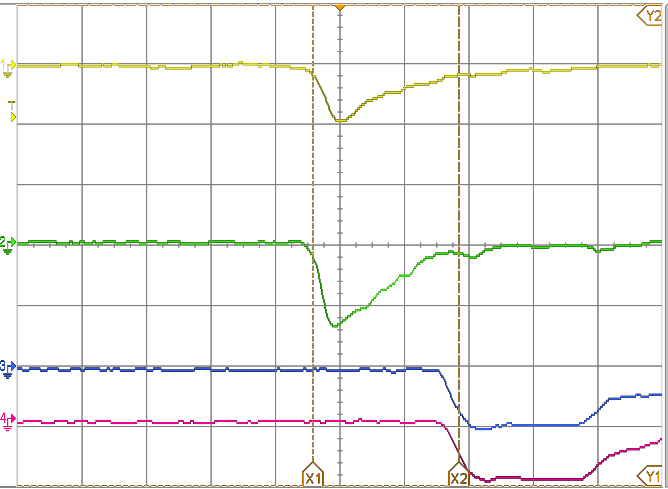} &
  \includegraphics[width=6cm]{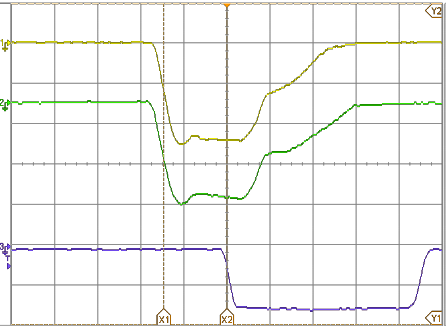} \\
  (a) & (b)
  \end{tabular}
  \caption{(a) Yellow, green: Scintillator outputs (\SI{100}{mV} scale); Blue, pink: CFD outputs (\SI{1}{V} scale). (b) Yellow, green: CFD outputs (\SI{200}{mV} scale); Purple: Coincidence signal (\SI{1}{V} scale).}
  \label{fig:cfdand}
\end{figure}

When the scintillators are set with their active (black) areas one on top of the other, cosmic particles will pass through and produce a signal in both scintillators. Each signal is then sent into the input of a Constant Fraction Discriminator (CFD). The latter is used to provide threshold triggering independent of the signal's peak height, while it also allows energy selection capability to avoid triggering on noise. The two CFD output signals are split and sent through a Coincidence port (AND gate) and onto an oscilloscope for observation. If both signals exceed the CFD voltage threshold ($V_{th}$), a Coincidence signal is created and sent to the Signal Counter. Each stage of the signal processing is shown on the scope in \cref{fig:cfdand}. 
%The coincidence signal is also fed to the oscilloscope, however this is not shown in \cref{fig:sci-signal-setup} to avoid cluttering.

\begin{figure}[htbp]
  \centering
  \begin{tabular}{cc}
  \includegraphics[width=6cm]{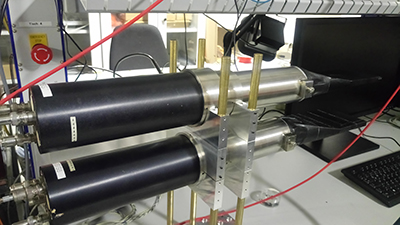} &
  \includegraphics[width=6cm]{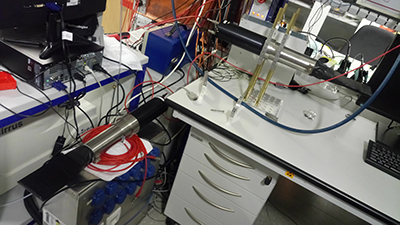} \\
  (a) & (b)
  \end{tabular}
  \caption{Scintillator positions for (a) signal measuring, (b) noise measuring.}
  \label{fig:scintipos}
\end{figure}

\begin{figure}[htbp]
  \centering
  \begin{tabular}{c}
  \includegraphics[width=9cm]{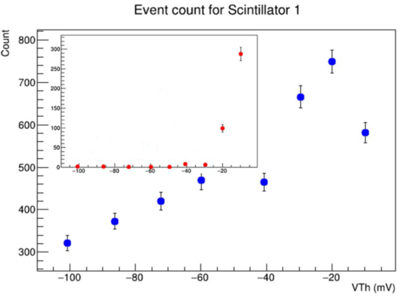} \\
  \includegraphics[width=9cm]{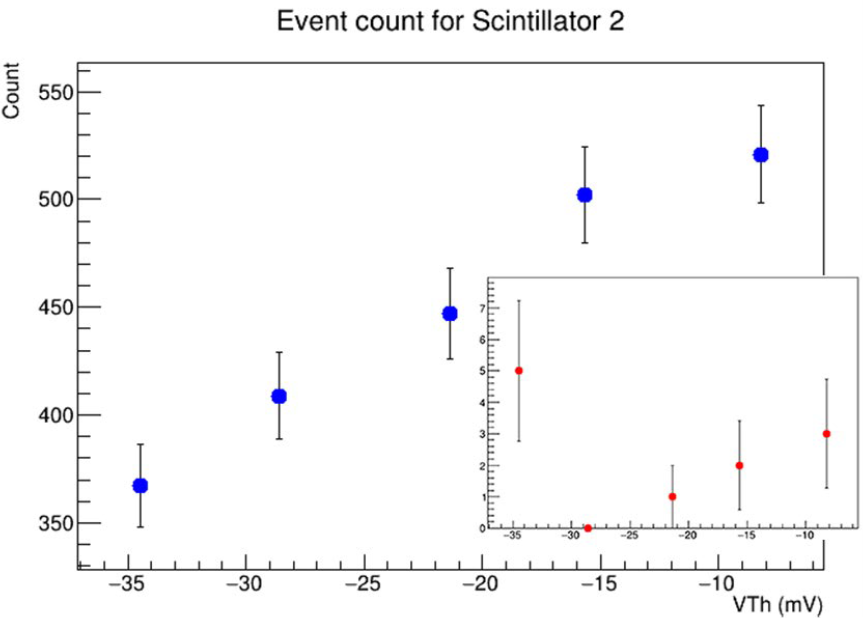} 
  \end{tabular}
  \caption{Signal-noise measurements for (top) Scintillator 1, (left) Scintillator 2. The noise is shown in the inserts for better optical effect. Poissonian errors were assumed for all points.}
  \label{fig:signal-noise}
\end{figure}

One of the tasks involved in setting the appropriate trigger is determining a functional CFD threshold voltage $V_{th}$ for both scintillators. To do that, one needs to compare the event count as opposed to the noise count for different $V_{th}$ values of the CFD. To determine the signal count, the scintillators are placed one above the other, as in \cref{fig:scintipos}a. To determine the noise count, the scintillators are placed with their active areas apart, as in \cref{fig:scintipos}b. During measurements for Scintillator 1, the Scintillator 2 CFD threshold was set at \SI{-24.65}{mV}, based on an initial optical evaluation of the signal-noise count. After determining the optimal $V_{th}$ for Scintillator 2 at \SI{-63.08}{mV}, this was then used during the measurements for Scintillator 1. For each $V_{th}$ setting, a 10-minute count was taken both for events and for noise. The measurement points for both scintillators are shown in \cref{tab:scinti}, while a graphical display is provided in \cref{fig:signal-noise}. The error bars reflect the Poissonian errors. In the case of Scintillator 2, the noise count at thresholds below \SI{-8.21}{mV} was incomprehensible, yielding low and high values between measurements and hence it was considered to be an unstable operative point.  The initial optical evaluation of the Scintillator 2 $V_{th}$ remained as optimal. So with these settings, the overall particle/trigger rate is about $\sim$ 1 hit per second or $\sim$ 45$\pm$7 hits/min. 

\begin{table}[htbp]
\centering
\begin{tabular}{ c c c }
\toprule
$V_{th}$ (mV) & Count & Noise \\ 
\midrule
-9.77 & 582 & 288 \\
-19.9 & 749 & 99 \\
-29.55 & 666 & 6 \\
-40.8 & 465 & 8 \\
-49.48 & 520 & 1 \\
-59.84 & 469 & 1 \\
-72.17 & 420 & 1 \\
-86.27 & 373 & 2 \\
-100.81 & 321 & 2 \\
\bottomrule
\end{tabular}
\quad
\begin{tabular}{ c c c }
\toprule
$V_{th}$ (mV) & Count & Noise \\ 
\midrule
-8.21 & 521 & 3 \\
-15.67 & 502 & 2 \\
-21.35 & 447 & 1 \\
-28.6 & 409 & 0 \\
-34.48 & 367 & 5 \\
\bottomrule
\end{tabular}
\caption{Event-Noise measurements for Scintillators 1 (left) and 2 (right)}%
\label{tab:scinti}
\end{table}

\begin{figure}[htbp]
  \centering
  \includegraphics[width=5.5cm]{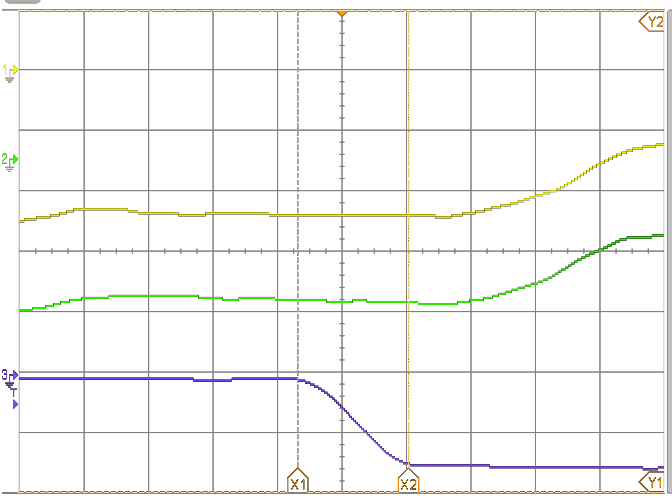}
  \caption{Trigger uncertainty.}
  \label{fig:trigdelay}
\end{figure}

An important parameter to consider is the time between the trigger and the actual occurrence of an event, or in other words, the trigger uncertainty. As shown in \cref{fig:trigdelay}, the falling edge of the coincidence signal (purple line) has a time delay. The larger this fall time is, the larger is the trigger uncertainty. This is due to a variety of reasons, including the time the PMT takes to multiply the charge, the length of the cables used, as well as the efficiency of the CFD and coincidence unit. Part of the trigger delay can be attributed to the peak distance between the input scintillator signals. A closer look at \cref{fig:cfdand}a can reveal that the two peaks are not perfectly aligned. 

\begin{figure}[htbp]
  \centering
  \begin{tabular}{cc}
  \includegraphics[width=5.5cm]{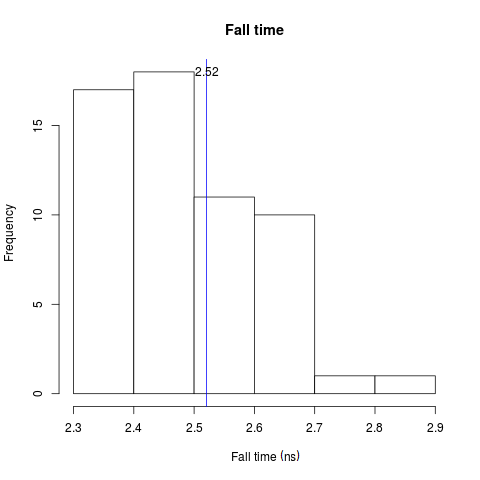} &
  \includegraphics[width=5.5cm]{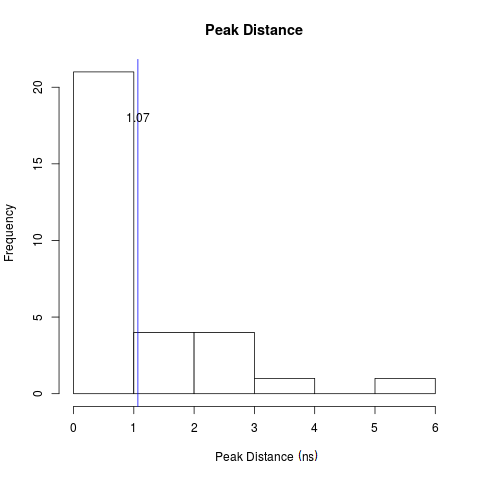} \\
  (a) & (b)
  \end{tabular}
  \caption{Fall time of the coincidence/trigger signal (a) and peak distance between the scintillator input signals (b).}
  \label{fig:falltime}
\end{figure}

In order to accurately measure the fall time and peak distance, about 32 signal frames of the oscilloscope were taken into account. The measurement points were plotted using the R programming environment in order to determine average values. Results are displayed in \cref{fig:falltime}. A more detailed review of all the measurement points can be found in \cref{sec:app:apv25}.

The average fall time was calculated at \SI{2.52}{ns}. The peak distance shows that on average, 42\% of the trigger uncertainty is due to the scintillators and the cables connecting them to the CFD input. A fall time of $\sim$\SI{2.5}{ns} seems to be a negligible uncertainty, but one needs to also check with the trigger latency of the APV25 chip. 

\begin{figure}[htbp]
  \centering
  \includegraphics[width=8cm]{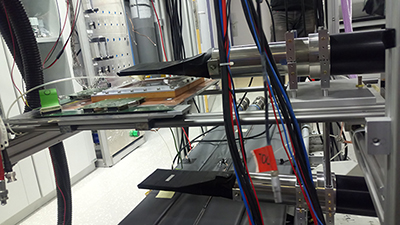}
  \caption{Final setup with the triple GEM detector and the scintillators.}
  \label{fig:gemtrig}
\end{figure}

Finally, the scintillators were mounted on the crane unit holding the GEM detector, while the coincidence signal was fed into the TCS receiver of the GeSiCA, as shown in \cref{sec:theory:apv25}, \cref{fig:readoutschematic}. The setup with the detector and the scintillators is shown in \cref{fig:gemtrig}. 

%------------------------------------------------------------------------------
\section{Latency}%
\label{sec:apv25:latency}
%------------------------------------------------------------------------------

As mentioned in \cref{sec:theory:apv25}, the trigger latency is the time within which a trigger decision is made and distributed to the digital elements. Because the APV25 operates in multi-mode (three consecutive samples of the signal, with the last one on the peak), a certain level of accuracy is required when placing the writer pointer. Since the latency is set in units of \SI{25}{ns}, a 10\% inaccuracy is expected due to the trigger uncertainty found in the previous section.

\begin{figure}[htbp]
  \centering
  \includegraphics[width=5.5cm]{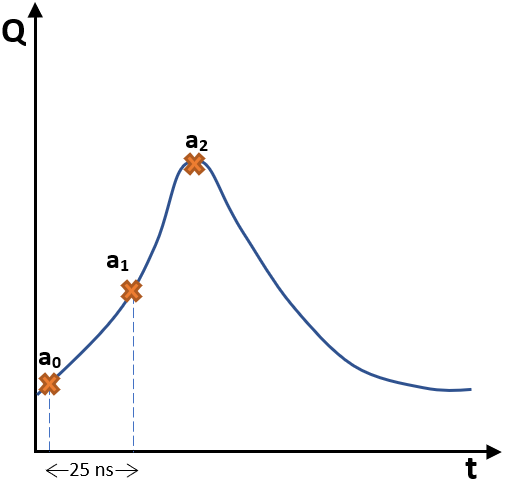}
  \caption{Signal sampling.}
  \label{fig:samples}
\end{figure}

\begin{figure}[htbp]
  \centering
  \begin{tabular}{cc}
  \includegraphics[width=6.8cm]{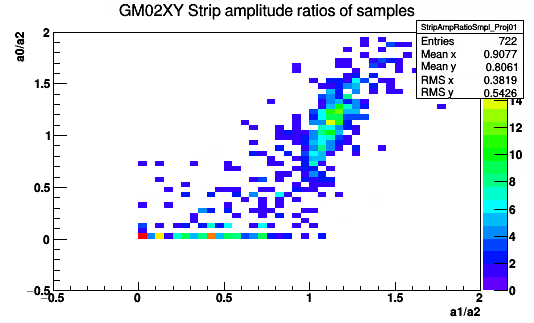} &
  \includegraphics[width=6.8cm]{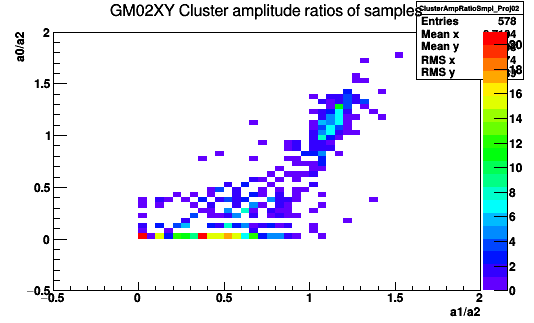} \\
  \includegraphics[width=6.8cm]{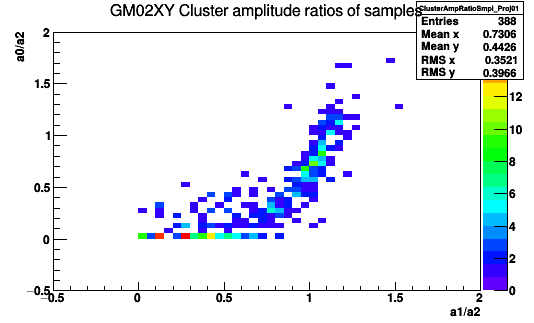} &
  \includegraphics[width=6.8cm]{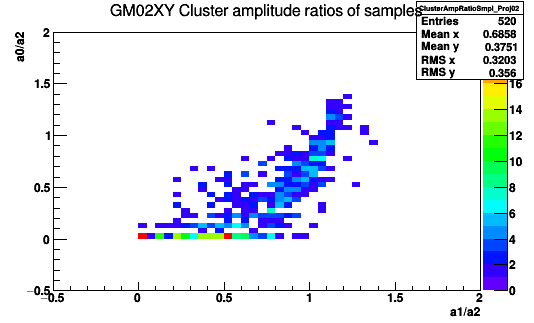} \\
  \includegraphics[width=6.8cm]{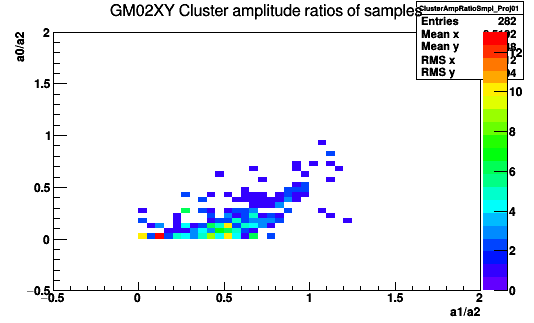} &
  \includegraphics[width=6.8cm]{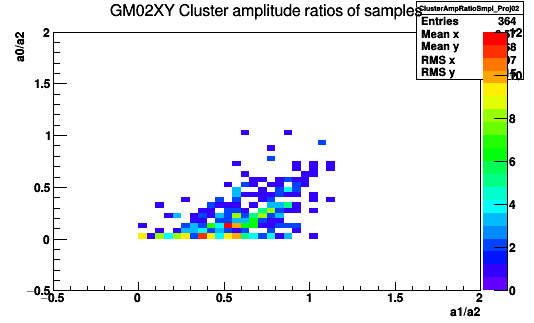} 
  \end{tabular}
  \caption{X coordinate (left) and Y coordinate (right) of latency 18 (top), latency 19 (middle) and  latency 20 (bottom).}
  \label{fig:latency}
\end{figure}

In order to check the accuracy of the writer pointer and subsequently the correct value for the trigger latency, the ratios of the three signal samples are plotted \cite{hoesgen2017tracking}. The three samples $a_0$, $a_1$ and $a_2$, taken as shown in \cref{fig:samples}, are used to build the ratios: $r_{0,2}=a_0/a_2$ and $r_{1,2}=a_1/a_2$. The plot should resemble a so-called banana shape, while limits for the two ratios should be: $[0,0.8]$ for $r_{0,2}$ and $[0.4,1.2]$ for $r_{1,2}$. 

Several latency values were tested, from 16 to 36. Plots for latency values of 18, 19 and 20 are displayed in \cref{fig:latency}. While the plot for latency 18 also resembles a banana shape, most of the entries exceed the expected ratio limits, which means most of the signal pulses are sampled at the falling edge. On the other hand, latency 20 displays most of its entries near $(0,0)$, which is the case when the pulse is sampled far before the peak and hence the resulting signal amplitude might be lower than the noise threshold and zero suppression takes effect. It is thus evident that latency 19 is the optimal setting.

%------------------------------------------------------------------------------
\section{Signal to Noise Ratio}%
\label{sec:apv25:snr}
%------------------------------------------------------------------------------

\begin{figure}[htbp]
  \centering
  \begin{tabular}{cc}
  \includegraphics[width=4cm]{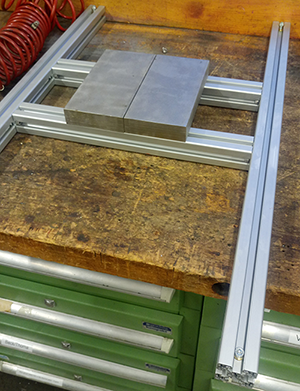} &
  \includegraphics[width=4cm]{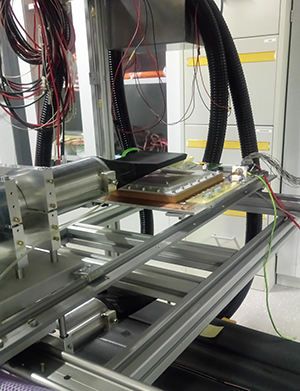} \\
  (a) & (b)
  \end{tabular}
  \caption{Shielding structure with steel bricks.}
  \label{fig:gemshield}
\end{figure}

The first cosmics measurement was conducted in January, 2020. The detector was powered at \SI{3900}{V} initially and additionally four more high voltage (HV) settings were used in following measurements, up to \SI{4100}{V} with a \SI{50}{V} step. Due to software restrictions, the measurement duration was cut at four hours for each measurement. Additionally, cosmics measurements were taken with the inclusion of steel bricks between the detector and the lower scintillator. The setup is shown in \cref{fig:gemshield}.

The steel bricks are used in order to shield the active detector medium from particles with momenta lower than that of the minimum ionizing particles. These particles deposit high amounts of energy in the medium, leading in large energy fluctuations (Landau tails) and hence, they affect the uniformity of our signal. The stopping power in the steel can prevent the passage of these particles through the second scintillator and hence they won't be triggered upon. In this sense, the steel bricks could also be placed above the detector. The reason this wasn't the choice is no other than simply that such a placement would have made it difficult to have physical and optical access to the detector board.  

In the following, a \textit{hit} is considered to be a single cosmic event, while a \textit{cluster} is defined as a group of strips impacted by the same hit. Therefore, the \textit{cluster size} (cS) is the number of strips correlated to one cosmic event. The \textit{cluster amplitude} refers to the signal strength of the event and it is calculated as such: all the hit strips on the detector plane are sorted by amplitudes first in a descending order, then beginning by the first hit strip, the algorithm searches if there is a hit strip next to it; if yes, it is assumed to be part of the cluster, otherwise a new cluster is created and the next hit strip is assumed to be part of the new one. Conventionally, a $3\sigma$ cut of the pedestal noise was taken on the single strip amplitude, while the cluster amplitude was filtered through a threshold of $5\sigma$ of the quadratic mean of the group of strip noises. Hence, a cluster of eg. two strips will go through a $3\sigma$ filter for each strip and then, the added amplitudes of the strips must exceed the $5\sigma$ barrier (in this sense, the amplitude of a cluster consisting of one strip will be filtered first through the $3\sigma$ threshold of its pedestal noise and, if it survives, through the $5\sigma$ one). The data shown are without steel protection, unless it is stated otherwise.  

\begin{figure}[htbp]
  \centering
  \begin{tabular}{cc}
  \includegraphics[width=7cm]{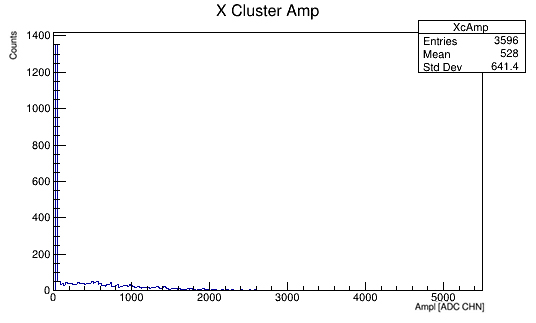} &
  \includegraphics[width=7cm]{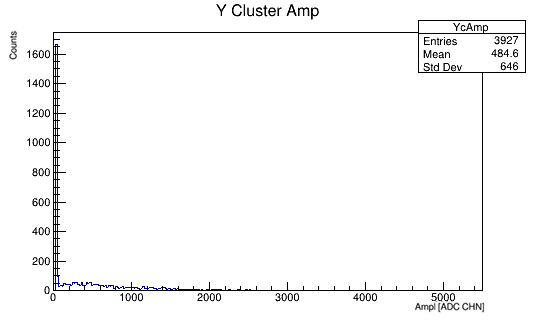} \\
  (a) & (b) 
  \end{tabular}
  \caption{Cluster amplitude (a) X and (b) Y.}
  \label{fig:apvclusterampcs1}
\end{figure}

\begin{figure}[htbp]
  \centering
  \begin{tabular}{cc}
  \includegraphics[width=7.5cm]{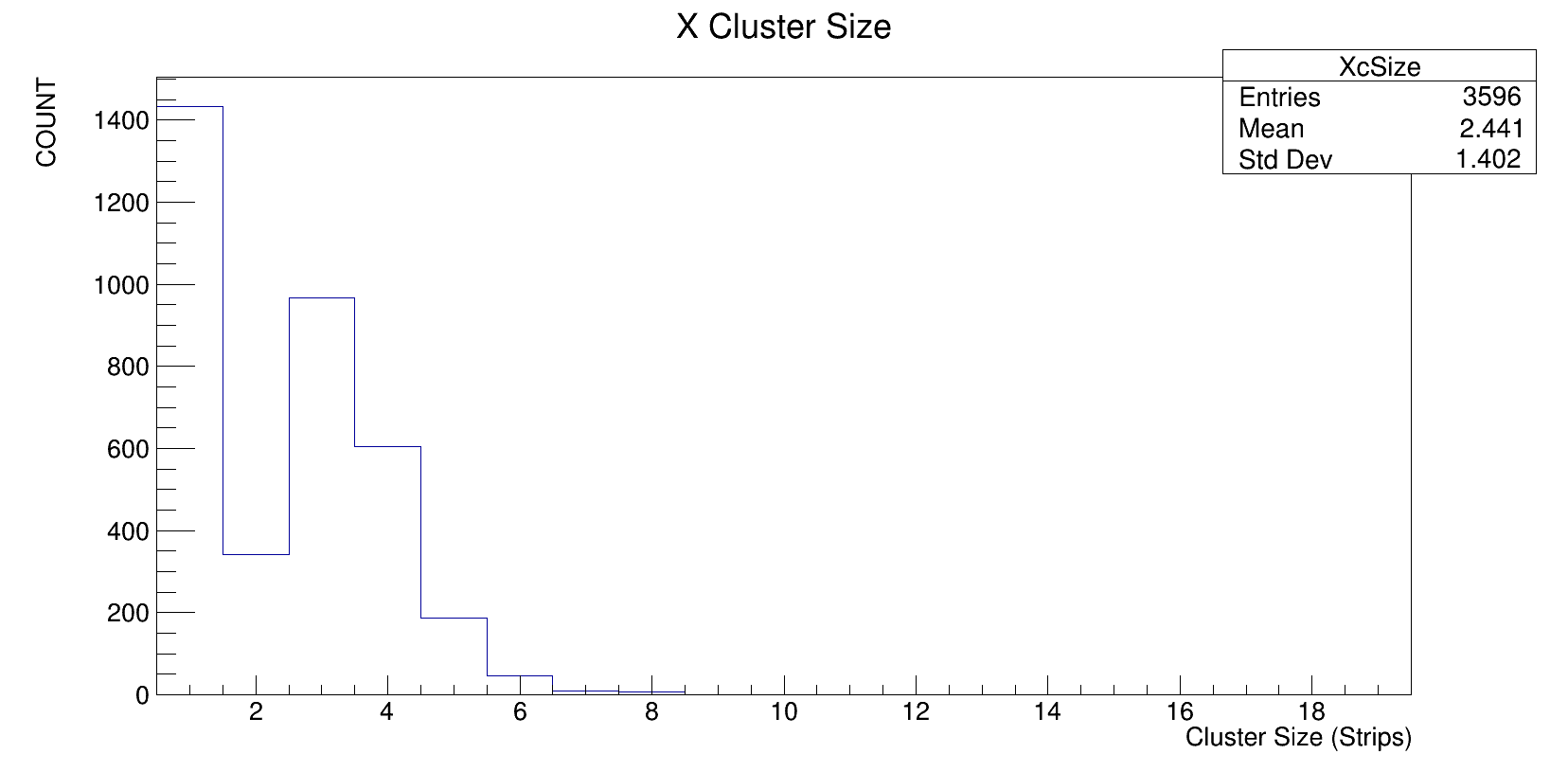} &
  \includegraphics[width=7.5cm]{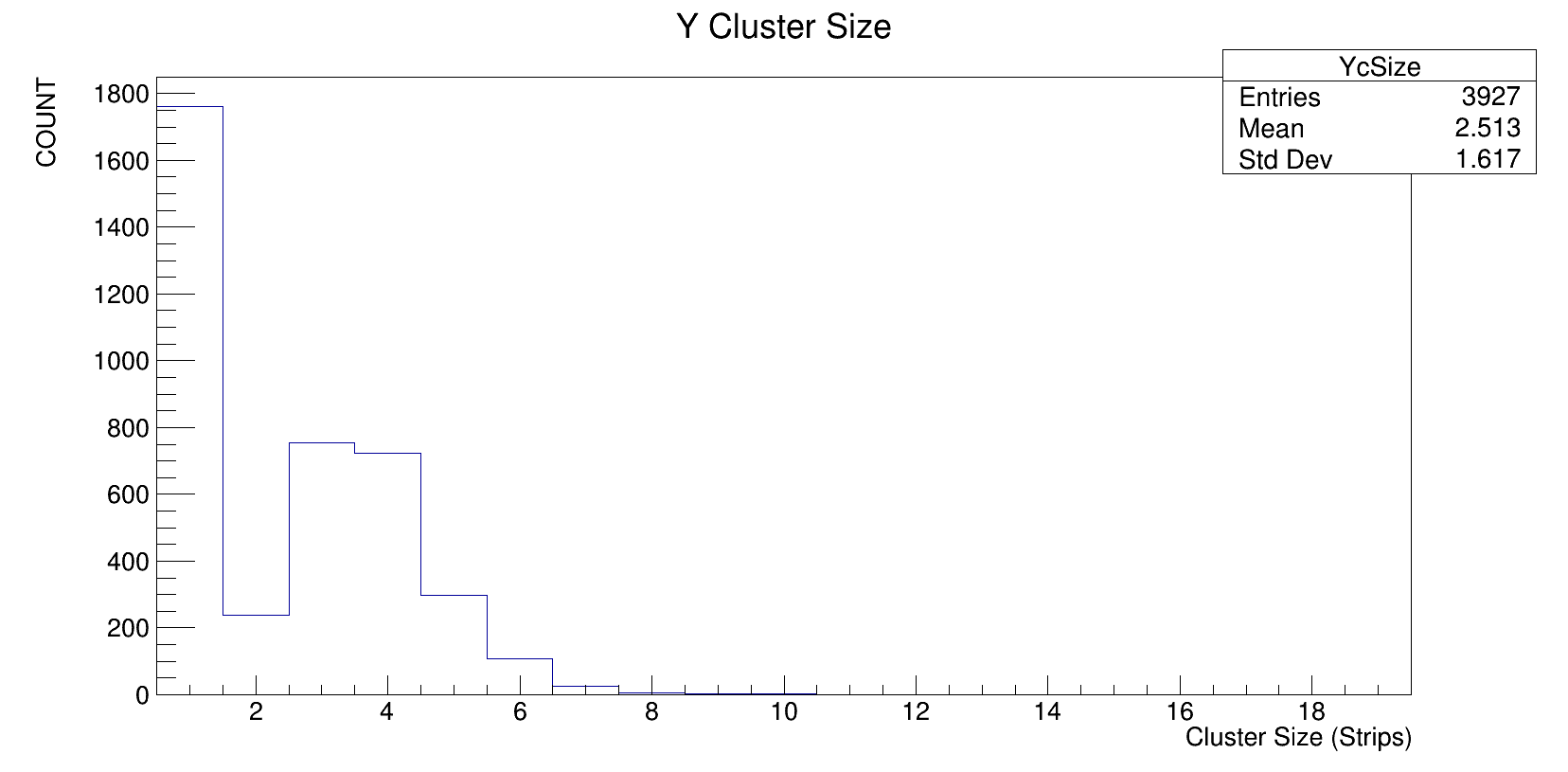} \\
  (a) & (b) 
  \end{tabular}
  \caption{Cluster size (a) X and (b) Y.}
  \label{fig:apvclustersizecs1}
\end{figure}

\begin{figure}[htbp]
  \centering
  \begin{tabular}{cc}
  \includegraphics[width=7cm]{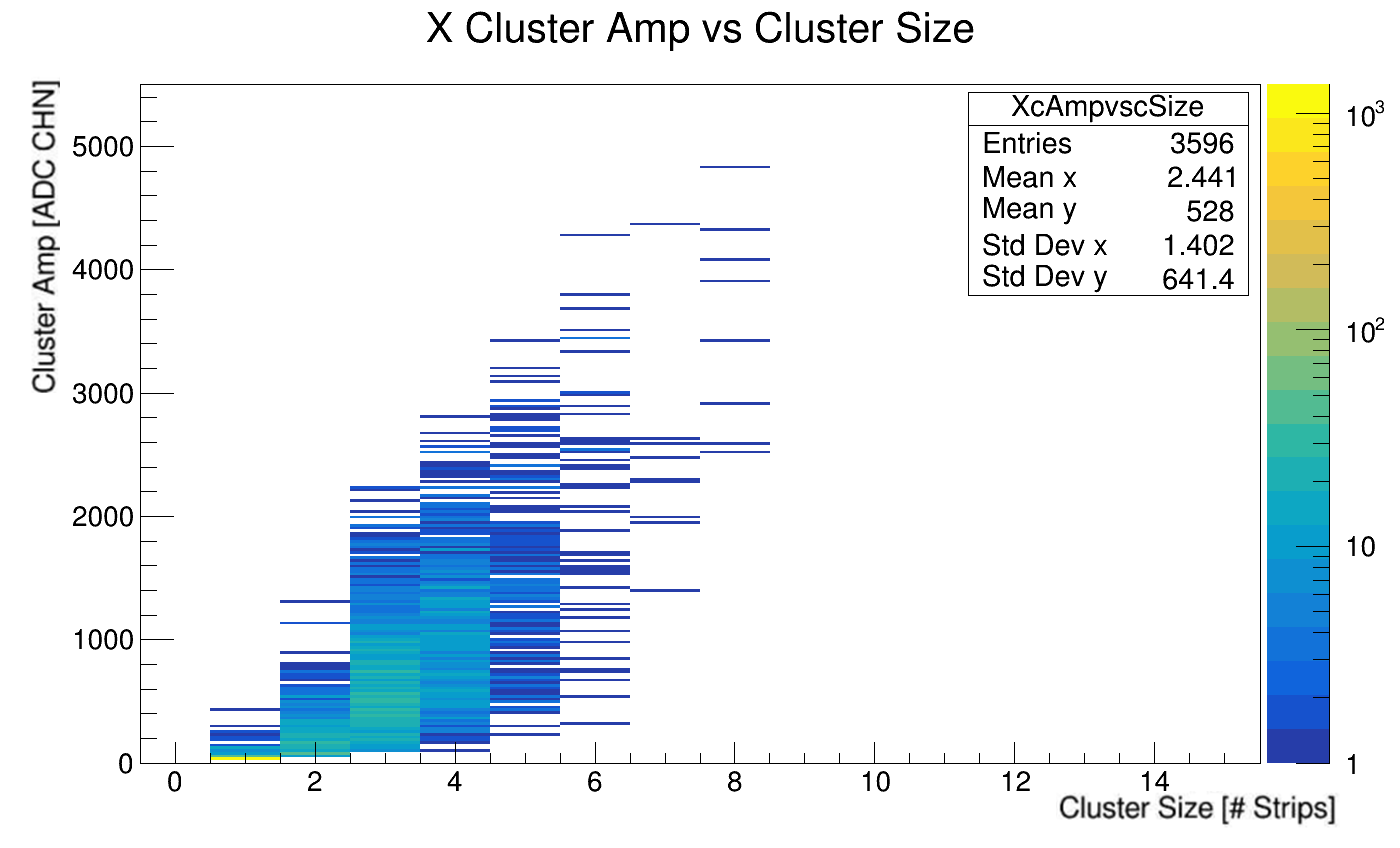} &
  \includegraphics[width=6cm]{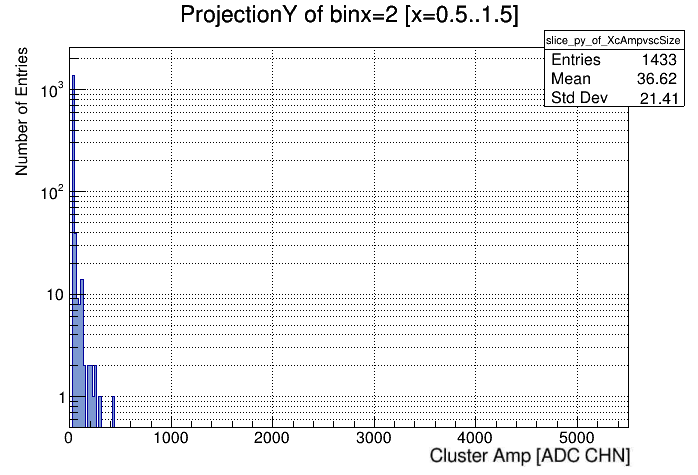} \\
  \includegraphics[width=7cm]{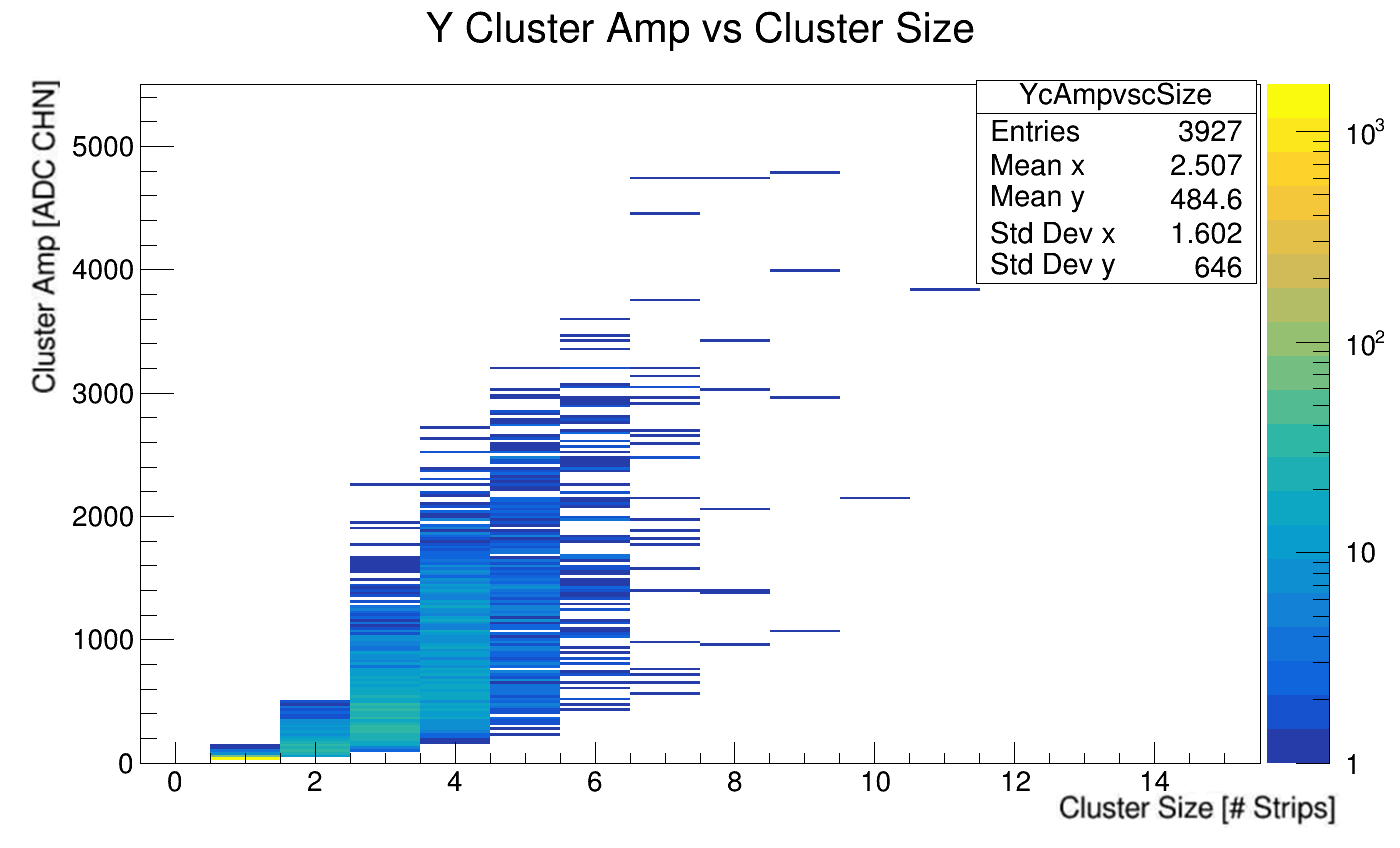} &
  \includegraphics[width=6cm]{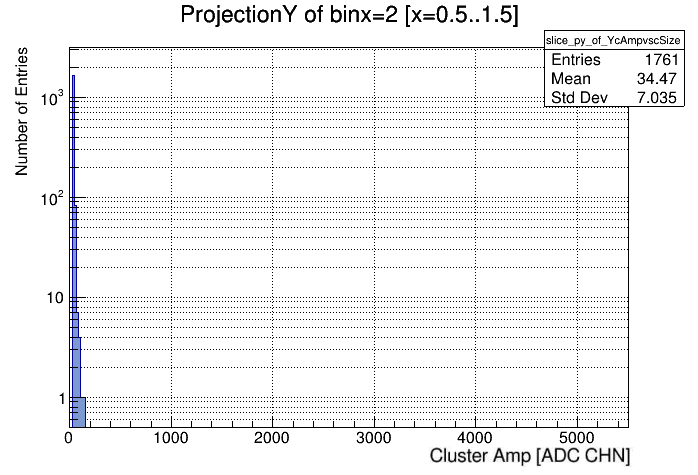} 
  \end{tabular}
  \caption{Cluster amplitude vs. cluster size for the X strips (top) and Y strips (bottom), plotted in a logarithmic scale for the z axis. The figures on the right provide a detailed view of the entries with cS$=$1 and their amplitudes.}
  \label{fig:apvclampvssize}
\end{figure}

In \cref{fig:apvclusterampcs1}, the cluster amplitude for the X and Y detector strips is displayed, while in \cref{fig:apvclustersizecs1}, we can see the corresponding cluster size. An accumulation of low amplitude events and events of cluster size of one can be immediately observed from a first look at the plots. Plotting the cluster amplitude against cluster size on a logarithmic scale for the Z axis can help further understand these low-amplitude events. In particular, one can see in \cref{fig:apvclampvssize} that the majority of these events are limited to a cluster size of one. Projections of the Y axis for the cS$=$1 bin are provided for each detector projection, showing that the low-amplitude/single-cluster events occupy almost a $\sim$ 40\% of all entries.

One argument that could be made is that since our clustering method creates a new cluster for each non-adjacent hit strip, many of the events that we have are in lateral angles, hence they might hit strips that are further apart and each of these strips would then count as a separate event of cluster size one. However, this is highly unlikely, because of the relatively small active area of the detector and because the placement of the scintillators as previously shown in \cref{fig:gemtrig} would only allow mostly perpendicular particle trajectories to be triggered upon. 

\begin{figure}[htbp]
  \centering
  \begin{tabular}{cc}
  \includegraphics[width=7cm]{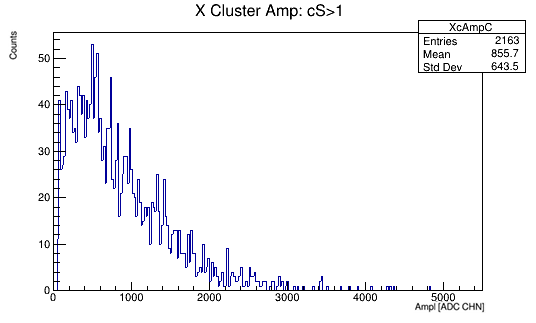} &
  \includegraphics[width=7cm]{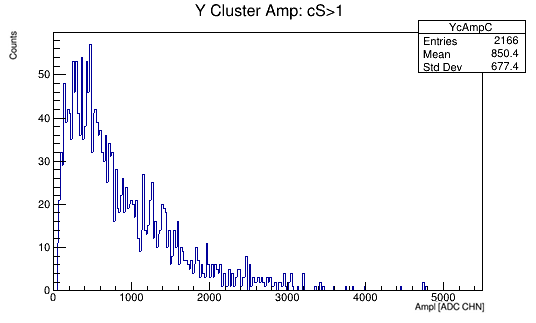} \\
  (a) & (b) 
  \end{tabular}
  \caption{Cluster amplitude (a) X and (b) Y. Cut of cS$>$1 applied.}
  \label{fig:apvclusterampcs2}
\end{figure}

\begin{figure}[htbp]
  \centering
  \begin{tabular}{cc}
  \includegraphics[width=7.5cm]{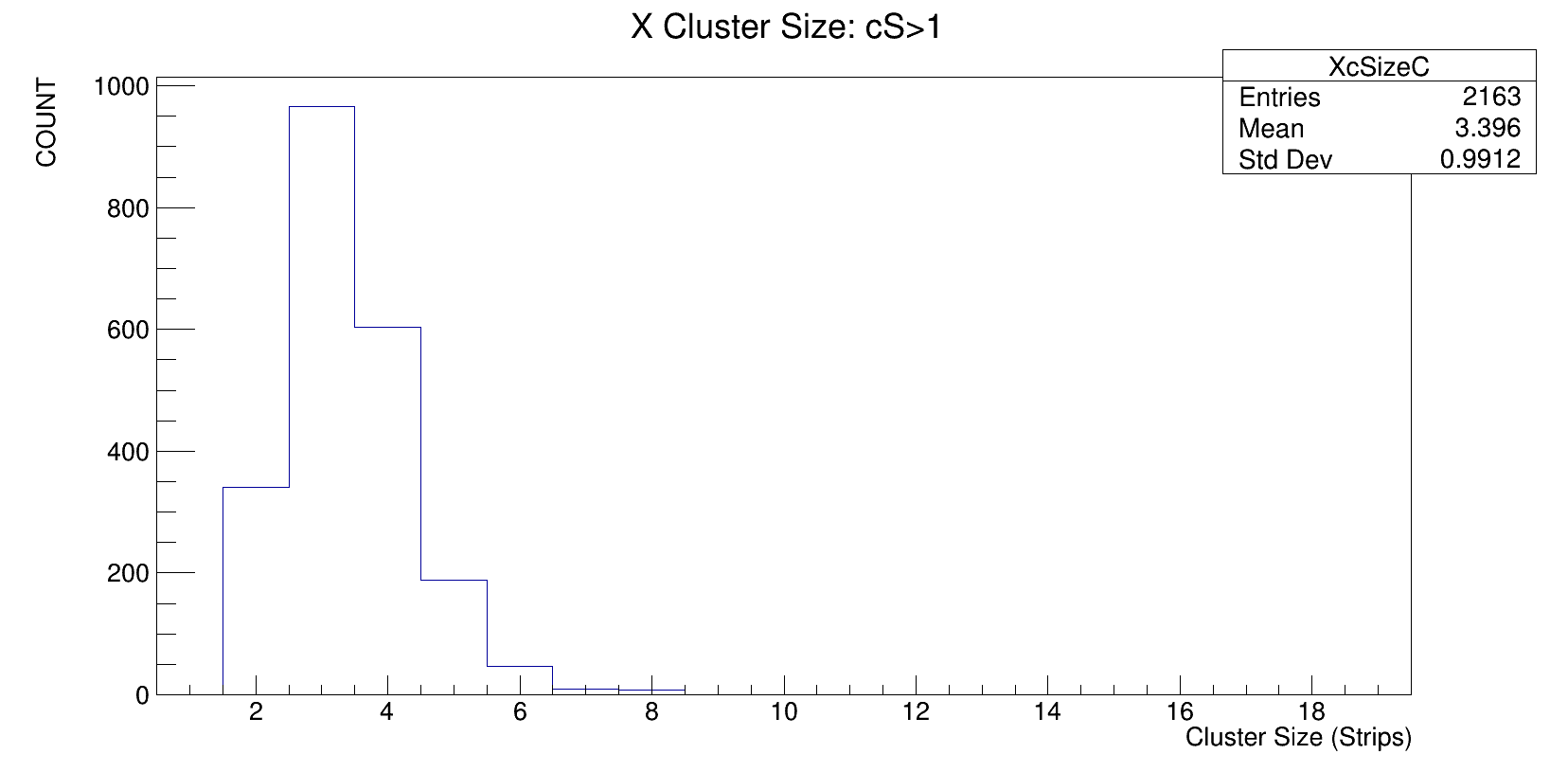} &
  \includegraphics[width=7.5cm]{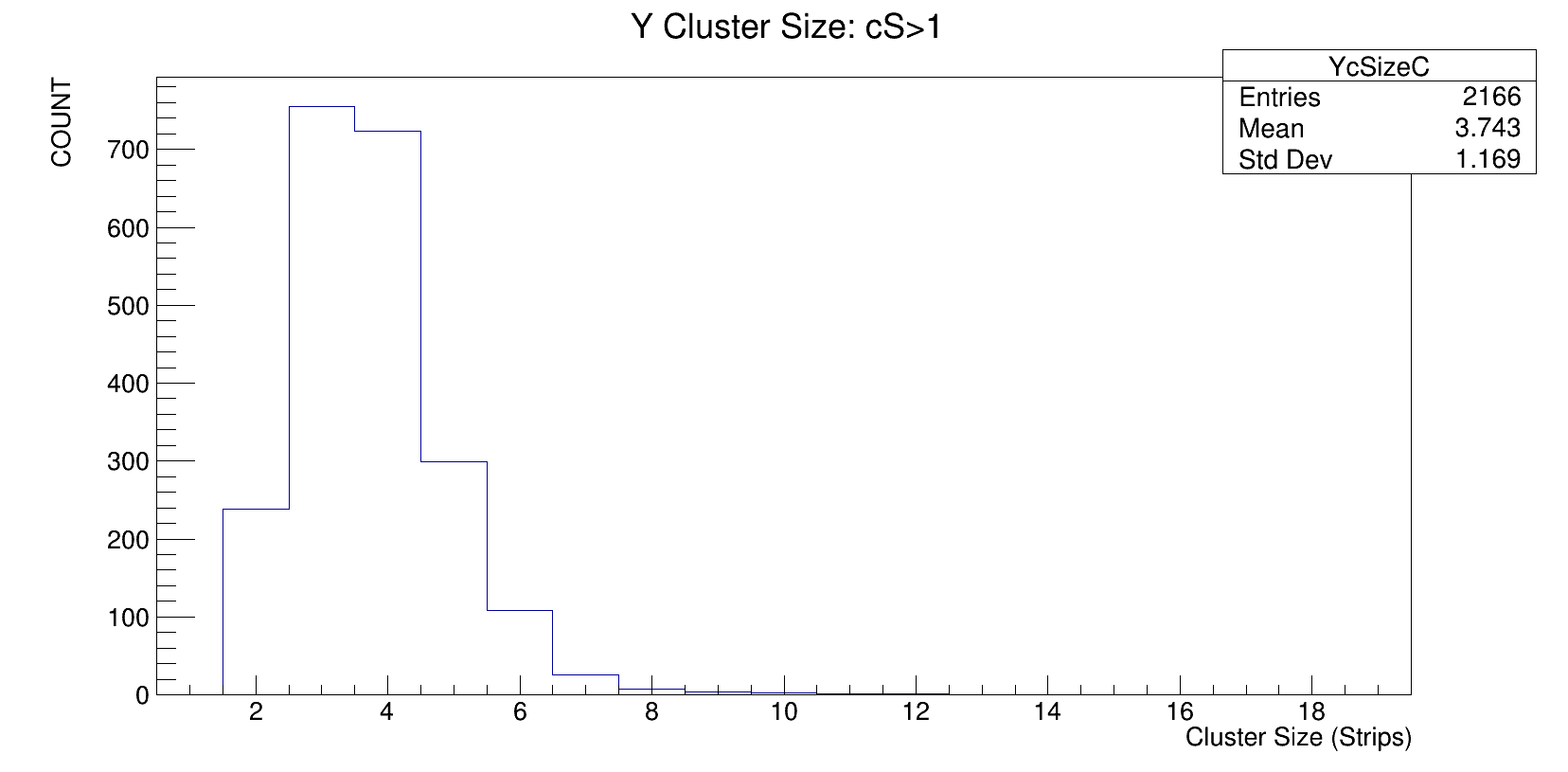} \\
  (a) & (b) 
  \end{tabular}
  \caption{Cluster size (a) X and (b) Y. Cut of cS$>$1 applied.}
  \label{fig:apvclustersizecs2}
\end{figure}

Because of their low amplitude, it can mostly be assumed that these events are simply random fluctuations on the baseline that managed to survive zero suppression. Furthermore, according to previous studies of the detector gain \cite{hoesgen2017tracking,,cozzi2019asic}, the probability of physics events affecting a single strip is fairly low. Therefore, it is safe to apply a cut of cS$>$1. The cluster amplitude for the two detector coordinates with the applied cuts is shown in \cref{fig:apvclusterampcs2}. Improvement is immediately evident, though the lack of events leads to a considerable reduction in statistics. In a similar manner, \cref{fig:apvclustersizecs2} shows the corresponding cluster size plots.

%\begin{figure}[htbp]
 % \centering
 % \begin{tabular}{cc}
 % \includegraphics[width=7cm]{figs/apv25/xcsize.png} &
 % \includegraphics[width=7cm]{figs/apv25/ycsize.png} \\
  %\includegraphics[width=7cm]{figs/apv25/xcsizec.png} &
  %\includegraphics[width=7cm]{figs/apv25/ycsizec.png} 
  %\end{tabular}
  %\caption{Cluster size. Left: X, Right: Y, Bottom: cS$>$1 cut.}
  %\label{fig:apvclustersize}
%\end{figure}

\begin{figure}[htbp]
  \centering
  \begin{tabular}{cc}
  \includegraphics[width=7cm]{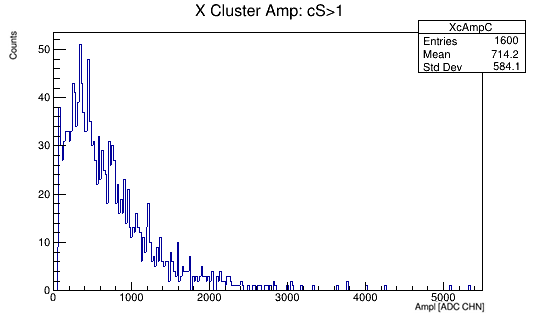} &
  \includegraphics[width=7cm]{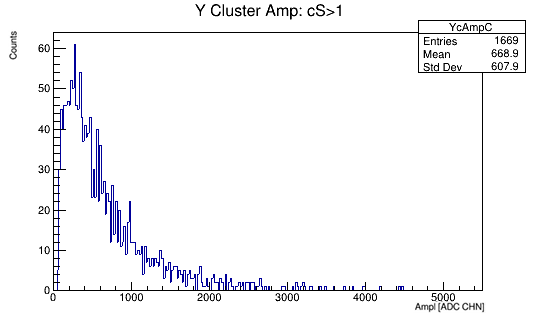} \\
  (a) & (b)
  \end{tabular}
  \caption{Cluster amplitude with steel shielding for the two detector coordinates (cS$>$1 cut also applied).}
  \label{fig:apvclusterampwithsteel}
\end{figure}

\begin{figure}
    \centering
    \begin{tabular}{cc}
        \includegraphics[width=7.2cm]{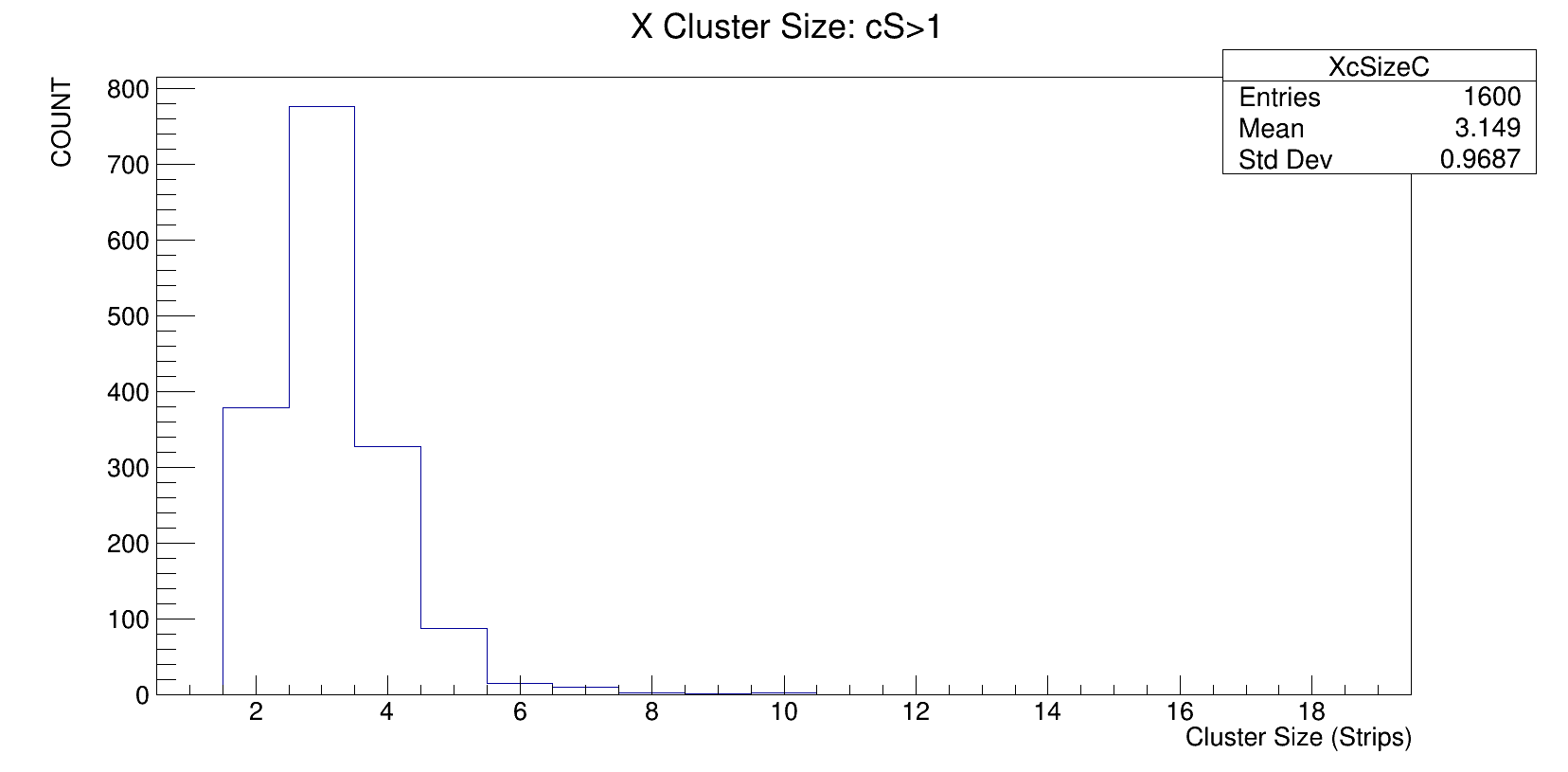} &
        \includegraphics[width=7.2cm]{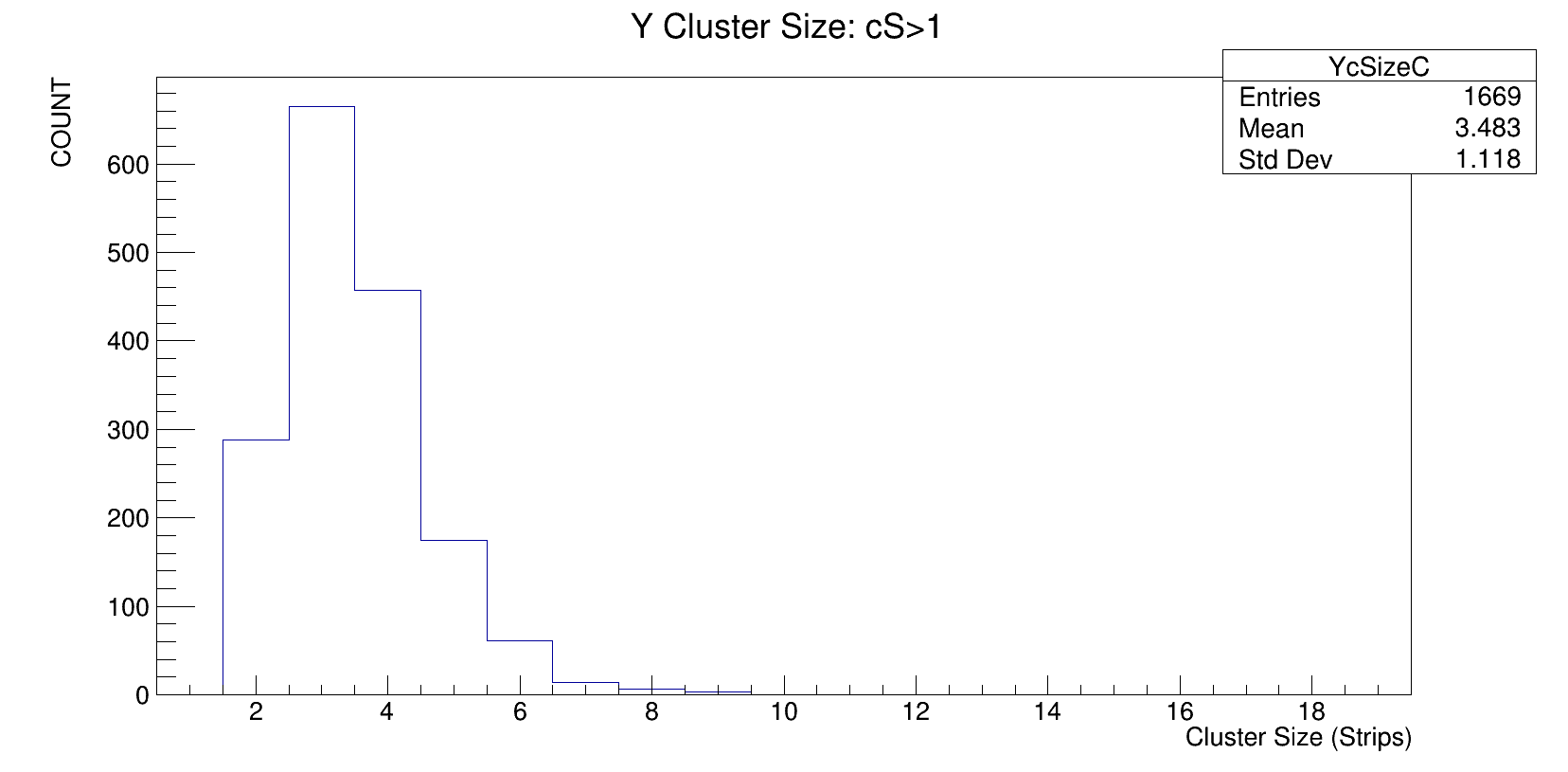} \\
         (a) & (b)
    \end{tabular}
    \caption{Cluster size with steel shielding for the two detector coordinates (cS$>$1 cut also applied).}
    \label{fig:apvclustersizewithsteel}
\end{figure}

In a next step, steel shielding was applied below the detector. \cref{fig:apvclusterampwithsteel,,fig:apvclustersizewithsteel} show the cluster amplitude and cluster size respectively after steel protection was applied.

\begin{figure}[htbp]
  \centering
  \begin{tabular}{cc}
  \includegraphics[width=7cm]{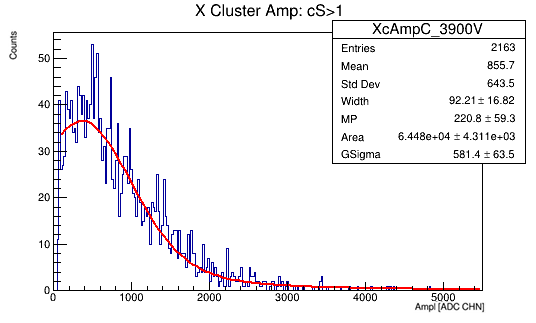} &
  \includegraphics[width=7cm]{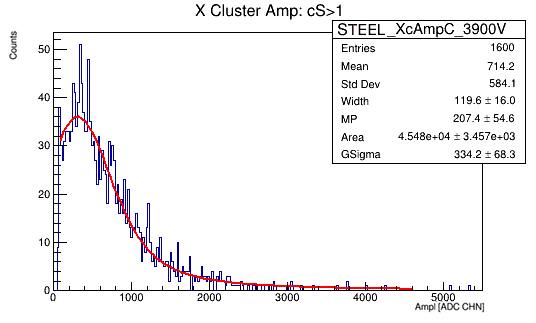} \\
  \includegraphics[width=7cm]{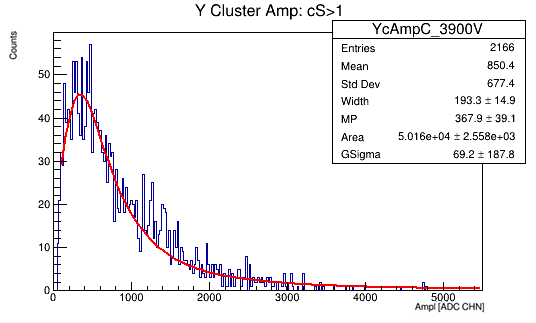} &
  \includegraphics[width=7cm]{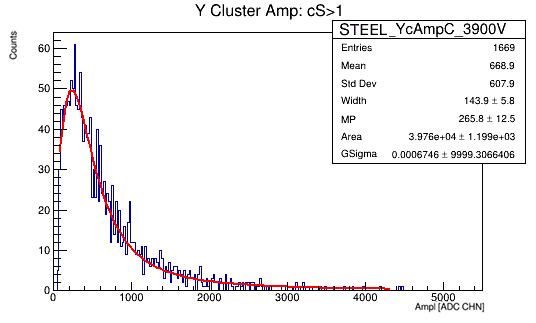} 
  \end{tabular}
  \caption{Gaussian-Landau fits on the X and Y projections of the cluster amplitude with steel shielding (right) and without (left).}
  \label{fig:apvratefits}
\end{figure}

At a first observation, we can see a reduction in entries at about $\sim$ 26\% in the X strips and about $\sim$ 22\% in the Y strips. It is also expected that the mean amplitude of the signal is reduced when using the steel bricks, because we shield against particles that deposit higher energies in the detector medium and hence the low-energy depositions dominate the data, which causes a reduction in the average signal amplitude that is obtained. In order to see whether there is actual improvement in the signal fluctuations, a convoluted Gaussian-Landau function was fitted on the cluster amplitude spectrums. In \cref{fig:apvratefits}, the fits for the data both with steel protection and without are shown side by side for comparison. The corresponding detector gain is \SI{3900}{V}. The generated \textit{p}-values under the $\chi^2$ test are displayed in \cref{tab:p-value}. 

The Gaussian-Landau fits were more successful for the data with steel (99\% probability that the obtained $\chi^2$ value is the minimum possible and thus the fit model highly represents the data). However, the fit models for the data without steel also exceed the 5\% significance level and thus they could also be deemed mostly successful. Despite this, some of the fit parameters display large fluctuations. 

The Gaussian Sigma can be used to determine whether the signal is improved with the steel protection. So in the case of the X strips, a $\sim$ 40\% improvement can be seen. In the case of the Y strips, the fit parameters cannot be relied upon and so, judging from the standard deviation of the data, a $\sim$ 10\% improvement is visible. Optically, one can observe a regression of the 'landau tails' and thus a reduction in highly energy-depositing particles. 

\begin{table}[htbp]
\centering
\begin{tabular}{ c c c }
\toprule
data & \textit{p}-value (X)  &  \textit{p}-value (Y) \\
\midrule
w/o steel & 0.3416 & 0.0602 \\
with steel & 0.9987 & 0.9926 \\
\bottomrule
\end{tabular}
\caption{Generated \textit{p}-values under the $\chi^2$ test for the Gaussian-Landau fits of the cluster amplitude data.}%
\label{tab:p-value}
\end{table}

%The null hypothesis is that the fit describes the data accurately. The test statistic is the chisquared which shows higher values in case less accuracy is achieved. 
%tmath: prob is the probability of the observed chisquared exceeding the value of given chi2 in Prob(chi2,ndf) by chance, even for a correct model/fit. 
%The reverse of this prob is that the observed chisquared for a correct model should be less than chi2.
%When prob is very low, there is a HIGH Probability that I should observe a chisquared which is less than the obtained one. Thus, my model is not entirely accurate. 

A complication that could be introduced with the steel bricks is the possibility of cutting on signal from minimum ionizing muons which is the major contribution from cosmics events. Theoretically, minimum ionizing muons need a range of at least $\sim$\SI{17}{cm} in steel in order to be completely absorbed by the material. The calculation of this value is thoroughly detailed in \cref{sec:app:range}. So a \SI{5}{cm} thick steel brick which is our case is highly unlikely to absorb these particles. In order to cross-check this experimentally, one can look at the most probable value (MPV) of the cluster amplitude distributions between using the steel protection and not. This would give access to the number of ionizations in the drift region and, indirectly, to the reduction in the signal that we derive. The MPV of the convoluted Gaussian-Landau fit was plotted for each detector HV. The expectation is that for the data with steel (blue points), the MPV will be slightly reduced (again, because higher deposited energies are masked from the medium and the most probable energy loss is hence shifted closer to that of minimum ionizing particles). This is true for HV of \SI{3900}{V} and \SI{4100}{V}, but not for the in-between HV settings, and in fact, in these cases, the MPV of the data with steel is actually increased (the Y-strip blue point at HV \SI{3950}{V} is due to a failed fit). A possible reason for this is that the data for each HV setting were taken on different days, thus the existence of different temperature and pressure conditions in the laboratory could have affected the results. Despite this, it can be surmised that particles with minimal ionization or of larger energy range are not significantly dampened from the steel shielding.

\begin{figure}[htbp]
  \centering
  \includegraphics[width=8cm]{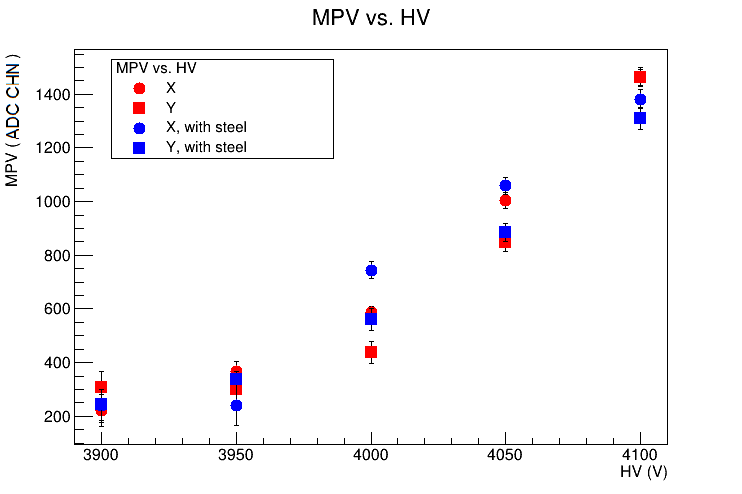}
  \caption{Most probable value (MPV) of the cluster amplitude against the detector gain. The Y-strip blue point at HV \SI{3950}{V} is due to a failed fit.}
  \label{fig:apvrate}
\end{figure}

\begin{figure}[htbp]
  \centering
  \begin{tabular}{cc}
  \includegraphics[width=7cm]{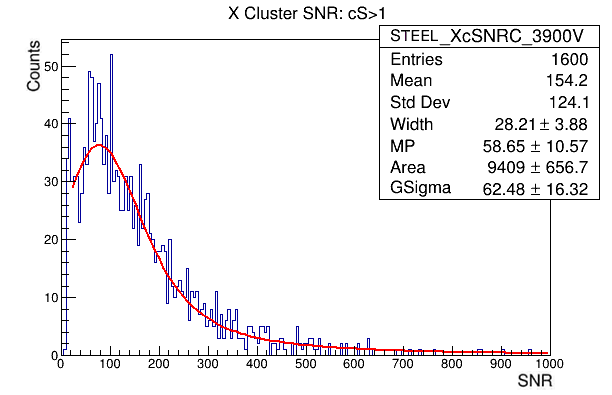} &
  \includegraphics[width=7cm]{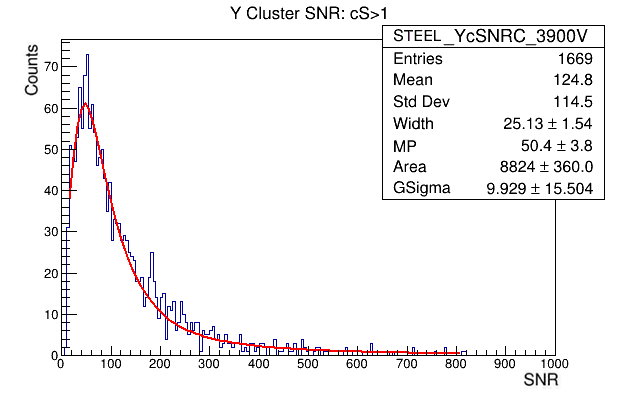} \\
  \includegraphics[width=7cm]{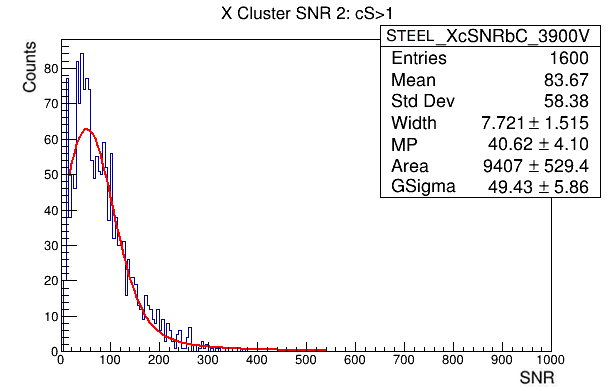} &
  \includegraphics[width=7cm]{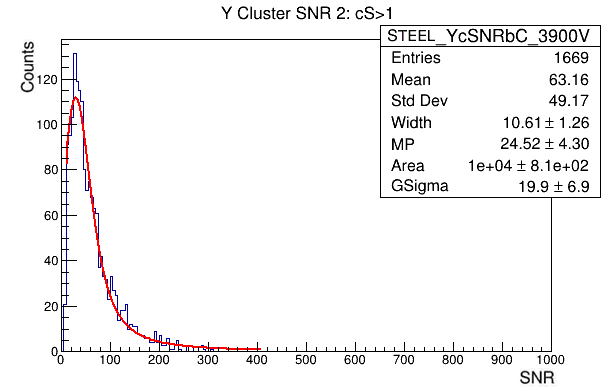} 
  \end{tabular}
  \caption{Gaussian-Landau Fits of the APV25 SNR 1 (top) and SNR 2 (bottom) plots from the X (left) and Y (left) detector strips.}
  \label{fig:apvsnr3900}
\end{figure}

It now remains to estimate the signal to noise ratio (SNR) of the APV25 data, which is the main and final part of the APV25 measurements. The SNR was calculated using two different methods, shown in \cref{eq:SNR1} and \cref{eq:SNR2}. 
For a single cluster $j$ consisting of $N$ strips, the SNR and SNR 2 values are given as:
\begin{equation}
    \label{eq:SNR1}
    SNR_j = \frac{C_A}{\sqrt{\sigma_i^2}}
  \end{equation}
  
\begin{equation}
    \label{eq:SNR2}
    SNR_{j,2} = \frac{C_A}{\sqrt{N}\sqrt{\sigma_i^2}} %= \frac{C_A}{\sqrt{N\sigma_i^2}}
  \end{equation}
where $\sigma_i^2$ is the quadratic sum of the noise over the strips of the cluster and $C_A$ is the cluster amplitude. Plots for both types of the SNR were taken and fitted with convoluted Gaussian-Landau functions. In \cref{fig:apvsnr3900}, the fitted plots are displayed, while the corresponding \textit{p}-values are shown in \cref{tab:p-value-snr}.  
  
\begin{table}[htbp]
\centering
\begin{tabular}{ c c c }
\toprule
data & \textit{p}-value (X)  &  \textit{p}-value (Y) \\
\midrule
SNR & 0.9424 & 0.9187 \\
SNR 2 & 0.0352 & 0.4628 \\
\bottomrule
\end{tabular}
\caption{Generated \textit{p}-values under the $\chi^2$ test for the Gaussian-Landau fits of the SNR and SNR 2 data.}%
\label{tab:p-value-snr}
\end{table}

\begin{figure}[htbp]
  \centering
  \begin{tabular}{c}
  \includegraphics[width=9cm]{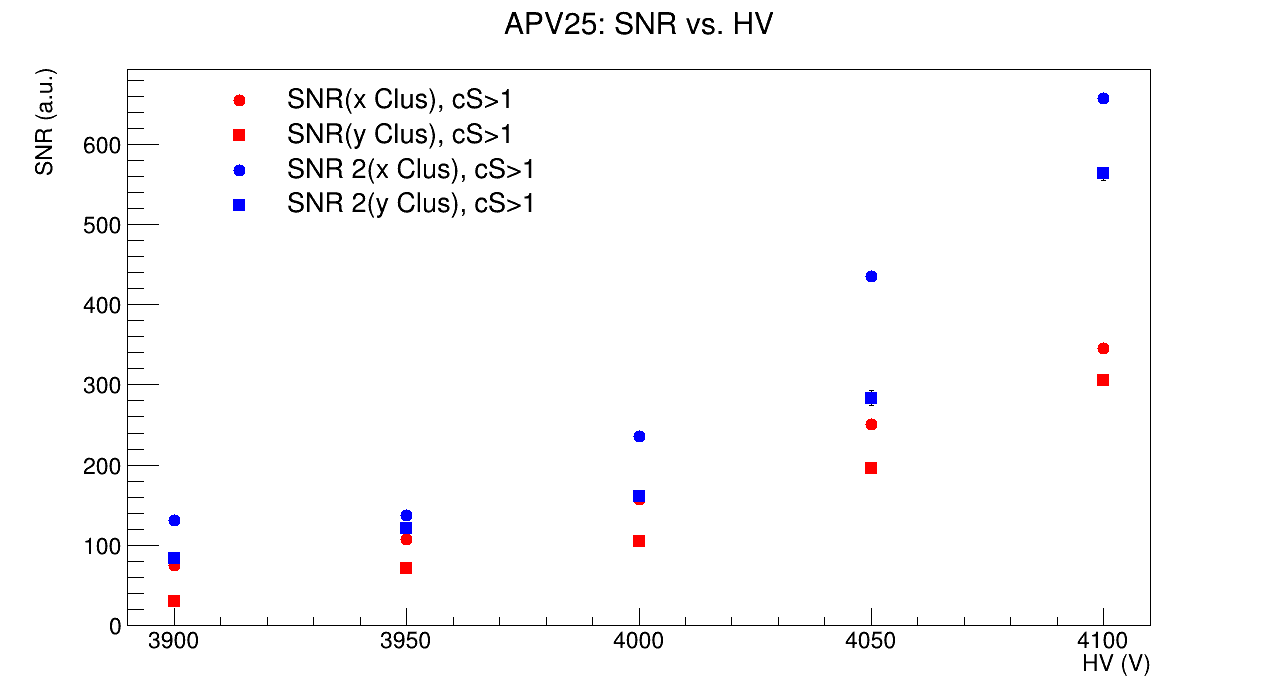} \\
  \includegraphics[width=9cm]{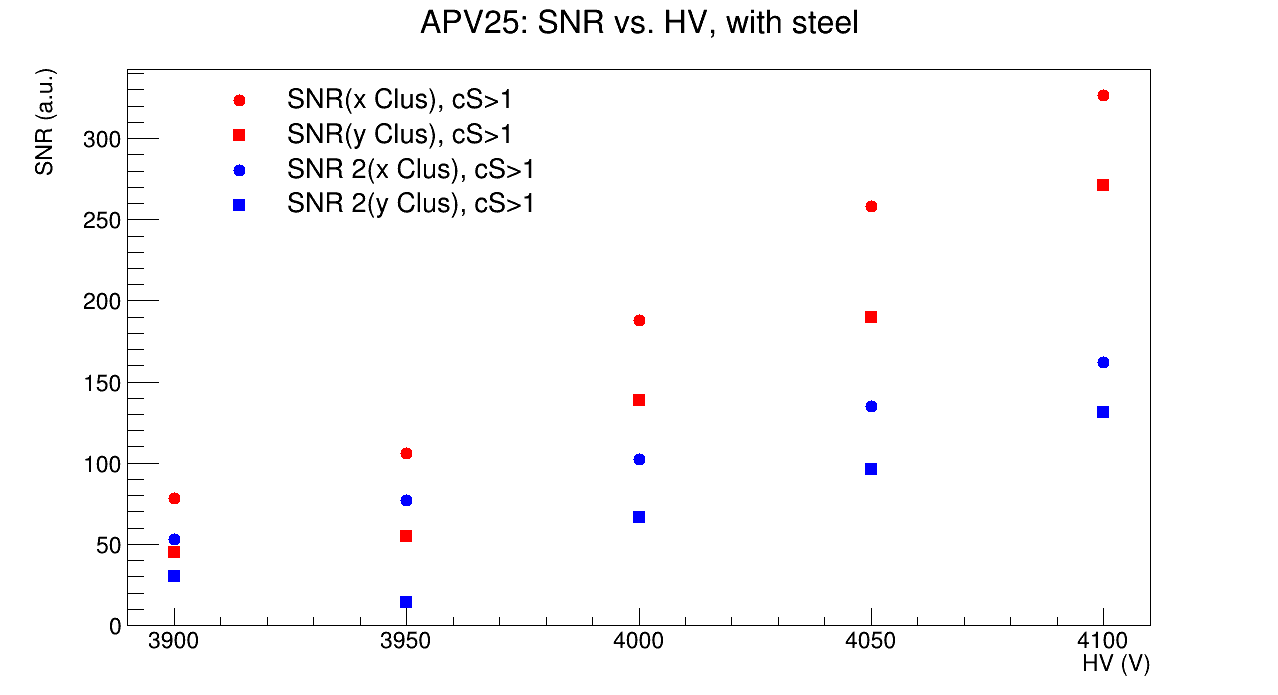}
  \end{tabular}
  \caption{Signal to Noise Ratio (top) without and (bottom) with steel shielding. The Gaussian-Landau means were used in these plots. The fit failed for the Y-strip data with steel shielding at HV$=$\SI{3950}{V}.}
  \label{fig:apvsnr}
\end{figure}

Again, the fit models seem to represent the data correctly, though the fit for the X strips for SNR 2 cannot be accepted under the $\chi^2$ test. On the other hand, the Y strips SNR demonstrates large fluctuations (Gaussian Sigma $=$ $9.9\pm15.5$). As a follow-up, the SNR and SNR 2 data of all the applied detector HV were fitted with Gaussian-Landau functions. For a detailed review of the quality of the fitting, see \cref{sec:app:p-value}. In most cases, the $\chi^2$ test succeeded and so the Gaussian-Landau means were used to plot the two types of the SNR with and without steel shielding against the detector high voltage. Because the convoluted function is generated numerically, the Gaussian-Landau means are found iteratively from the fit, while their error is obtained as equal to 5\% of the Landau width, which is the step used to search for the mean in the iterative process. The results are shown in \cref{fig:apvsnr}. 

The type-1 SNR (red points) is only slightly lower when steel shielding is used, while the SNR 2 goes through a considerable scale reduction. It has been observed that the signal amplitude is only slightly affected by the steel shielding. Additionally, if this was the cause for the scale reduction it would be expected that the same would happen for the type-1 SNR. The noise in both cases is also derived from the same pedestal measurement, so it cannot be considered as the culprit. The only thing that remains is the cluster size, which is the main difference between SNR and SNR 2. 

For the average detector gain of \SI{4000}{V} and with steel shielding applied, the X strips show an SNR of 188.18$\pm$1.35 and an SNR 2 of 102.26$\pm$0.16, while the Y strips show an SNR of 138.804$\pm$1.074 and an SNR 2 of 66.32$\pm$0.19. This yields an average X-Y SNR of 163.49$\pm$0.86 and an average X-Y SNR 2 of 84,29$\pm$0.12.

%could indicate the presence of a large common noise source during the measurements with the steel bricks. The blue point at HV$=$39\SI{50}{V} of the Y projection data with steel deviates from the general curve due to a failed fit.

These results have to eventually be compared with corresponding values from the VMM chip. In order to achieve that, the noise performance of the VMM chip was thoroughly studied, as it will be demonstrated in the following chapters. 
%==============================================================================
\chapter{VMM3a Noise Performance: First Tests}
\label{sec:vmmstg1}
\begin{figure}[htbp]
  \centering
  \begin{tabular}{cc}
  \includegraphics[width=5.5cm]{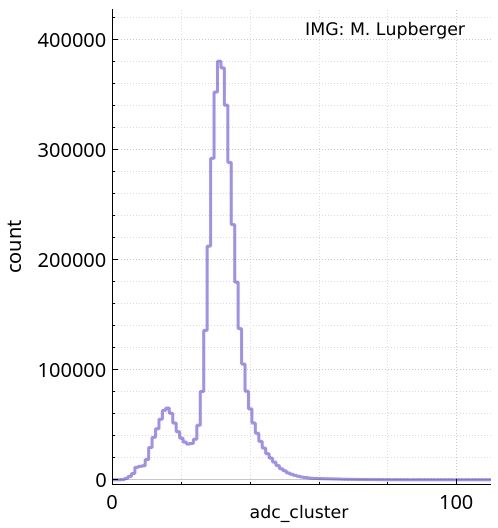} &
  \includegraphics[width=5.5cm]{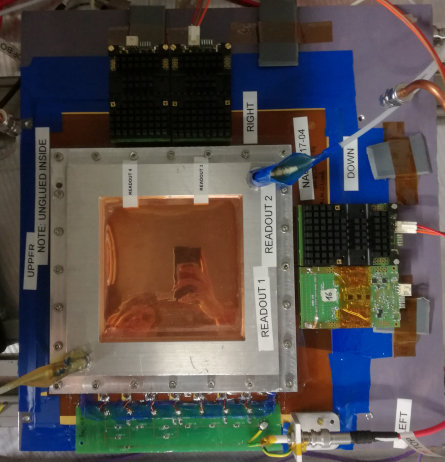} \\
  (a) & (b)
  \end{tabular}
  \caption{(a) First data obtained with the VMM: FE-55 spectrum, and (b) Setup with the VMMs on the tripe GEM detector [M. Lupberger, private communication].}
  \label{fig:imglup}
\end{figure}

%==============================================================================
In May 2019, the first data obtained with the VMM chip were produced at HISKP using a FE-55 source on a triple GEM detector. The goal was to measure a signal to noise ratio as it was similarly done for the APV25 chip. In this case, the signal was defined as the mean ADC cluster of a triggered event from the FE-55 source, while the mean ADC cluster of random events was used as pedestal noise. The results are shown in \cref{fig:imglup}a, while the setup used is depicted in \cref{fig:imglup}b. Both images were provided by M. Lupberger.

However, a major flaw of the recorded data was the absence of low-ADC cluster values, which is the result of the internal global VMM threshold being set too high to record any considerable noise. A high threshold is generally desired in the normal operation of the detectors, however, it is counter-productive when the goal is to measure the noise. Also, as a self-triggered chip, the VMM needs a relatively high threshold, as otherwise, because of the self-triggering, the output bandwidth would be filled with noise data. On the other hand, adjusting the global threshold introduces its own obstacles, since each VMM channel has a different baseline and hence, requires a different threshold. At the time of the measurement, no fine threshold adjustment for individual channels was available and even a few channels in the noise data would be sufficient to fill the output bandwidth of the chip. Therefore, in order to avoid low-threshold channels dominating the data, threshold equalization was necessary. 

The first part of this chapter gives a thorough explanation of the threshold examinations of individual channels that were carried out towards the goal of adjusting the VMM global threshold. Problems that arose from this approach called for a more detailed study of the VMM noise using the oscilloscope, which is the main topic discussed in the latter section.
%------------------------------------------------------------------------------
\section{Threshold Scans}%
\label{sec:vmmstg1:thres}
%------------------------------------------------------------------------------

\begin{figure}[htbp]
  \centering
  \begin{tabular}{cc}
  \includegraphics[width=4cm]{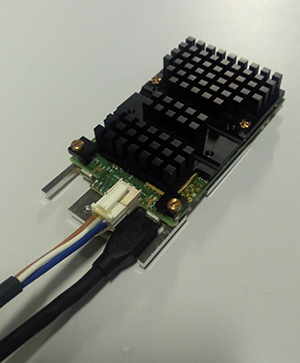} &
  \includegraphics[width=4cm]{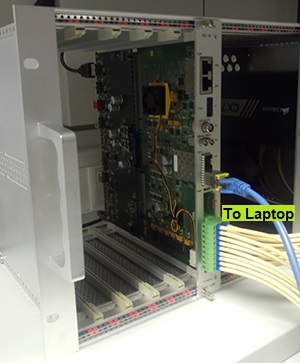} \\
  (a) & (b)
  \end{tabular}
  \caption{The VMM hybrid (a) and the SRS FEC \& Adapter Card (b).}
  \label{fig:setup1}
\end{figure}

\begin{figure}[htbp]
  \centering

  \includegraphics[width=12cm]{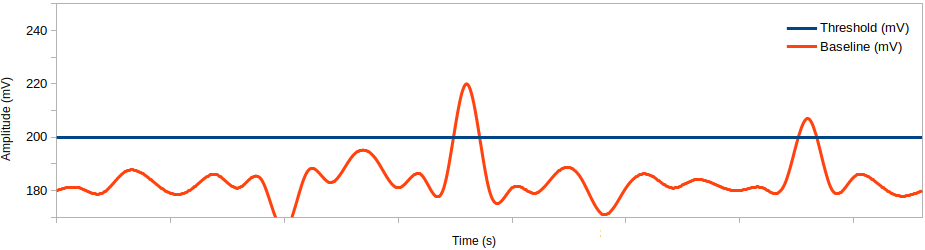}
  \caption{Schematic view of a threshold scan over the baseline. The blue line represents the threshold, which is to be set at varying positions around the baseline. The fluctuating baseline (orange line) will only yield few events when the threshold is further from it and the event number will increase as the threshold line closes in. }
  \label{fig:thresbase}
\end{figure}

The initial setup used was with the VMM powered from the FEC through the HDMI connection between the adapter card and the hybrid (see \cref{fig:setup1}). In order to determine the threshold of an individual channel, the first step was to determine its baseline. The VSCI option to directly read the analog baseline (in ADC CHN) of a channel at the monitoring output was used for this purpose. Subsequently, a threshold scan over the baseline value would help define the mean and RMS of the baseline. More specifically, the global threshold DAC was adjusted close to the ADC value of the baseline, as displayed in \cref{fig:thresbase}, while using the Wireshark network analyzer to count the noise rate coming from the hybrid.

\begin{figure}[htbp]
  \centering
  \begin{tabular}{cc}
  \includegraphics[width=8cm]{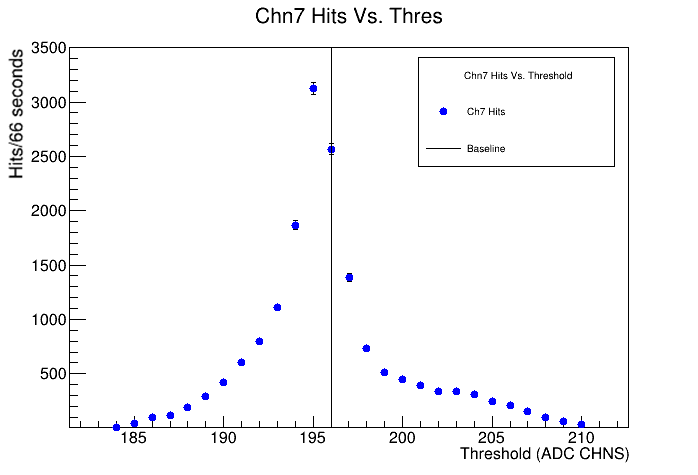} &
  \includegraphics[width=8cm]{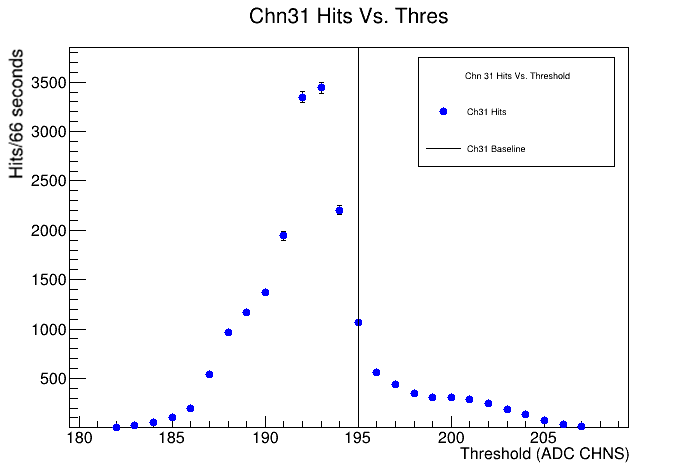} 
  \end{tabular}
  \caption{Threshold scan for channel 7 (left) and channel 31 (right) of the VMM.}
  \label{fig:thres7-31}
\end{figure}
\begin{figure}[htbp]
  \centering
  \begin{tabular}{cc}
  \includegraphics[width=8.3cm]{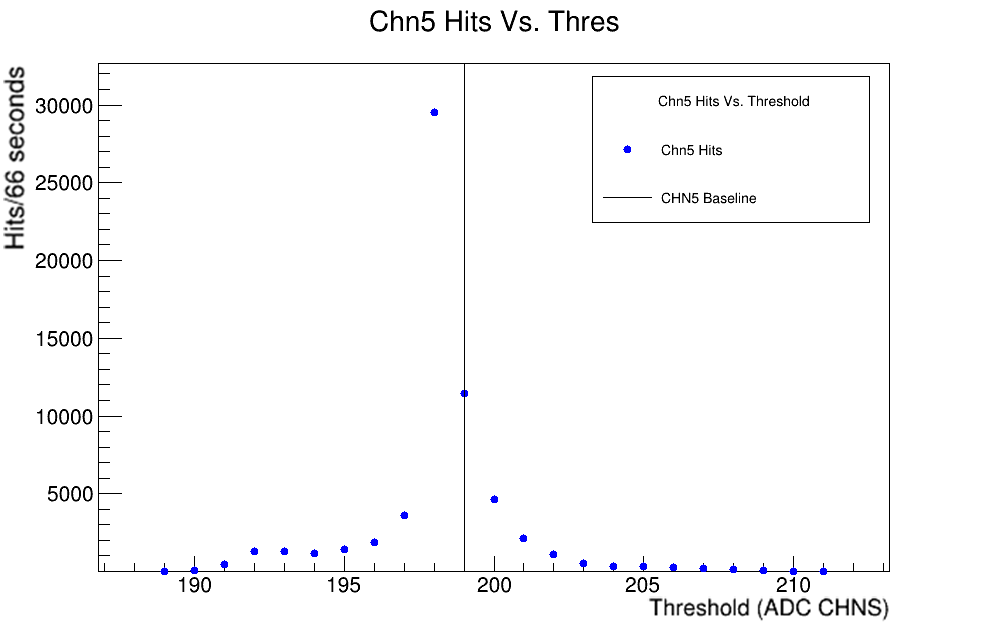} &
  \includegraphics[width=8cm]{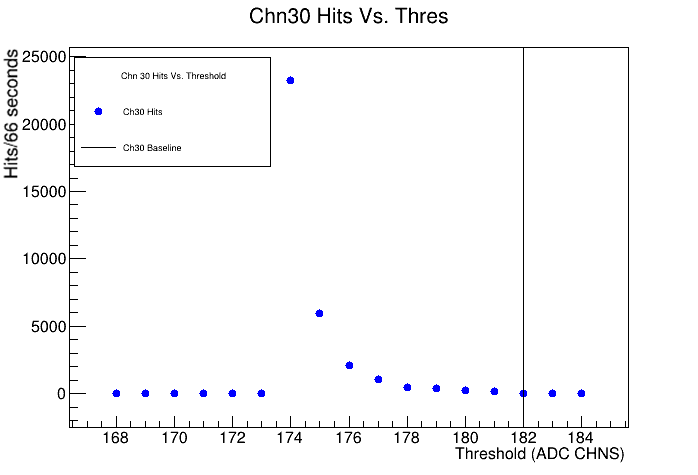}
  \end{tabular}
  \caption{Threshold scan for channel 5 (left) and channel 30 (right) of the VMM.}
  \label{fig:thres5-30}
\end{figure}
\begin{figure}[htbp]
  \centering
  \begin{tabular}{cc}
  \includegraphics[width=8cm]{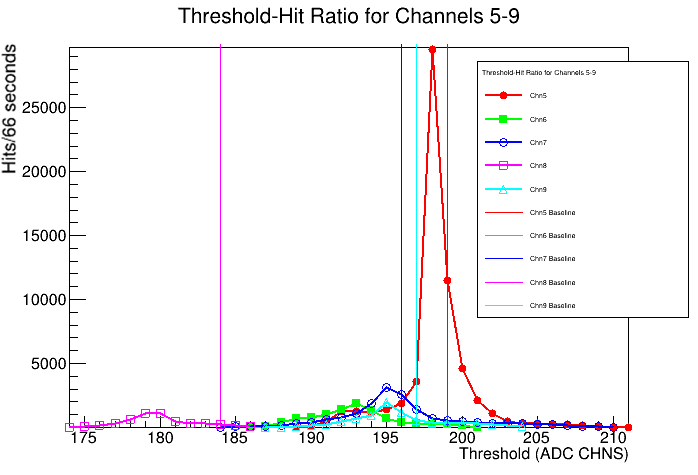} &
  \includegraphics[width=8cm]{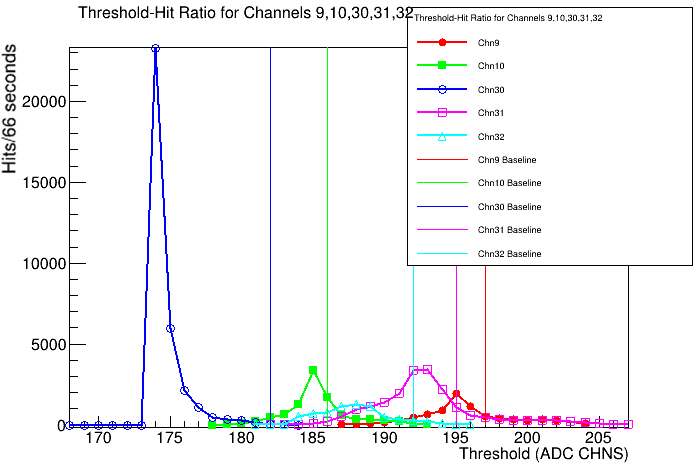}
  \end{tabular}
  \caption{Threshold scan for channels 5 to 9 (left) and for channels 9, 10, 30, 31, 32 (right) of the VMM.}
  \label{fig:thresall}
\end{figure}

The expectation was that the noise rate would surge to a maximum when the threshold matches the baseline and then it would slowly decline, similarly to the behavior shown in \cref{fig:thres7-31}. Even though this plot looks to be fine at a first look, there are still many issues, some of which were discovered by examining more VMM channels. Even though the maximum deviates from the baseline, the difference is minimum and could be attributed to statistical errors. However, since the threshold can only be adjusted in steps of 1 ADC CHN, there are only a few points available close to the maximum and hence, it is impossible to generate sufficient statistical analysis.

%Better: Investigation of baseline rms with osci showed that even for highest gain, rms <~ 1 mV, which is the step size of the threshold. So the baseline can be missed + at this test: from oszi: a lot of external noise sources

In other cases, extreme behaviors like the ones in \cref{fig:thres5-30} were observed. Here, the noise rate reaches extremely high maxima, while in the case of channel 30, the rate also abruptly plummets to zero and the analog baseline is recorded much further away from the maximum.In \cref{fig:thresall}, the threshold scans for multiple channels are plotted together for comparison. A later investigation of the baseline RMS with an oscilloscope showed that even for the highest VMM gain, the RMS is less than or almost equal to \SI{1}{mV}, which is the step size of the threshold. Therefore, the actual position of the baseline can be overstepped and missed. In addition, the oscilloscope observations revealed the presence of a lot of external noise. 
%This could indicate two possibilities: 1) channels 5 and 30 are open channels or affected by the presence of a large noise source, and 2) the threshold step is too large and hence, the maximum rate was overstepped in all the other recorded channels. The second option was found to be the correct one.

It is evident that threshold equalization is difficult to achieve under these circumstances. A better understanding of the VMM noise was required and hence, this led to the next step, which was to use an oscilloscope in order to take measurements of the baseline of individual channels. 

%------------------------------------------------------------------------------
\section{Measurements with the Oscilloscope}%
\label{sec:vmmstg1:osci}
%------------------------------------------------------------------------------

\begin{figure}[htbp]
  \centering
  \begin{tabular}{ccc}
  \includegraphics[width=3.25cm]{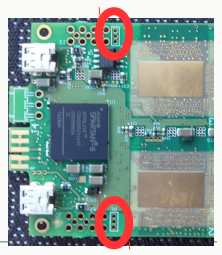} &
  \includegraphics[width=4cm]{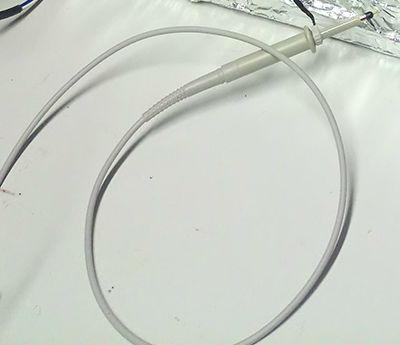} &
  \includegraphics[width=5cm]{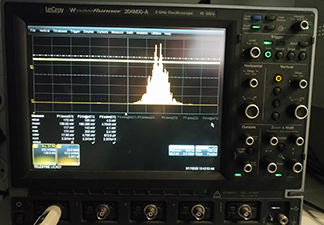}
  \end{tabular}
  \caption{MO pad on the hybrid (left), oscilloscope probe (middle) and the baseline of channel 0 measured on the oscilloscope (right).}
  \label{fig:oszi}
\end{figure}

As mentioned in \cref{sec:theory:vmm}, the baseline of the VMM can be accessed through the MO pad on the hybrid. From there the analog shaper output can be sent onto an oscilloscope screen using an oscilloscope probe, as shown in \cref{fig:oszi}. In the same figure on the right, we see the oscilloscope during an on-going measurement of the baseline of channel 0 of the VMM, which is the channel that was used during this test. The oscilloscope that was used is a 2 GHz Tektronix-MSO56.

\begin{figure}[htbp]
  \centering
  \includegraphics[width=7cm]{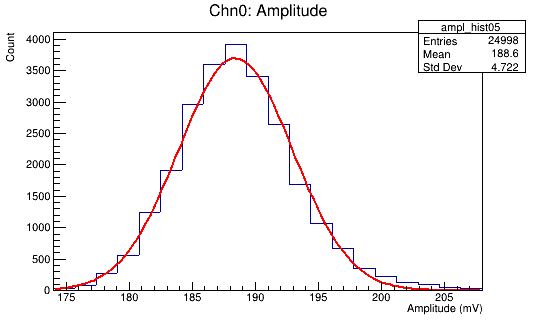} 
  \caption{Gaussian fit on the derived amplitude data of channel 0 of the VMM at a gain of \SI{9}{mV}/fC.}
  \label{fig:base9}
\end{figure}

%each of x samples [for you: samples are the number of enties in the oszi data of one baseline measurement]
The method followed was to save the channel-0 baseline data from random scope frames, each of $\sim$\SI{25000}{samples}, at all 8 gains of the VMM (0.5 - \SI{16}{mV}/fC.) In continuation, the data was projected onto the Y axis and fitted with Gaussian functions in order to derive the RMS value in mV, as shown in \cref{fig:base9}. The RMS was then plotted against the increasing gain. 

\begin{figure}[htbp]
  \centering
  \includegraphics[width=8cm]{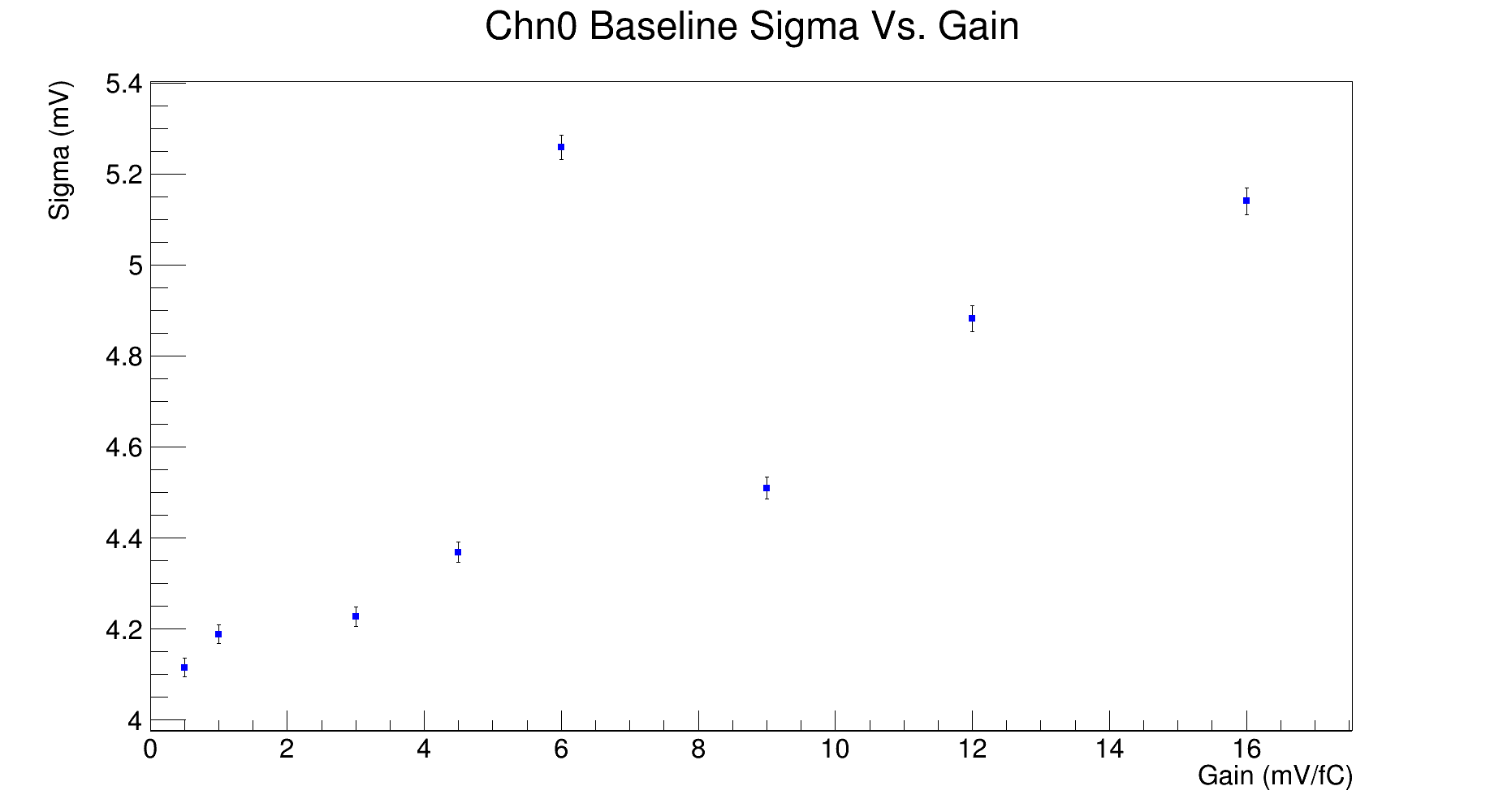} 
  \caption{Channel 0 baseline RMS vs. VMM gain. The data for the RMS at \SI{6}{mV}/fC showed a very noisy behavior and hence the extremely high value.}
  \label{fig:baseline1}
\end{figure}

As it can be seen in \cref{fig:baseline1}, the results derived from this practice are far from desirable. As a first note, the baseline RMS is almost four times higher than previous similar measurements done by M. Lupberger in the Gaseous Detector Development (GDD) group \cite{lupbergerenc}. The data for gain \SI{6}{mV}/fC also demonstrated a strange noise behavior. 

\begin{figure}[htbp]
  \centering
  \includegraphics[width=8cm]{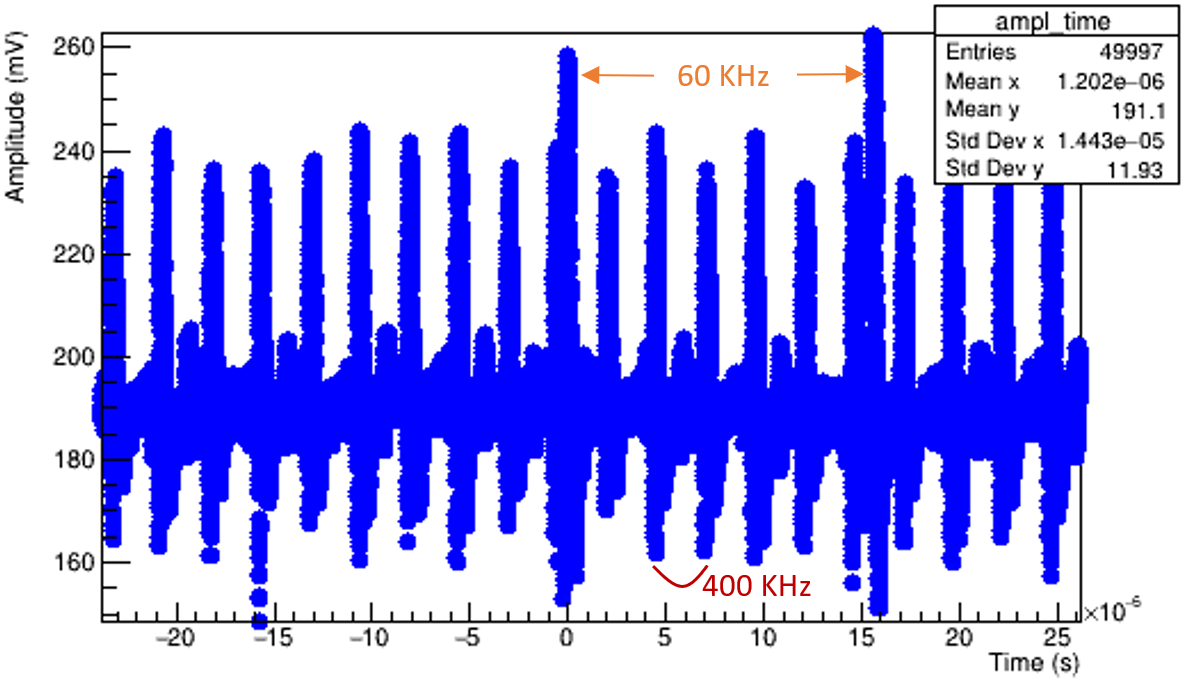} 
  \caption{Channel 0 baseline with external noise (Gain: \SI{0.5}{mV}/fC).}
  \label{fig:noisespikes}
\end{figure} 

As it was realized later, the baseline is severely dominated from external noise sources. Namely, noise from the measuring setup, which includes the hybrid components, the oscilloscope probes, the back-end electronics, as well as noise from other existing instruments in the lab. A constant 160 MHz noise signal was found to be related to the Address in Real Time (ART) clock, which is a clock specific to the MicroMegas operation in the ATLAS experiment and is not needed in our setup. Therefore it was switched off from the VMM control panel. The major external noise spikes were a \SI{60}{KHz} one and a \SI{400}{KHz} one, which did not appear to be synchronized. They are both shown in \cref{fig:noisespikes}.

\begin{figure}[htbp]
  \centering
  \begin{tabular}{cc}
  \includegraphics[width=7.1cm]{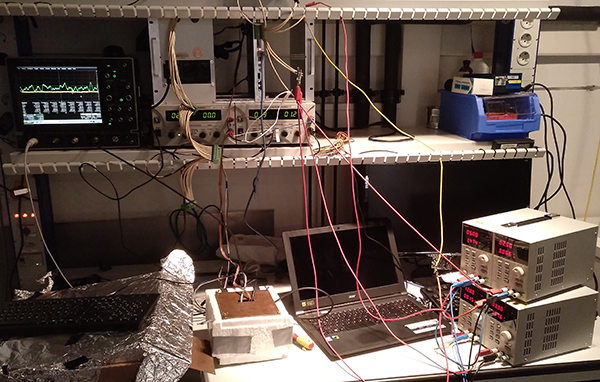} &
  \includegraphics[width=3.3cm]{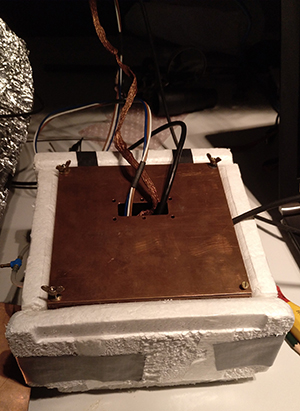} \\
  \end{tabular}
  \caption{New setup with externally powered VMM placed inside a Faraday Box.}
  \label{fig:newsetup}
\end{figure}

\begin{figure}[htbp]
  \centering
  \includegraphics[width=8cm]{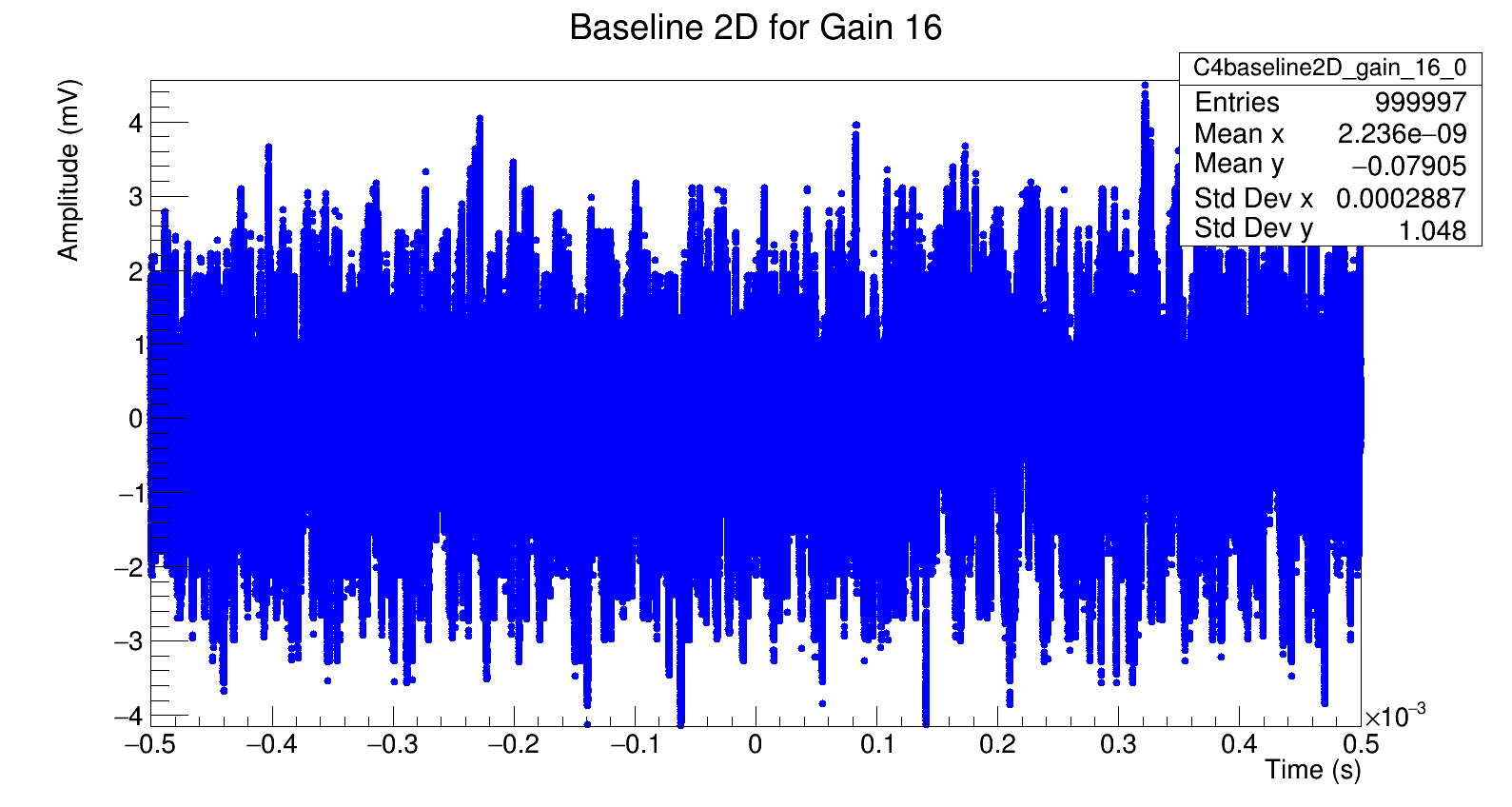} 
  \caption{Channel 4 baseline after noise removal, (Gain: \SI{16}{mV}/fC).}
  \label{fig:baseline2}
\end{figure}

The \SI{400}{KHz} noise source was eventually found to be a malfunctioning ATX power supply in the back-end electronics, so it was removed by using an external power supply for the VMM. As a follow-up, the entire setup was rearranged in order to suppress interferences. In specific, the VMM was placed inside a Faraday box, which successfully removed the \SI{60}{KHz} noise. Further improvements included replacing the standard oscilloscope probe with BNC cables and an active probe which was soldered to the MO pad. In order to keep the hybrid cool, a mini fan was placed on top of the Faraday Box. The renewed setup along with a close view of the Faraday box is shown in \cref{fig:newsetup}. The final baseline RMS that was obtained is shown in \cref{fig:baseline2}. In this example, channel 4 was monitored, while the gain was set at \SI{16}{mV}/fC. Additionally, the number of samples is now close to $\sim$10$^6$.

Additionally, with the hybrid off, the oscilloscope with only the BNC cable connected demonstrates an RMS value of about \SI{0.14}{mV}, which is still less than the VMM baseline RMS value for the lowest gain setting ($\sim$ \SI{0.2}{mV}). The pickup noise from the hybrid electronics can also be determined individually for each channel/gain setting by measuring the baseline RMS at data acquisition (ACQ) on and off. We provide results both with ACQ on and off and also, the BNC cable noise is quadratically subtracted from final results.

\begin{figure}[htbp]
  \centering
  \includegraphics[width=8.5cm]{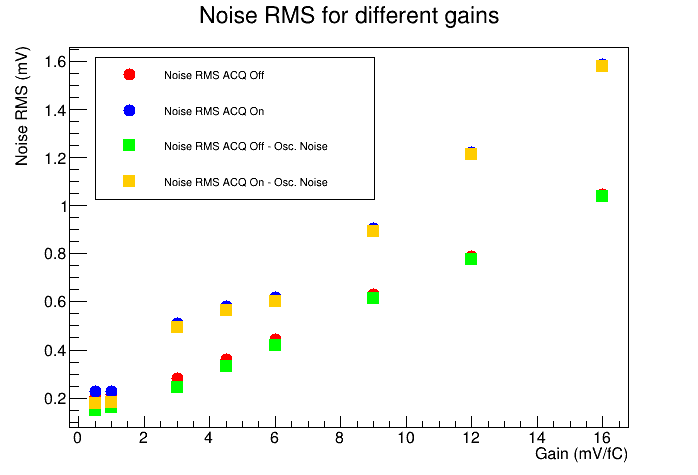} 
  \caption{Channel 4 baseline RMS vs. gain.}
  \label{fig:baseline3}
\end{figure}

The scale of improvement was rather impressive--the baseline was reduced to about four times less than previous values, coming to be superior to the GDD results. However, one can come ever so close to excluding 100\% of noise sources, so the value that we obtain for the baseline RMS (and for the ENC) can be considered as an upper limit for the real VMM hybrid noise. Finally, a new RMS vs. gain measurement was done, producing the results shown in \cref{fig:baseline3}. Data with the ACQ contribution and the oscilloscope noise are also displayed. The increase with gain is mostly linear, aside from the lower gains, where the oscilloscope noise seems to have a stronger impact. The statistical errors generated from the analysis were negligible and hence, they were not added to the final plot. 

So far, the baseline RMS was obtained in mV. A calibration was needed in order to convert it into a number of electrons, or in other words, characterise it in terms of the ENC (see \cref{sec:theory:noise}). The main idea used to achieve this was to induce a pulse of known charge into the VMM channel and then measure the amplitude of the signal at the MO output, as it was similarly done for the baseline. Hence, the ENC can be calculated as explained in \cite{lupbergerenc} and \cite{spieler2002pulse} and shown in \cref{eq:ENC}. 
\begin{align}
\label{eq:ENC}
\begin{split}
\frac{U_{pulse}}{Q_{pulse}}=\frac{U_{RMS}}{Q_{noise}} 
\\
Q_{noise}\rightarrow ENC[e^-] 
\\
ENC[e^-] = \frac{\Delta Q_{in}}{e}\cdot \frac{U_{RMS}}{U_{pulse}} 
\end{split}
\end{align}
%where $U_{RMS}$ is the RMS of the analog baseline and $U_{pulse}$ the amplitude of the induced pulse. 

\begin{figure}[htbp]
  \centering
  \begin{tabular}{cc}
  \includegraphics[width=7.5cm]{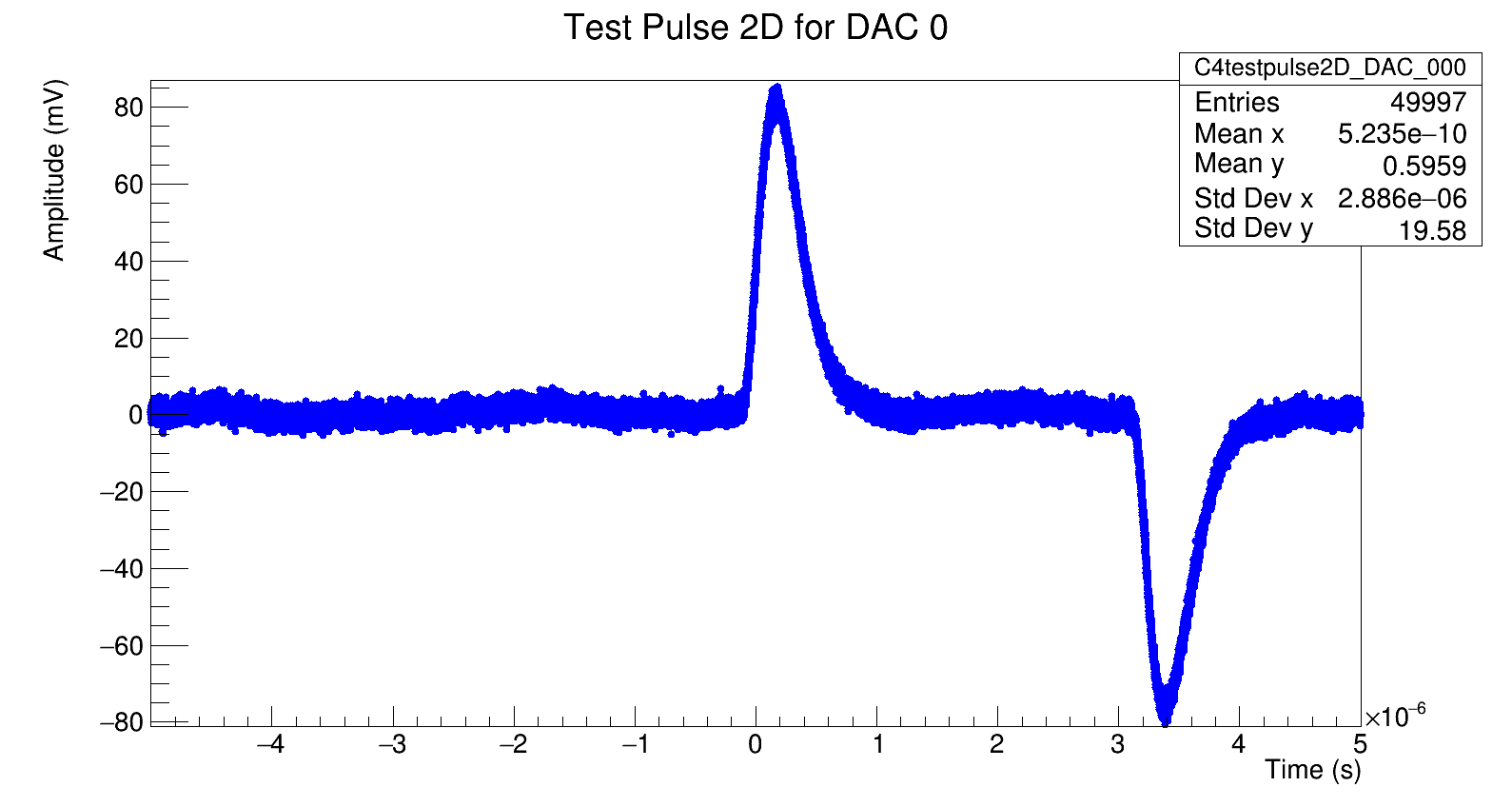} &
  \includegraphics[width=7.5cm]{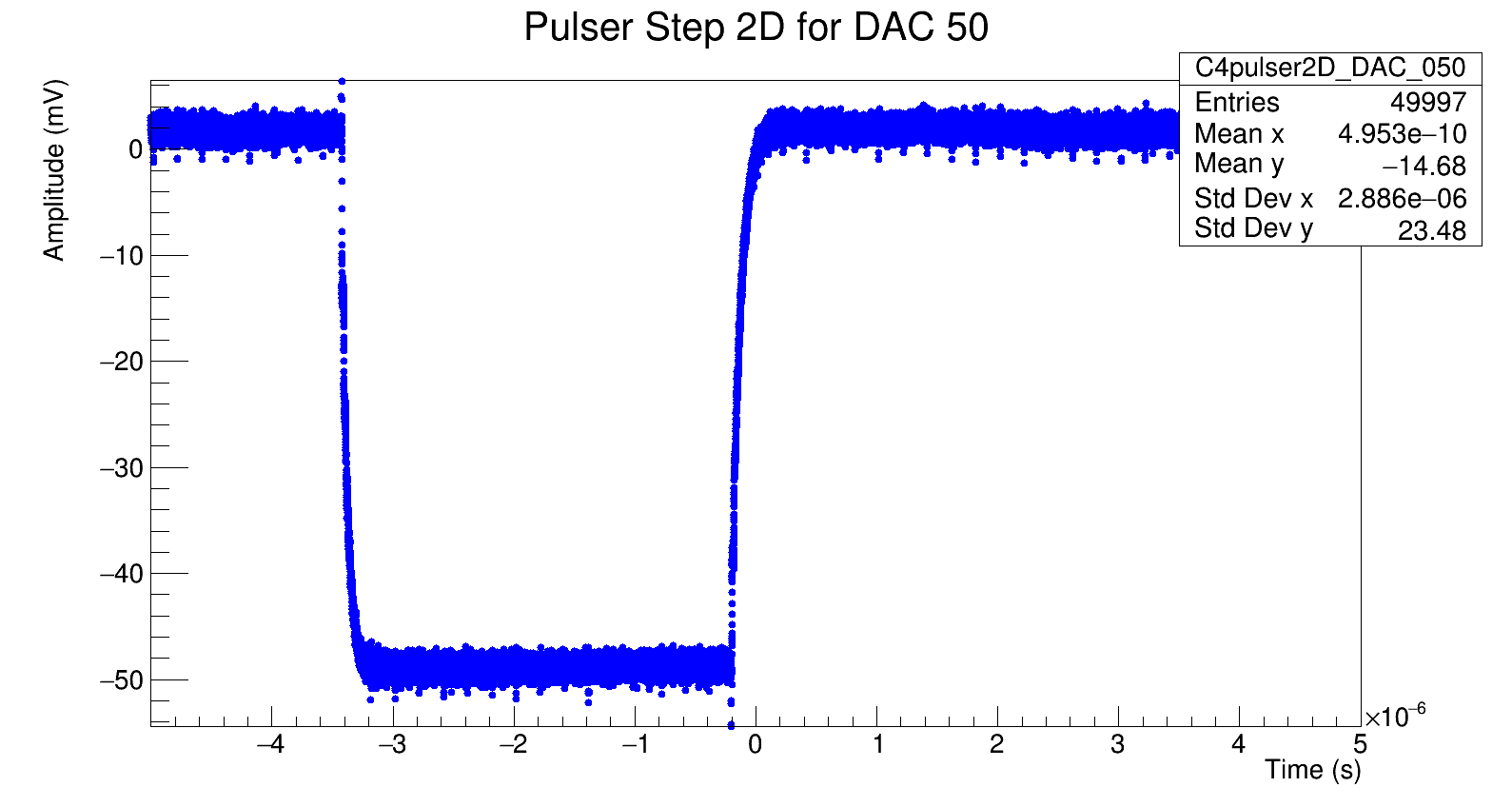} \\
  (a) & (b)
  \end{tabular}
  \caption{Example of a 2D histogram of (a) a test pulse, (b) the pulser step.}
  \label{fig:tpulse}
\end{figure}

The internal pulse circuit of each VMM channel can be used to send a test pulse of input charge $\Delta Q_{in}$ and injection capacitance $C_s=0.$\SI{3}{pF} (see \cref{fig:pcircuit}). Both the pulse amplitude, $U_{pulse}$, and the RMS of the analog baseline, $U_{RMS}$, can be routed to the MO via configuration of the VMM and be calculated. The potential difference, $\Delta U$, of the generated pulse can be set from the VMM DAC and be calculated at the MO as well. Hence, the input charge is known through:
\begin{figure}[htbp]
  \centering
  \includegraphics[width=7cm]{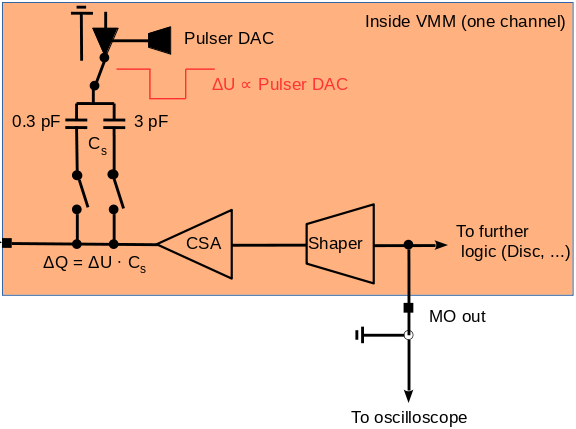} 
  \caption{The test pulse circuit integrated in each VMM channel [M. Lupberger, private communication].}
  \label{fig:pcircuit}
\end{figure}
\begin{equation}
    \label{eq:DQin}
    \Delta Q_{in}=C_s\cdot\Delta U
\end{equation}
With $e=1.602\times10^{-19}$ C and $C_s=0.$\SI{3}{pF}, \cref{eq:ENC} can finally be written as: 
\begin{equation}
    \label{eq:ENCfinal}
    ENC[e^-] = (1.8645\times10^6)\cdot\frac{U_{RMS}}{U_{pulse}}\cdot\Delta U 
\end{equation}
In order to calculate the ENC, the $U_{pulse}$ and $U_{RMS}$ were obtained for each gain setting. An example of a test pulse diagram is shown in \cref{fig:tpulse}a. The pulse amplitude was calculated by applying a zoom on the pulse's peak so as to obtain an average of the peak height. The mean of the baseline was then subtracted from the peak height, giving the pulse amplitude. Additionally, the test pulse histograms were used to calculate the baseline RMS value with ACQ on. This was done by scaling the histogram to a region without the pulse and then deriving the Gaussian sigma of the corresponding Y-projection histogram.  

\begin{figure}[htbp]
  \centering
  \begin{tabular}{cc}
  \includegraphics[width=7.5cm]{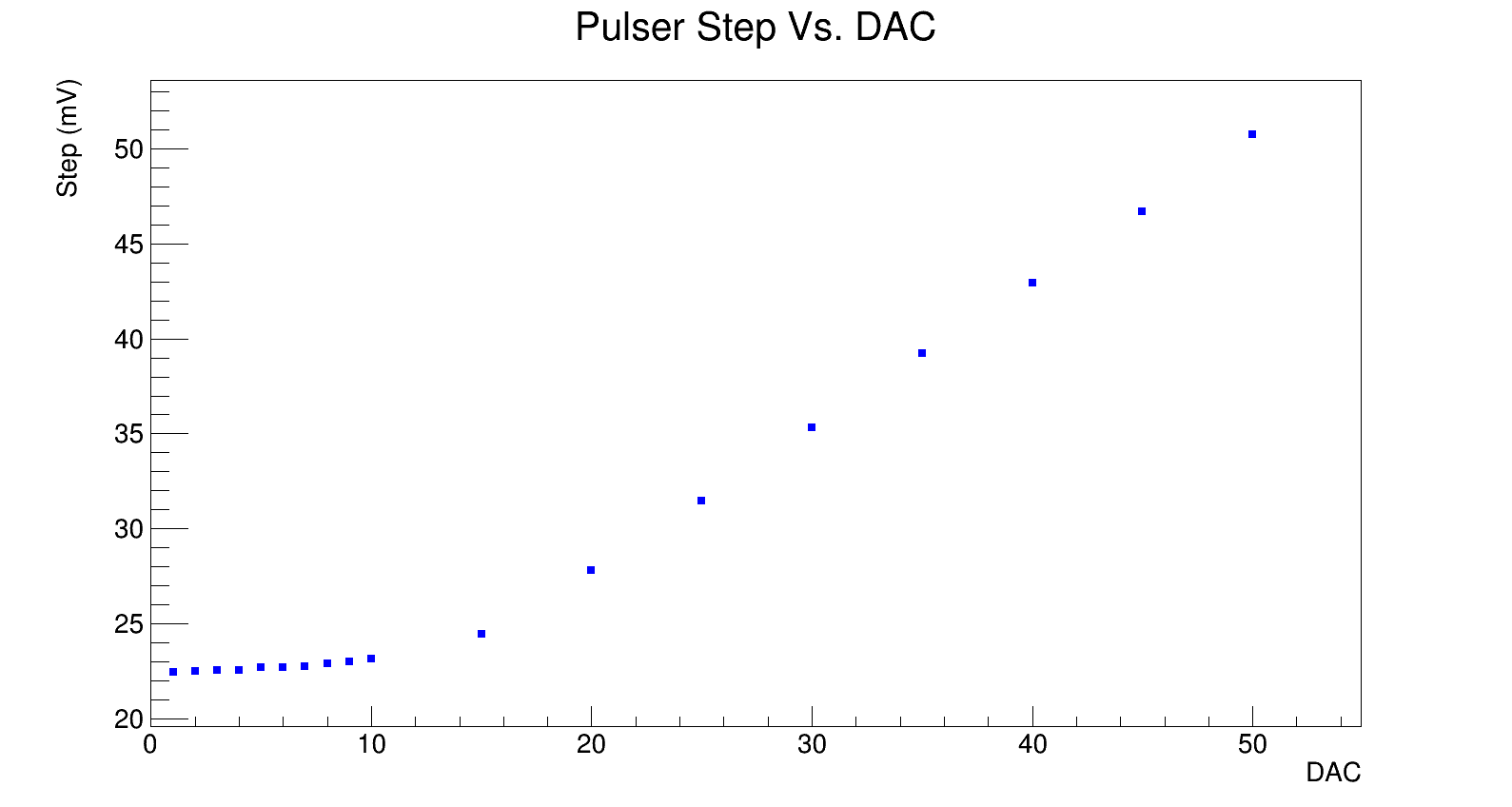} &
  \includegraphics[width=7.5cm]{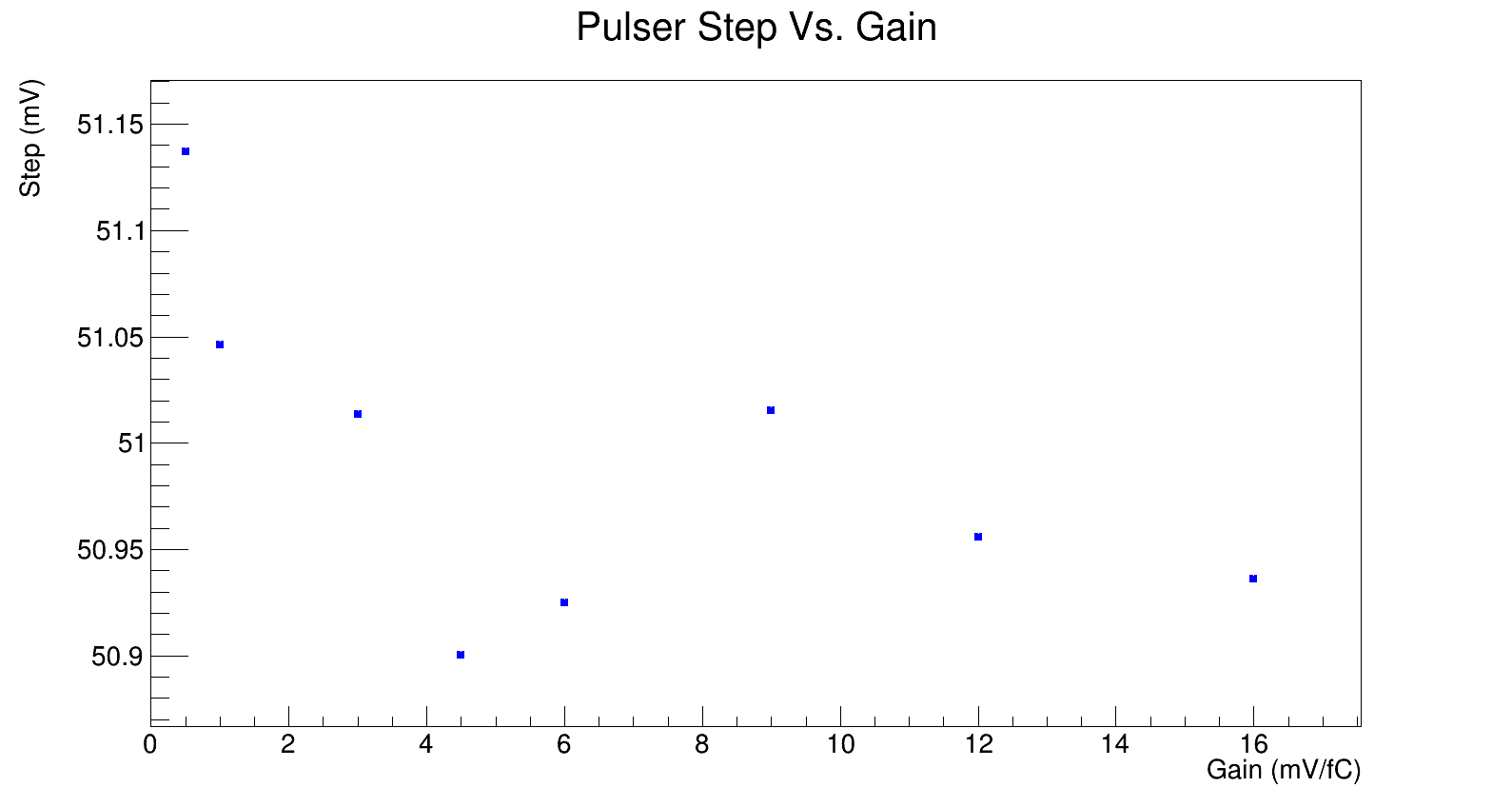} \\
  (a) & (b)
  \end{tabular}
  \caption{(a) Pulser step calibration from DAC to mV, (b) pulser step (mV) against the gain increase.}
  \label{fig:pulser_step}
\end{figure}

On the other hand, the value of $\Delta U$ was determined as a constant through a series of calibrations. In order to calculate the pulser step, a histogram such as the one shown in \cref{fig:tpulse}b was derived. The low- and high-amplitude regions were then plotted and fitted separately with a Gaussian to derive an average of the low and high level of the step. Hence, their difference would be the desired value. It was necessary to understand the relation between the pulser DAC and the voltage step in mV. This practice is shown in \cref{fig:pulser_step}a. The curve becomes linear after an onset of DAC $\sim$ 15. For the measurements, the pulser step value was set at DAC $=$ 50. Another important check was between the pulser step in mV and the gain increase, shown in \cref{fig:pulser_step}b. As one can see, the fluctuations across different gains are minimal with a mean value of \SI{50.99}{mV} and a standard deviation of \SI{0.07}{mV} (0.1\% relative standard error). Additionally, no systematic effect is visible.

\begin{figure}[htbp]
  \centering
  \begin{tabular}{cc}
  \includegraphics[width=7.5cm]{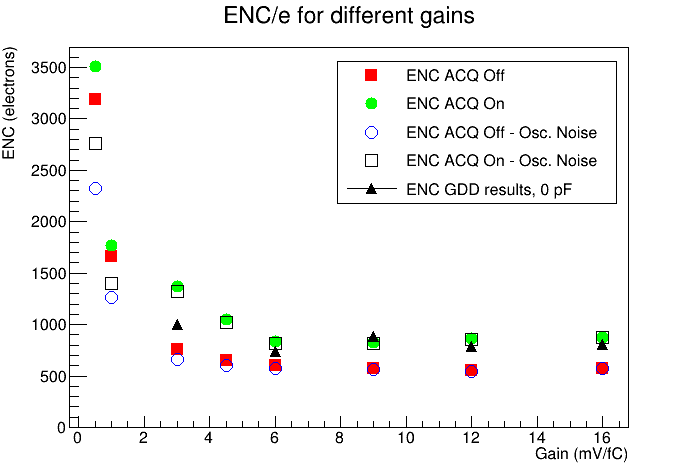} &
  \includegraphics[width=7.5cm]{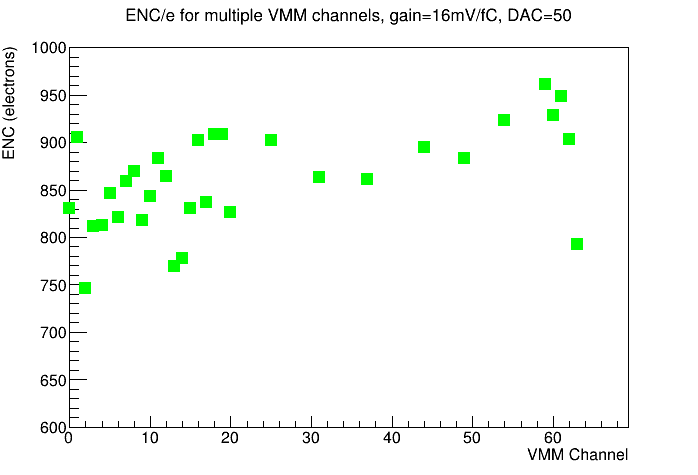} \\
  (a) & (b)
  \end{tabular}
  \caption{(a) ENC measurement for channel 4 against the gain increase. Results are compared with previous GDD results (black triangles). The ACQ contribution and oscilloscope noise are also demonstrated. (b) ENC measurement for multiple ENC channels at gain \SI{16}{mV}/fC.}
  \label{fig:encfinal}
\end{figure}

The final ENC results are shown in \cref{fig:encfinal}a and b. In the first one, the channel 4 ENC is plotted against the chip's gain increase. Again, points were taken both with ACQ on and off and with the oscilloscope noise removed. The black triangle points provide a relative comparison to the former GDD results. In the GDD measurement, points for the lower gains are missing, since the oscilloscope that was used lacked a high enough resolution. This could mean that the data between the two measurements either agree or they were improved due to a better shielding of the noise at the HISKP. The second figure displays the ENC for multiple VMM channels at the highest gain setting. A minimal ENC variation is observed between channels. Since the statistical errors for the RMS and pulse amplitude were negligible, the pulser step variation was used to calculate errors for the ENC. However, this only amounted to an additional noise of 3-4 electrons, which can be barely seen in these plots.

At this stage, results were only preliminary, since there was no thorough investigation done regarding the different elements of the setup, such as the internal test pulse unit and the temperature effects. The lack of an automated procedure also made it difficult and time-consuming to derive enough measurement samples for error estimation. Aside from that, the results lacked completeness, since only a few channels of the VMM at specific parameter settings could be studied at a time and since no tests with external input capacitance were done. Therefore, it was necessary to develop an automation framework with which to fulfill the above requirements. 

%==============================================================================
\chapter{Development of an Automatic VMM Noise Measurement System}
\label{sec:script}
%==============================================================================

In order to minimize the duration and error factors of the VMM3a noise measurements, a systematic method of data taking and processing was developed. The main goal was to unify the procedures involved in the VMM noise measurement, as mentioned in \cref{sec:vmmstg1}. In brief summary, these were: a) configuration of the VMM3a readout chip through the VMM slow control software, b) visualization and acquisition of the signal waveform from an oscilloscope, c) calculative analysis of the derived data in order to measure the noise RMS in mV and subsequently convert it to electrons. The easiest way to implement this was by adding a new calibration routine in the already existent calibration module of the VSCI. The main challenge that was encountered during this effort was establishing communication between a Linux-operated laptop and an oscilloscope. While the VSCI is based on C++, the oscilloscope responds to its own language based on Standard Commands for Programmable Instruments, or SCPI commands. SCPI commands vary among oscilloscope models, thus specific knowledge of the SCPI commands for the Tektronix-MSO56 oscilloscope that was used was required. This was obtained from \cite{lecroy2005osci}. Additionally, a third-party program was used to translate C++ commands into the oscilloscope language and back, found at \cite{sharples2008vxi11} and based on the VXI-11 Ethernet Protocol. In the following, a description is given of the calibration routine that was written for the above purpose, as well as of the individual programming components included. 

%------------------------------------------------------------------------------
\section{The RMS Calibration Routine}%
\label{sec:script:rms-routine}
%------------------------------------------------------------------------------

\begin{figure}[htbp]
  \centering
  \includegraphics[width=10cm]{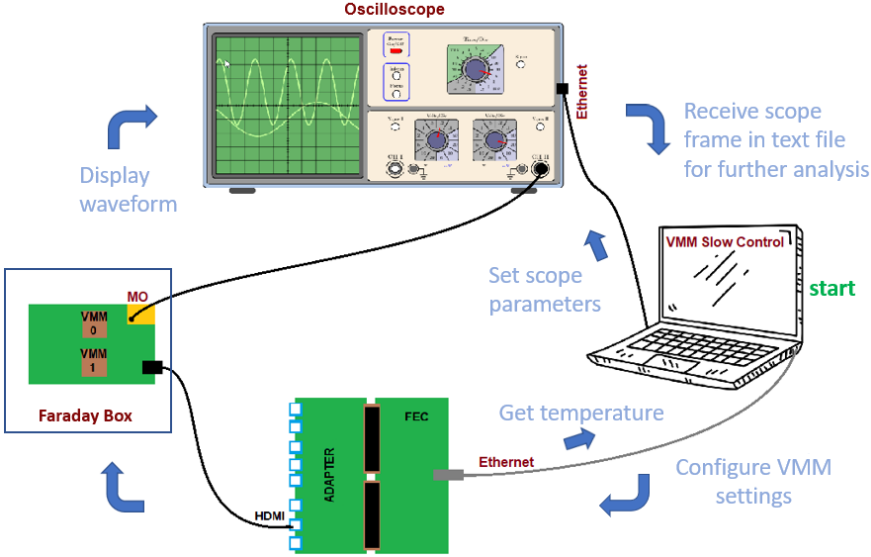}
  \caption{Schematic of one measurement cycle. The VMM is configured through the VSCI and parameters such as the channel, gain, peaktime and temperature are recorded. Communication to the oscilloscope is established, parameters like the vertical scale are set. The waveform of either the rms, test pulse or pulser step is displayed on the oscilloscope. Several scope frames of the displayed waveform are saved into the laptop, where data analysis commences.}
  \label{fig:meas-cycle}
\end{figure}

The added calibration routine configures the VMM parameters and settings for RMS data taking and calls on a framework of executables which 1) establish connection to the oscilloscope, 2) transfer the data of several scope frames, 3) analyze and plot the data using the \texttt{C++/ROOT} programming environment. Once the RMS is taken, the VMM is configured again and data acquisition is set on to derive test pulse and pulser step data. For each case, similar executables are called on for communication with the oscilloscope and data processing. The above procedures constitute one measurement cycle, which is schematically demonstrated in \cref{fig:meas-cycle}.

In a full-scale measurement, this cycle runs repeatedly in an iterative process in order to cover measurements with all VMM channels, gains and peaktimes. The parameters as well as the temperature of the VMM are recorded for each measurement loop. The final RMS, test pulse and pulser step values along with their errors are each statistically derived from the analysis of 10 scope frames. The ENC is then calculated as described in \cref{sec:vmmstg1}, while additional calculations are carried out for the removal of the oscilloscope noise. All these values are stored in a text file which is later used to plot the results for the individual VMM parameters. For instance, for a measurement of VMM 0, channel 0, lowest gain and highest peaktime, the line entry in the final text file is shown split in \cref{tab:entry1,,tab:entry2}. The "off" and "on" labels signify data with ACQ on and off, while the "osz" label signifies rms values without the oscilloscope noise. All rms entries, as well as the pulse amplitude, the pulser step and the ACQ contribution are expressed in mV; the ENC is expressed in electrons, while the temperature is calculated in \SI{}{\celsius}. The gain and peaktime are recorded as integers and not with their actual values.

\begin{table}[htbp]
\centering
\resizebox{1.\columnwidth}{!}{\begin{tabular}{|c|c|c|c|c|c|c|c|c|c|}
\toprule
vmm$\_$id & channel & gain & peaktime & rms$\_$off & rms$\_$off$\_$osz & rms$\_$off$\_$er & rms$\_$on & rms$\_$on$\_$osz & rms$\_$on$\_$er \\
\midrule
0 & 0 & 0 & 0 & 0.3208 & 0.2887	& 0.1648 & 0.4063 &	0.3814 & 0.0573 \\
\bottomrule
\end{tabular}}
\caption{An entry line in the final text file containing the results from the RMS calibration, part 1.}
\label{tab:entry1}
\end{table}

\begin{table}[htbp]
\centering
\resizebox{1.\columnwidth}{!}{\begin{tabular}{|c|c|c|c|c|c|c|c|c|c|}
\toprule
pulse$\_$ampl & p$\_$ampl$\_$er & enc$\_$off & enc$\_$off$\_$er & enc$\_$on & enc$\_$on$\_$er & ACQ$\_$contr & tempr & $\Delta U$ & $\Delta U\_$er \\
\midrule
5.7555 & 0.3173 & 2935.13 &	1387.07	& 4055.07 & 1888.11 & 0.2996 & 56.2162 & 29.5917 & 13.5146 \\
\bottomrule
\end{tabular}}
\caption{An entry line in the final text file containing the results from the RMS calibration, part 2.}
\label{tab:entry2}
\end{table}

As mentioned at the beginning of the chapter, the main part of the automation system was establishing control and communication with the oscilloscope. This was done by utilizing the VXI-11 Ethernet Protocol and the SCPI commands of the Tektronix-MSO56 oscilloscope that was used in the setup. 
%------------------------------------------------------------------------------
\section{The SCPI Language}%
\label{sec:script:scpi}
%------------------------------------------------------------------------------

SCPI commands are ASCII-based commands \cite{scpi2017r&s}, which are designed to remotely communicate with measurement instruments; in our case, an oscilloscope. They are separated into two categories: 
\begin{itemize}
    \item the set commands,
    \item the query commands, or queries.
\end{itemize}
A set command may simply set a parameter of the measurement instrument or oscilloscope. For instance, the command "\texttt{C4:VDIV 50 MV}" sets the vertical scale of channel 4 to \SI{50}{mV} per division. The quotation marks are used here to enclose the command and are not part of its syntax. On the other hand, query commands request information from the oscilloscope. Thus, the query "\texttt{C4:VDIV?}" requests the oscilloscope's vertical division in mV. Queries can be distinguished from set commands since their syntax always includes a \textbf{?} character following their header. Both of the example commands mentioned are specific to the oscilloscope used in the current setup and can be found in \cite{lecroy2005osci}. 

%------------------------------------------------------------------------------
\section{The VXI-11 Ethernet Protocol}%
\label{sec:script:vxi11}
%------------------------------------------------------------------------------

Devices with LAN extensions for instrumentation, or LXI-enabled devices, provide communication protocols over an Ethernet connection. The VXI-11, designed by the VXIbus Consortium in 1995 \cite{vxibusvmebus}, is one of these protocols. It is a Transmission Control and Internet Protocol (TCP/IP) specification that allows communication with LXI instruments and provides mechanisms to send ASCII strings (or SCPI commands) to control them. 

In order to establish communication with a device over the network, the VXI-11 protocol makes use of the Open Network Computing/Remote Procedure Call (ONC/RPC) standard, which is based on TCP/IP. With RPC, the system
controller or the laptop, which is defined as the RPC client, uses remote functions in the measurement instrument or in our case, the oscilloscope, which is defined as the RPC server. These functions are defined in the RPC Language (RPCL) definition provided by the VXI11 specification. Therefore one can use a protocol generator (rpcgen) to create client and server stubs in the C language \cite{franksen2000vxi11,kopp2006linux}. 

A simple solution to use this protocol from a Linux-operated laptop was created by Steve D. Sharples in 2006 \cite{sharples2008vxi11}. The program specifications can be found in the link provided in the citation. Basically, the program contains a numbers of intuitive and interactive functions that one can use to send and receive data from an ethernet-enabled device, which can be easily located through its IP address.

%==============================================================================
\chapter{VMM3a Noise Performance: Systematic Measurements}
\label{sec:vmmstg2}
%==============================================================================

With the automatic procedure, systematic measurements of the VMM noise became attainable. However, before moving on, it was important to understand certain parameters which are essential to the calculations--namely, the efficiency of the internal VMM test pulse injector and the temperature impact. These matters are discussed in  \cref{sec:vmmstg2:pulser,,sec:vmmstg2:temp}, respectively. Finally, \cref{sec:vmmstg2:enc} of this chapter summarizes the full-scale of ENC measurements that were executed, expanding across 10 VMMs in total.

%------------------------------------------------------------------------------
\section{Test Pulse Circuit Efficiency}%
\label{sec:vmmstg2:pulser}
%------------------------------------------------------------------------------

In order to calculate the ENC correctly, it was essential to certify the quality of the internal VMM test pulses which are used. This matter can be approached by a number of aspects, such as the linearity of the pulse amplitude with the pulser step, the accuracy of the input charge and hence of the internal capacitor $C_s$, as well as the pulse's behavior with the ADC offset.

At this point it should be mentioned that previously taken measurements were performed without the VMM analog output buffers enabled. Enabling these buffers, which are summarized in \cref{tab:buffers}, is essential in order to achieve a stable and reliable output signal, especially for large amplitudes, as communicated by ATLAS VMM experts.

\begin{table}[htbp]
\centering
\begin{tabular}{|c|c|}
\hline
\textbf{Buffer} & \textbf{Description} \\ 
\hline
\texttt{sbft} & Enables TDO analog output buffer.\\
\hline
\texttt{sbfp} & Enables PDO analog output buffer.\\
\hline
\texttt{sbfm} & Enables MO analog output buffer. \\
\hline
\texttt{sbmx} & Routes analog monitor to PDO output.\\
\hline
\end{tabular}
\caption{VMM analog output buffers \cite{polychronakos2018vmm3a}.}
\label{tab:buffers}
\end{table}

\begin{figure}[htbp]
  \centering
  \begin{tabular}{cc}
  \includegraphics[width=9cm]{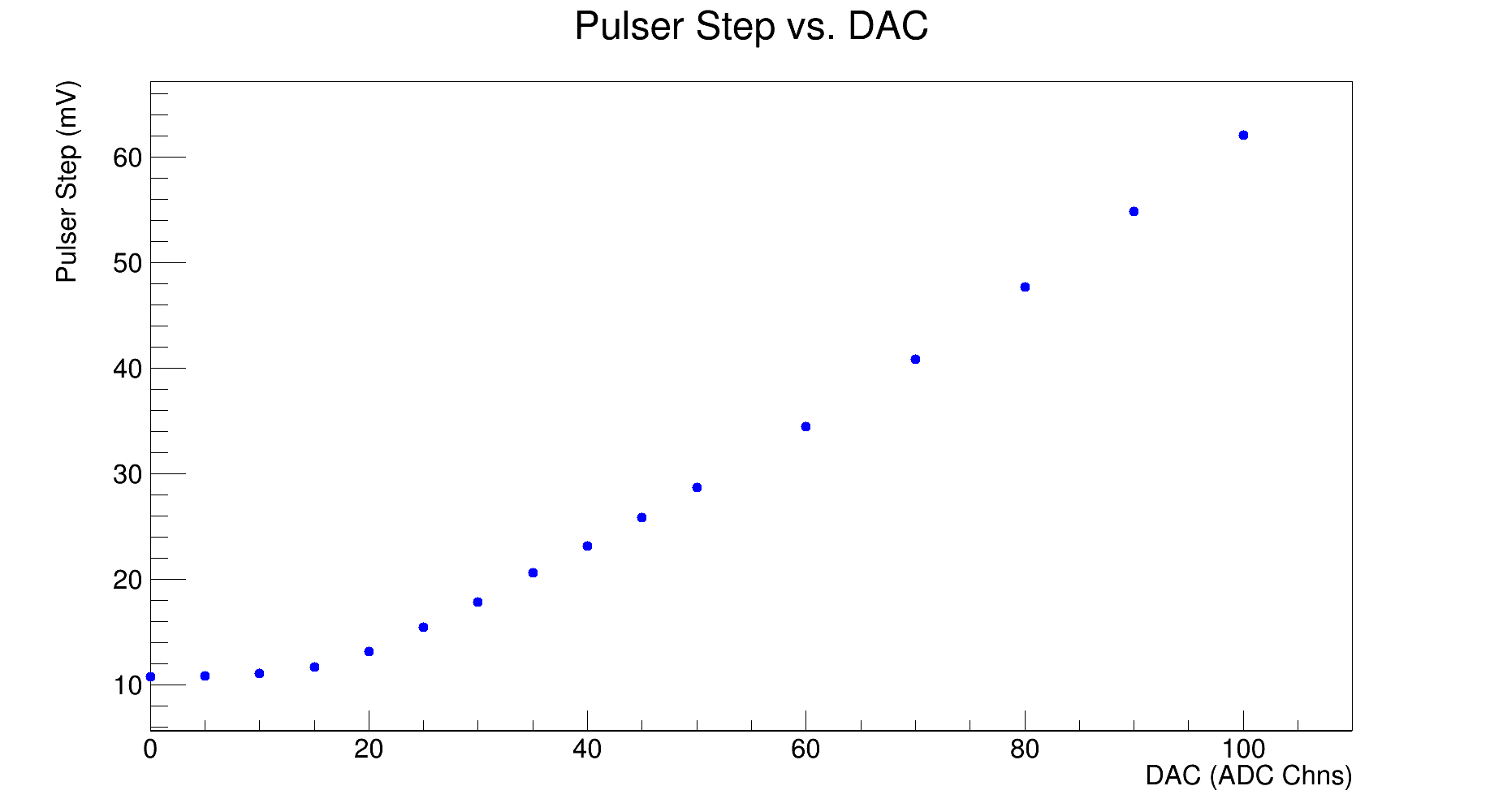}
  \end{tabular}
  \caption{Pulser step calibration from DAC to mV, new buffers applied.}
  \label{fig:stepcal}
\end{figure}

As a follow-up, a new pulser-DAC/step-in-mV calibration was done. The method is the same as the one described in \cref{sec:vmmstg1:osci}. The result is shown in \cref{fig:stepcal}. Linearity is achieved after an onset of DAC $\sim$ 20. At DAC 50, the step now amounts to $\sim$ \SI{28}{mV}. With \SI{0.3}{pF} test capacitance, this amounts to a total input charge of 52\SI{434.45}{e} (\cref{eq:DQin}).

\subsection{The pulse amplitude vs. the pulser step}

\begin{figure}[htbp]
  \centering
  \begin{tabular}{cc}
  \includegraphics[width=8cm]{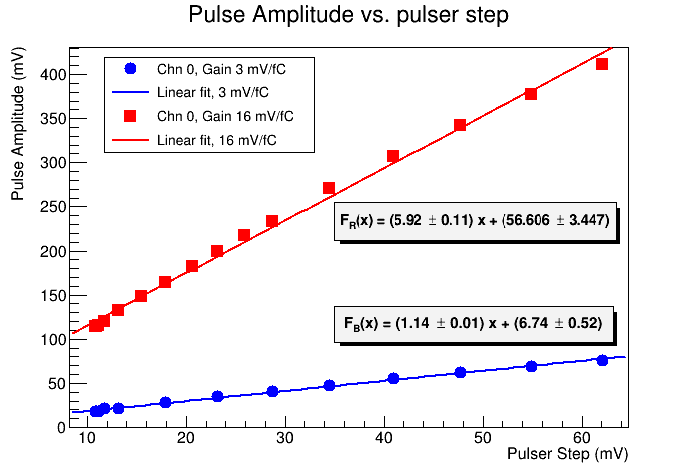}
  \end{tabular}
  \caption{Pulse amplitude/Pulser step linearity.}
  \label{fig:pulsedac}
\end{figure}

In order to check the behavior of the test pulse with different pulser steps, the pulse amplitude was measured for two different gains of the VMM (channel 0, peaktime \SI{200}{ns}). This measurement is shown in \cref{fig:pulsedac}. From this plot it can be induced that the pulse amplitude is more or less linearly coupled to the pulser step with slight saturation effects observable for the highest gain and amplitude. 

Additionally, the scaling with the gain is correct. Between gain \SI{3}{mV}/fC and \SI{16}{mV}/fC, a scale factor of 16/3 $\sim$ 5.3 is expected. From the slopes of the fitting curves in \cref{fig:pulsedac}, one measures a scaling factor of: 
$$ scale = \frac{5.92}{1.14}\approx 5.19$$
Hence, the pulse amplitude vs. the pulser step behavior is sufficiently close to the expected standard. 

\subsection{Test Charge}

\begin{figure}[htbp]
  \centering
  \begin{tabular}{cc}
  \includegraphics[width=11cm]{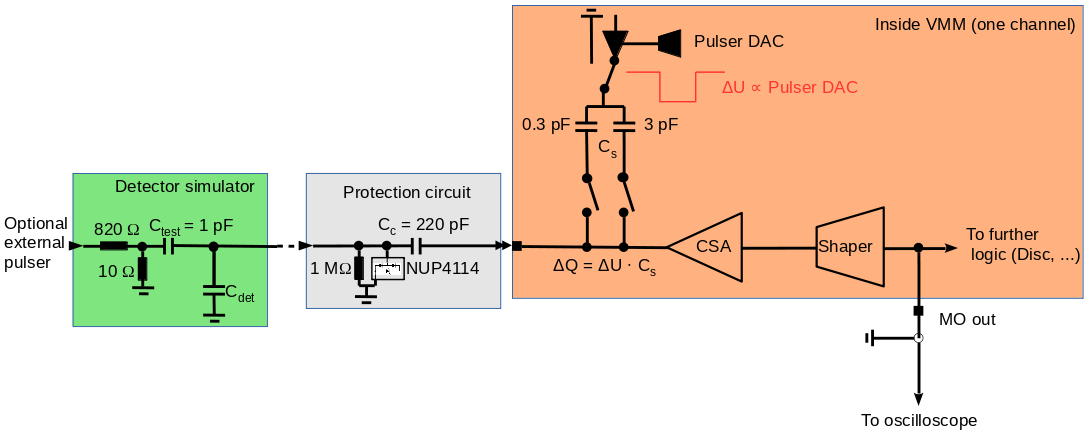}
  \end{tabular}
  \caption{VMM pulser circuit with external capacitance and protection circuit, [M. Lupberger, private communication].}
  \label{fig:pcircuit2}
\end{figure}

In order to verify the internal VMM test capacitance of \SI{0.3}{pF}, external capacitors were connected to the VMM, as shown in \cref{fig:pcircuit2}, in order to send external pulses. In this figure, a \SI{1}{pF} (1\%) external capacitance of a detector simulator is used. This detector simulator was designed by Chistrian Honisch and was used in the measurements with external capacitance which are going to be shown later in this chapter. Additional external capacitors were of \SI{5.6}{pF} (10\%) and \SI{10}{pF} (10\%). Unlike the detector simulator's capacitance which has a connector, the other two capacitors were soldered directly to the input in order to avoid influences from the protection circuit. 

The philosophy behind this method is that if the test charge of verified external capacitances results in the same pulse amplitude as the pulse amplitude of internal VMM test pulses, then this can verify that the internal test capacitance is indeed \SI{0.3}{pF} and help define the precision of this value. 

\begin{figure}[htbp]
  \centering
  \begin{tabular}{cc}
  \includegraphics[width=8.5cm]{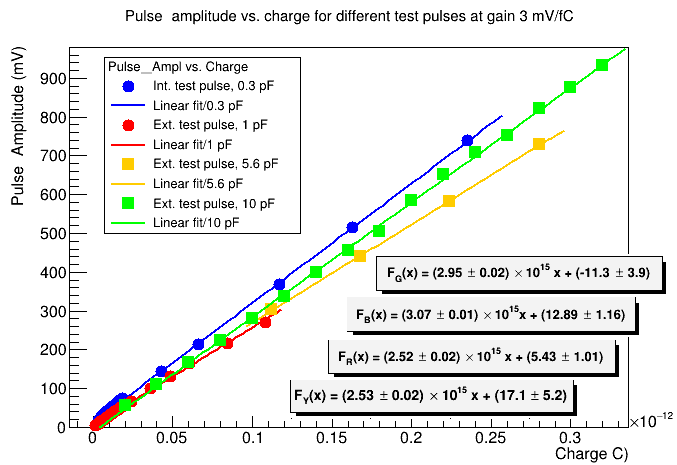}
  \end{tabular}
  \caption{Pulse amplitude-Charge relation for internal and external test pulses.}
  \label{fig:amplcharge}
\end{figure}

In \cref{fig:amplcharge} the pulse amplitude is plotted against the input charge for all external capacitors as well as the internal capacitor for gain \SI{3}{mV}/fC of the VMM. Furthermore, the resulting data were fitted with linear polynomial functions. The functions equations are displayed in the figure. Function $F_G(x)$ corresponds to the \textbf{G}reen curve, function $F_B(x)$ corresponds to the \textbf{B}lue curve and likewise for the rest of the curves. 

The slopes of the curves display the gain of the VMM more or less correctly ($\sim$ $3\times 10^{15}$ mV/C). As we saw, the external capacitances have a precision of 10\% and 1\%.  For the internal capacitance, we expect a precision also in the 1\% range. It has also been observed that the amplitude of the test pulse does not fluctuate much between different channels, which means that the channel-to-channel variation of the internal capacitor is also in the 1\% range. It should be noted that the external capacitance of the detector simulator is influenced from the protection circuit and the detector capacitance, as can be seen in \cref{fig:pcircuit2}. This is not the case for the other two external capacitors. Hence, the detector simulator capacitance might be the one with the highest precision, but also the one with the least reliable connection method. Calculatively, from this graph, a less than 9\% variation was found for the internal capacitance of $\sim$ \SI{0.3}{pF}. 

\subsection{Internal ADC measurement}

\begin{figure}[htbp]
  \centering
  \begin{tabular}{cc}
  \includegraphics[width=7.3cm]{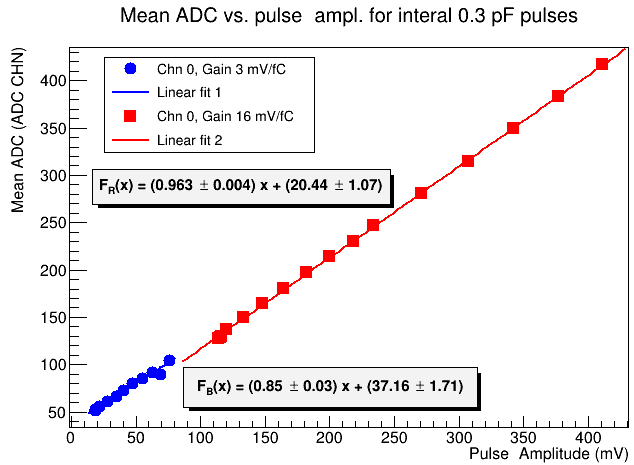} &
  \includegraphics[width=7.3cm]{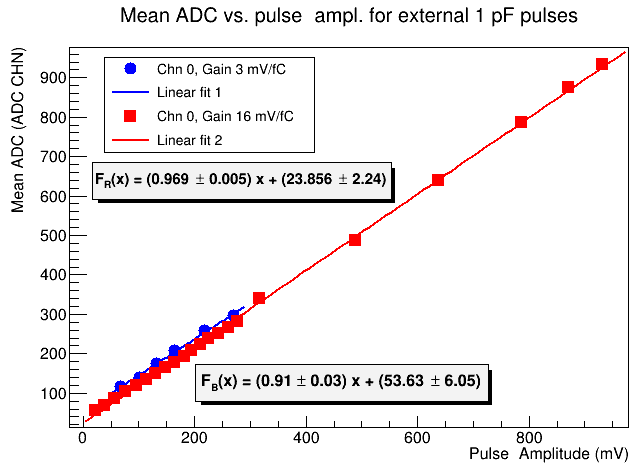} \\
  \includegraphics[width=7.3cm]{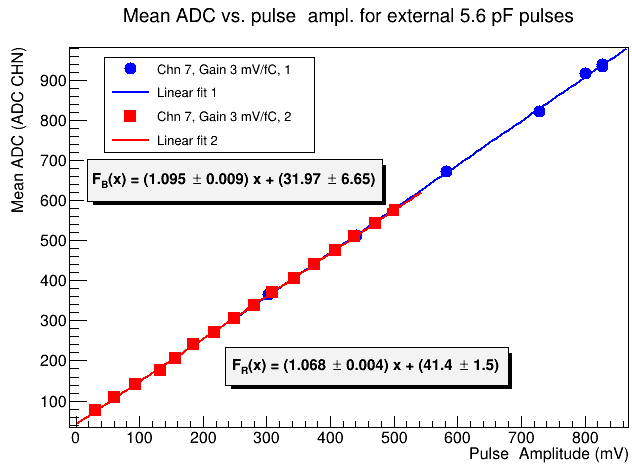} &
  \includegraphics[width=7.3cm]{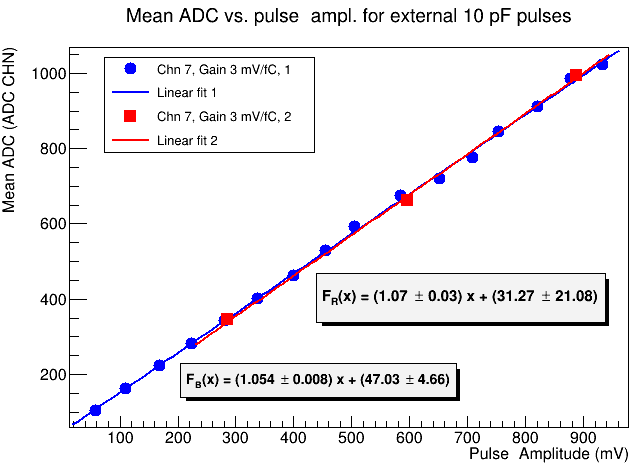} \\
  \end{tabular}
  \caption{Mean ADC-Pulse amplitude relation for internal \SI{0.3}{pF} pulses (top left), external \SI{1}{pF} pulses (top right), external \SI{5.6}{pF} pulses (bottom left), external \SI{10}{pF} pulses (bottom right).}
  \label{fig:adcpulse}
\end{figure}

Another important certification is the behavior of the pulse amplitude with the internal ADC. A non-linear behavior, for example, would indicate issues with the MO output. In order to cross-check this, the mean ADC vs. the pulse amplitude was plotted for all capacitances mentioned in the previous subsection, across different values of the gain and VMM channel. The result from this practice is shown in \cref{fig:adcpulse}. 
In specific, the top figures compare the behaviors between \SI{0.3}{pF} internal capacitance (left) and \SI{1}{pF} external capacitance (right), while they both combine points for gains 3- and \SI{16}{mV}/fC of channel 0 and peaktime \SI{200}{ns} of the VMM. The bottom figures display behaviors for the external capacitances of \SI{5.6}{pF} (left) and \SI{10}{pF} (right), this time combining points from two different measurements with the same settings (gain \SI{3}{mV}/fC, channel 7, peaktime \SI{200}{ns}). All points were fitted with linear polynomials, the functions of which are all displayed in the figures.   

As it can be observed, the ADC offset shows a 1:1 scaling with the pulse amplitude with less than a 10\% error across different parameters. This is expected because the ADC has a range of 1024, which is used to measure a test pulse on the baseline at about \SI{180}{mV} until the maximum voltage of the chips of \SI{1200}{mV}. The baseline of the ADC itself can be configured, such that the intercept of the fits is an arbitrary number.

\subsection{ENC dependence}

\begin{figure}[htbp]
  \centering
  \begin{tabular}{cc}
  \includegraphics[width=11cm]{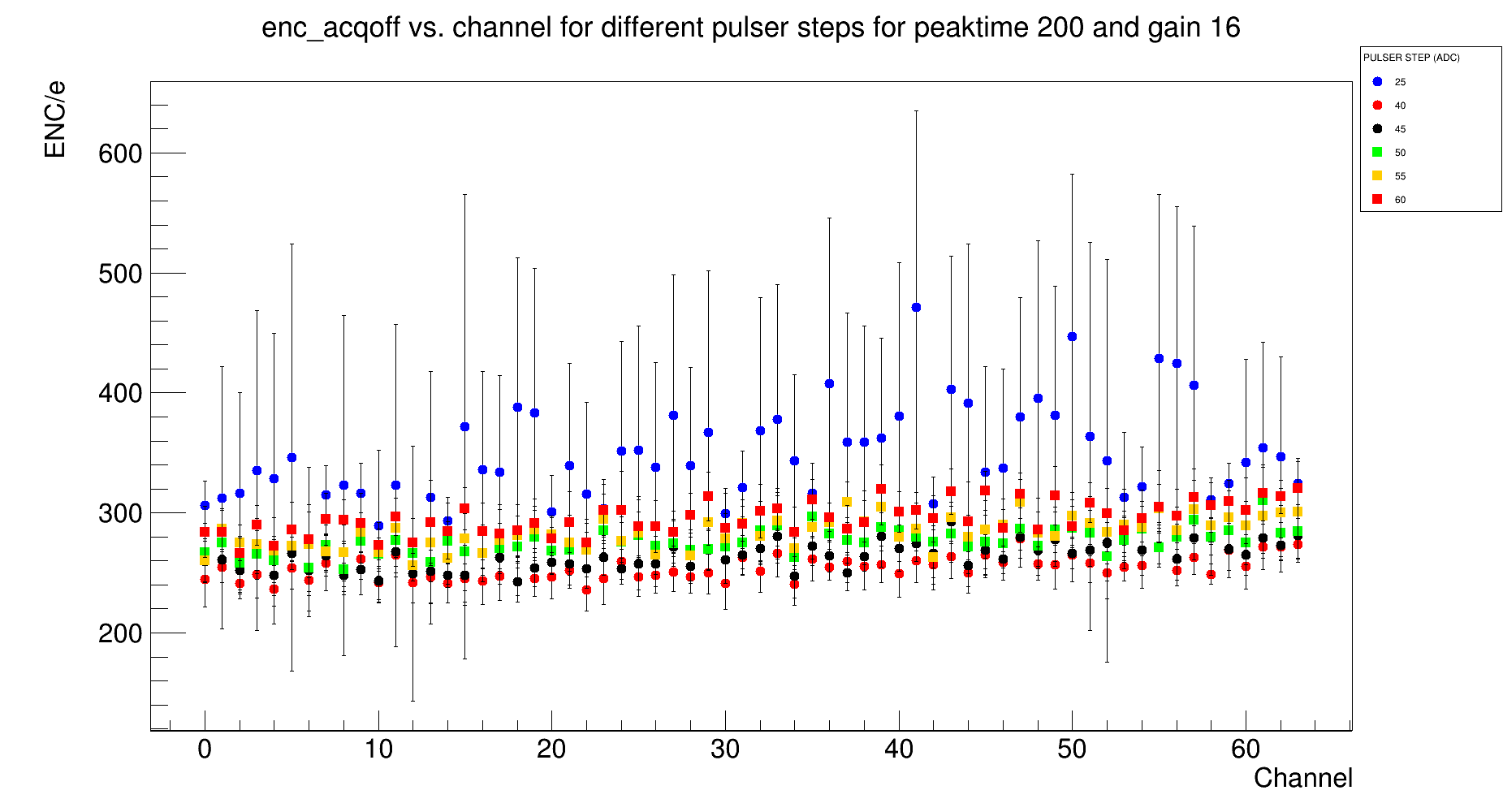}
  \end{tabular}
  \caption{ENC vs. channel for different pulser steps. No visible effect.}
  \label{fig:encstep}
\end{figure}

Finally, it remained to observe whether there was any impact of the test charge on the ENC. For this purpose, the ENC was calculated across different voltage steps for all channels of a single VMM at gain \SI{16}{mV}/fC and peaktime \SI{200}{ns}. The result is displayed in \cref{fig:encstep}. 

The measurements for different test charges are compatible within their error bars and no systematic dependence between the ENC and the pulser step size can be observed. The input charge was varied by sending test pulses induced by different step sizes. For low step sizes, however, the measurement became unstable, which is reflected in the larger error bars. 

%new buffers, show new pulser step/amplitude. shows linearity of pulser dac, hence efficiency. 
%show ENC vs pulser step additionally?
%testing of the input charge, M.Lup.

%------------------------------------------------------------------------------
\section{Temperature Effects}%
\label{sec:vmmstg2:temp}
%------------------------------------------------------------------------------

\begin{figure}[htbp]
  \centering
  \begin{tabular}{c}
  \includegraphics[width=11cm]{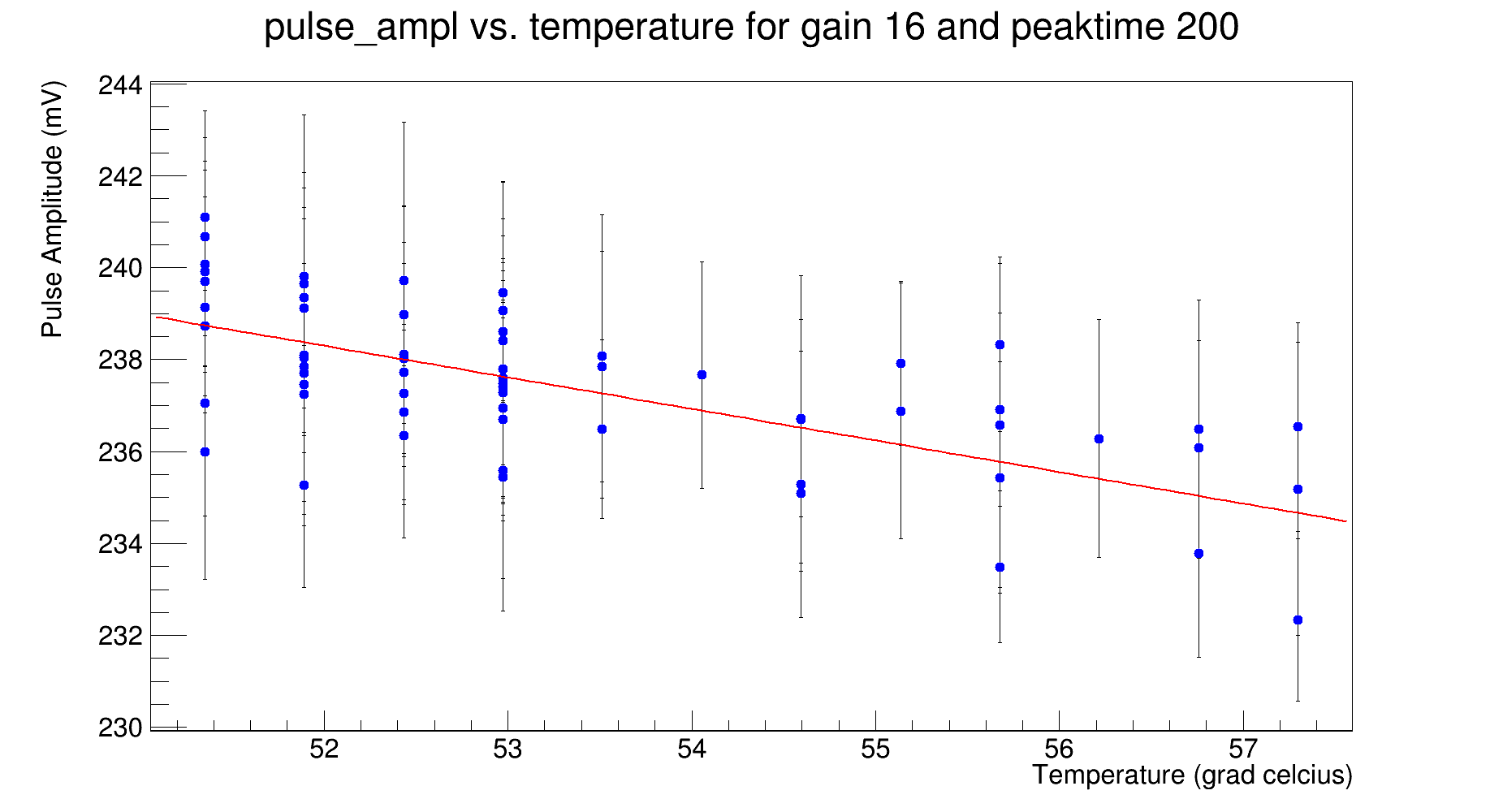}
  \end{tabular}
  \caption{Pulse amplitude vs. VMM core temperature. A 3\% variation is observed.}
  \label{fig:temp}
\end{figure}

As mentioned in \cref{sec:vmmstg1:osci}, the VMM measurement setup is cooled using a fan. However, there are still fluctuations of the temperature observed, which could be due to the change of temperature in the lab as well as due to the varying position of the fan. This fluctuation spreads from $\sim$ \SI{51}{\celsius} to \SI{57}{\celsius} and can be seen in \cref{fig:temp}. 
%fit: f(x) = (-0.686 +/- 0.155)x + (273.992 +/- 8.286)
As mentioned earlier, the temperature is measured in each measurement cycle and hence in this figure, for a single gain and peaktime, the displayed points correspond to all 64 channels. According to this plot, the pulse amplitude shows a $\sim$ 3\% variation with the temperature. 

\begin{figure}[htbp]
  \centering
  \begin{tabular}{cc}
  \includegraphics[width=7.5cm]{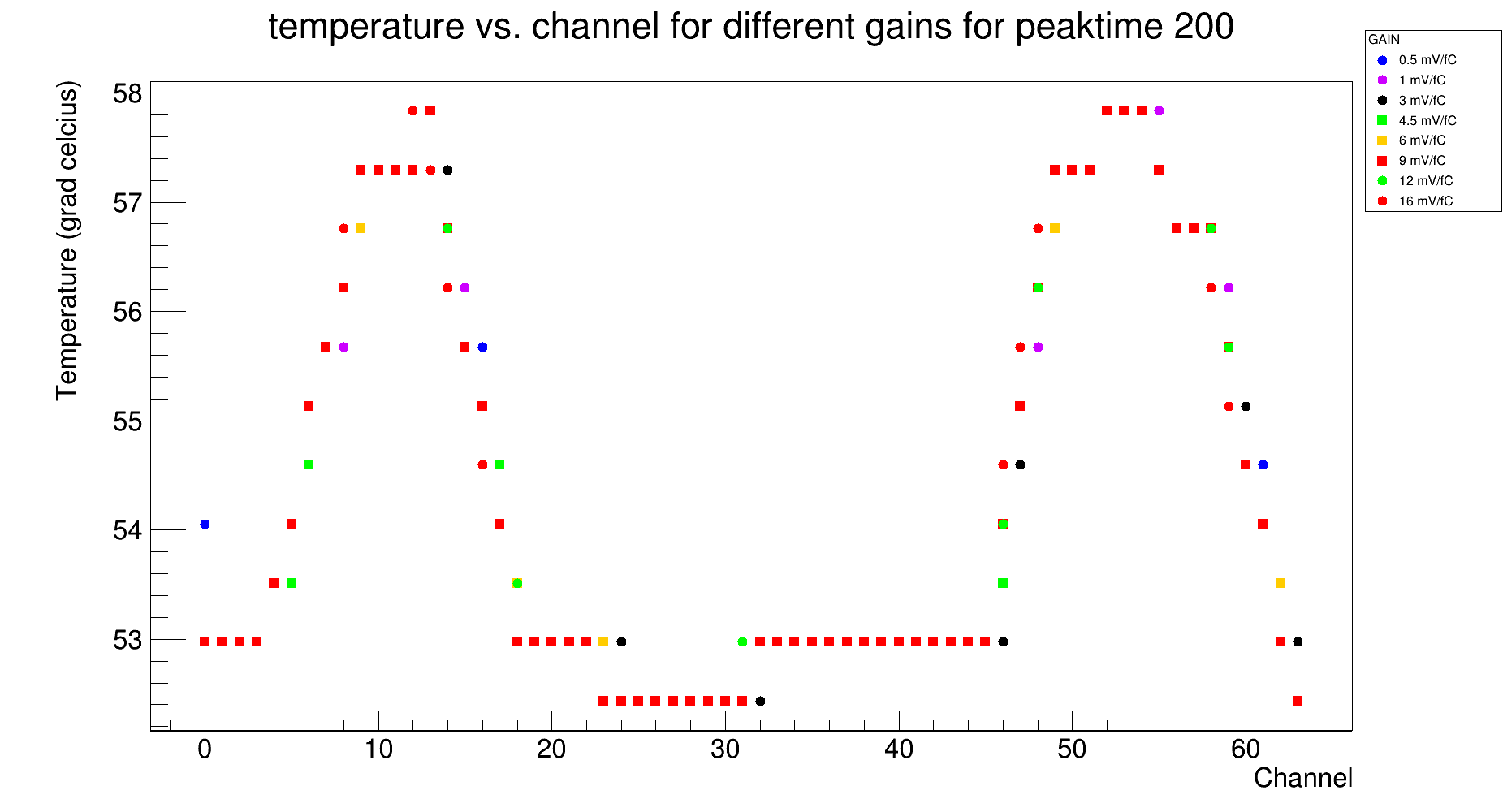} & \includegraphics[width=7.5cm]{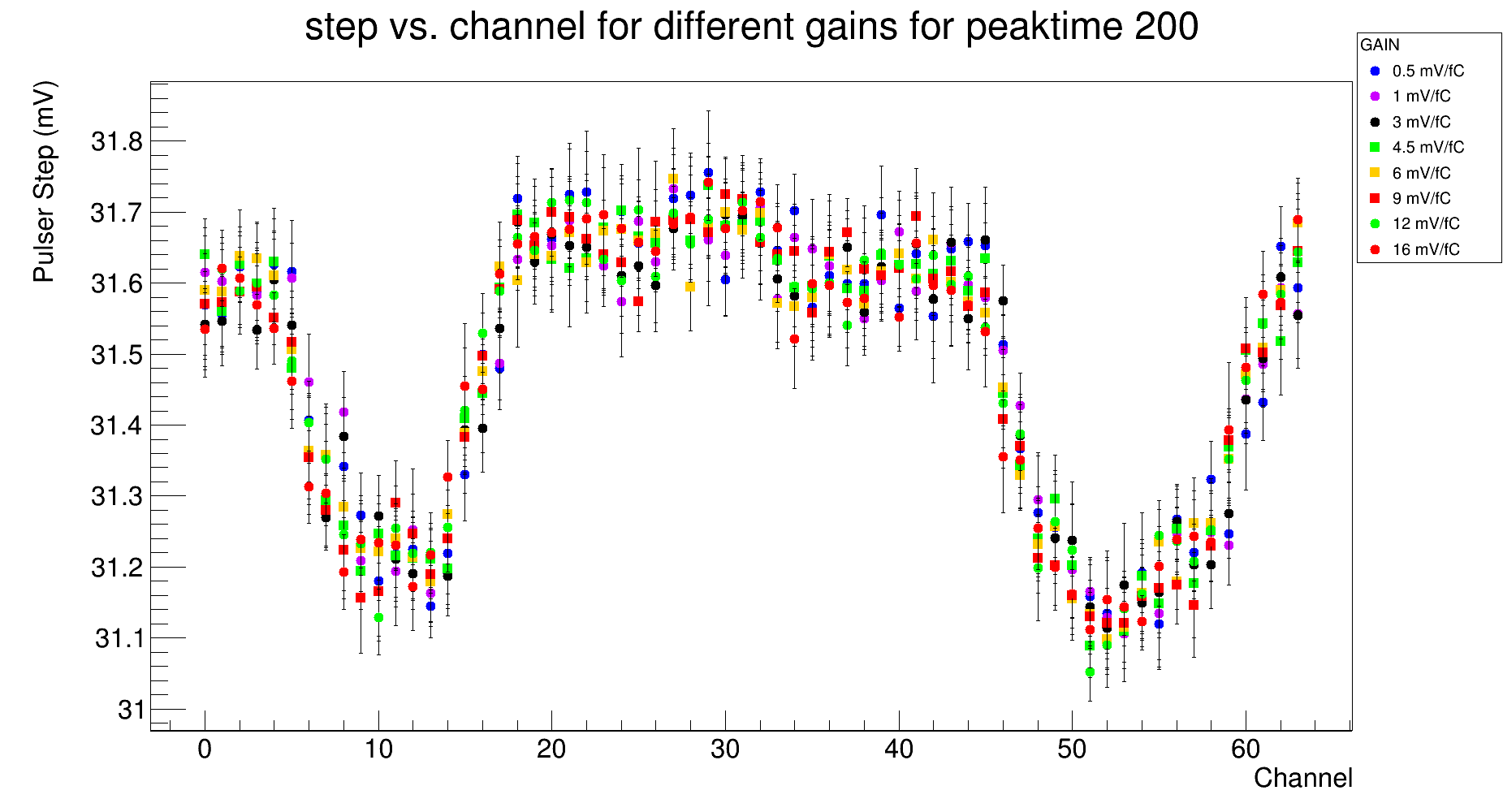} \\
  \includegraphics[width=7.5cm]{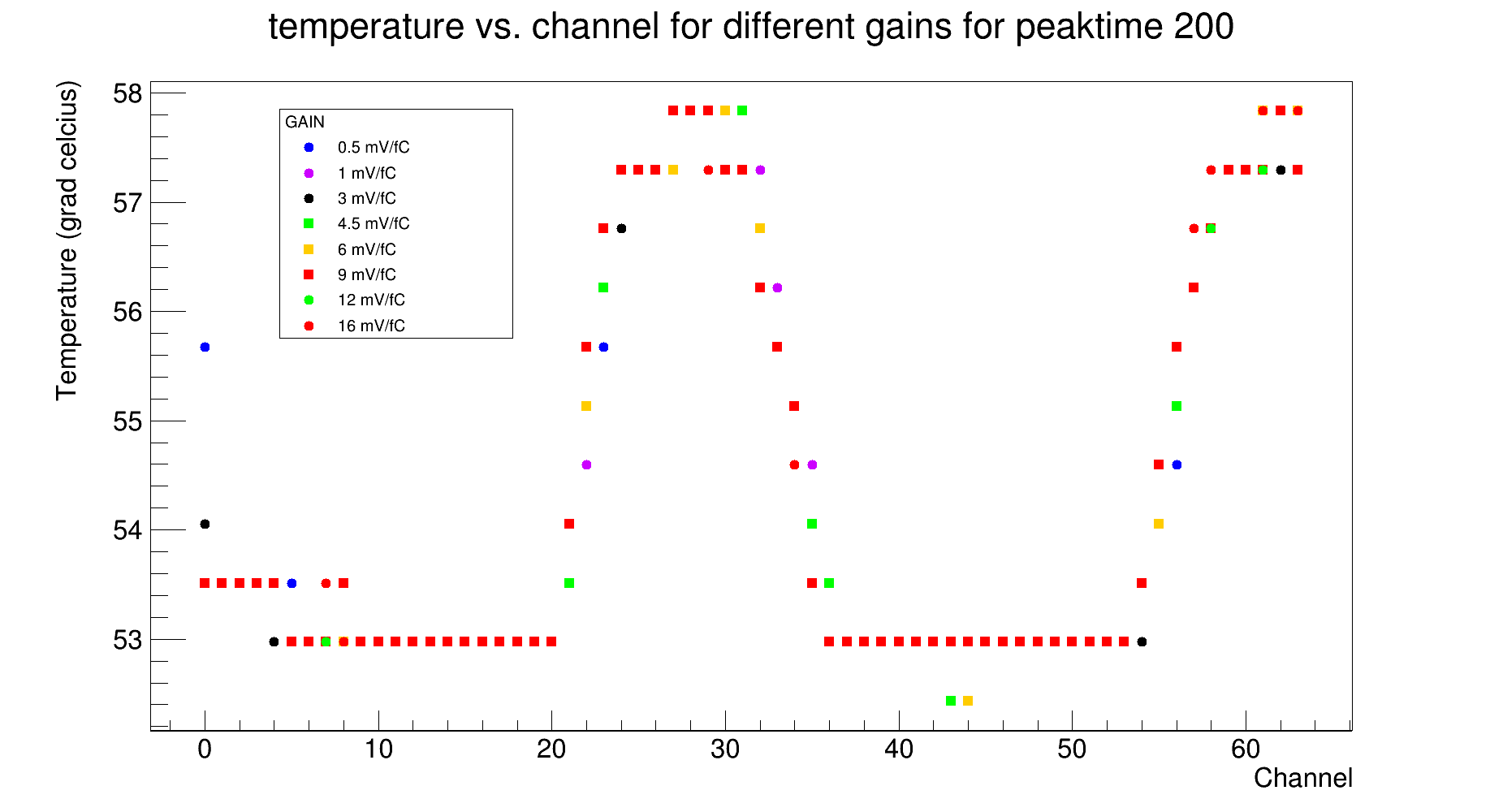} & \includegraphics[width=7.5cm]{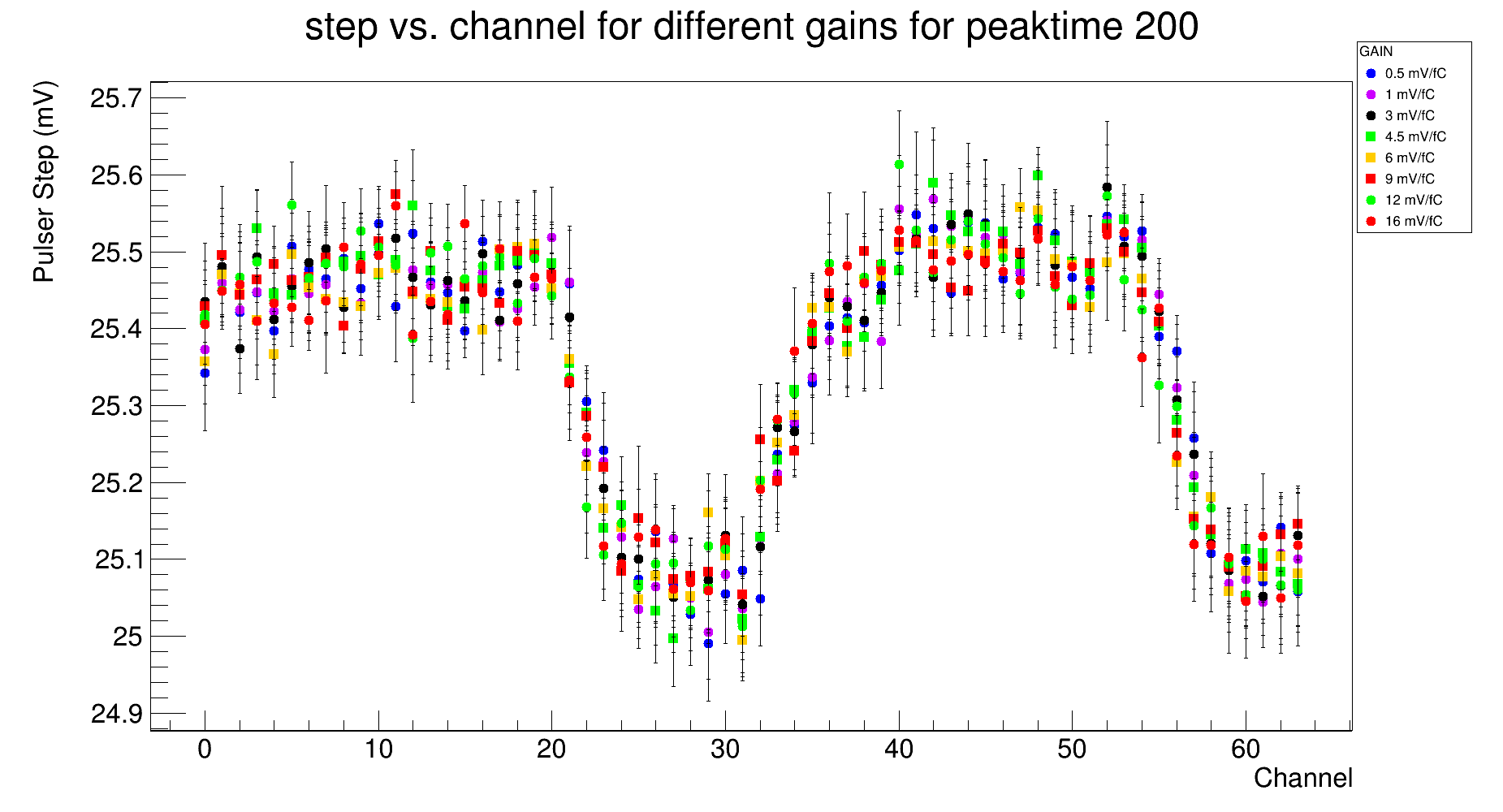}
  \end{tabular}
  \caption{Extreme measurements: Temperature vs. channel (left), Pulser step vs. channel (right), pulser step (ADC CHN): 55 (top), 45 (bottom).}
  \label{fig:extreme_45-55}
\end{figure}

In a more detailed evaluation of this effect, it was discovered that it is not the pulser amplitude by itself, which depends on the temperature, but the pulser step and therefor, the test charge. This can be clearly seen in \cref{fig:extreme_45-55}, where in some extreme measurements, the observed behavior occurred. In this figures, one sees the temperature (left) and the pulser step (right) plotted for all VMM channels for a pulser DAC set to 55 (top) and 45 (bottom) ADC CHN. In the case of the temperature, a $\sim$ 9\% variation is observed, while the pulser step is affected by $\sim$ 1.6\%. 

\begin{figure}[htbp]
  \centering
  \begin{tabular}{cc}
  \includegraphics[width=9.5cm]{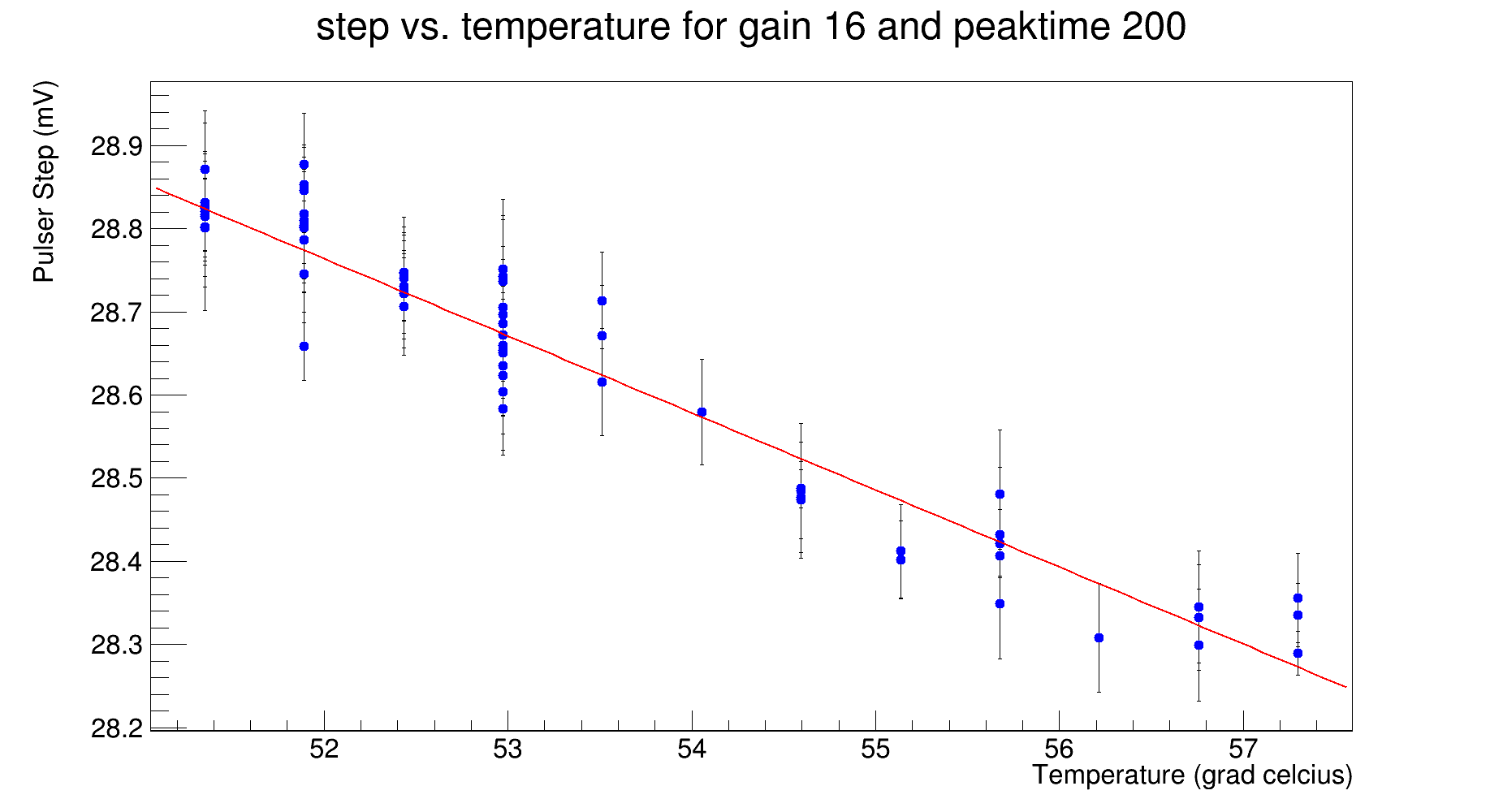}
  \end{tabular}
  \caption{Pulser step vs. temperature, 0.5\% effect.}
  \label{fig:pstep_vs_temp}
\end{figure}

Finally, for a complete overview of the temperature effect, the pulser step was plotted against the temperature for pulser DAC 50 ($\sim$ \SI{30}{mV}). A preliminary fit on \cref{fig:pstep_vs_temp} shows a $\sim$ 0.5\% overall effect.  

%f(x) = (-0.092+/-0.003)x + (33.582 +/- 0.182)

\begin{figure}[htbp]
    \centering
    \includegraphics[width=9.5cm]{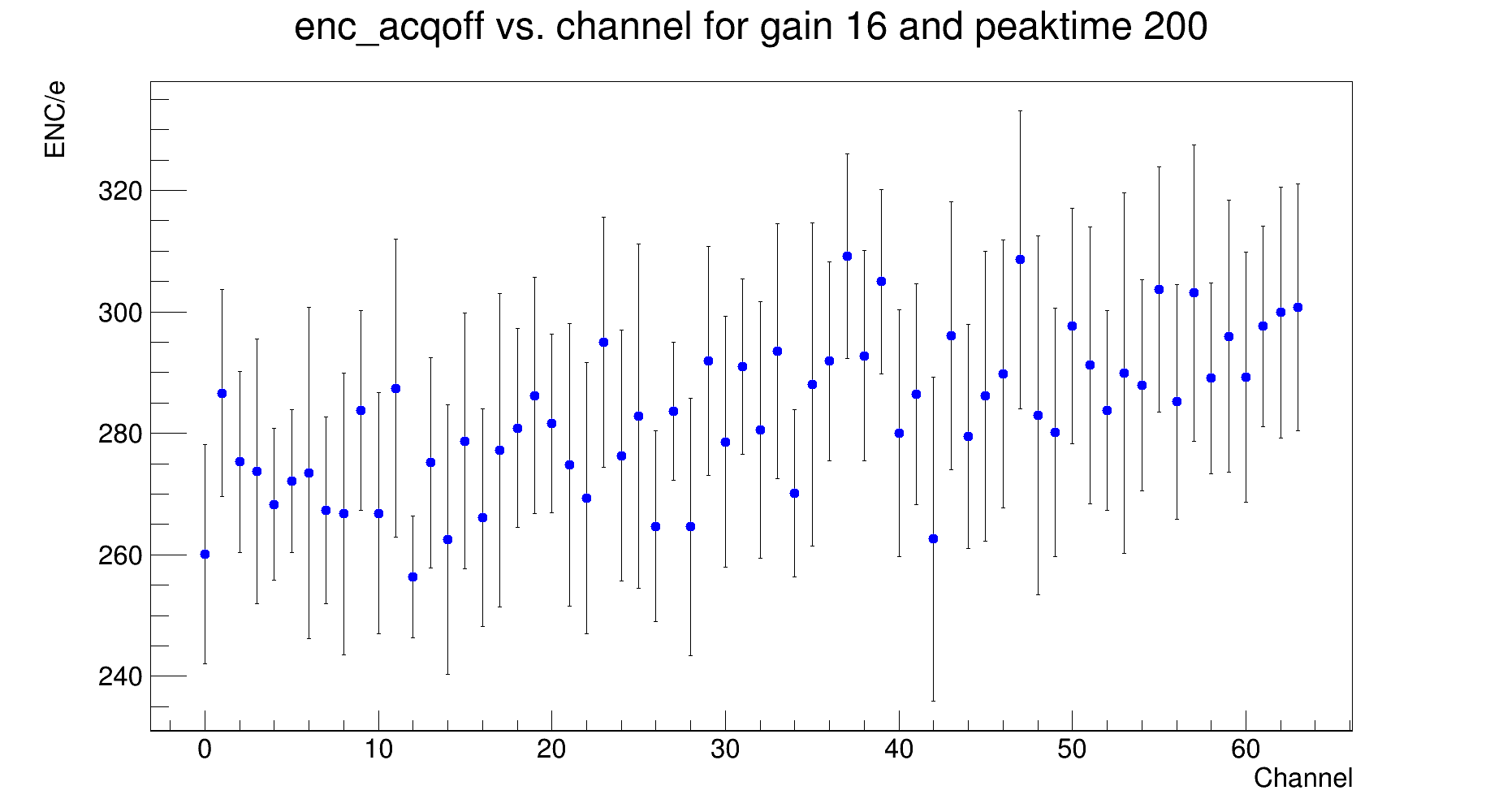}
    \caption{ENC vs. channel for pulser step (ADC CHN): 55 (extreme measurement).}
    \label{fig:encextreme}
\end{figure}

The conclusion of this study is that the known imput charge slightly varies with temperature. However, as the ENC does not depend on the input charge, the ENC is not affected by the temperature, which can be seen from \cref{fig:encextreme}. The ENC is plotted for the same measurement, however, compared to \cref{fig:extreme_45-55}, the effects of temperature variations have vanished. Despite this, for the systematic ENC measurement, all three quantities that enter the calculation of the ENC (baseline RMS, pulser amplitude and pulser step size) have to be measured for each set of configuration, which significantly increases the duration of data taking. 

%Show a figure of the ENC for the same run as with the extreme temperature variation (I hope there is no effect visible).

%------------------------------------------------------------------------------
\section{Systematic ENC Measurements}%
\label{sec:vmmstg2:enc}
%------------------------------------------------------------------------------
 
\begin{figure}[htbp]
  \centering
  \begin{tabular}{cc}
  \includegraphics[width=5cm]{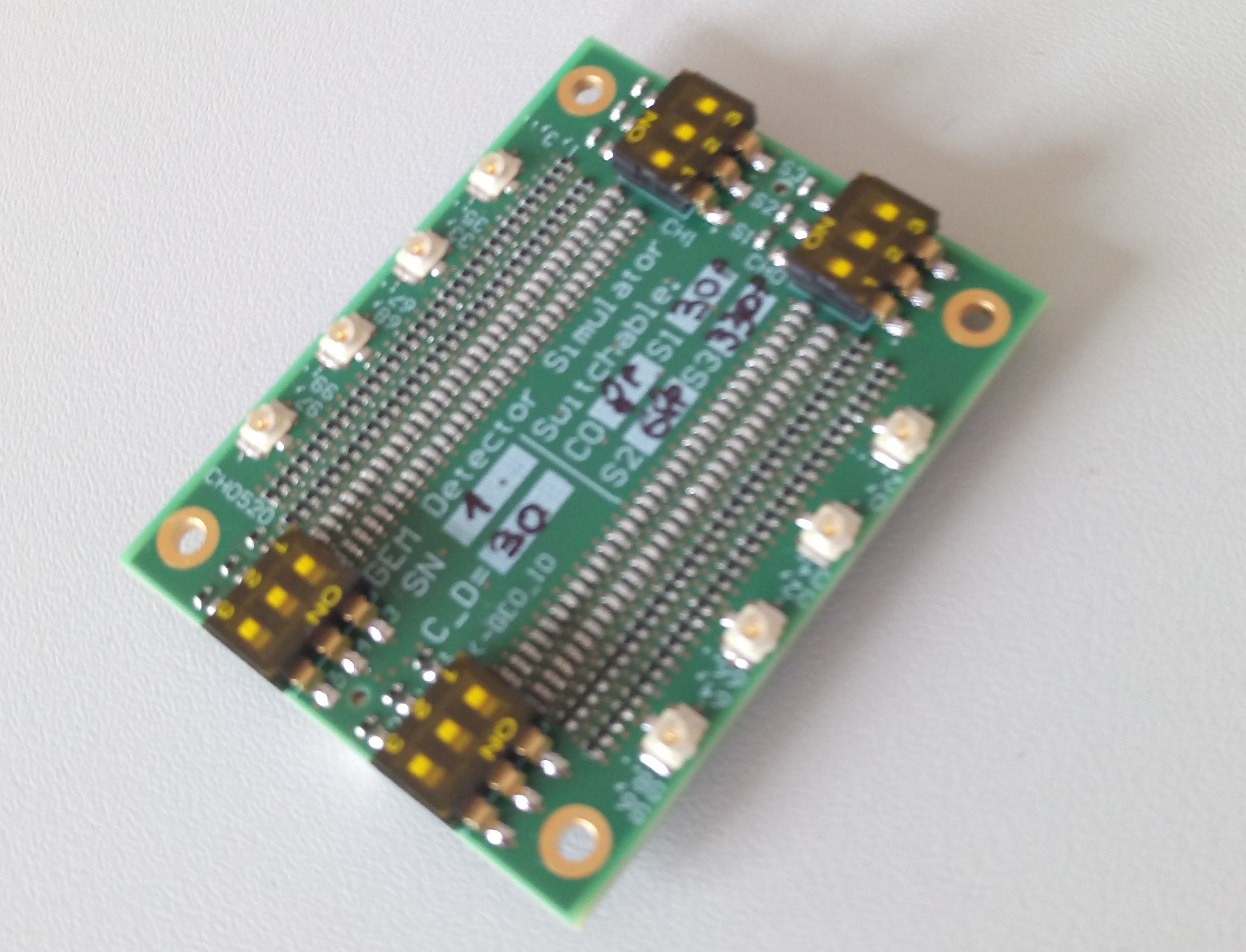} & \includegraphics[width=5.7cm]{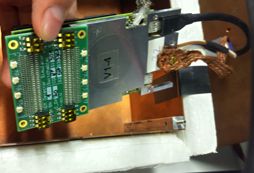} \\
  (a) & (b)
  \end{tabular}
  \caption{Detector simulator, designed by Christian Honisch.}
  \label{fig:ext_cap}
\end{figure}

Having concluded all the necessary checks and with an automated measuring process at hand, it was finally possible to obtain systematic ENC measurements. In total, 10 VMM chips were tested, both with and without an external capacitance. In the case of the external capacitance, the detector simulator shown in \cref{fig:ext_cap} was used, which was designed by Christian Honisch. With the detector simulator PCB, a \SI{30}{pF} external capacitance can be applied on all channels of the VMM. In more detail, by simple connection of the PCB, the default capacitance of \SI{30}{pF} is applied to all channels except for the first or last two, depending on the orientation of the detector PCB. These channels can be either 0 and 1 or 62 and 63, depending on which VMM of the hybrid the oscilloscope is connected to. For these channels, the default capacitance is \texttt{C0} $\rightarrow$ \SI{8}{pF}, while an additional six external capacitances can also be applied to them, from \SI{68}{pF} to \SI{428}{pF}. The different capacitances can be applied by turning the yellow switches than can be seen in \cref{fig:ext_cap}a on and off, while the specific values corresponding to each switch are written on the top, specifically: \texttt{S1} $\rightarrow$ \SI{22}{pF} (not \SI{30}{pF} as is written), \texttt{S2} $\rightarrow$ \SI{68}{pF}, \texttt{S3} $\rightarrow$ \SI{330}{pF}.  \cref{tab:switches} summarizes the complete scope of external capacitances for channels 0,1/62,63 with their switch combinations. 

\begin{table}[htbp]
\centering
\resizebox{0.4\columnwidth}{!}{\begin{tabular}{cc}
\toprule
\textbf{Capacitance} & \textbf{Switch Combination} \\ 
\midrule
\SI{8}{pF} & \texttt{C0}\\
\SI{30}{pF} & \texttt{C0}+\texttt{S1}\\
\SI{76}{pF} & \texttt{C0}+\texttt{S2}\\
\SI{98}{pF} & \texttt{C0}+\texttt{S1}+\texttt{S2}\\
\SI{338}{pF} & \texttt{C0}+\texttt{S3}\\
\SI{360}{pF} & \texttt{C0}+\texttt{S1}+\texttt{S3}\\
\SI{406}{pF} & \texttt{C0}+\texttt{S2}+\texttt{S3}\\
\SI{428}{pF} & \texttt{C0}+\texttt{S1}+\texttt{S2}+\texttt{S3}\\
\bottomrule
\end{tabular}}
\caption{Summary of external capacitances and their PCB switch combinations for channels 0,1/62,63. While the default capacitance  \texttt{C0} does not physically represent a switch on the PCB, it is used here in order to fully comprehend the capacitance values.}
\label{tab:switches}
\end{table}

\subsection{Measurement Log}

Despite the existence of an automated procedure, testing all parameters and all external capacitances for all 10 VMMs was not a viable option since a full-reference measurement takes about 72 hours and since the adjustment of the external capacitance is still something that must be done manually. Hence, a complete measurement exists only for VMM 0 and VMM 1, with the exception that for VMM 0 not all values of the input capacitance were tested. For the rest of the chips, only standard parameters were taken. A complete overview of the measurement log is shown in \cref{tab:log}. 

\begin{table}[htbp]
\centering
\resizebox{1.0\columnwidth}{!}{\begin{tabular}{|c|c|}
\hline
\textbf{VMM ID} & \textbf{Comments} \\
\hline
VMM 0 & full-reference tests with 0, 8 and 30 pF
cap. \\
\hline
VMM 1 & full-ref. tests with 0, 8 and 30 pF
cap.; for channels 62, 63: caps 30 - 428 pF also tested \\
\hline
VMM 2,3,4,5,7,9,10,11 & ref. tests for gains 3, 6, 16 mV/fC, caps 0, 8, 30 pF \\
\hline
\end{tabular}}
\caption{A detailed log of the VMM ENC systematic measurements.}
\label{tab:log}
\end{table}

\subsection{VMM 1 Results}

In this section, a complete overview of the VMM 1 ENC results is provided. While all available gain and peaktime combinations were tested, in the following results a certain focus is applied on measurements with peaktime \SI{200}{ns} and gain \SI{16}{mV}/fC, since these are the settings for which the VMM has the lowest ENC and hence it is most sensitive to interesting effects. 
\begin{figure}[htbp]
  \centering
  \begin{tabular}{c}
  \includegraphics[width=11.5cm]{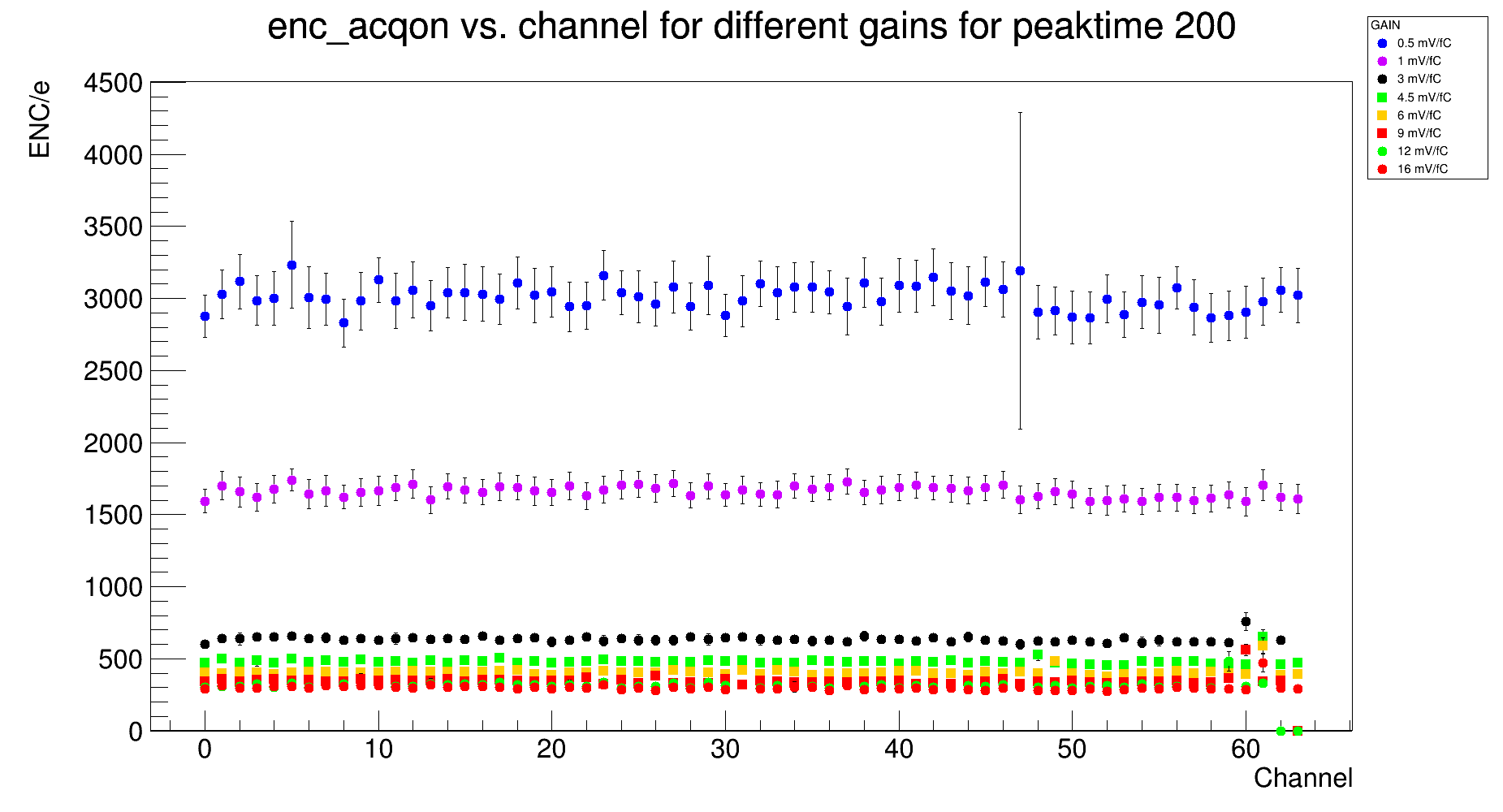} \\
  \includegraphics[width=11.5cm]{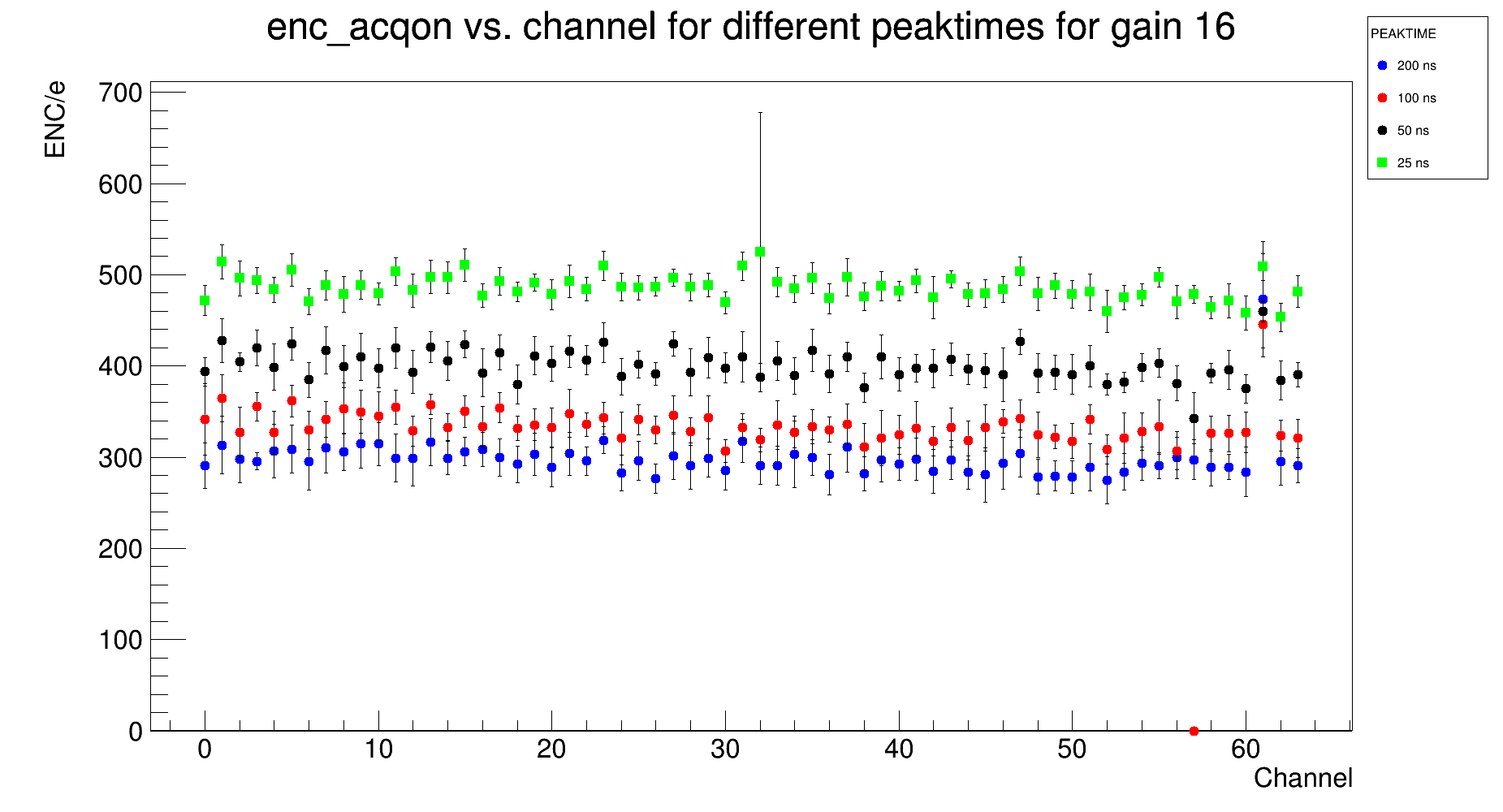} 
  \end{tabular}
  \caption{Top: ENC, ACQ on, vs. channel for different gains and peaktime \SI{200}{ns}. Bottom: ENC, ACQ on, vs. channel for different peaktimes and gain \SI{16}{mV}/fC. Channel 61 is likely damaged. Missing points are set to zero. Large errors occur due to losses in connection to the oscilloscope.}
  \label{fig:vmm1_channel_ENC_0pF}
\end{figure}
Additionally, measurements with gain \SI{6}{mV}/fC are also demonstrated in some cases, since it is a setting which is typically used in applications (largest dynamic range for standard detector gains). Plots with different peaktimes are also provided in \cref{sec:app:vmm-enc}.

Beginning with \cref{fig:vmm1_channel_ENC_0pF}, the ENC for ACQ on is plotted against the increasing VMM channel for different gains (top figure) and peaktimes (bottom figure), using no external capacitance. So it can be seen that with no input capacitance and with data acquisition on, the ENC for the highest gain and peaktime is around $\sim$ 300 electrons. Omitting the lowest gains, the variation of the ENC among different gains at peaktime \SI{200}{ns} is around $\sim$ 300 electrons. Alternatively, the ENC variation among different peaktimes at gain \SI{16}{mV}/fC is around $\sim$ 200 electrons. 

The points with large error bars are due to temporary losses of connection to the oscilloscope, while some missing points were set at zero by default. Missing points occur due to technical issues of the oscilloscope, eg. it gets stuck in a single process. This can be usually prevented if the oscilloscope screen display is set off during long-time measurements and also with adjusting the priority of the VSCI program to the highest on the PC. 

\begin{figure}[htbp]
  \centering
  \begin{tabular}{c}
  \includegraphics[width=11.5cm]{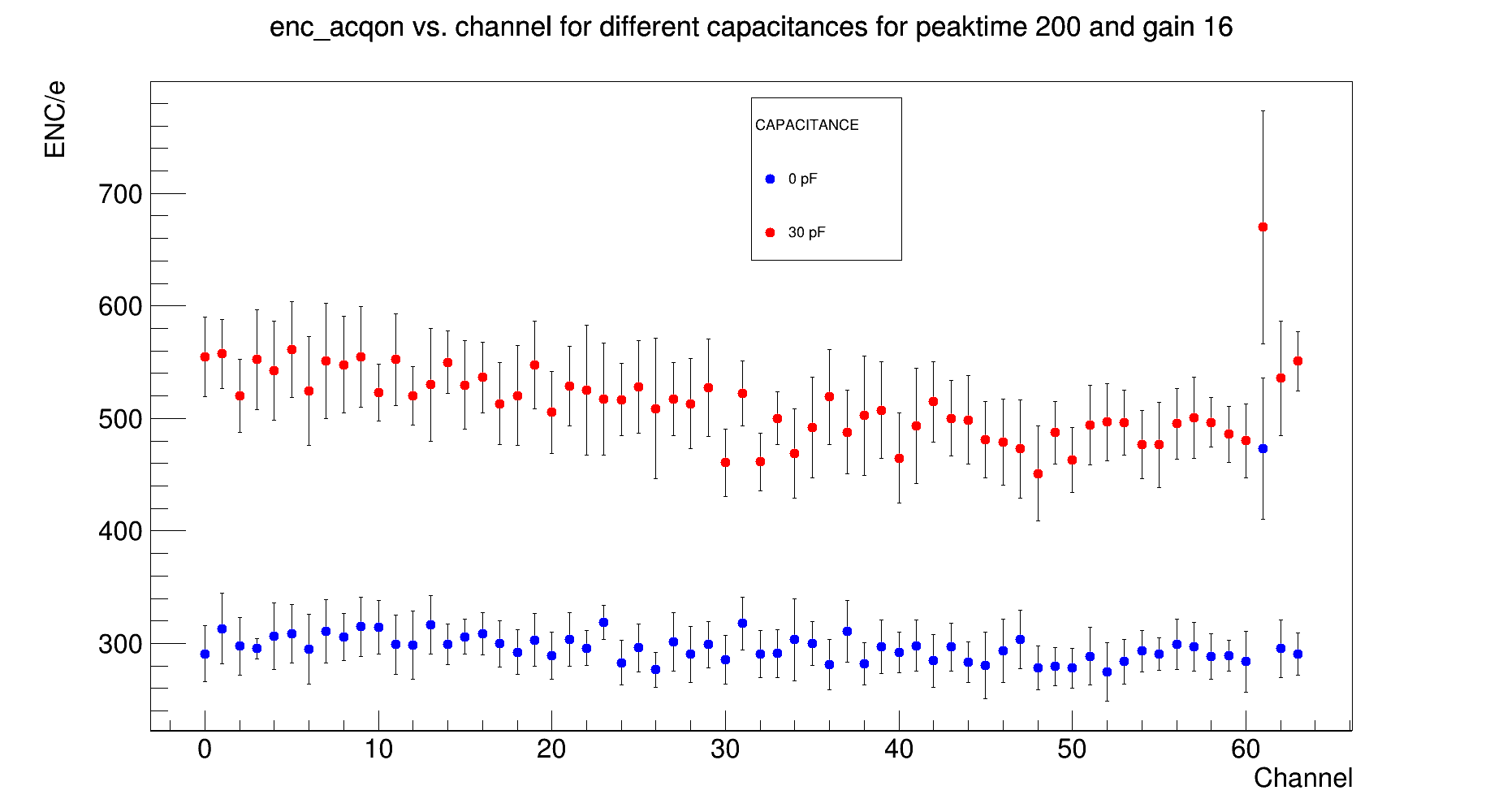} \\
  \includegraphics[width=11.5cm]{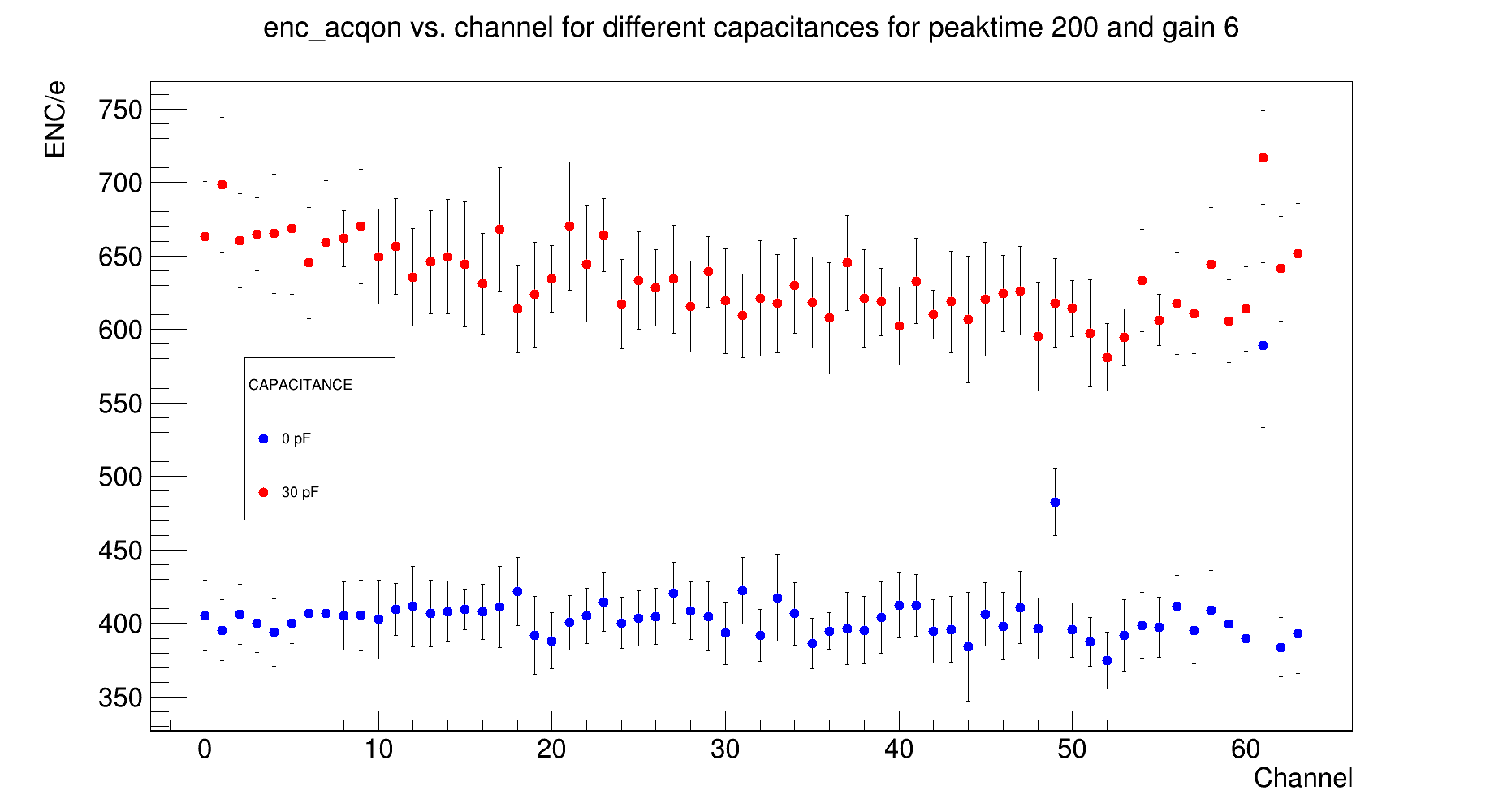}
  \end{tabular}
  \caption{ENC, ACQ on, vs. channel with and without external capacitance for peaktime \SI{200}{ns}, gain \SI{16}{mV}/fC (top) and gain \SI{6}{mV}/fC (bottom).  Channel 61 is likely damaged. In the bottom figure, the blue point at Channel 49 is likely due a random faulty measurement.}
  \label{fig:vmm1_channel_ENC_0-30pF}
\end{figure}

Additionally, in \cref{fig:vmm1_channel_ENC_0-30pF}, the ENC is shown for measurements both with and without input capacitance for the highest gain and peaktime (top) and also for gain \SI{6}{mV}/fC (bottom). It is evident from these plots that an external capacitance of \SI{30}{pF} increases the ENC (with ACQ on) for about $\sim$ 250 electrons. For gain \SI{6}{mV}/fC, the ENC in both examined cases is only $\sim$ 100 electrons higher. In the case of channel 61, the extreme result is most probably due to permanent damage on the channel, since this behavior is consistent among all measurements (with and without capacitance, both gains). 

\begin{figure}[htbp]
  \centering
  \begin{tabular}{c}
  \includegraphics[width=11.5cm]{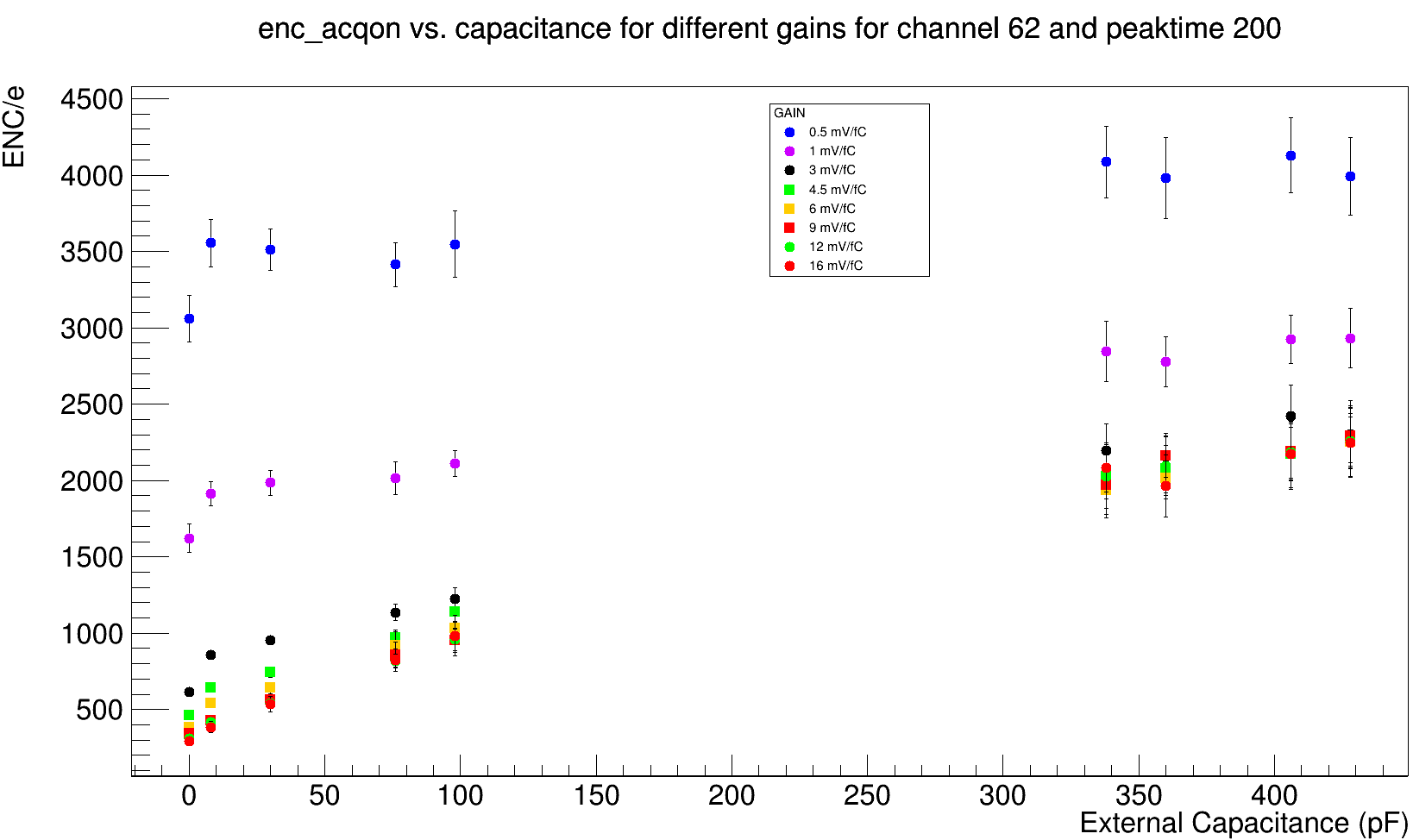} \\
  \includegraphics[width=11.5cm]{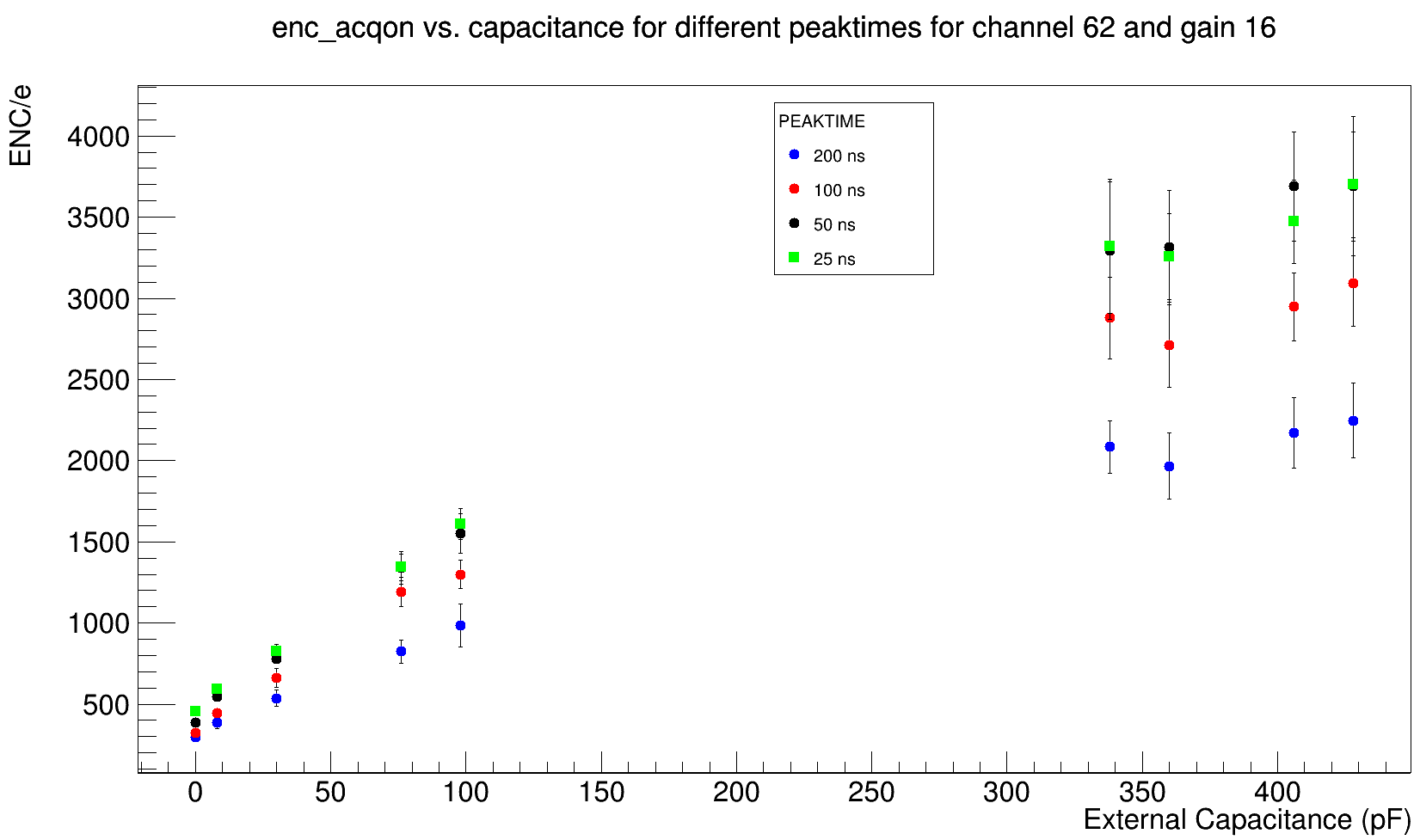} \\
  \end{tabular}
  \caption{ENC, ACQ on, vs. capacitance for channel 62 for different gains (top) and different peaktimes (bottom).}
  \label{fig:vmm1_chn62_cap_ENC}
\end{figure}

As mentioned, higher input capacitances were only available for two channels of the VMM; in this case, channels 62 and 63. Since results for both channels were similar, only results for channel 62 are provided here. Channel 63 can be found in the appendix. In \cref{fig:vmm1_chn62_cap_ENC}, the ENC (ACQ on) is plotted against the increasing input capacitance for different gains (top figure) and peaktimes (bottom figure). As it will be seen in some following graphs, results for gain \SI{6}{mV}/fC are quite similar and thus they are not provided here. In these figures, the observable trends deviate from linearity due to a protection capacitor placed in the protection circuit of the hybrid (see \cref{fig:pcircuit2}) \cite{lupbergerenc}. This capacitor amounts to \SI{470}{pF} and has to also be taken into account in order for linearity to be restored. The ENC can reach up to about $\sim$ 5000 electrons. 

\begin{figure}[htbp]
  \centering
  \begin{tabular}{c}
  \includegraphics[width=11.5cm]{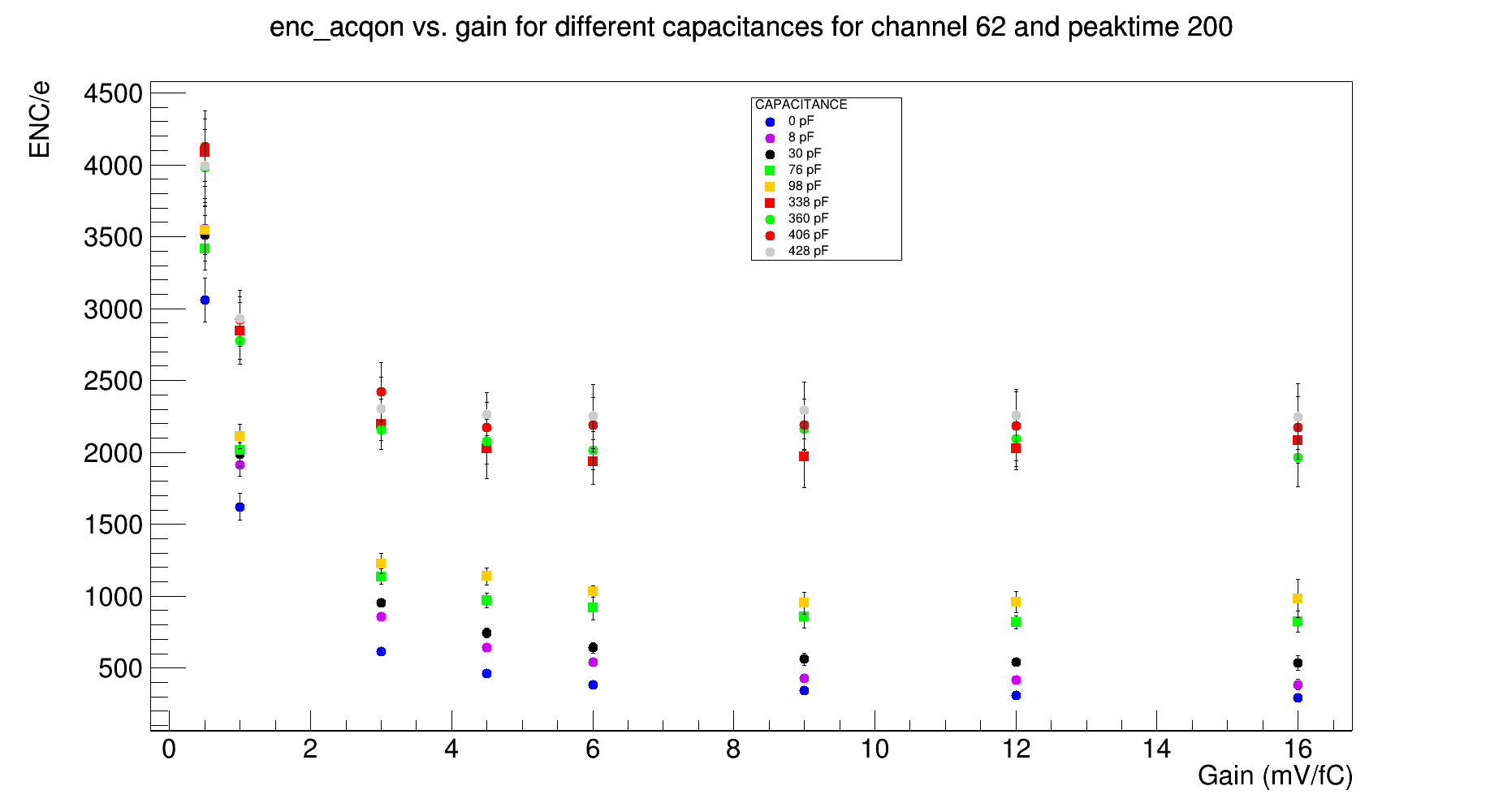} \\
  \includegraphics[width=11.5cm]{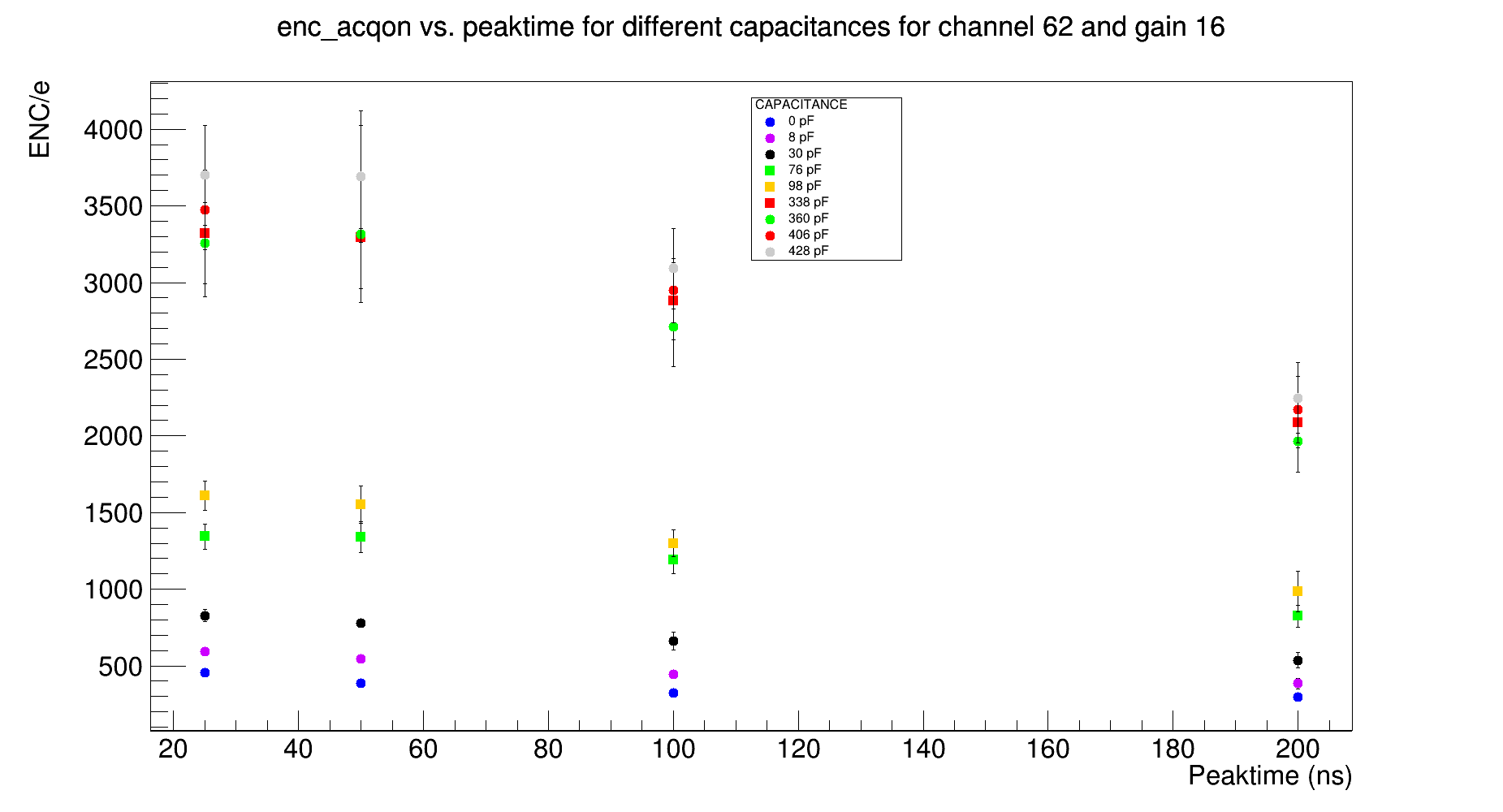}
  \end{tabular}
  \caption{ENC, ACQ on, vs. gain (top) and vs. peaktime (bottom) for channel 62 for different capacitances.}
  \label{fig:vmm1_chn62_gain-peaktime_ENC}
\end{figure}

\cref{fig:vmm1_chn62_gain-peaktime_ENC} shows the ENC (ACQ on) of channel 62 plotted against the increasing gain and peaktime for all different capacitances. It can be seen from the top figure that the results down to gain \SI{6}{mV}/fC are quite similar and thus \cref{fig:vmm1_chn62_cap_ENC} (bottom) can be used as reference for all these gains. On the other hand, results for lower peaktimes show a greater variation. More elaborate plots for lower peaktimes are again provided in the appendix.

\subsection{Multiple VMM Results}

In this section, results for the total of 10 VMMs that were measured are presented in comparison. 

\begin{figure}[htbp]
  \centering
  \begin{tabular}{c}
  \includegraphics[width=13cm]{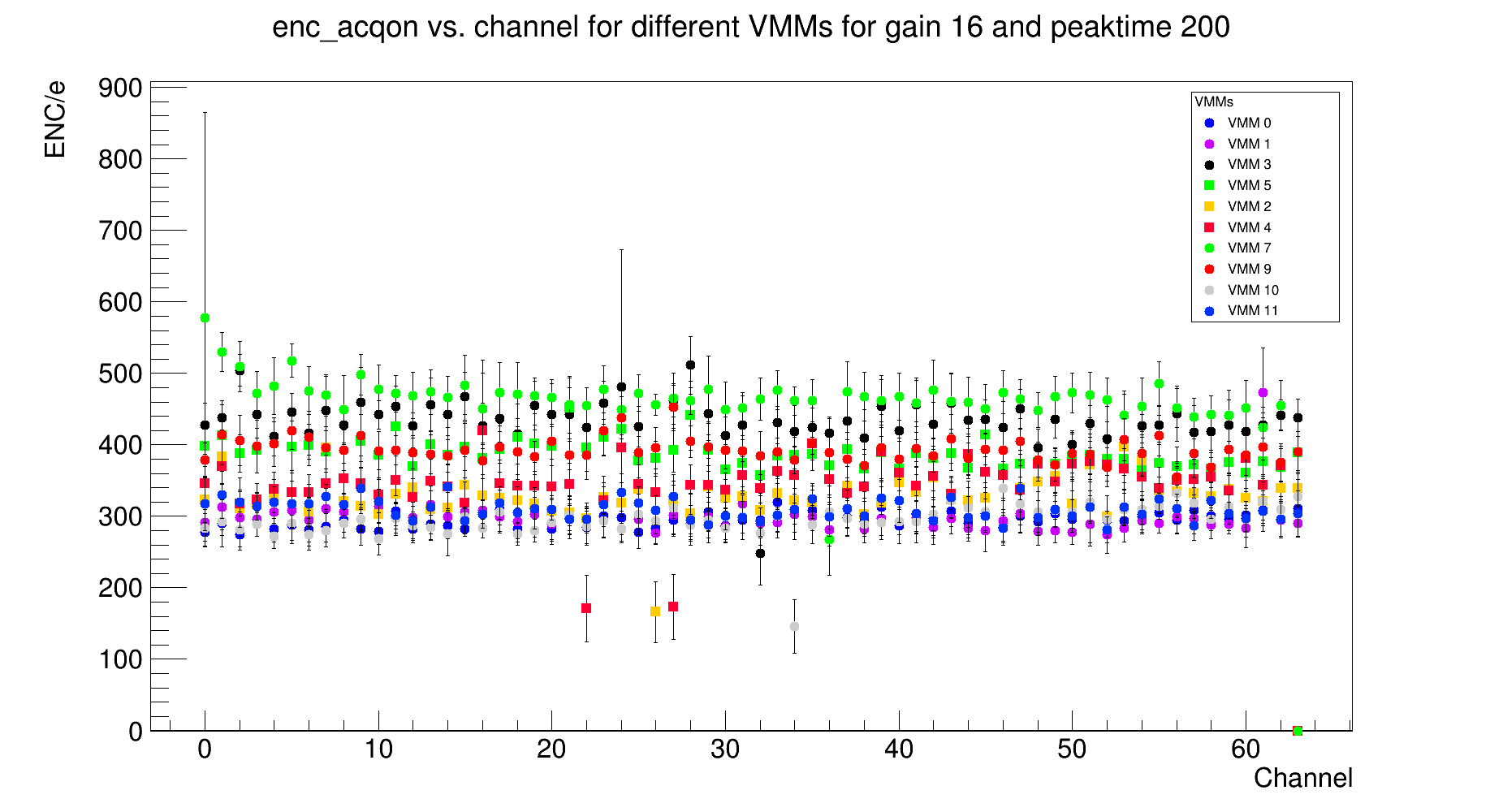} \\
  \includegraphics[width=13cm]{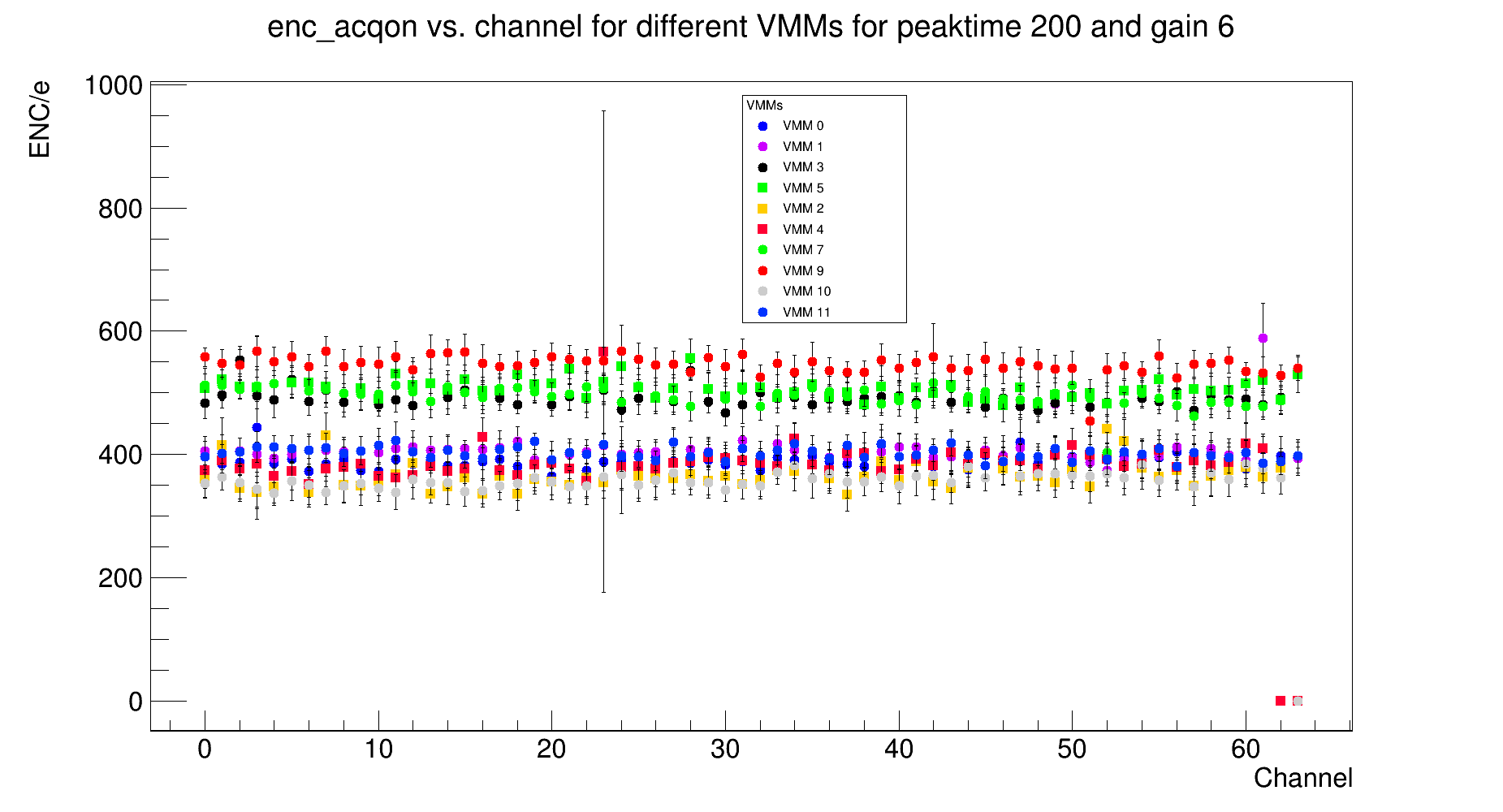}
  \end{tabular}
  \caption{ENC, ACQ on, vs. channel for all VMMs, peaktime \SI{200}{ns} and gain \SI{16}{mV}/fC (top), gain \SI{6}{mV}/fC (bottom). A $\sim$ 16\% effect can be observed. Points with large error bars are due to random faulty measurements. Missing points are set to zero by default.}
  \label{fig:ENC_allvmms}
\end{figure}

Specifically, \cref{fig:ENC_allvmms} shows the ENC (ACQ on) plotted against all channels for all VMMs at peaktime \SI{200}{ns} and gains \SI{16}{mV}/fC and \SI{6}{mV}/fC. In total, among all VMMs the ENC shows a $\sim$ 16\% variation. Several factors have to be taken into account in order to understand these deviations. One of them is that as it was seen in \cref{sec:theory:vmm}, \cref{fig:vmmsrs}, the VMM hybrid has two inserts for the data cable. Normally, the insert furthest from the power connector was used. However, for VMMs 2,4,10 and 11 this data connection proved very noisy and so the firmware was changed in order to use the data connector closest to the power connector. Additionally, for some measurements the data cable was changed and the grounding was different. Again, temporary losses in connection and technical issues can lead to large error bars or even missing points in the measurements. The latter were set to zero by default. 

\begin{figure}[htbp]
  \centering
  \begin{tabular}{c}
  \includegraphics[width=13cm]{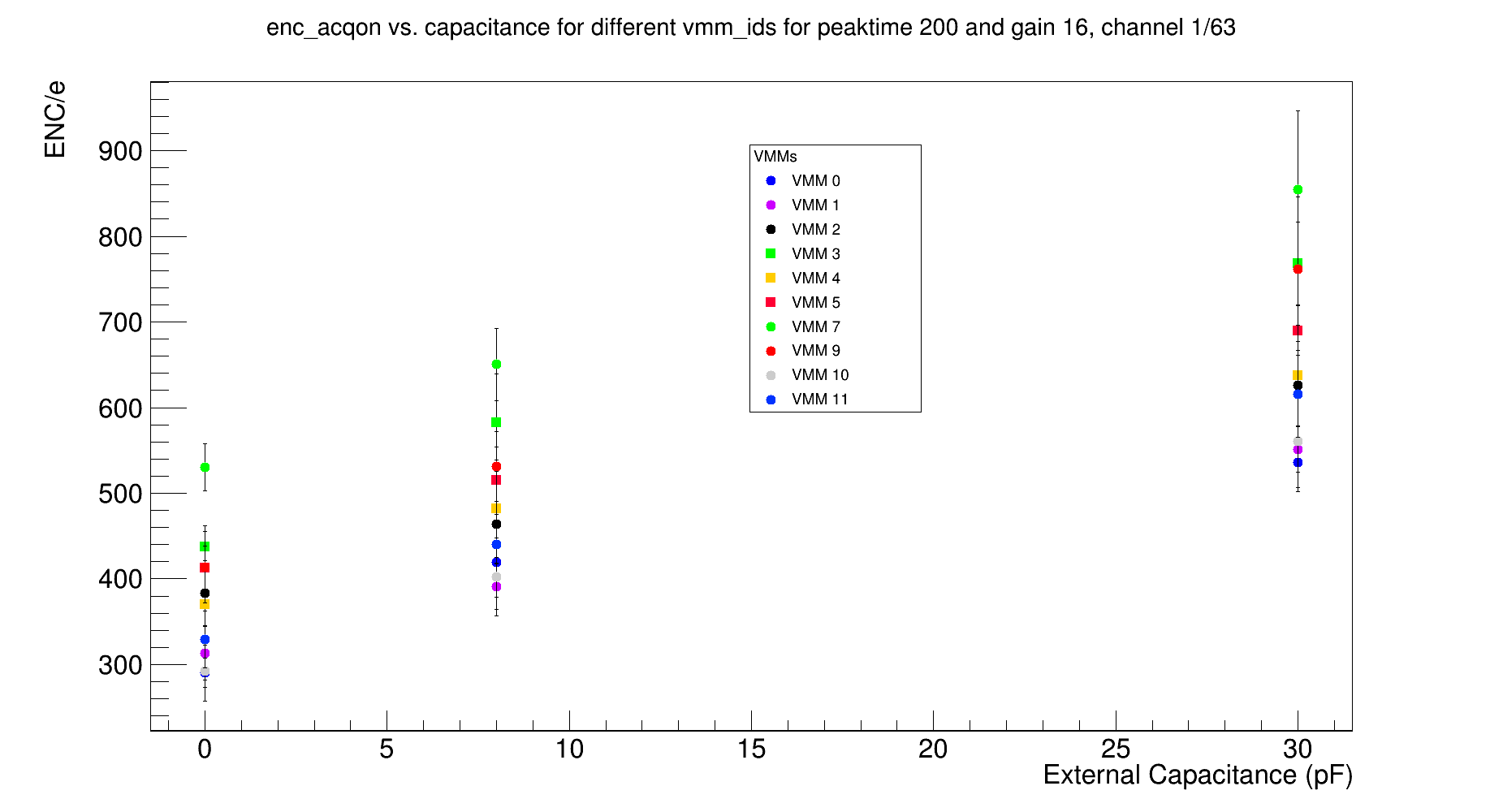} \\
  \includegraphics[width=13cm]{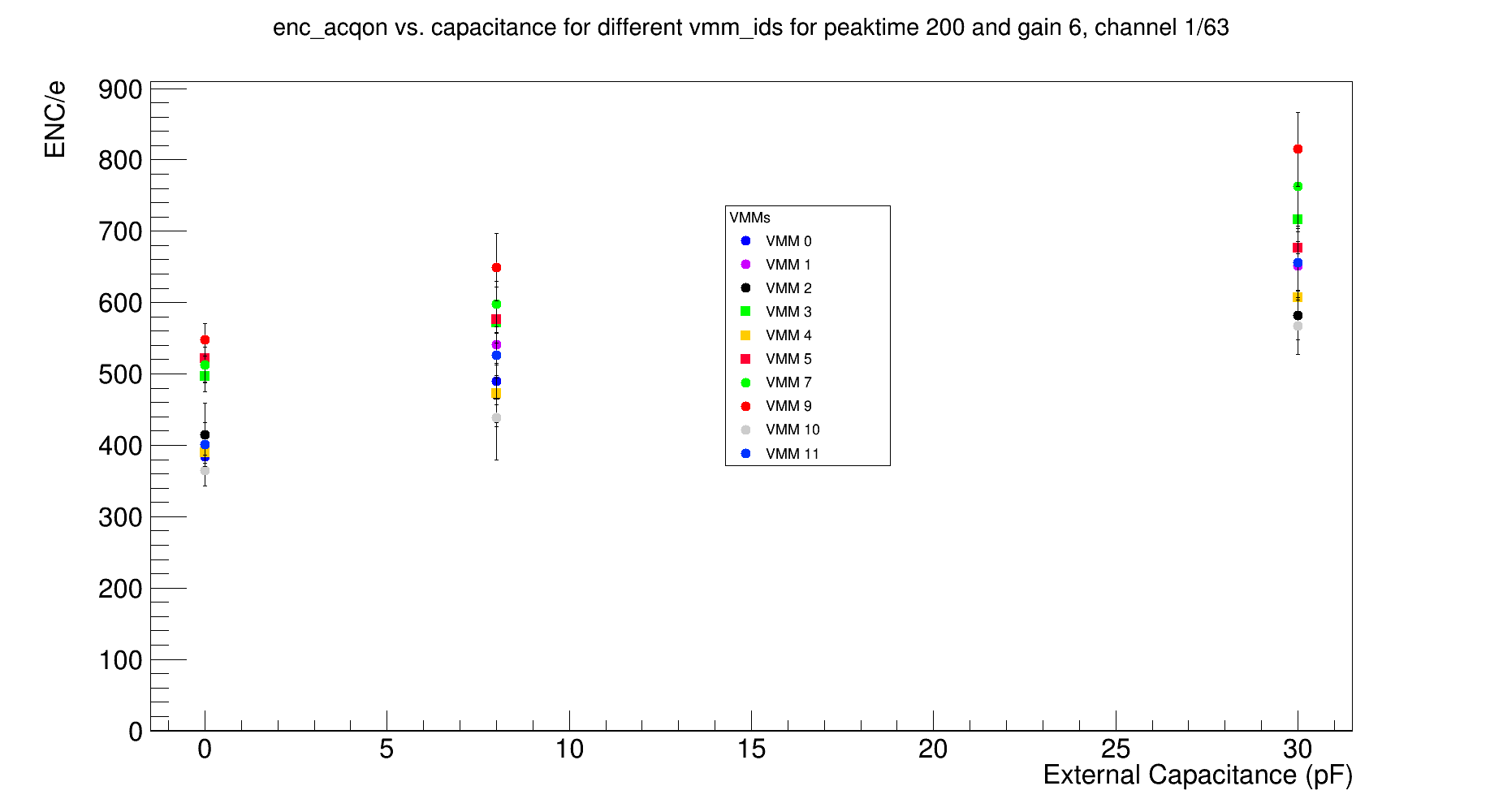}
  \end{tabular}
  \caption{ENC, ACQ on, vs. capacitance for all VMMs, peaktime \SI{200}{ns}, and gain \SI{16}{mV}/fC (top), gain \SI{6}{mV}/fC (bottom). Channel 63 is measured for the odd-numbered VMMs, while channel 1 is measured for the even-numbered VMMs.}
  \label{fig:ENC_cap_allvmms}
\end{figure}

Additionally, for all VMMs, for channels 0, 1 (or 62, 63) measurements were also taken for external capacitance of \SI{8}{pF} and \SI{30}{pF}. Plots for the ENC (ACQ on) against the external capacitance for all VMMs can be found in \cref{fig:ENC_cap_allvmms}. Results are shown both for gain \SI{16}{mV}/fC  and gain \SI{6}{mV}/fC. In these figures, for the odd-numbered VMMs, channel 63 is the one plotted, while for the even-numbered VMMs, channel 1 is the displayed channel. Again, the measuring of alternate channels is due to the geometry of the detector simulator. Similarly, in this case, a $\sim$ 16\% effect was calculated. The 250-electron difference between using the detector simulator and not using it remains stable for most of the VMMs.  

\subsection{Comparison to Relevant Works}

Finally, in order to understand the quality of these results, comparisons to previous similar measurements were made. Namely, measurements from the ATLAS and GDD groups. It should be noted that in these works, results were shown for only a limited set of parameters, which are shown in the following plots, while this measurement campaign scanned the whole parameter space.  

\begin{figure}[htbp]
  \centering
  \begin{tabular}{c}
  \includegraphics[width=13cm]{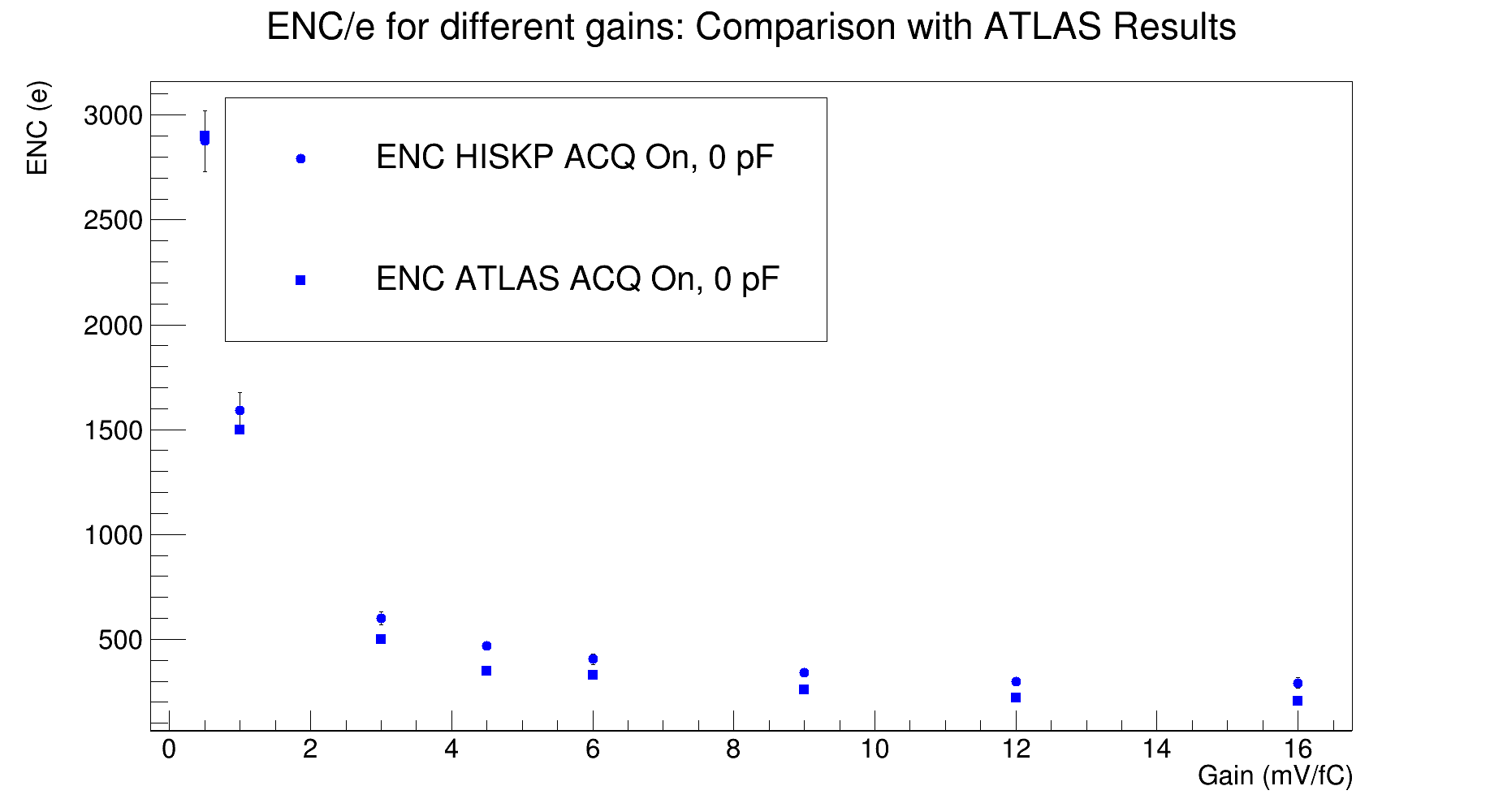} \\
  \includegraphics[width=13cm]{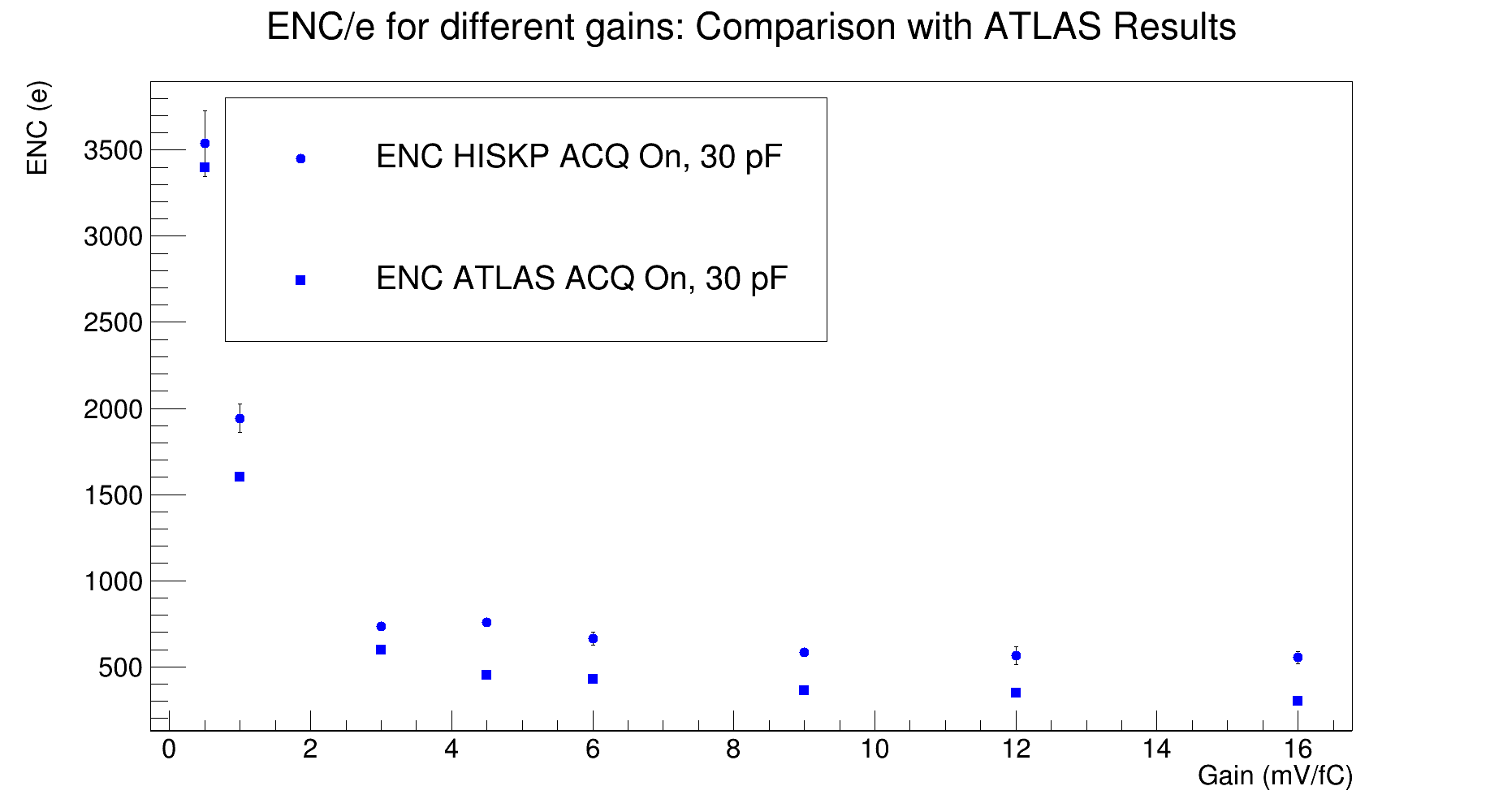}
  \end{tabular}
  \caption{Comparison of ENC results between measurements at ATLAS \cite{iakovidis2018vmm3} and measurements in this work for \SI{0}{pF} (top) and \SI{30}{pF} (bottom) input capacitance. Peaktime: \SI{200}{ns}.}
  \label{fig:atlas}
\end{figure}

In \cref{fig:atlas}, the comparison to the ATLAS results \cite{iakovidis2018vmm3} is provided. In specific, the ENC (ACQ on) is plotted against the VMM gain for \SI{0}{pF} and \SI{30}{pF} external capacitance. The peaktime is set to \SI{200}{ns}. The ENC as measured in this work is only slightly higher than the one from ATLAS. This minimal variation is justifiable, since the ATLAS team tested the VMM on different readout boards than the RD51 hybrid. In these readout boards, the space and density of electronic components is smaller. In particular, the power electronics are further away from the VMM chips, while on the hybrid they are placed on the backside. This could potentially add more noise to our results. 

\begin{figure}[htbp]
  \centering
  \begin{tabular}{c}
  \includegraphics[width=13cm]{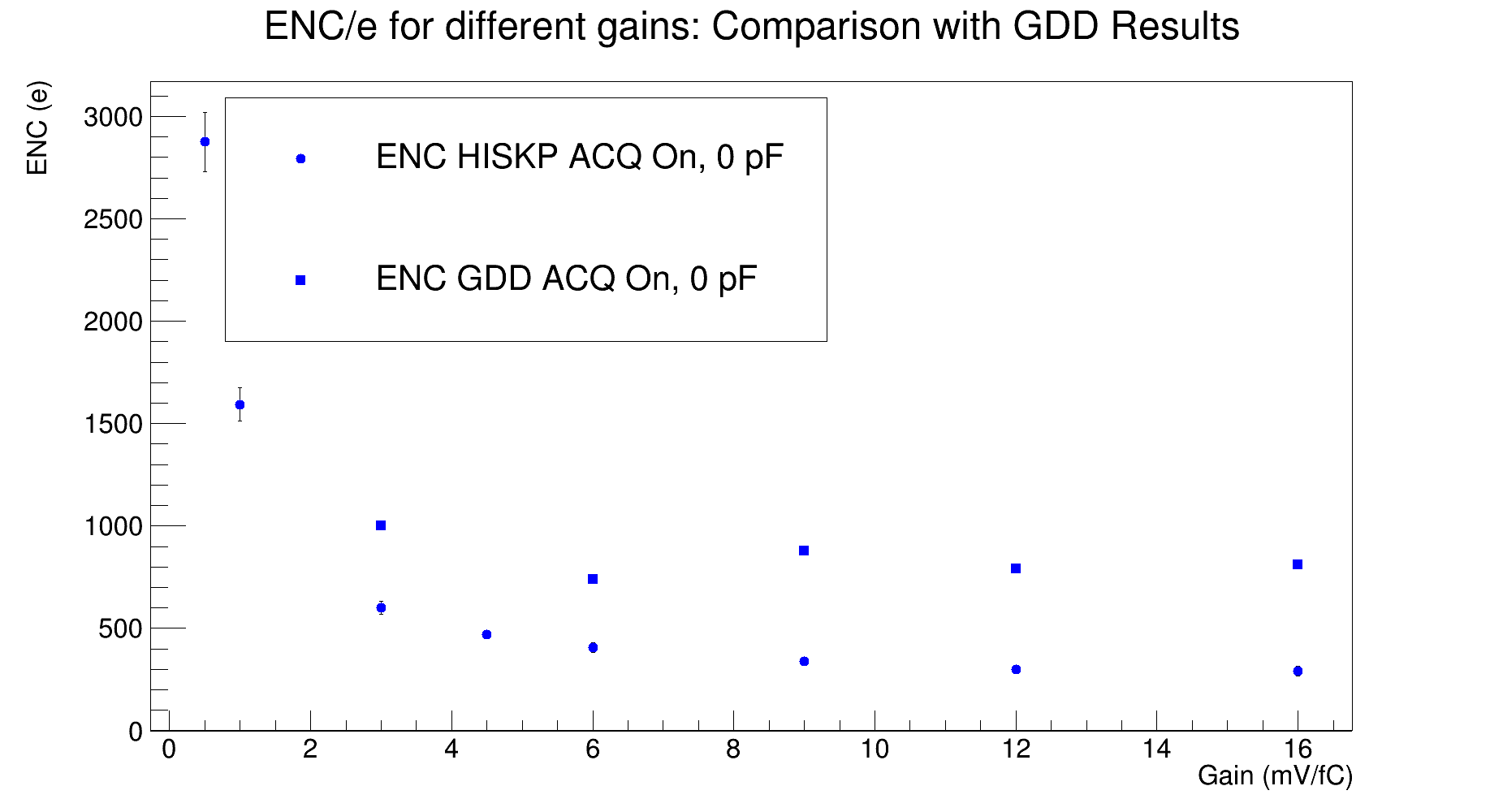}
  \end{tabular}
  \caption{Comparison of ENC results between measurements at GDD \cite{lupbergerenc} and HISKP for \SI{0}{pF} input capacitance. Peaktime: \SI{200}{ns}.}
  \label{fig:gdd}
\end{figure}

To compare to a more similar setup, the results of the GDD group \cite{lupbergerenc}, where the VMM was also tested on the hybrid, are provided in \cref{fig:gdd}. Results are shown for no input capacitance and peaktime \SI{200}{ns}. In this case, the results in this work show a great improvement with an almost $\sim$ 50\% reduction in the ENC at high gains. A comparison between points at the lowest gains was not possible, since the oscilloscope used at the GDD lacked the necessary resolution to measure at those gains. This improvement can be explained from the fact that the GDD team did not apply the necessary buffers (as shown at the beginning of this chapter, while additionally no appropriate care was given to external noise sources, and the Faraday box and connection to the oscilloscope were not as efficient. In this work, a much more improved setup was used, while also better control was taken regarding the external and internal noise sources. 

%==============================================================================
\chapter{Conclusions}
\label{sec:vmmsn}
%==============================================================================

An investigation was conducted regarding the performance of an APV25 and a VMM3a readout chip with the prospect of using them on a triple-GEM detector front-end readout board. 

In the case of the APV25 chip, the signal-to-noise ratio was evaluated for various high voltage settings in the detector. For a detector HV of \SI{4000}{V}, the average SNR between the X and Y strips was calculated at 163.49$\pm$0.86. A second relation for the SNR was also derived, its main difference being that it divides the cluster amplitude by the square root of the number of strips in the cluster. This was calculated at an average of 84.29$\pm$0.12 for the X and Y strips.

Regarding the VMM3a chip, a direct signal-to-noise measurement was not achievable at the time, since as a self-triggered chip, closer examinations of its noise levels needed to be executed first. For this reason, the ENC of several VMM chips was determined including all possible parameters of the chip (channel, gain, peaktime) and also with varying input capacitances. In order to implement this in a systematic way, an automatic noise measurement framework was developed. For the highest gain and peaktime, the ENC was calculated at about $\sim$300 electrons $+$ 8 electrons/pF, while for a gain of \SI{6}{mV}/fC, which is a typical operative setting in detector readouts, an ENC of 400 electrons $+$ 8 electrons/pF. For these values, a $\sim$10\% statistical variation was calculated between measurements of one VMM with the same settings, while a $\sim$16.5\% variation applies for measurements across different VMMs. 

\begin{figure}[htbp]
    \centering
    \begin{tabular}{c}
        \includegraphics[width=8cm]{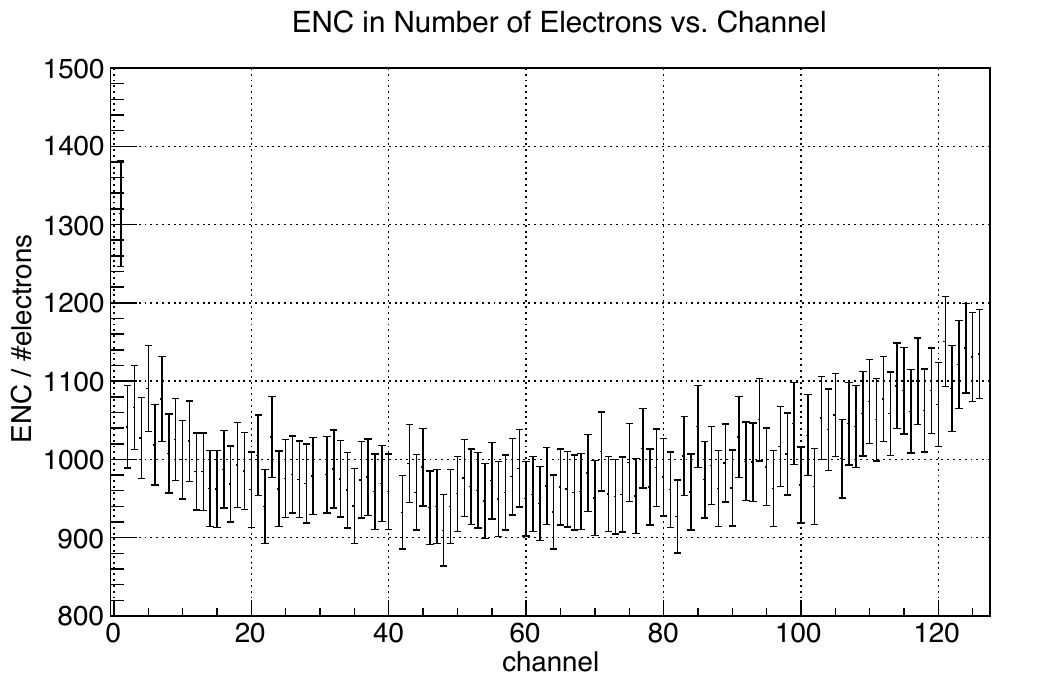} 
        %\includegraphics[width=3.95cm]{figs/vmmsn/IMG_20210208_162233.jpg}\\
         %(a) & (b)
    \end{tabular}
    \caption{APV25 EN, gain \SI{4.6}{mV}/fC, peaktime \SI{50}{ns}, input capacitance \SI{30}{pF} \cite{floethner}.}
    \label{fig:karl}
\end{figure}

\begin{figure}[htbp]
    \centering
    \begin{tabular}{c}
        \includegraphics[width=9cm]{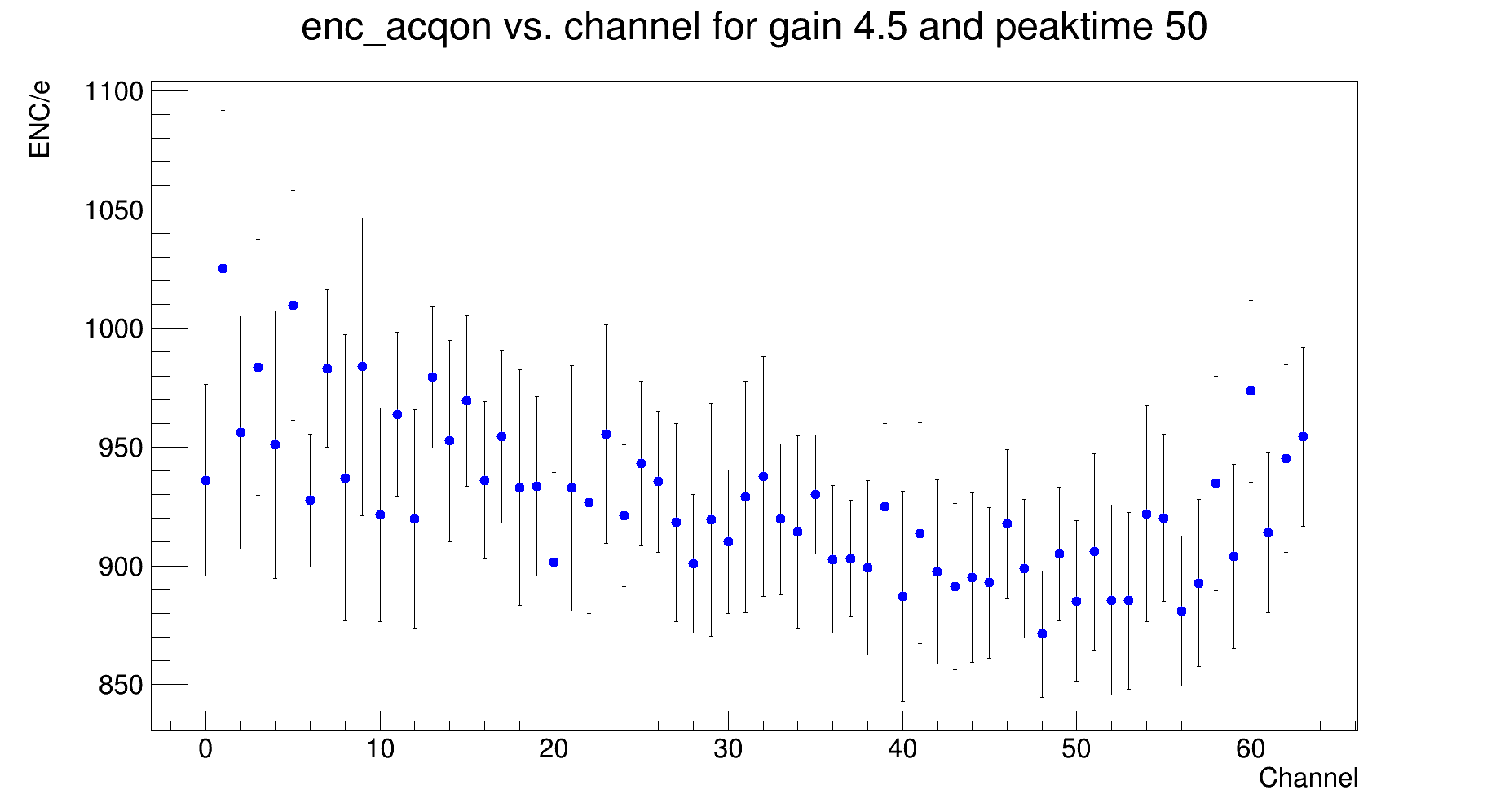}   
        %\includegraphics[width=3.95cm]{figs/vmmsn/IMG_20210208_162233.jpg}\\
         %(a) & (b)
    \end{tabular}
    \caption{VMM 1 ENC with ACQ on, gain \SI{4.5}{mV}/fC, peaktime \SI{50}{ns}, input capacitance \SI{30}{pF}.}
    \label{fig:enccomp}
\end{figure}

Since a direct comparison of the two readout chips in terms of the signal-to-noise ratio is not possible at this time, the VMM ENC that is calculated in this work can be compared instead to the APV25 ENC that was derived in this work \cite{floethner} by Karl Jonathan Flöthner, shown in \cref{fig:karl}. This measurement was also with the detector simulator PCB capacitance of \SI{30}{pF}. As shown in \cref{sec:theory:apv25}, the APV gain is \SI{4.6}{mV}/fC, with a peaktime of \SI{50}{ns}. To compare this to the VMM, we use the VMM settings of gain \SI{4.5}{mV}/fC and peaktime also \SI{50}{ns}, with input capacitance of \SI{30}{pF}. This can be seen in \cref{fig:enccomp}. 

On average, the VMM ENC for the above settings is roughly 1000 electrons, while for the APV it is around 1200 electrons. So it can be observed that generally the VMM demonstrates a slightly lower noise performance.

\begin{figure}[htbp]
    \centering
    \begin{tabular}{c}
        \includegraphics[width=5cm]{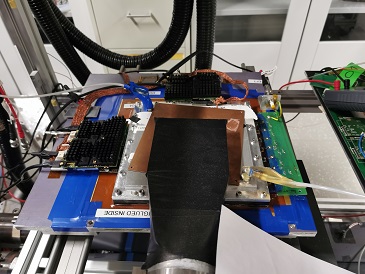}   
        %\includegraphics[width=3.95cm]{figs/vmmsn/IMG_20210208_162233.jpg}\\
         %(a) & (b)
    \end{tabular}
    \caption{Setup for cosmics measurement with VMM hybrids installed on a Triple-GEM detector.}
    \label{fig:vmmsnrsetup}
\end{figure}

Of course, a more reliable comparison would be with the SNR values. The setup for cosmics measurements with the VMMs installed on a triple-GEM detector is already at place, as shown in \cref{fig:vmmsnrsetup}. While pedestal noise can be derived with the VMM automatic measurement framework that was developed in this work, there is a lack of appropriate software for data taking with cosmics for the present. Temporary solutions like the use of network analyzer programs such as Wireshark could be used to derive the first data sets and estimate the VMM SNR. 

% Uncomment the following command to get references per chapter.
% Put it inside the file or change \include to \input if you do not want the references
% on a separate page
% \printbibliography[heading=subbibliography]

%------------------------------------------------------------------------------
% Use biblatex for the bibliography
% Add bibliography to Table of Contents
% Comment out this command if your references are printed for each chapter.
\printbibliography[heading=bibintoc]

%------------------------------------------------------------------------------
% Include the following lines and comment out \printbibliography if
% you use BiBTeX for the bibliography.
% If you use biblatex package the files should be specified in the preamble.
% \KOMAoptions{toc=bibliography}
% {\raggedright
%   \bibliographystyle{../refs/atlasBibStyleWithTitle.bst}
%   % \bibliographystyle{unsrt}
%   \bibliography{./thesis_refs,../refs/standard_refs-bibtex}
% }

%------------------------------------------------------------------------------
\appendix
% \part*{Appendix}
% Add your appendices here - don't forget to also add them to \includeonly above
%------------------------------------------------------------------------------
\chapter{The APV25 Trigger Uncertainty}
\label{sec:app:apv25}
%------------------------------------------------------------------------------

The fall time and peak distance of the APV25 cosmics trigger signal was measured using the setup in \cref{sec:theory:apv25}. A total of 32 scope frames were studied for this purpose. In \cref{tab:falltime} shown below, all measured variables are expressed in units of ns. The $x_{min1}$ and $x_{min2}$ signify the position of minimum (since the signal is negative) for peaks 1 and 2 respectively. An average of \SI{2.52}{ns} fall time and \SI{1.07}{ns} peak distance were found.

\begin{table}[htbp]
\centering
\resizebox{.42\columnwidth}{!}{\begin{tabular}{ c c c c c }
\toprule
Scope Frame & $x_{min1}$ & $x_{min2}$ & Peak distance & Fall time \\ 
\midrule
1 & -32.0 & -31.4 & 0.6 & 2.5\\
2 & -30.5 & -31.4 & 0.9 & 2.5\\
3 & -30.9 & -30.9 & 0.0 & 2.3\\
4 & -31.4 & -31.4 & 0.0 & 2.7\\
5 & -31.4 & -31.4 & 0.0 & 2.3\\
6 & -32.2 & -31.9 & 0.3 & 2.5\\
7 & -30.6 & -30.5 & 0.1 & 2.3\\
8 & -30.6 & -30.6 & 0.0 & 2.5\\
9 & -31.3 & -31.3 & 0.0 & 2.3\\
10 & -31.4 & -30.3 & 1.1 & 2.7\\
11 & -31.4 & -30.5 & 0.9 & 2.5\\
12 & -30.9 & -31.9 & 1.0 & 2.5 \\
13 & -30.2 & -30.8 & 0.6 & 2.7 \\
14 & -30.5 & -32.0 & 1.5 & 2.5 \\
15 & -32.3 & -31.8 & 0.5 & 2.6 \\
16 & -29.9 & -32.1 & 2.2 & 2.6 \\
17 & -30.5 & -33.2 & 2.7 & 2.5 \\
18 & -31.6 & -32.3 & 0.7 & 2.5\\
19 & -32.3 & -32.0 & 0.3 & 2.6\\
20 & -31.9 & -31.7 & 0.2 & 2.6\\
21 & -31.6 & -32.6 & 1.0 & 2.3\\
22 & -30.5 & -32.4 & 1.9 & 2.7\\
23 & -30.6 & -33.0 & 2.4 & 2.4\\
24 & -34.5 & -29.0 & 5.5 & 2.3\\
25 & -32.8 & -32.1 & 0.7 & 2.5\\
26 & -32.0 & -32.3 & 0.3 & 2.6\\
27 & -30.5 & -31.4 & 0.9 & 2.6\\
28 & -34.1 & -30.9 & 3.2 & 2.3\\
29 & -32.0 & -33.0 & 1.0 & 2.5\\
30 & -31.6 & -31.1 & 0.5 & 2.4\\
31 & -33.0 & -31.0 & 2.0 & 2.4\\
32 & -32.5 & -31.4 & 1.1 & 2.5\\
\bottomrule
\end{tabular}}
\caption{Measurements of the peak distance and fall time of the APV25 trigger signal}%
\label{tab:falltime}
\end{table}

%------------------------------------------------------------------------------
\chapter{Calculation of the Range of Minimum Ionizing Muons in Steel}
\label{sec:app:range}
%------------------------------------------------------------------------------
\begin{figure}[htbp]
  \centering
  \begin{tabular}{cc}
  \includegraphics[width=7.3cm]{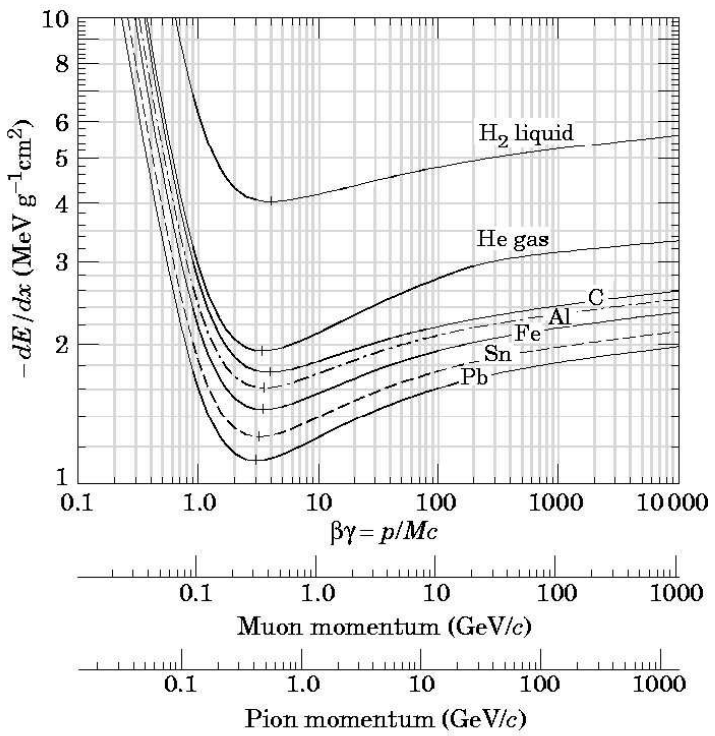} &
  \includegraphics[width=6cm]{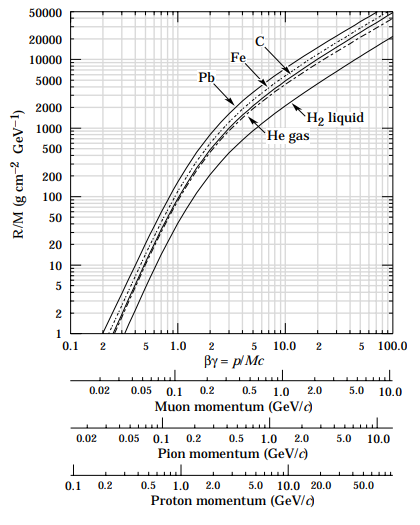} \\
  (a) & (b)
  \end{tabular}
  \caption{(a) Energy loss and (b) Range of heavy charged particles in various materials \cite{groom2000passage}. }
  \label{fig:range}
\end{figure}

From \cref{fig:range}a, it is evident that the velocity of minimum ionizing muons is around $\beta\gamma = 3.5$. One can use this to find their range $R/M$ in steel from \cref{fig:range}b, where $M$ is the mass of the incident particle (the muon, in our case). Considering steel to be a mixture of carbon and iron, this range is around 1300-1600 $g$ $cm^{-2}$ $GeV^{-1}$. To derive the range in cm, one needs to multiply this value with the mass of the muon, $M=0.105$ $GeV/c^2$ and divide by the density of steel, $\rho_{steel}\approx8.05$ $g/cm^3$.  
$$R (cm) = \frac{\frac{R}{M}\cdot M}{\rho_{steel}} =
\frac{1300\times 0.105}{8.05} cm \approx 17 cm $$

%------------------------------------------------------------------------------
\chapter{\textit{P}-Values of the SNR fits for the APV25}
\label{sec:app:p-value}
%------------------------------------------------------------------------------

\begin{table}[htbp]
\centering
\begin{tabular}{ c c c }
\toprule
HV (V) & $p_{SNR}$ & $p_{SNR2}$ \\ 
\midrule
3900 & 0.068 & 0.137 \\
3950 & 0.695 & 0.005 \\
4000 & 0.879 & 0.685 \\
4050 & 0.663 & 0.044 \\
4100 & 0.887 & 0.100 \\
\bottomrule
\end{tabular}
\quad
\begin{tabular}{ c c c }
\toprule
HV (V) & $p_{SNR}$ & $p_{SNR2}$ \\ 
\midrule
3900 & 0.511 & 0.927 \\
3950 & 0.971 & 0.398 \\
4000 & 0.662 & 0.415 \\
4050 & 0.789 & 0.007 \\
4100 & 0.202 & 0.191 \\
\bottomrule
\end{tabular}
\caption{Generated \textit{p}-values under the $\chi^2$ test for the Gaussian-Landau fits of the SNR and SNR 2 data of the X (left) and Y (right) projection without steel shielding.}%
\label{tab:p-value-snr2}
\end{table}

\begin{table}[htbp]
\centering
\begin{tabular}{ c c c }
\toprule
HV (V) & $p_{SNR}$ & $p_{SNR2}$ \\ 
\midrule
3900 & 0.942 & 0.035 \\
3950 & 0.545 & 0.237 \\
4000 & 0.710 & 0.051 \\
4050 & 0.928 & 0.668 \\
4100 & 0.152 & 0.0005 \\
\bottomrule
\end{tabular}
\quad
\begin{tabular}{ c c c }
\toprule
HV (V) & $p_{SNR}$ & $p_{SNR2}$ \\ 
\midrule
3900 & 0.918 & 0.462 \\
3950 & 0.204 & 0.015 \\
4000 & 0.986 & 0.535 \\
4050 & 0.348 & 0.577 \\
4100 & 0.371 & 0.234 \\
\bottomrule
\end{tabular}
\caption{Generated \textit{p}-values under the $\chi^2$ test for the Gaussian-Landau fits of the SNR and SNR 2 data of the X (left) and Y (right) projection with steel shielding.}%
\label{tab:p-value-snr3}
\end{table}

%------------------------------------------------------------------------------
\chapter{VMM ENC Systematic Measurements: More Results}
\label{sec:app:vmm-enc}
%------------------------------------------------------------------------------

As mentioned in \cref{sec:vmmstg2:enc}, even though ENC measurements with all gain-peaktime combinations were done, it would be excessive to include all of them. However, some more of the results, mainly for peaktimes other than \SI{200}{ns}, are provided here. In the first section, results from VMM 1 are displayed, while in the second section additional plots with all VMMs are shown. 

\section{VMM 1}

\begin{figure}[htbp]
  \centering
  \begin{tabular}{c}
  \includegraphics[width=13cm]{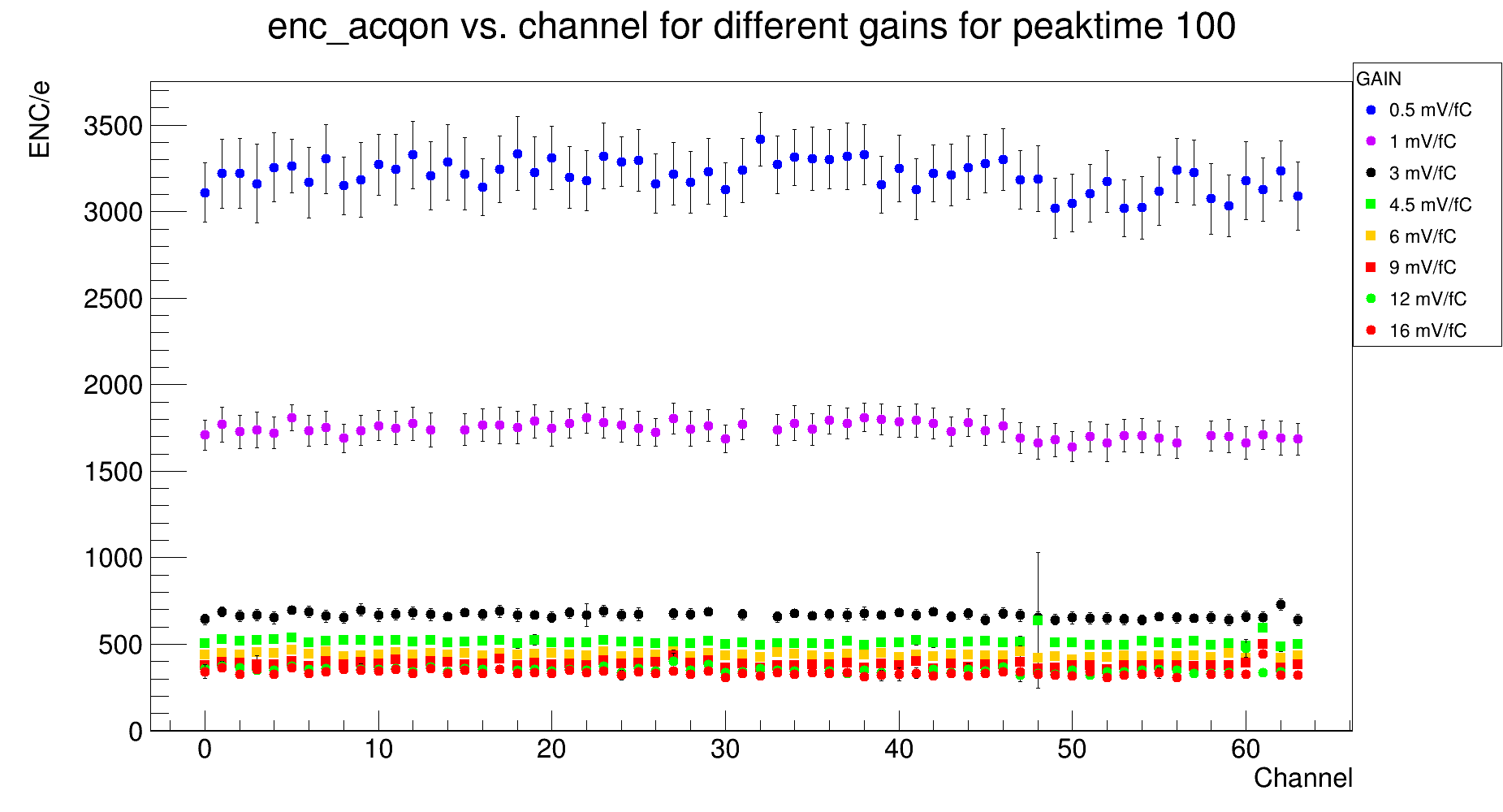} \\
  \includegraphics[width=13cm]{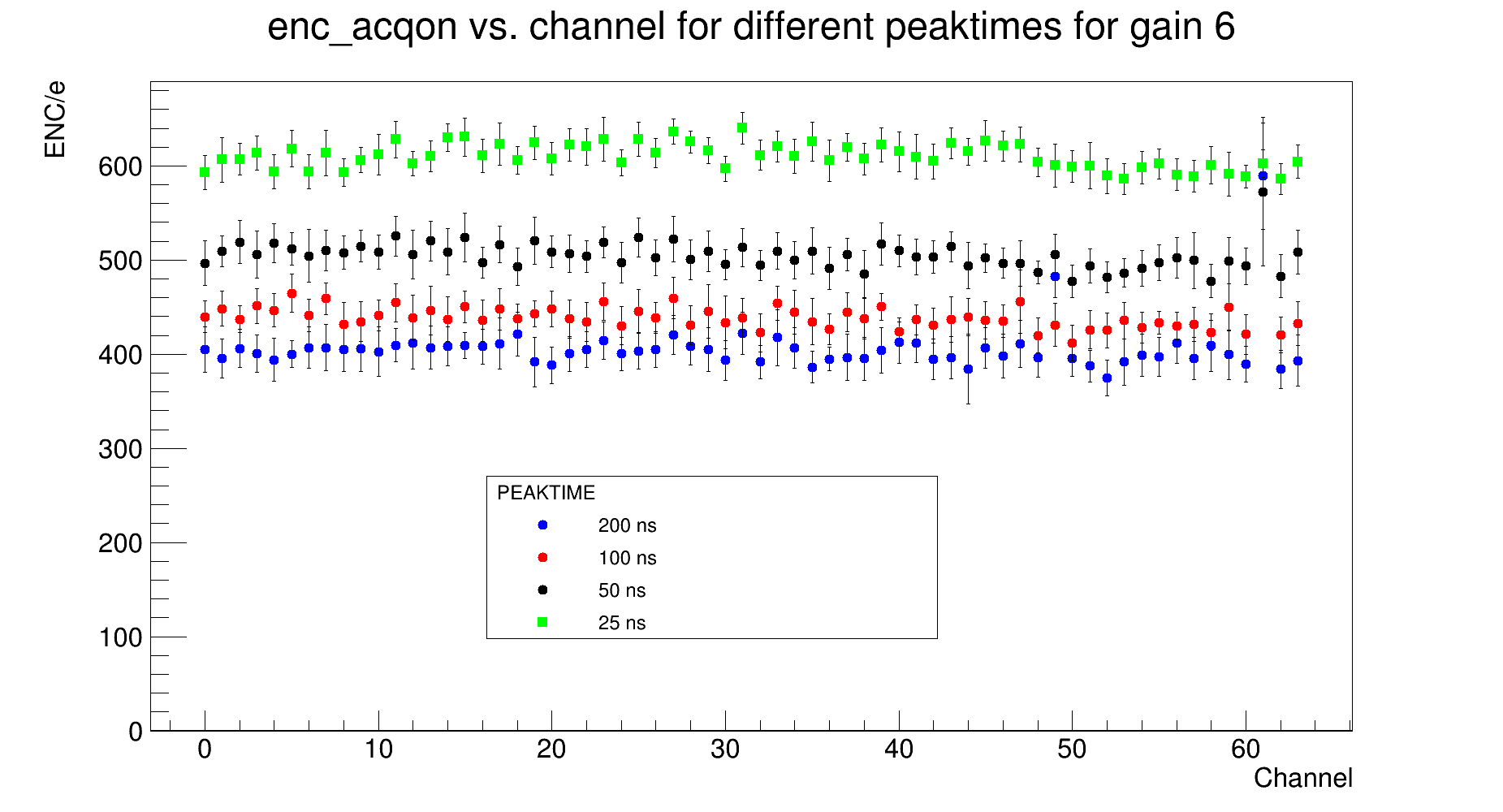} 
  \end{tabular}
  \caption{Top: ENC, ACQ on, vs. channel for different gains and peaktime \SI{100}{ns}. Bottom: ENC, ACQ on, vs. channel for different peaktimes and gain \SI{6}{mV}/fC. Channel 61 is likely damaged. Missing points are set to zero. Large errors occur due to losses in connection to the oscilloscope.}
  \label{fig:vmm1_channel_ENC_0pF_2}
\end{figure}

Similarly to \cref{fig:vmm1_channel_ENC_0pF} in \cref{sec:vmmstg2:enc}, in \cref{fig:vmm1_channel_ENC_0pF_2}, the ENC (ACQ on) is plotted against the channels for different peaktimes and different gains, but in each case, a different gain and peaktime are used as constants respectively. So in this way, we can see the ENC variation among different gains at peaktime \SI{100}{ns} this time, or alternatively, the ENC variation among different peaktimes at gain \SI{6}{mV}/fC. Omitting the lowest gains, the variation of the ENC among different gains at peaktime \SI{100}{ns} is around $\sim$ 300 electrons. Alternatively, the ENC variation among different peaktimes at gain \SI{6}{mV}/fC is around $\sim$ 200 electrons.

\begin{figure}[htbp]
  \centering
  \begin{tabular}{c}
  \includegraphics[width=13cm]{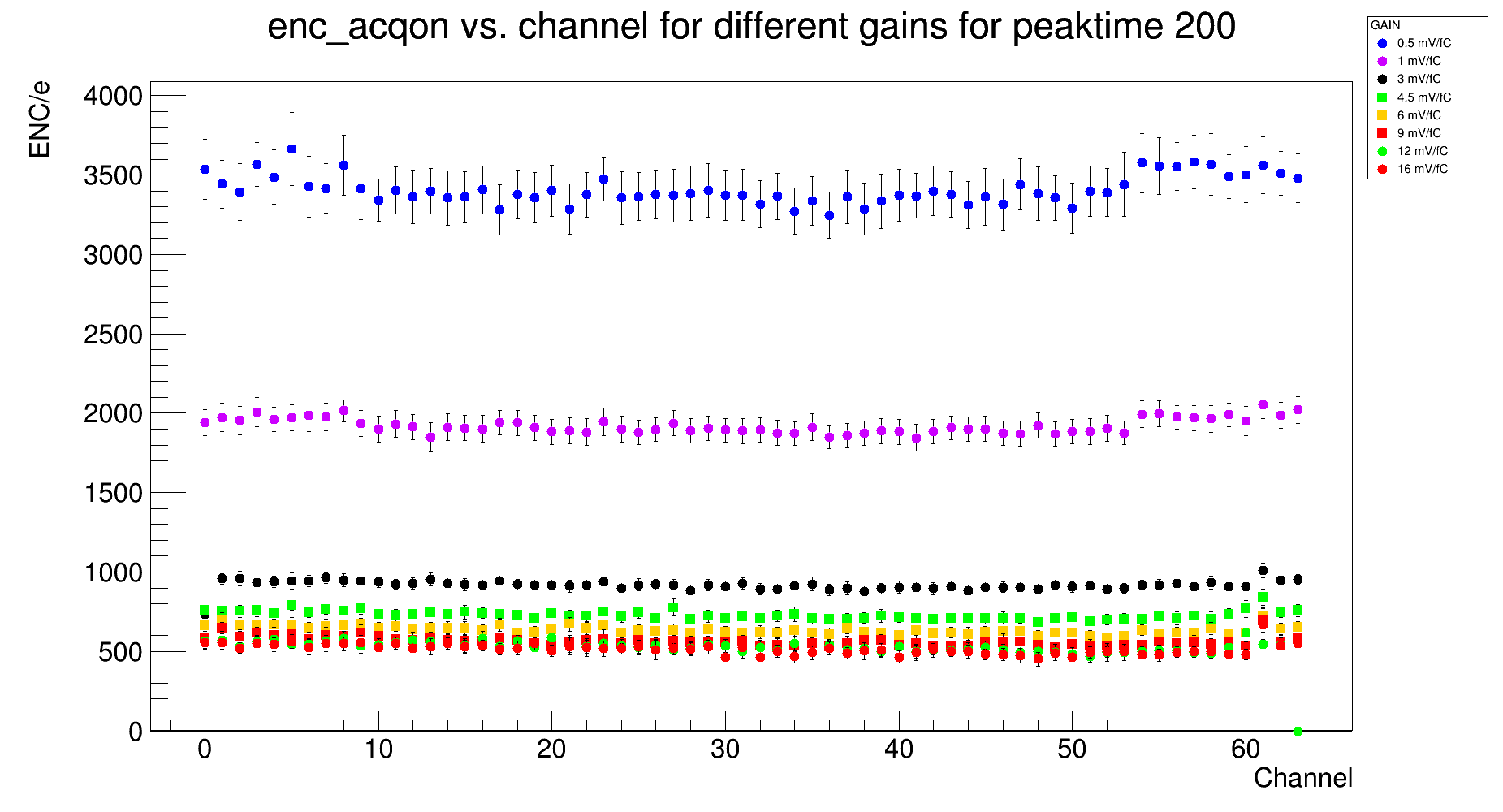} \\
  \includegraphics[width=13cm]{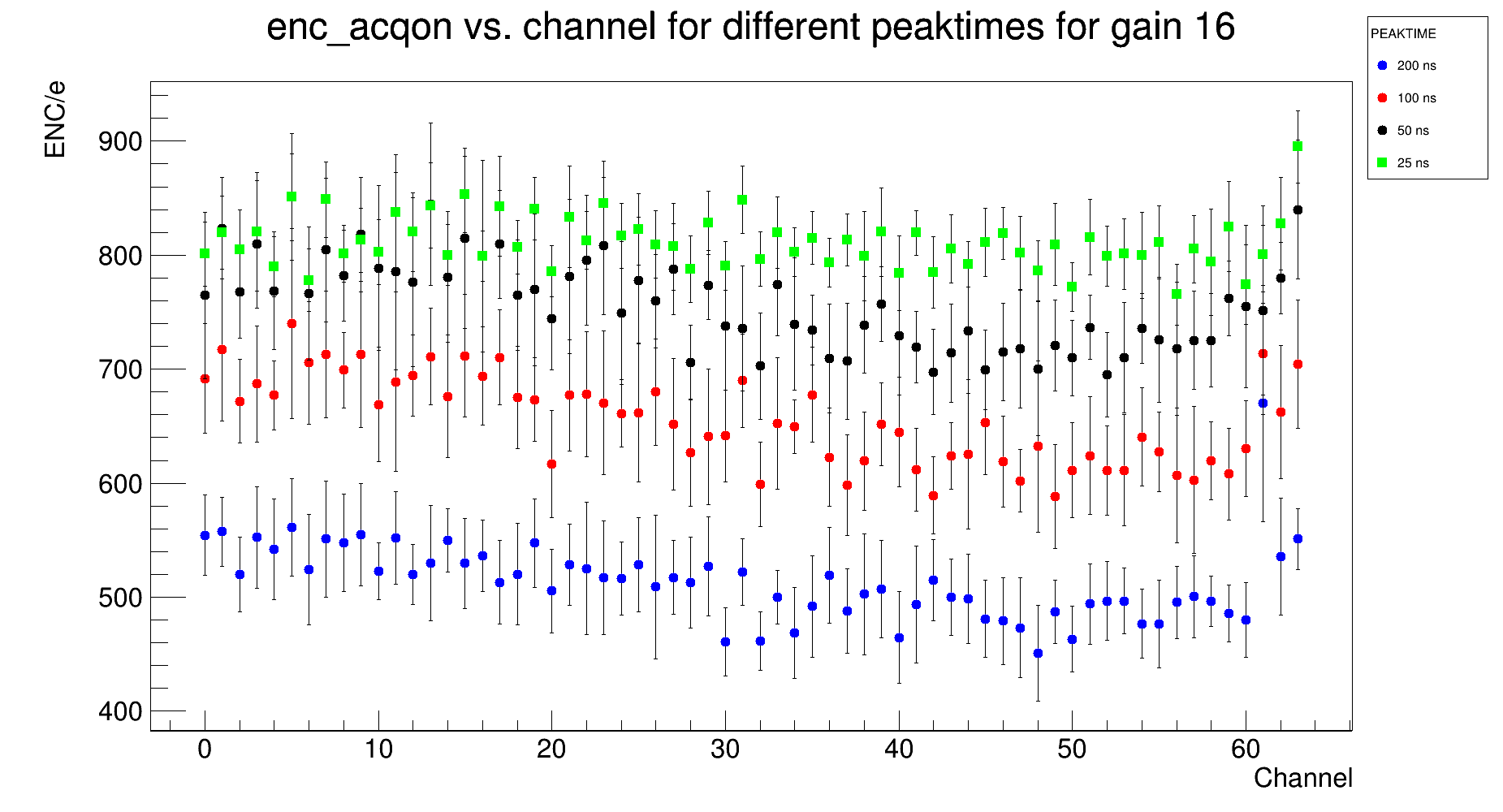} 
  \end{tabular}
  \caption{ENC, ACQ on, vs. channel for different gains (top) and peaktimes (bottom) for external capacitance of \SI{30}{pF}.}
  \label{fig:vmm1_channel_ENC_30pF}
\end{figure}

In \cref{fig:vmm1_channel_ENC_30pF}, the results for \SI{30}{pF} input capacitance are shown extensively for all gains at peaktime \SI{200}{ns} and for all peaktimes at gain \SI{16}{mV}/fC. As a reminder, in \cref{sec:vmmstg2:enc}, only a comparison between results at \SI{0}{pF} and \SI{30}{pF} input capacitance at the highest settings was provided (\cref{fig:vmm1_channel_ENC_0-30pF}),  while in this case a special focus is given to the cumulative results for \SI{30}{pF} external capacitance. 

\begin{figure}[htbp]
  \centering
  \begin{tabular}{c}
  \includegraphics[width=11.5cm]{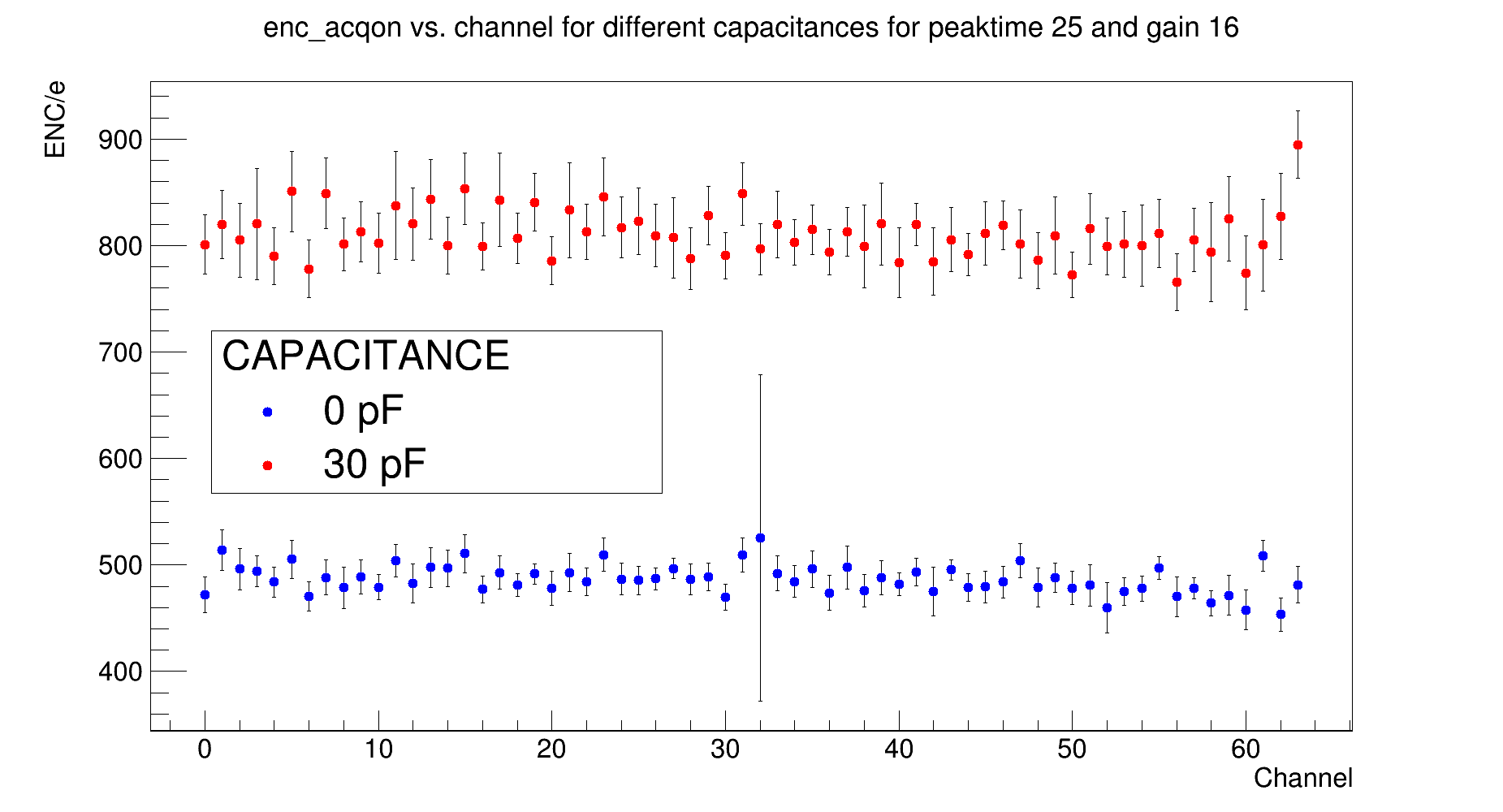} \\
  \includegraphics[width=11.5cm]{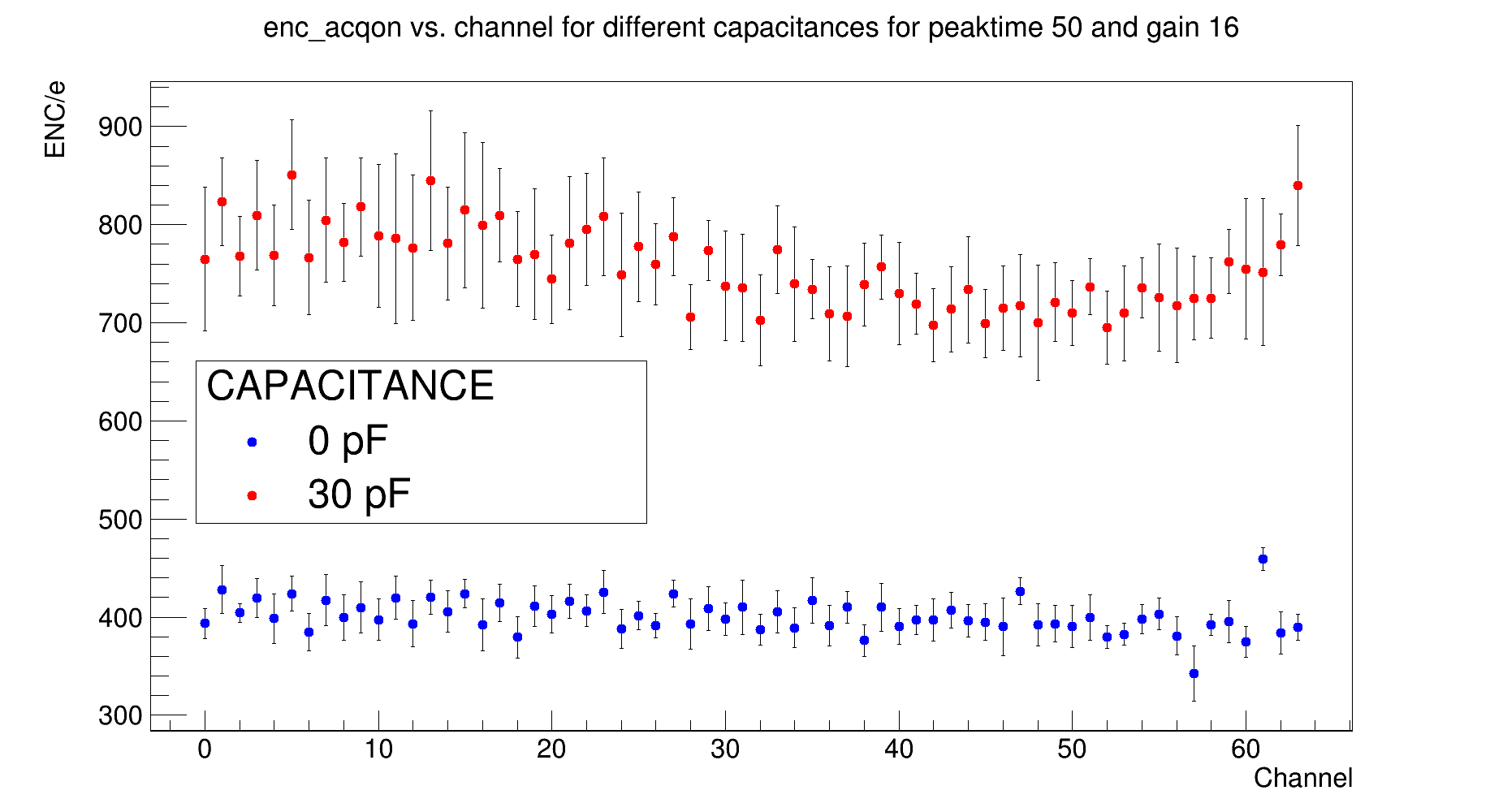} \\
  \includegraphics[width=11.5cm]{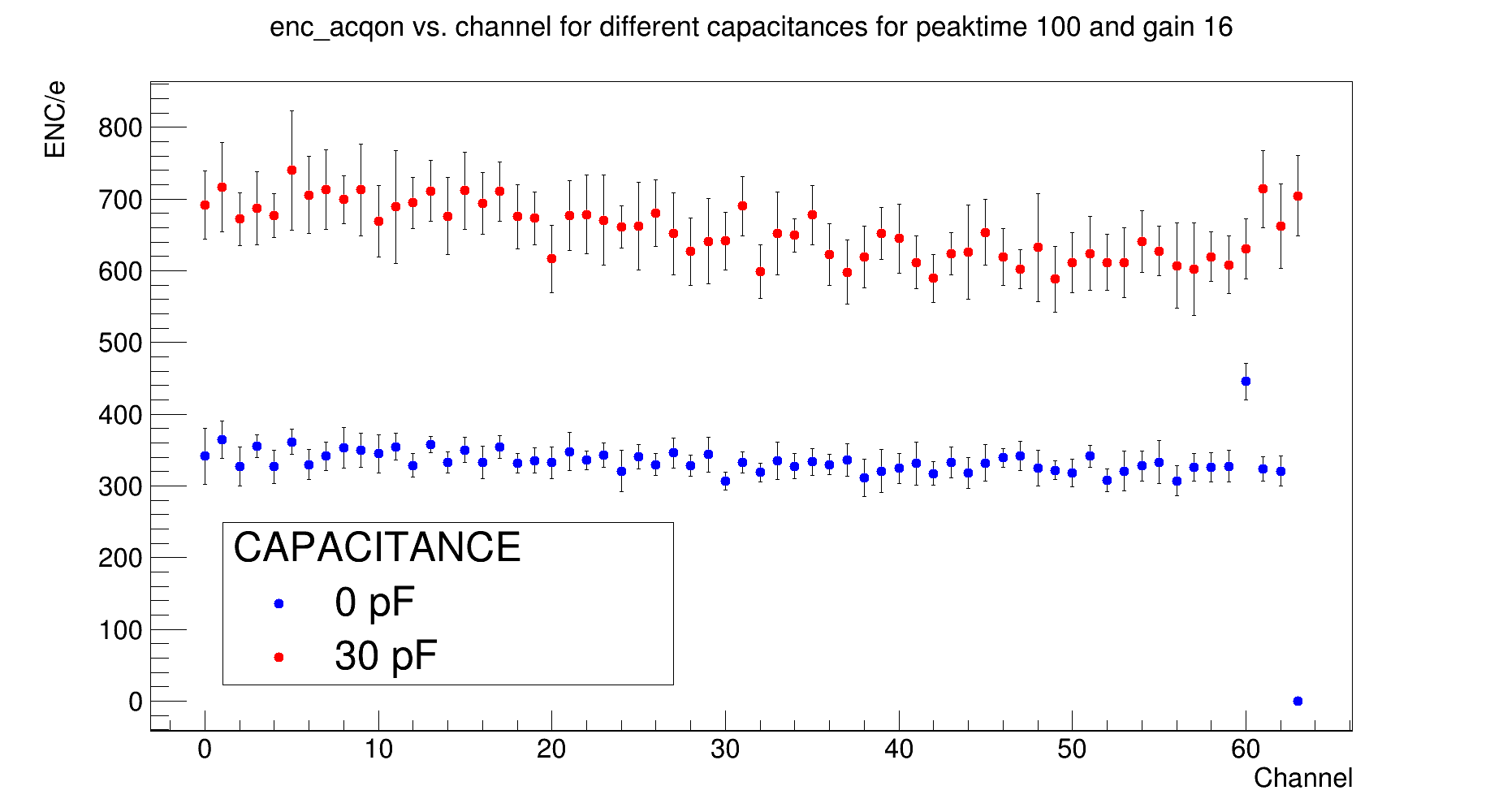}
  \end{tabular}
  \caption{ENC, ACQ on, vs. channel with and without external capacitance for peaktimes \SI{25}{ns} (top), \SI{50}{ns} (middle) and \SI{100}{ns} (bottom), gain \SI{16}{mV}/fC. Missing points are set to zero while large error bars occur from brief losses in connection to the oscilloscope.}
  \label{fig:vmm1_channel_ENC_0-30pF_2}
\end{figure}

Again, regarding the comparison between input capacitances (as done in \cref{fig:vmm1_channel_ENC_0-30pF}), \cref{fig:vmm1_channel_ENC_0-30pF_2} shows similar results for peaktimes \SI{25}{ns}, \SI{50}{ns} and \SI{100}{ns} (gain \SI{16}{mV}/fC). Channel 61 is most likely damaged as it can be understood from earlier plots as well, even though that in this case, its measurement for \SI{30}{pF} capacitance seems to agree with the rest of the channels.

\begin{figure}[htbp]
  \centering
  \resizebox{\columnwidth}{!}{\begin{tabular}{cc}
  \includegraphics[width=9cm]{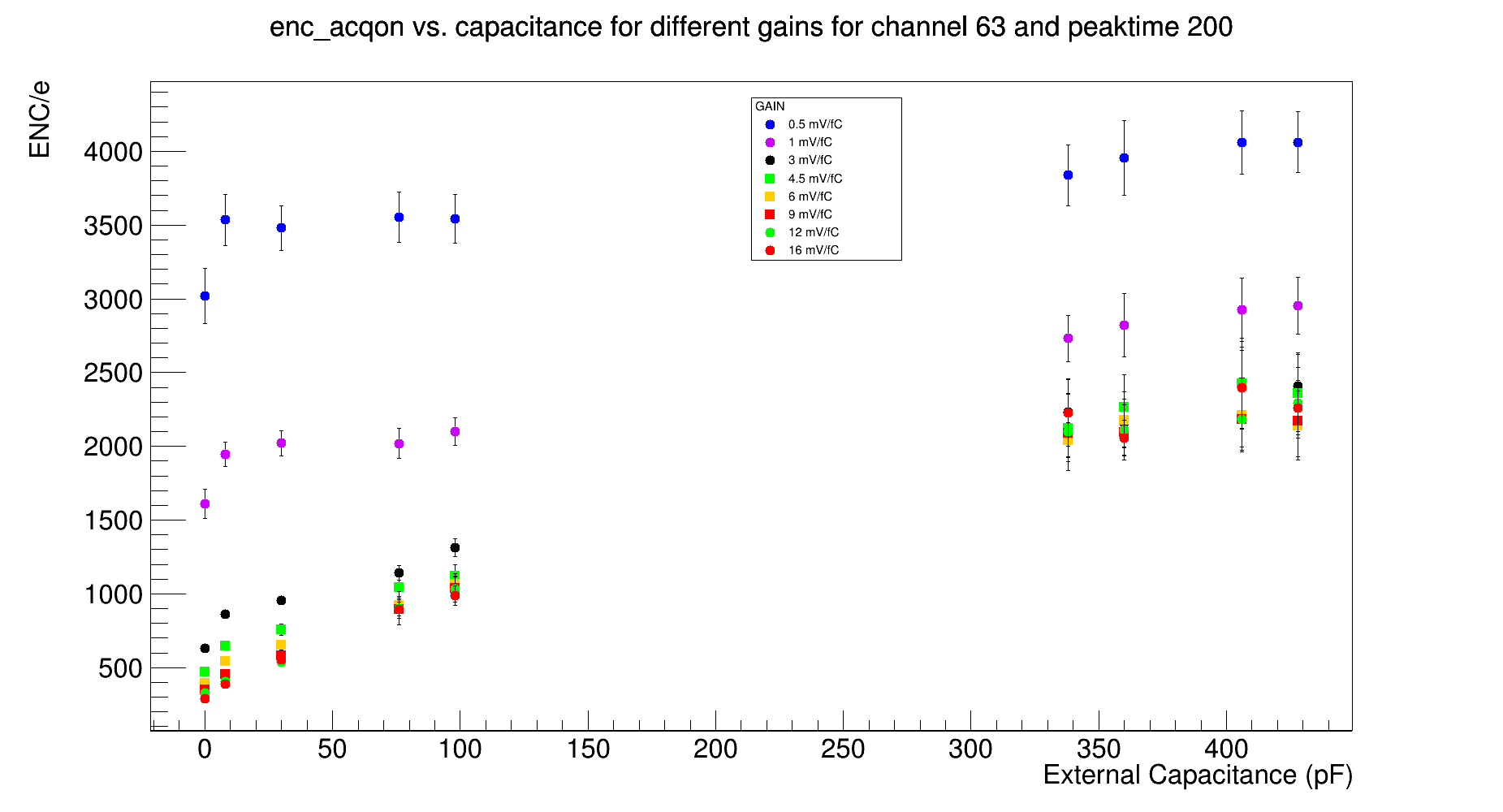} &
  \includegraphics[width=9cm]{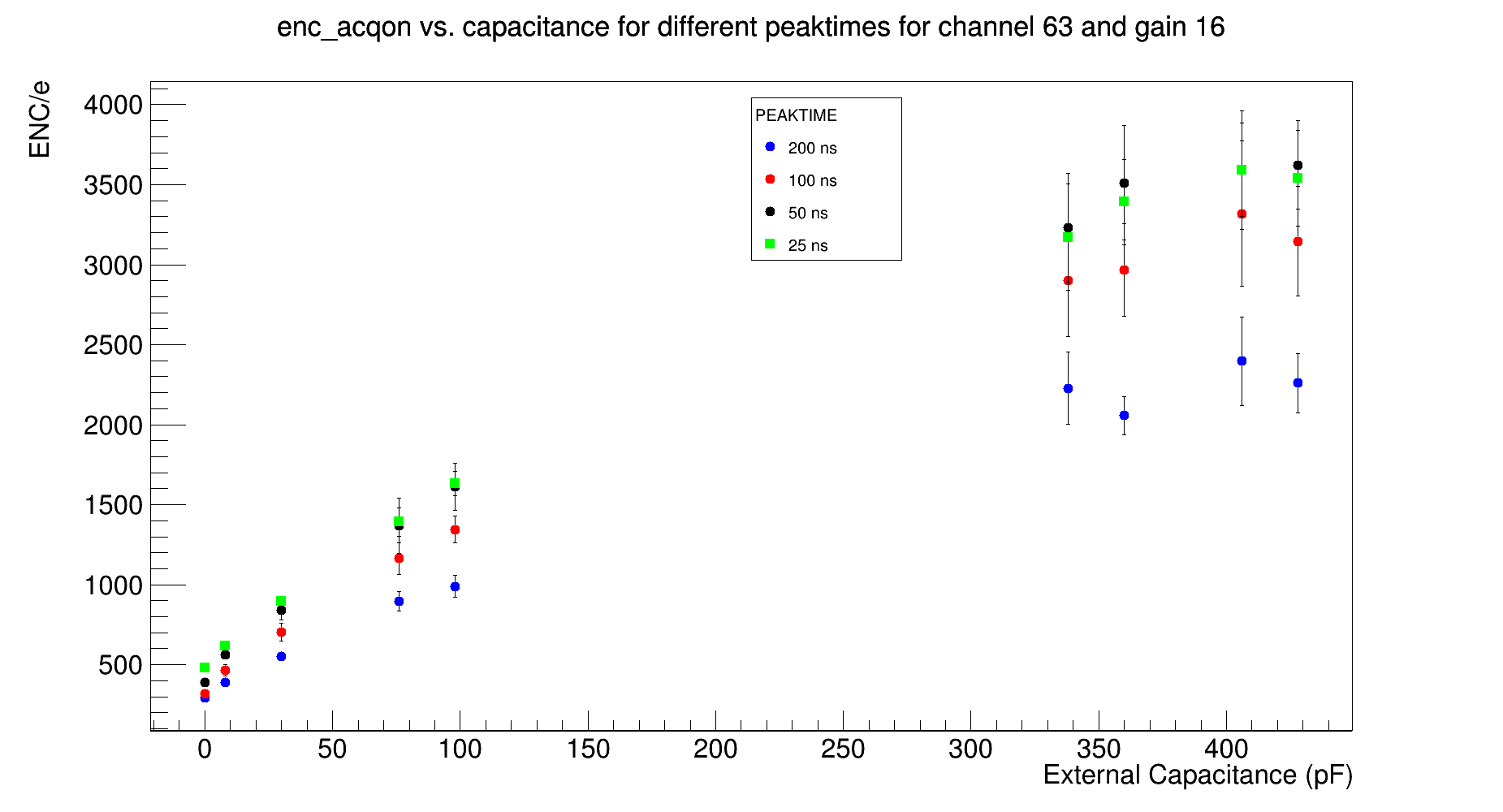} \\
  \end{tabular}}
  \caption{ENC, ACQ on, vs. capacitance for channel 63 for different gains (left) and different peaktimes (right).}
  \label{fig:vmm1_chn63_cap_ENC}
\end{figure}

\begin{figure}[htbp]
  \centering
  \resizebox{\columnwidth}{!}{\begin{tabular}{cc}
  \includegraphics[width=9cm]{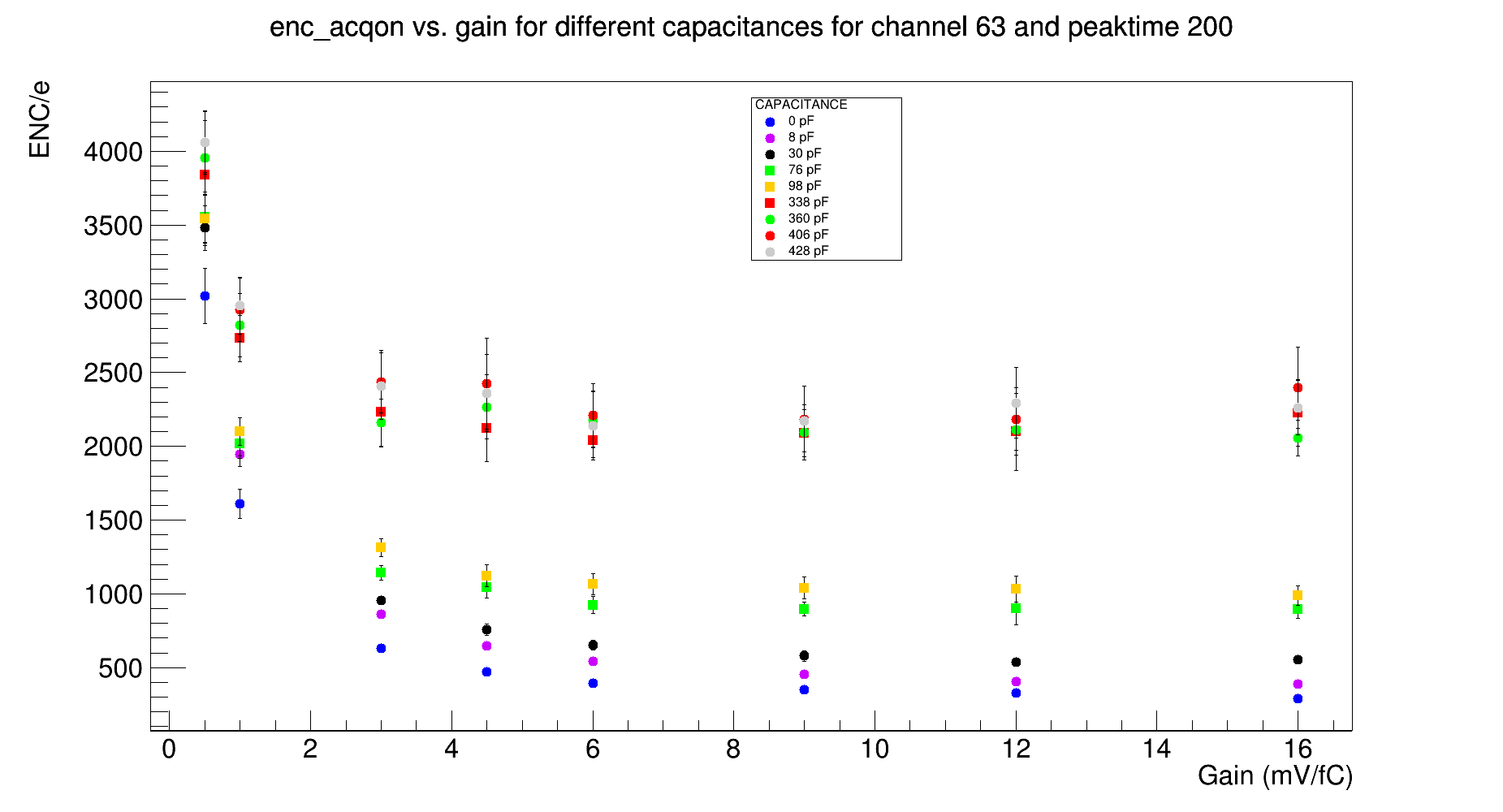} &
  \includegraphics[width=9cm]{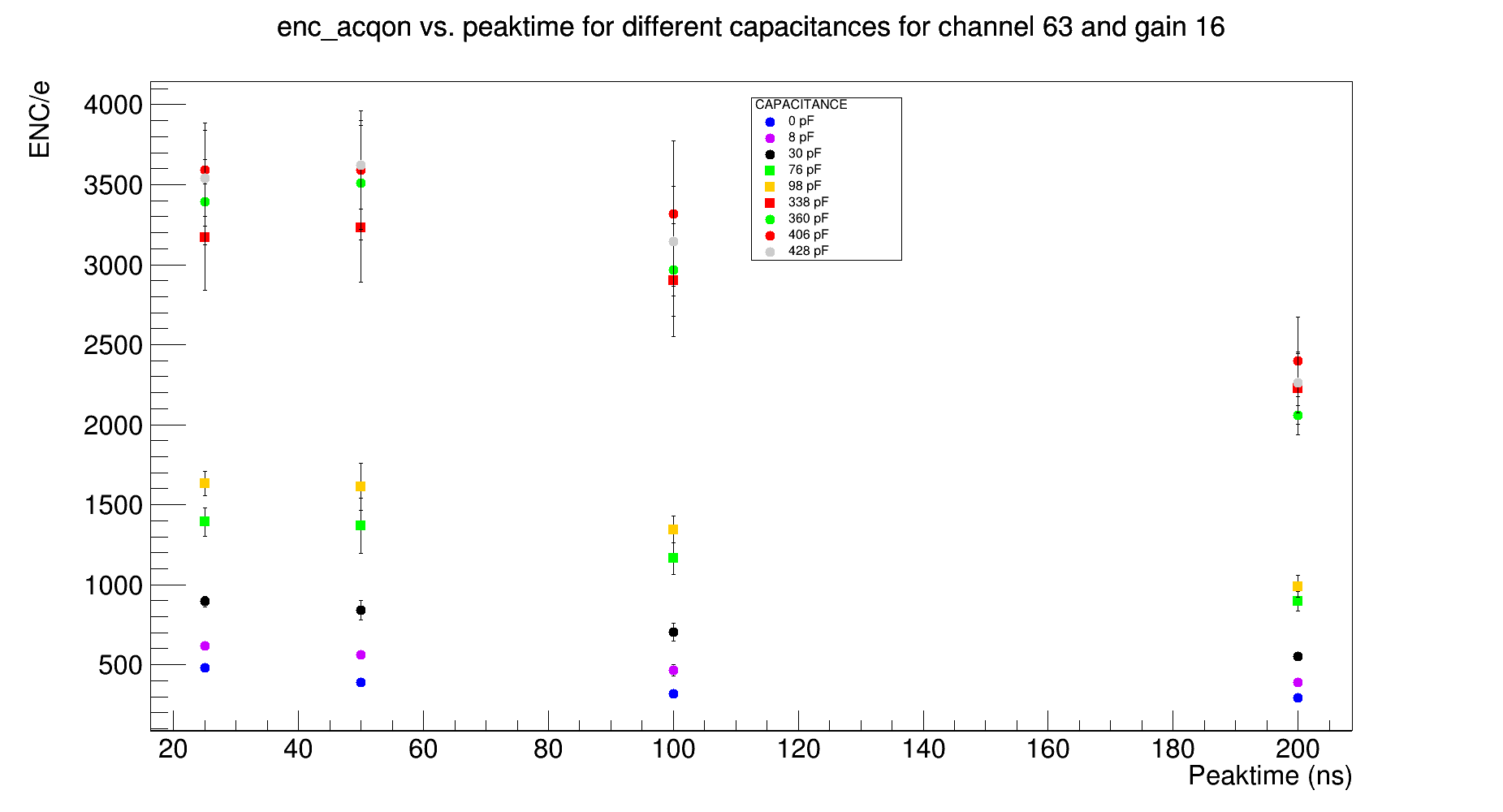} \\
  \end{tabular}}
  \caption{ENC, ACQ on, vs. gain (left) and vs. peaktime (right) for channel 63 for different capacitances.}
  \label{fig:vmm1_chn63_gain-peaktime_ENC}
\end{figure}

\begin{figure}[htbp]
  \centering
  \resizebox{\columnwidth}{!}{\begin{tabular}{cc}
  \includegraphics[width=9cm]{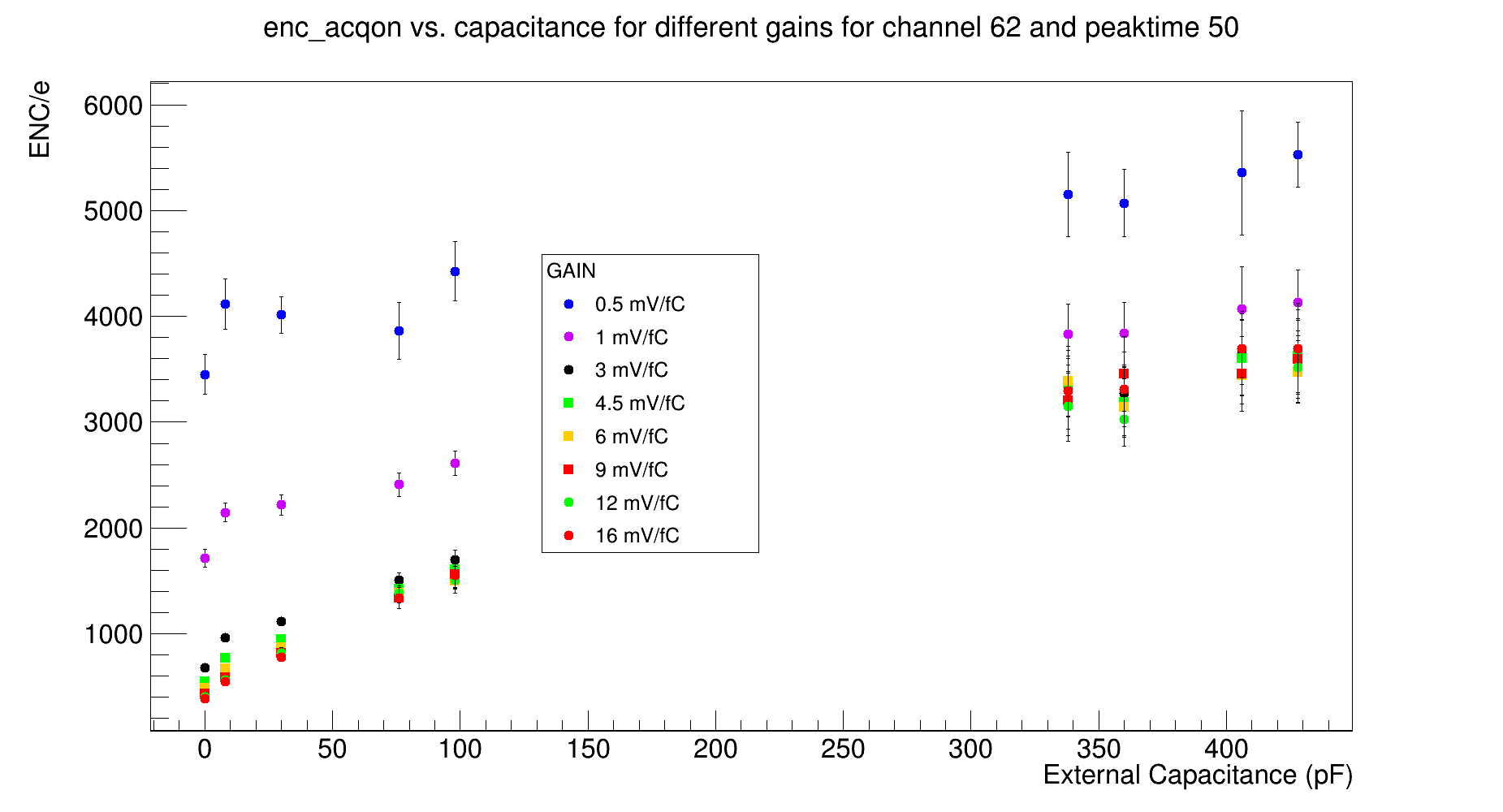} &
  \includegraphics[width=9cm]{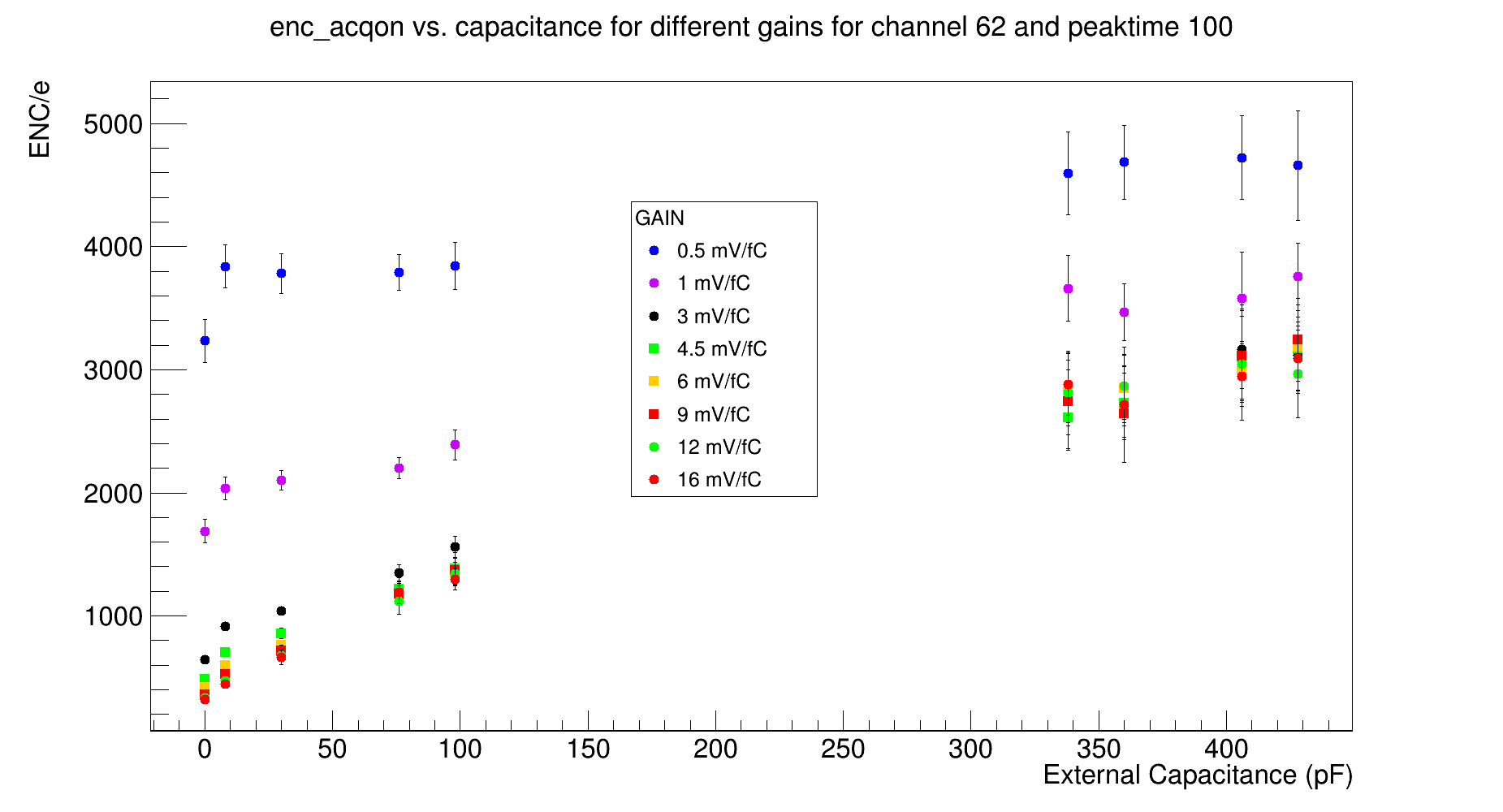} \\
   \includegraphics[width=9cm]{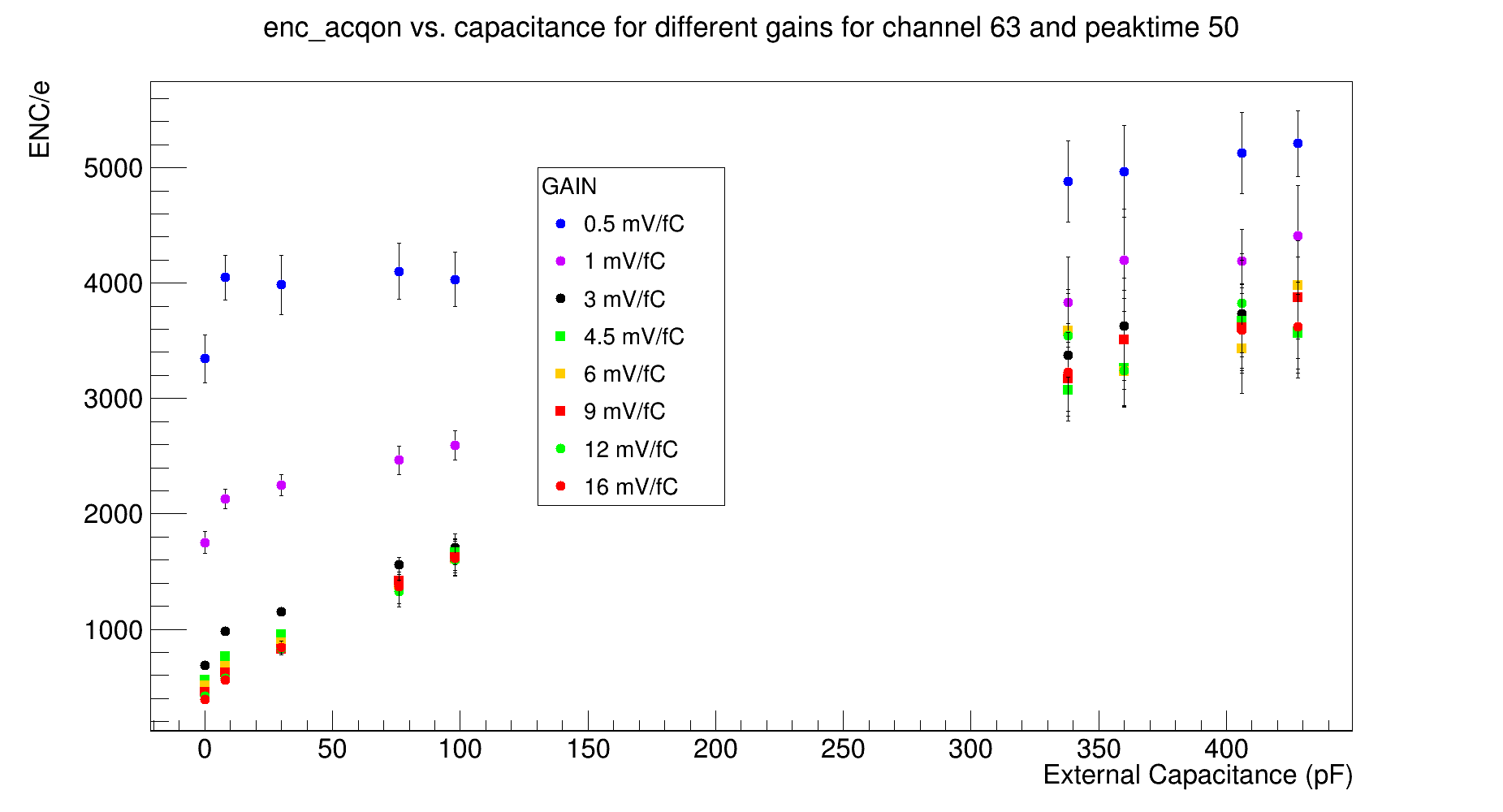} &
  \includegraphics[width=9cm]{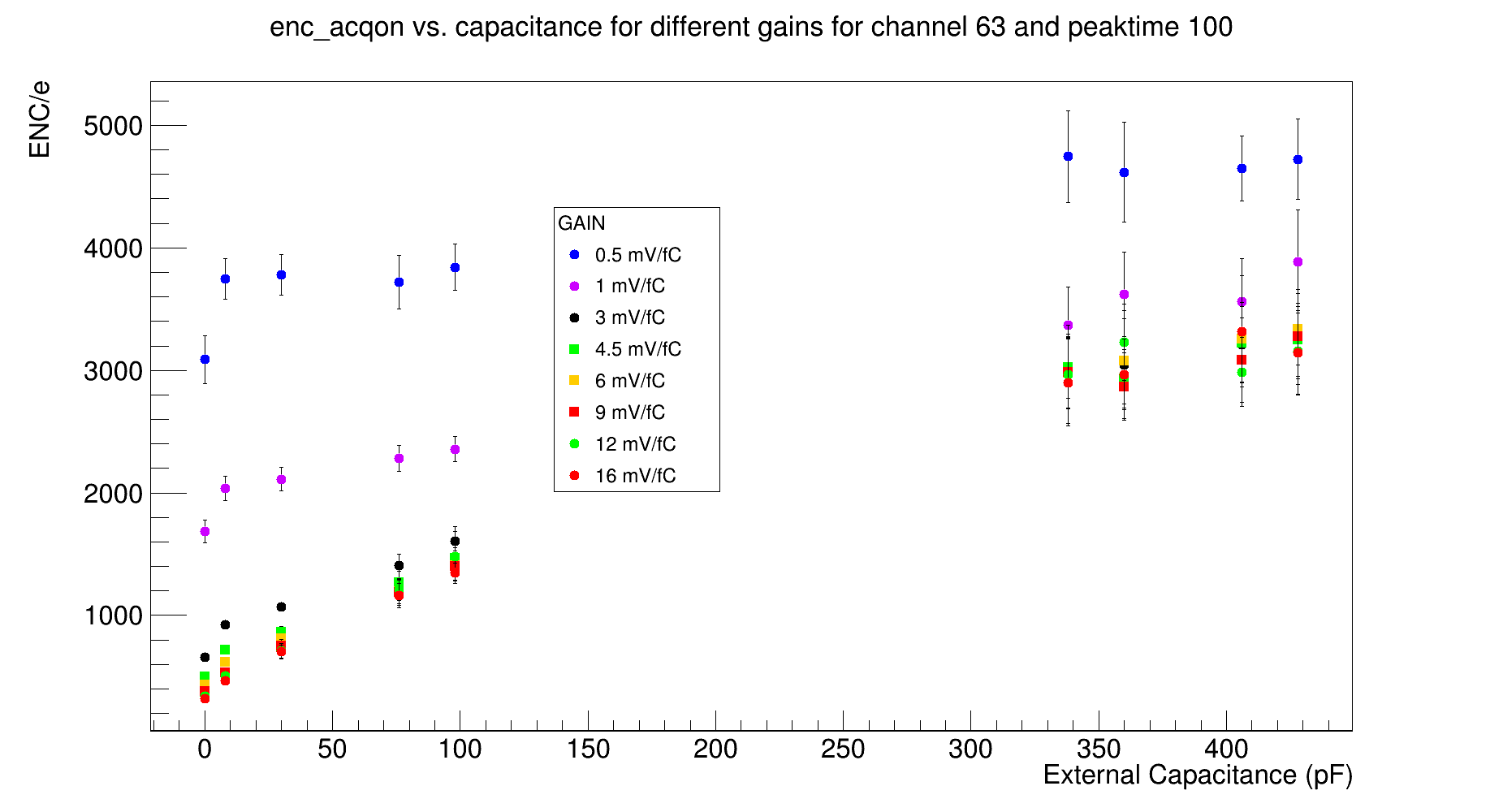}
  \end{tabular}}
  \caption{ENC, ACQ on, vs. capacitance for channel 62 and 63 for gains \SI{16}{mV}/fC\SI{16}{mV}/fC and different peaktime settings.}
  \label{fig:vmm1_channel_62-63_peak50-100}
\end{figure}

As mentioned, more capacitances were also tested for channel 62 and 63 of VMM 1. Results for channel 62 were shown in the main chapter, while results for channel 63 are provided here. Specifically, \cref{fig:vmm1_chn63_cap_ENC} shows the ENC against the external capacitance for different gains and peaktimes, while \cref{fig:vmm1_chn63_gain-peaktime_ENC} shows the ENC against the gain and peaktime for different capacitances. \cref{fig:vmm1_channel_62-63_peak50-100} shows results of the ENC against the input capacitance both for channels 62 and 63 for additional peaktime settings (\SI{50}{ns} and \SI{100}{ns}).

\section{Multiple VMMs}

Similarly to \cref{fig:ENC_cap_allvmms} in \cref{sec:vmmstg2:enc}, results for multiple VMMs and different capacitances are also shown here for channel 0/62, for both gains \SI{16}{mV}/fC and \SI{6}{mV}/fC. Channel 0 is measured for the even-numbered VMMs, while channel 62 is measured for the odd-numbered VMMs. 

\begin{figure}[htbp]
  \centering
  \begin{tabular}{c}
  \includegraphics[width=13cm]{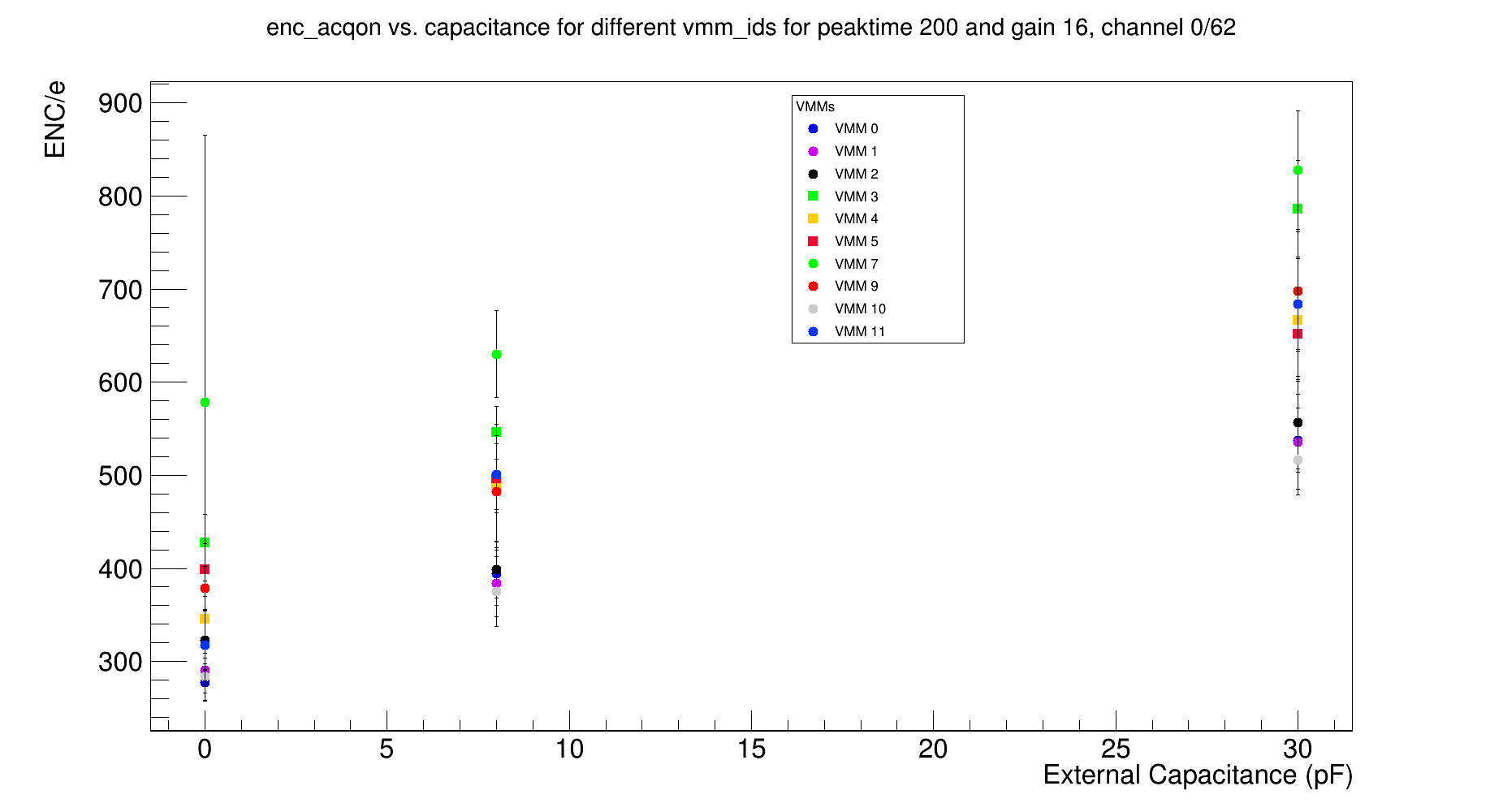} \\
  \includegraphics[width=13cm]{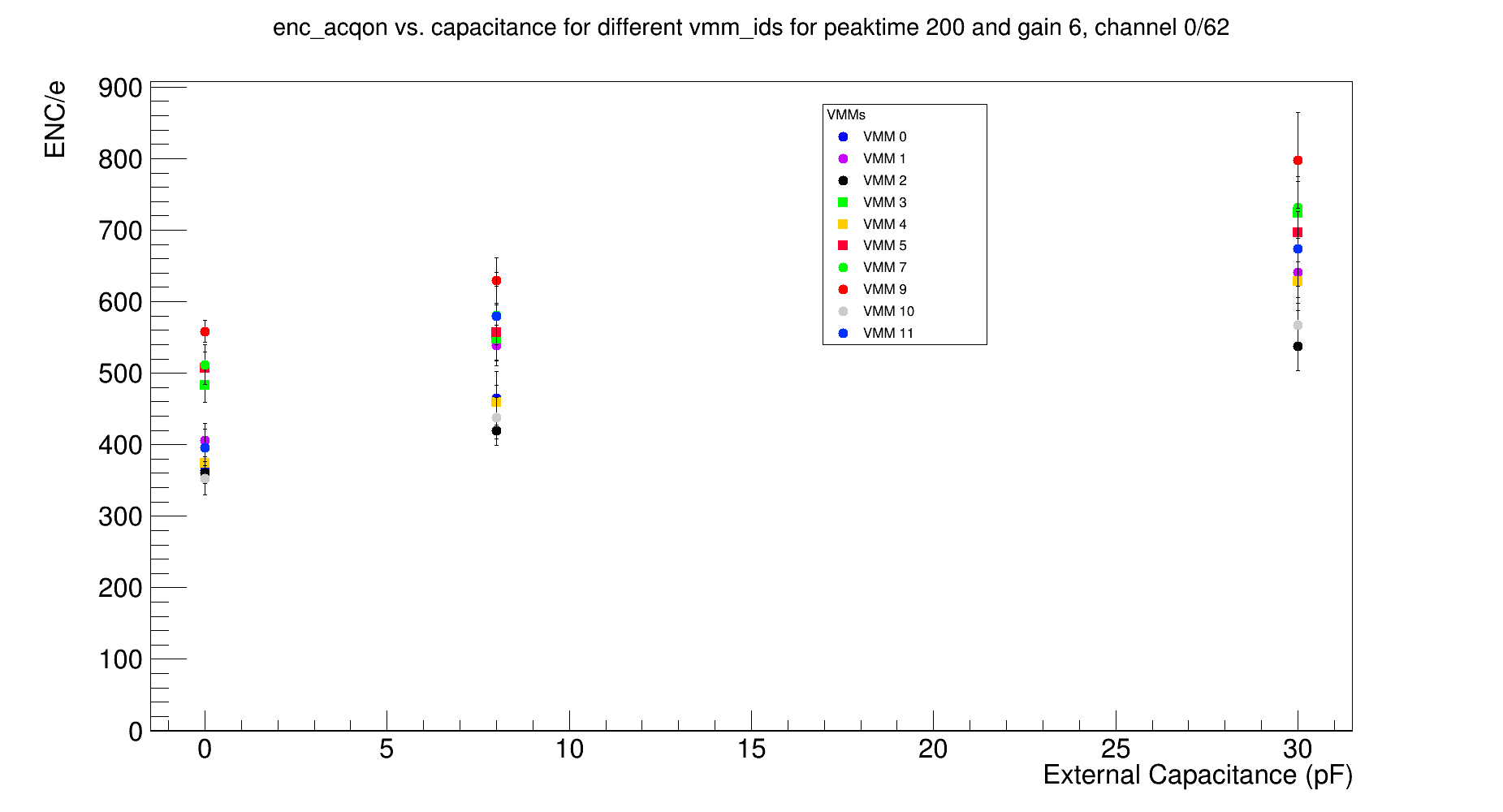}
  \end{tabular}
  \caption{ENC, ACQ on, vs. capacitance for all VMMs, peaktime \SI{200}{ns}, and gain \SI{16}{mV}/fC (top), gain \SI{6}{mV}/fC (bottom). Channel 62 is measured for the odd-numbered VMMS, while channel 0 is measured for the even-numbered VMMs.}
  \label{fig:ENC_cap_allvmms_0-62}
\end{figure}
%\printbibliography[heading=subbibliography]

%------------------------------------------------------------------------------
% Declare lists of figures and tables and acknowledgements as backmatter
% Chapter/section numbers are turned off
\backmatter

\listoffigures
\listoftables

%------------------------------------------------------------------------------
% Print the glossary and list of acronyms
% \printglossaries

%------------------------------------------------------------------------------
% You could instead add your acknowledgements here - don't forget to
% also add them to \includeonly above
% \include{thesis_acknowledge}

%------------------------------------------------------------------------------
% CV needed when you submit your PhD thesis
% \input{thesis_cv}

\end{document}